\documentclass[nofootinbib, aps, twocolumn,  longbibliography]{revtex4-1}
\usepackage{amsmath}
\usepackage{amssymb}
\usepackage{graphicx}
\usepackage[usenames,dvipsnames]{xcolor}
\usepackage{tikz}
\usepackage{pgffor}
\usepackage{verbatim}
\usepackage{bbm}
\usepackage{float}
\usepackage{mathrsfs}
\usepackage[colorlinks, breaklinks=true,linkcolor=red, citecolor=blue, linktocpage=true]{hyperref}

\newcommand{\<}{\langle}
\renewcommand{\>}{\rangle}

\newcommand{\mac}{\mathcal}

\begin{document}

\title{Anyon condensation in the string-net models}

\author{Chien-Hung Lin}
\author{Fiona J. Burnell}
\affiliation{School of Physics and Astronomy, University of Minnesota, Minneapolis, Minnesota 55455, USA}

\date{\today}
\begin{abstract}
We study condensation of abelian bosons in string-net models, by constructing a family of Hamiltonians that can be tuned through any such transition.  We show that these Hamiltonians admit two exactly solvable, string-net limits: one deep in the uncondensed phase, described by an initial, uncondensed string net Hamiltonian, and one deep in the condensed phase, described by a final, condensed string net model.  
We give a systematic description of the condensed string net model in terms of the uncondensed string net and the data associated with the condensing abelian bosons.
Specifically, if the uncondensed string net is described by a fusion category $\mathcal{C}$, we show how the string labels and fusion data of the fusion category  $\mathcal{\tilde{C}}$
describing the condensed string net can be obtained from that of $\mathcal{C}$ and the data describing the string oeprators that create the condensing boson.  
This construction generalizes previous approaches to anyon condensation in string nets, by allowing the condensation of arbitrary abelian bosons, including chiral bosons in string nets constructed from (for example) Chern-Simons theories, which describe time-reversal invariant bilayer states.  This gives a method for obtaining the full data for string nets without explicit time-reversal symmetry from such bilayer models.  
We illustrate our approach with several examples.  

\end{abstract}

\maketitle

\maketitle

\section{Introduction}

The universal, low-energy properties of gapped phases of quantum matter are described using two principles: symmetry and topological order.    
Considerable effort in recent years has gone into expanding our understanding of the resulting geneaology of quantum phases that cannot be described by the Landau paradigm of spontaneously broken symmetries, unveiling many new intriguing possibilities for strongly interacting systems.
Among the earliest notable exceptions to Landau's framework are topologically ordered phases in 2+1 dimensions,\cite{AndersonRVB,WenTO,KitaevToric,KitaevHoneycomb}, 
which harbor emergent point-like particles (known as anyons) with fractional statistics.

The long-ranged properties of topologically ordered phases are captured by a mathematical structure known as a unitary modular tensor category (UMTC) \cite{KitaevHoneycomb,FrolichTQFT,WangBook,BondersonThesis,Bakalov,Kong2017}, which describes the rules governing fusion and braiding of point-like excitations.  Thus our knowledge of the possible topologically ordered phases -- much like our knowledge of the possible symmetry groups --  is quite be complete.   Given this, it is natural to ask which phases can, in principle, be related by second-order phase transitions?  

In the case of topological order, this question is closely related to the question of which topological orders are related by so-called anyon condensation transitions (see \cite{BurnellReview} for a brief review).  Such transitions were first studied in the context of conformal field theory\cite{ConformalZoo,MooreSeibergNatural,Schellekens89,Gepner89,GoddardCoset1,GoddardCoset2,SCHELLEKENSFixed,FUCHSFixed,FuchsCoset}, and they have been discussed for general UMTC's in the mathematical literature \cite{MugerTSB,BruguieresTSB,Muger,KIRILLOV2002183,Davydov13,Hung15,KongCondensation,Kawahigashi21}.  Refs. \cite{BaisGaugeTheories,TSBPRL,BaisMathy,SlingerlandBais,Eliens} described how, in the context of 2+1 dimensional topologically ordered phases, these transitions physically correspond to processes in which emergent bosons condense; the topological order obtained is then a direct consequence of the new, condensed, vacuum.  

Anyon condensation has proven useful in understanding not only the structure of topological phases\cite{Bombin,NeupertBernevig,WenBarkeshliLong,VerstraeteCondense,SchuchCondensation,HuWan22}, but also when they admit gappable boundaries\cite{BravyiKitaev,KitaevKong,LevinBdy,KongCondensation,BaisBoundaries,Beigi11, BaisChiralBoundaries,YanWanCond,KapustinAbelianBdies,WangWenBdy,LanWangWen,GaneshanBdy,KongGaplessEdges,KongModular}, and how to create non-abelian topological orders\cite{DrinfeldGauging,GuWan14,HungWan14,GarciaCond,TeoTwist,LongQPaper,ChengAPS,Heinrich16,FradkinTwist,TarantinoTwist} or topological defects\cite{FuKane,LindnerPara,ClarkePara,ChengPara,BarkeshliQi,BombinTwist,WenGenon,BarkeshliAbelian1,BarkeshliAbelian2,BarkeshliBraiding,HughesSantos} from abelian ones.
Moreover, the possibility of condensing anyons to change a topological order also opens up the door for novel second-order critical points which may not have analogues in conventional symmetry-breaking transitions\cite{TSBLong,Gilsetal,GilsIsing,Schulz16,SchulzSU2,MotrunichSET,DSemToTC,ArdonneCond,WenBarkeshli,WB2,SchuchFibonacciCondense}.  Recently, it has been observed that anyon condensation can also be used to study certain dynamical processes in open quantum systems and quantum codes \cite{BaoAltman,Kesserling22}.

In studying anyon condensation, it is useful to have a lattice Hamiltonian that can be tuned between the two phases in question.  This establishes beyond a doubt that a direct transition between the two topological orders can occur, and enables a variety of analytical and numerical approaches to be used to study both the corresponding phase transitions, and verify the above description of the condensed phase \cite{Gilsetal,TSBShort,TSBLong,VidalToric,VidalToric2,SchulzSU2,VerstraeteCondense,SchuchCondensation}.   Lattice models of anyon condensation are also useful for constructing Hamiltonians realizing symmetry-enriched topological orders \cite{Heinrich16,ChengAPS}.

The present work focuses on a family of 2D topological orders known as Drinfeld centers, which are believed to be the most general class of (bosonic) topological orders compatible with gapped boundaries \cite{KitaevKong,KongCondensation,LinLevinstrnet,freed2020gapped}.   These can be realized by commuting projector lattice models known as string nets \cite{LevinWenstrnet,LinLevinstrnet,Kong2012,LanWen13,KitaevKong,LakeWu,HahnWolf,LinLevinBurnell}.  The string net construction begins not from the UMTC describing the anyon model, but from a pivotal fusion category $\mac{C}$, which describes the Hamiltonian and ground states.  The full topological order (i.e. UMTC) is exhibited by studying so-called string operators, which realize point-like anyonic excitations at their end-points.  A number of works have previously studied anyon condensation in these models \cite{Gilsetal,TSBShort,TSBLong,VidalToric,VidalToric2,SchulzSU2,Schulz16}.  
However, the literature so far has focused primarily on transitions which condense a particular type of boson within these string net models, corresponding to excitations of only the plaquette term in the string net Hamiltonian.   For such condensation processes a general prescription exists\cite{TSBLong} to modify the string-net Hamiltonian by adding a term which can drive anyon condensation; in the limit that this term is very large the Hamiltonian reduces to a new string net model realizing the topological order of the condensed phase, constructed from a sub-category of the original fusion category $\mathcal{C}$.  In other words, the data for the string net model of the condensed phase follows straightforwardly from that of the uncondensed phase.  (The relation between the string operators of the condensed and uncondensed phases, however, is not quite so straightforward).  

Here, we will describe a general formalism that describes condensation of arbitrary {\it abelian bosons} in string-net models.\footnote{An abelian boson is simply a boson that has a unique fusion outcome with any other anyon in the theory.}   First, we describe an extended version of the string net construction, obtained by extending the Hilbert space using an approach similar to that of Ref. \cite{WuTails}, albeit tailored to simplify the description of the condensed phase.  Within this extended Hilbert space, we construct a family of model Hamiltonians that can be tuned through a transition involving the condensation of any abelian boson, and outline the topological order expected for the resulting condensed phase.   Our modified model has the advantage that deep in the condensed phase, a general prescription can be given to identify both the low-energy effective Hilbert space and the ground state.   We describe how the Hilbert space of the condensed phase can be described by a new, effective,  set of string labels (i.e. a new effective fusion category $\tilde{\mathcal{C}}$), whose relationship to the original label set can be calculated explicitly.  We further show that the ground state of the condensed phase is also a string-net ground state, described by the data of the new fusion category $\tilde{\mathcal{C}}$.  In this regime, our Hamiltonian acts like the regular string net Hamiltonian on the new label set.  
Moreover, we obtain an explicit expression for the fusion data of $\tilde{\mathcal{C}}$ in terms of $\mathcal{C}$ and the condensing bosons.

The relationship between the topological order of the uncondensed and condensed anyon models can be 
characterized by the fate of  the anyons of the original topological order in this condensed vacuum.  First, anyons that braid non-trivially with any of the condensing bosons become confined, and are absent from the topological order of the condensed theory.  Second, two anyons that are related by fusion with one of the condensing bosons must be identified in the condensed phase, meaning that topologically speaking, they correspond to the same excitation.  Finally,  certain non-abelian anyons can split into multiple distinct anyon types after condensation.  For condensation of abelian bosons, these relationships are well known in conformal field theory, where they go by the name of central extensions\cite{ConformalZoo}, and the $S$ and $T$ matrices of the final topological orders can be computed explicitly.
However, as noted by Ref. \cite{Eliens}, the task of computing the full topological data - namely $F$ and $R$ matrices - is significantly more challenging.
Our approach does not easily give us access to the full topological data of the anyon model -- indeed, though the confinement, identification, and splitting of anyons in the final topological order is apparent from the form of our Hamiltonian, in our approach the associated topological data is inferred only indirectly, through the emergence of the new string net data $\tilde{\mathcal{C}}$, which in turn implies a new set of anyon- creating string operators.  However, 
we show that our approach to string net condensation does allow one to straightforwardly compute the data of the fusion category underlying the condensed phase, as we demonstrate explicitly in a number of examples.

An interesting application of the condensation transitions studied here is that they can take a string net with an explicit time-reversal symmetry (of the type described in the original work by Levin and Wen\cite{LevinWenstrnet}) into one that does not admit a naive time-reversal transformation (described in detail in Ref. \cite{LinLevinBurnell}).  One example that we will discuss in detail is the transition from SU(2)$_4 \times \overline{\text{SU(2)}}_4$ to U(1)$_3 \times \overline{\text{SU(2)}}_4$, for which $\mathcal{C} = SU(2)_4$ has all of the symmetries assumed by Ref. \cite{LevinWenstrnet}, while $\tilde{C} = TY_3$ does not.

The paper is organized as follows. In Sec. \ref{Sec:snmodel}, we review some basics of general string-net ground states, and introduce the extended string-net Hilbert space that we use to study anyon condensation.
In Sec. \ref{Sec:CondHam}, we introduce a family of modified string net Hamiltonians, which can be tuned across a transition condensing an arbitrary abelian boson.
We describe the effective Hilbert space deep in the condensed phase in Sec. \ref{Sec:newH}, where we discuss how the string types in $\tilde{\mathcal{C}}$ are related to those in $\mathcal{C}$.   In Sec. \ref{Sec:newsnmodel}, we study the condensed ground state, and show that it is indeed a string net.  In particular, we show how the new ground state allows one to describe the fusion data of $\tilde{\mathcal{C}}$, verify that this fusion data is indeed consistent, and argue that the full Hamiltonian projected into the condensed Hilbert space is indeed the associated string net Hamiltonian. 
We illustrate our construction with concrete examples in Sec. \ref{section:exp}. A number of technical details are elaborated on in the appendices.

\section{Extended String-net models } \label{Sec:snmodel}
In this section, we introduce the extended string-net models that we will use for our models of anyon condensation.

\subsection{Review: Generalized string net models} \label{section:psi}

We begin by reviewing the string-net construction.  Here we use the generalized string-net construction of Ref. \cite{LinLevinBurnell} (see also Refs. \cite{LinLevinstrnet,LanWen13,KitaevKong}), since the symmetries assumed in the original construction\cite{LevinWenstrnet} are not always present in the condensed phase.

We defer a discussion of string-net Hamiltonians to Sec. \ref{modifiedstring}, and here focus on the string-net Hilbert space, together with its ground states and certain excited states.

\subsubsection{The string-net ground state} \label{Sec:StrOld}

The string net model consists of a Hamiltonian whose ground state(s) obey certain special properties, which we now describe.  
These string-net ground states live in a Hilbert space of {\it string-net configurations}, each of which is defined  on an  oriented, trivalent graph.   (Though the string-net Hamiltonians are defined on the honeycomb lattice, this lattice structure is not necessary to describe the string-net ground states.)  Throughout this work, we use the convention that \emph{all strings are oriented upward}, i.e. the orientation vector has positive projection onto the $\hat{y}$ direction.  We therefore require this projection to be non-zero, such that our strings cannot have horizontal tangent vectors. 

The string net configuration is obtained by assigning to each edge $i$  a label (or string type) $a_i$. 
The combinations of string types $\{ (a,b; c)\}$ that are allowed to meet at a vertex is dictated by a set of branching rules-- i.e. if $(a,b;c)$ is among the branching rules, then the vertices 
\begin{equation}
	\raisebox{-0.22in}{\includegraphics[height=0.5in]{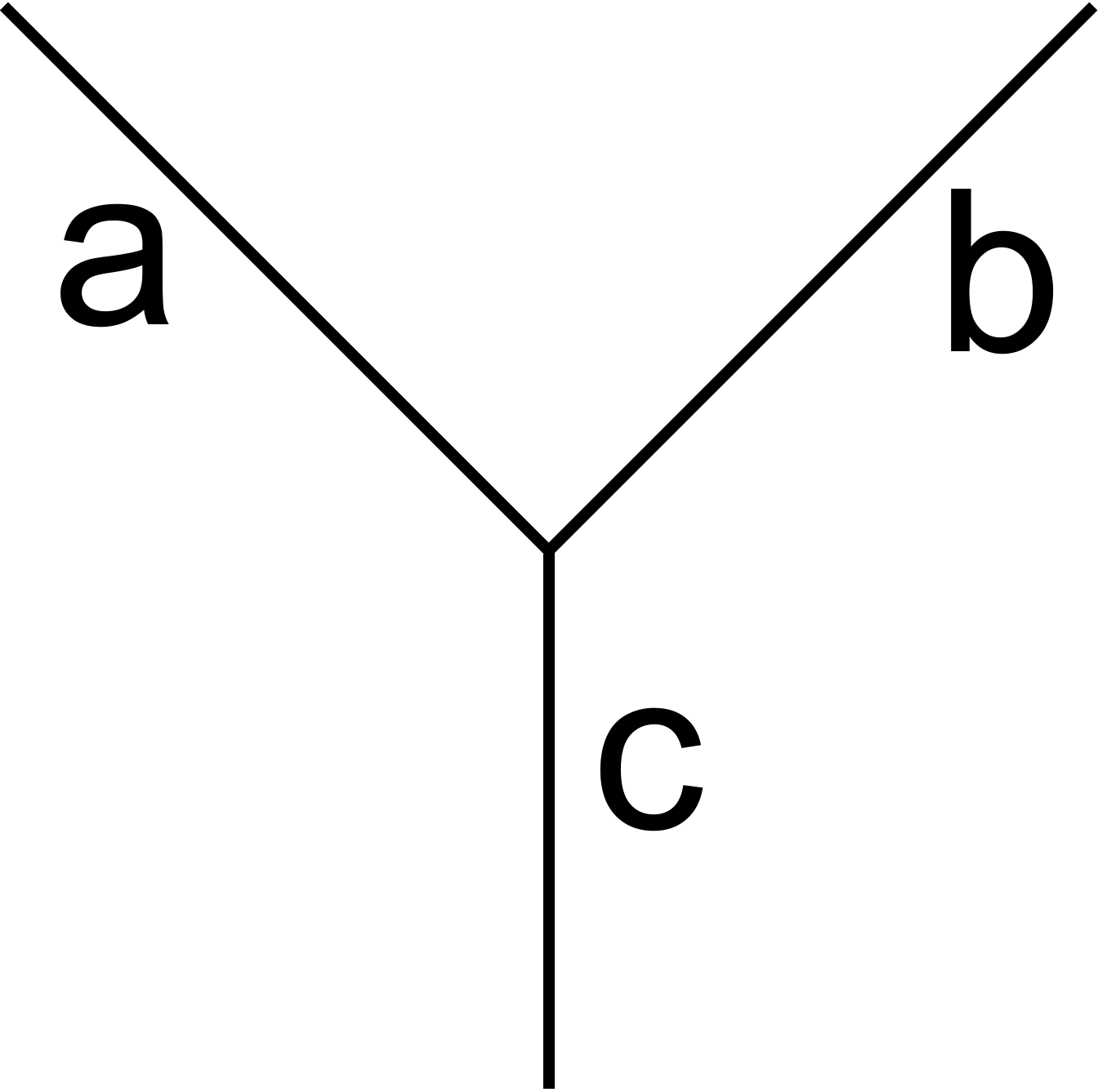}} \qquad
	\raisebox{-0.22in}{\includegraphics[height=0.5in]{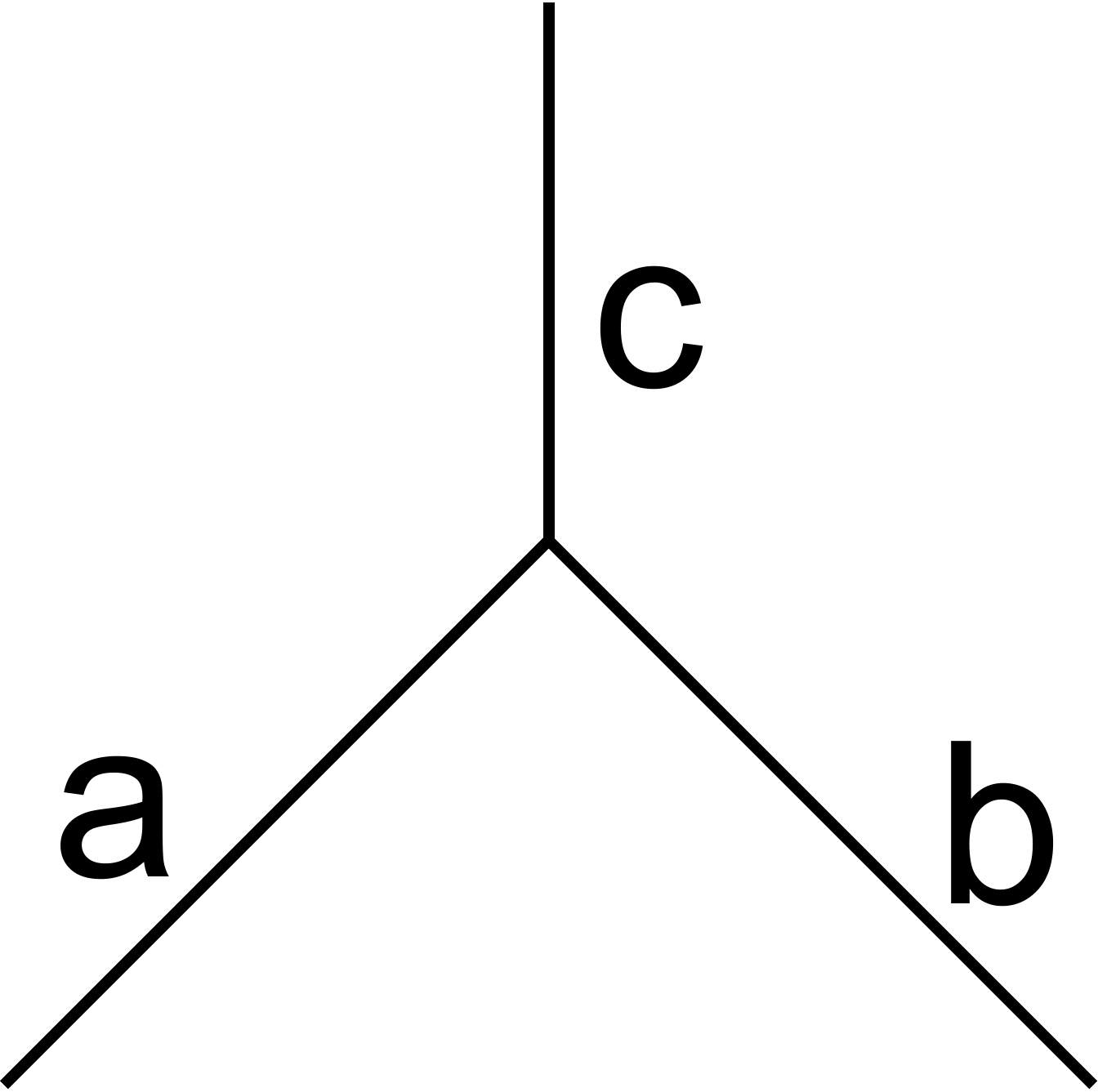}}
	\label{}
\end{equation}
are allowed.  
The set of all string-net configurations which satisfy the above branching rules form an orthonormal basis for the string-net Hilbert space $\mathcal{H}$.  
We will call those states in this string-net Hilbert space {\it string net states}.

In order to be able to define a string-net ground state, our label set and branching rules must satisfy certain conditions.  
First, defining  
\begin{equation}
	\delta^{ab}_c= \left\{
  \begin{array}{lc}
	  1, & \text{if } \{a,b;c\} \text{ is allowed}\\
	  0, & \text{otherwise}.
  \end{array}
\right.  ,
	\label{branching}
\end{equation}
the branching rules must satisfy:
\begin{equation}
\sum_e \delta^{ab}_e \delta^{ec}_d = \sum_f \delta^{bc}_f \delta^{af}_d \ .
\end{equation}
It follows that if $(a,b;c)$ is allowed, then so are  $(\bar{c}, a; \bar{b})$, $(b, \bar{c}; \bar{a})$, and $(\bar{b} , \bar{a}, \bar{c})$.  

Second, our label set must contain a null label, which we denote $0$, and depict diagrammatically with a dashed line.  This label is trivial in the sense that edges carrying this null label can be added to or deleted without changing the physical state (i.e. a null labeled edge is physically equivalent to having no edge at all).  Note that we will use $0$ to denote the trivial string label, and $\mathbf{1}$ to denote a trivial anyon.  
Finally, for each string type $a$, we require that our label set contains a dual string type $\bar{a}$, such that the branching rules must contain $(a,\bar{a};0)$ and $(\bar{a},a; 0)$ (but not $(a,b;0)$ for any $b \neq \bar{a}$). The null string $0$ is self-dual, $0=\bar{0}$.

The string-net ground state  $|\Phi \rangle$ is described by two sets of parameters: a set of complex numbers $F^{abc}_{def}$, known as the $F$ symbols, depending on 6 string types $a,b,\dots,f$, and a positive number $d_a$ for each string type $a$, often called its quantum dimension.  These determine the relative coefficients of different string net configurations in the ground state, via the relations:
\begin{subequations}
\begin{align}
	\Phi\left(\raisebox{-0.22in}{\includegraphics[height=0.5in]{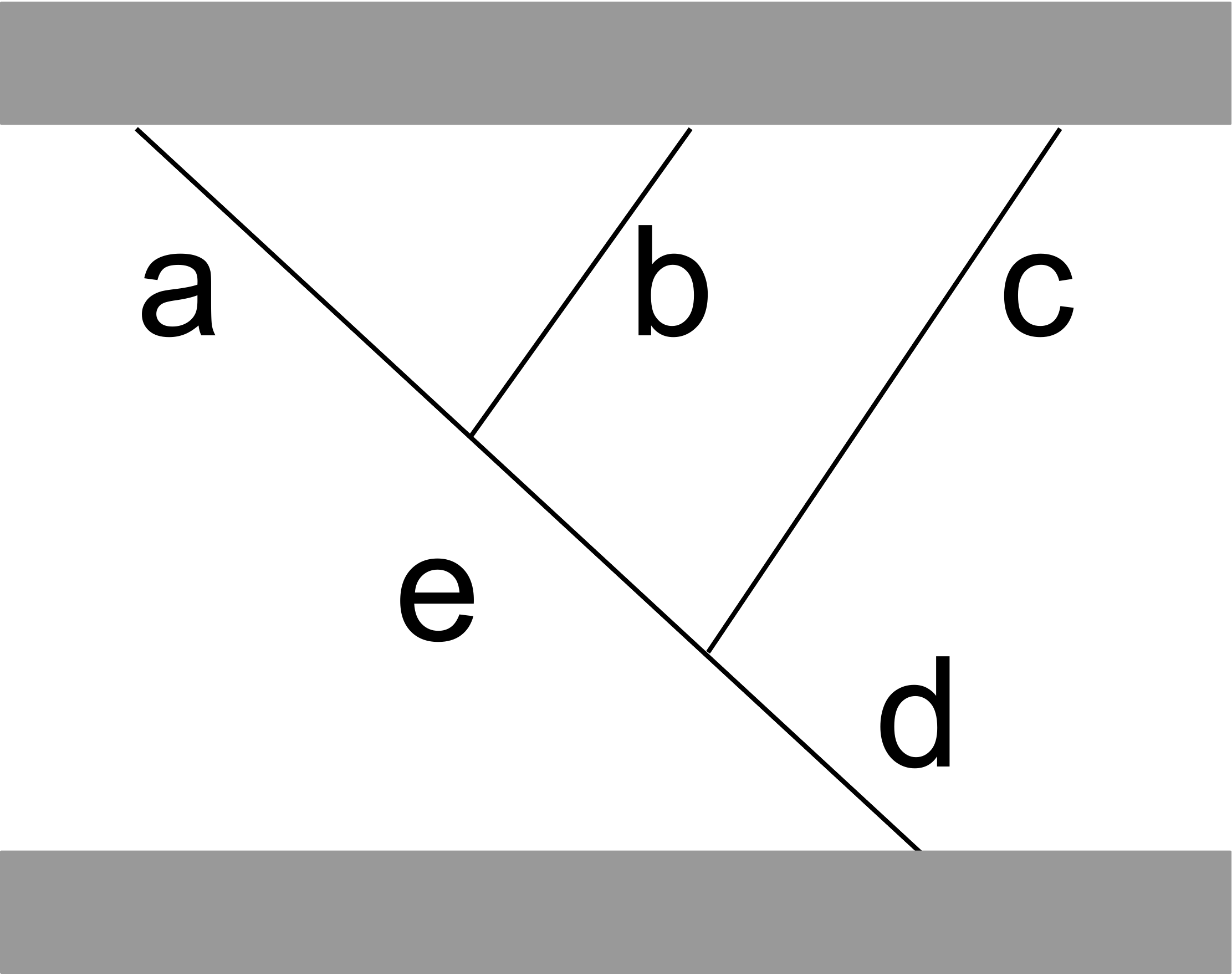}} \right)&=\sum_f F^{abc}_{def}\Phi \left(\raisebox{-0.22in}{\includegraphics[height=0.5in]{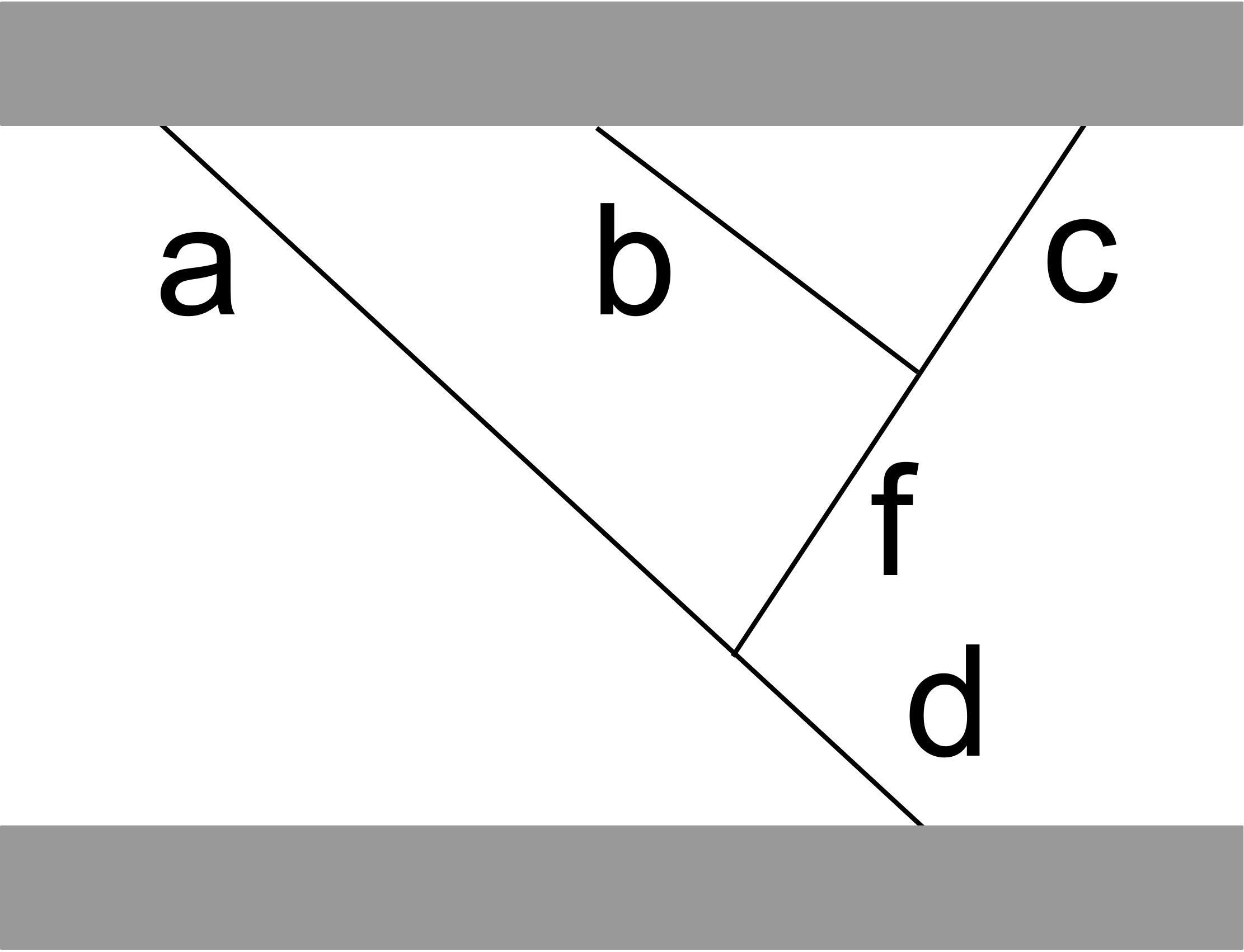}} \right) \label{1a}\\
	\Phi\left(\raisebox{-0.22in}{\includegraphics[height=0.5in]{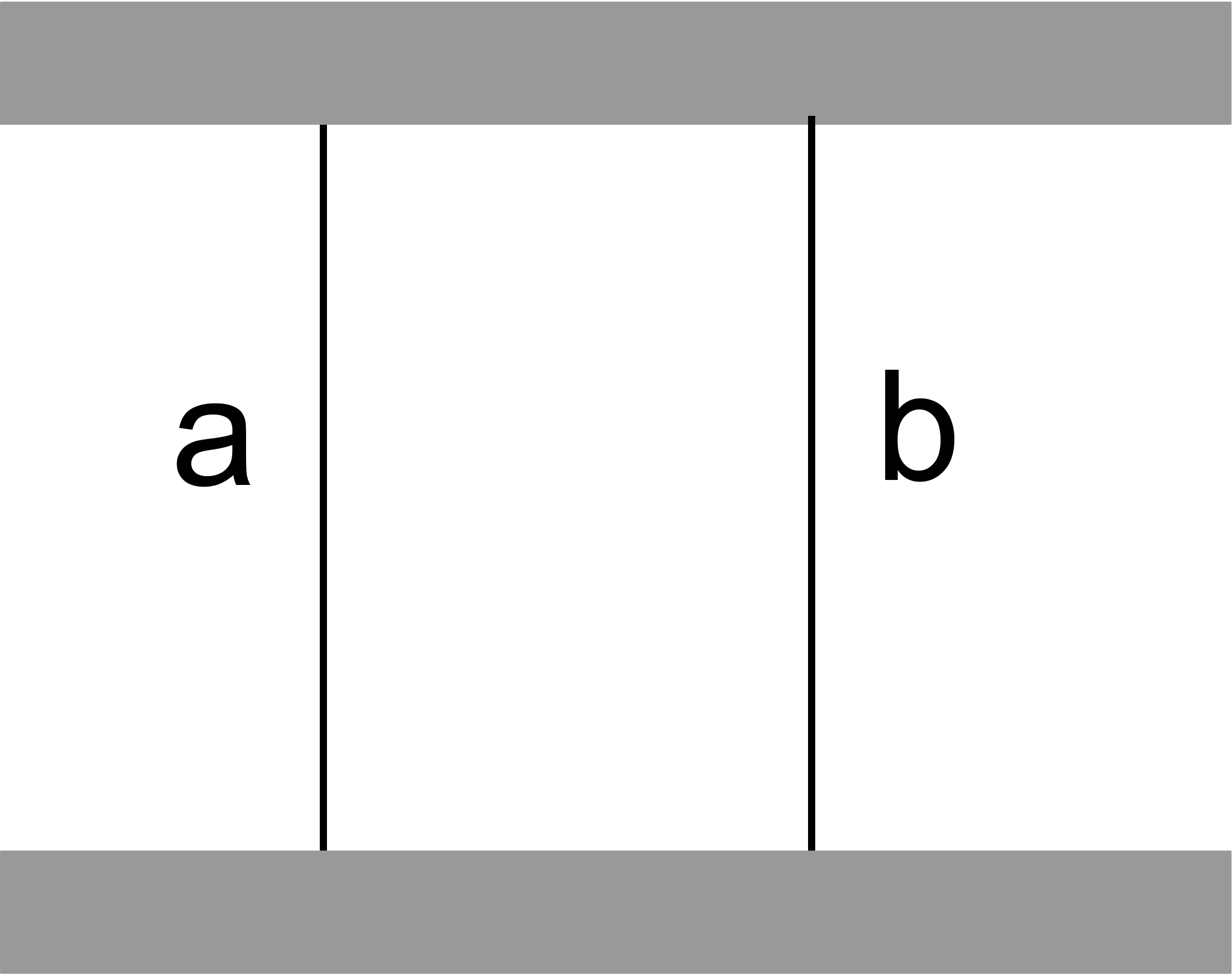}} \right)&=\sum_c  \sqrt{\frac{d_c}{d_a d_b}} 
	\Phi \left(\raisebox{-0.22in}{\includegraphics[height=0.5in]{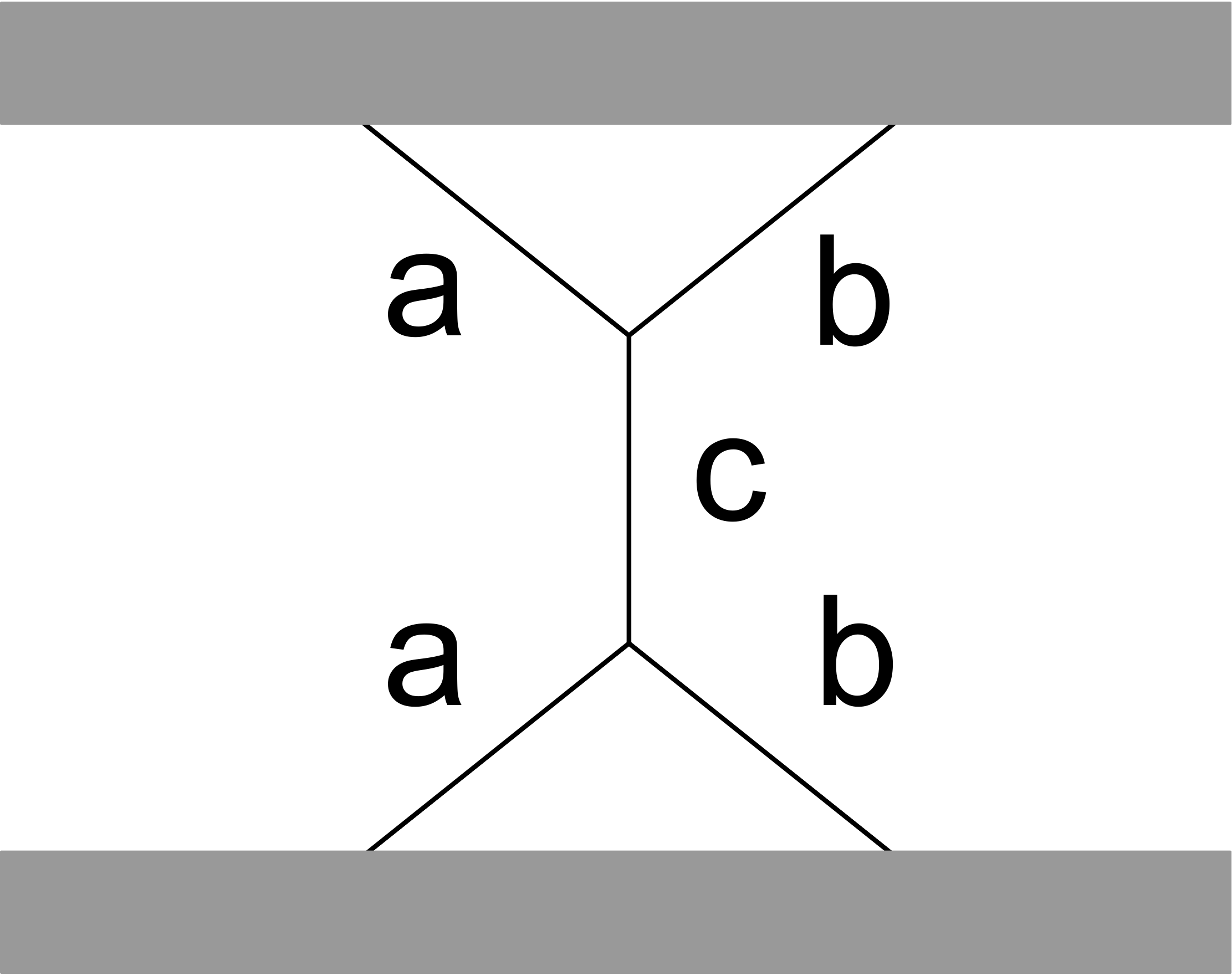}} \right) \label{1b} \\
	\Phi\left(\raisebox{-0.22in}{\includegraphics[height=0.5in]{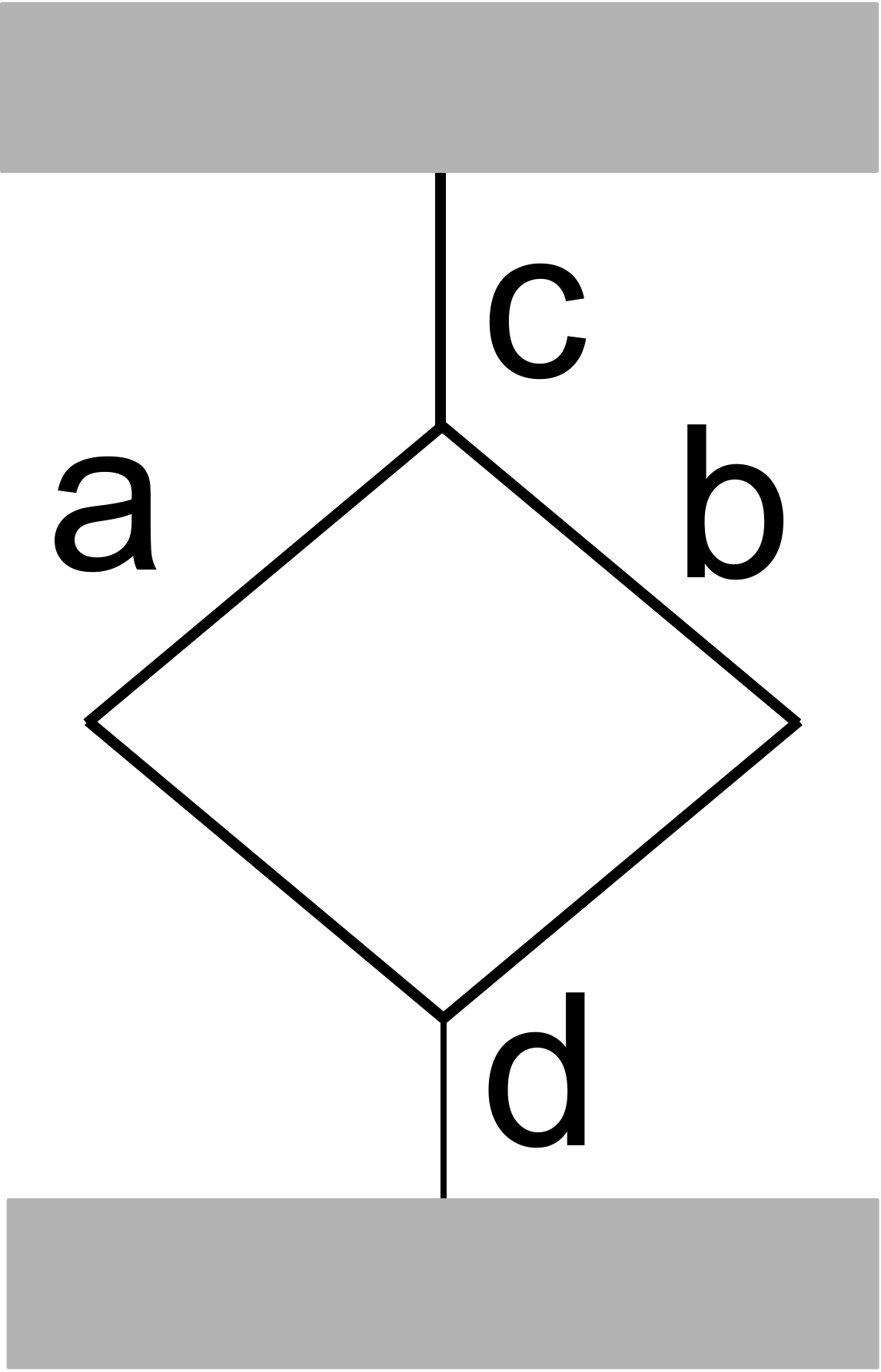}} \right)&=\delta_{c,d} \sqrt{\frac{d_a d_b}{d_c}}
	\Phi\left(\raisebox{-0.22in}{\includegraphics[height=0.5in]{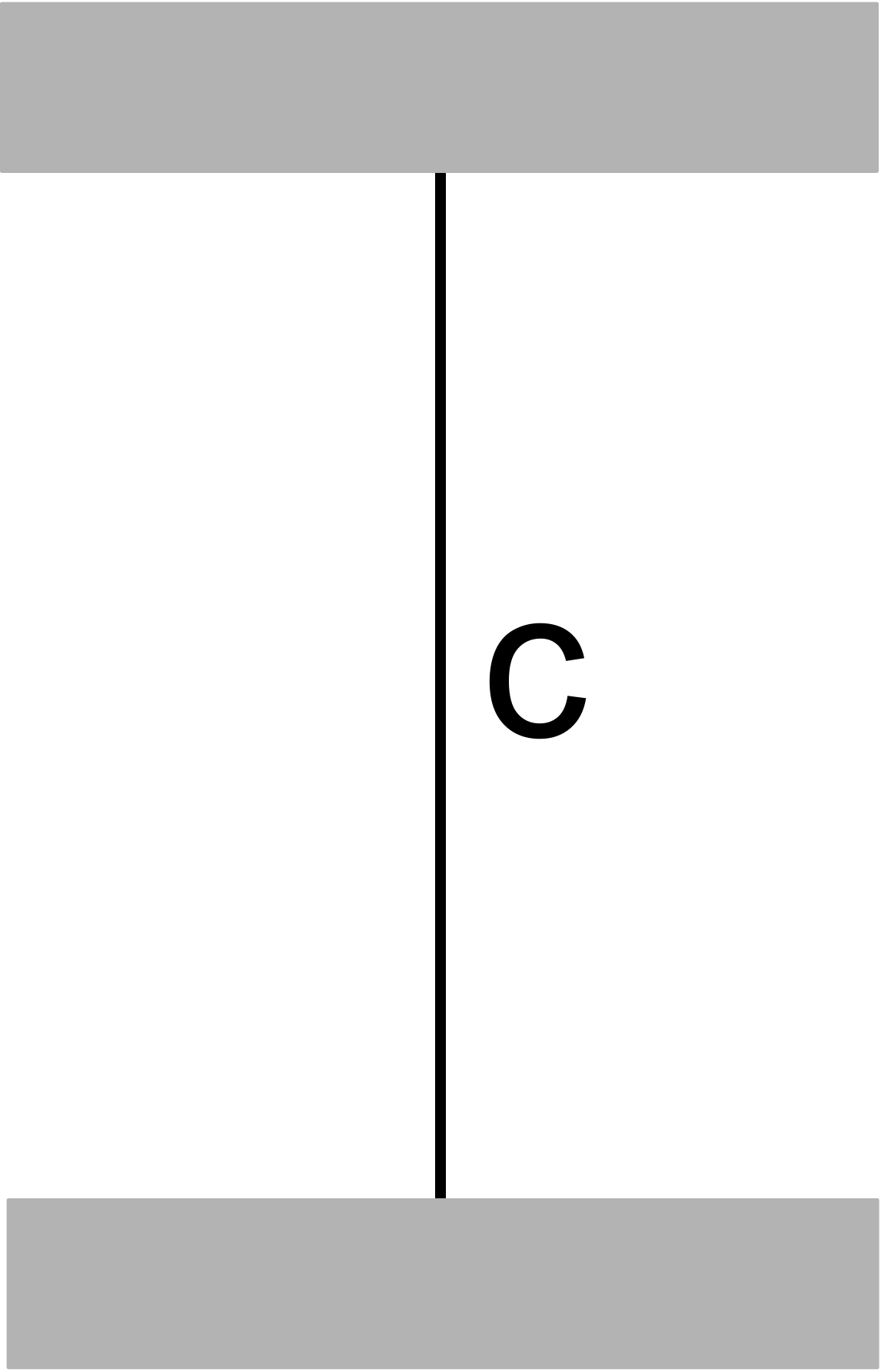}} \right) \label{1c}\\
	\Phi\left(\raisebox{-0.22in}{\includegraphics[height=0.5in]{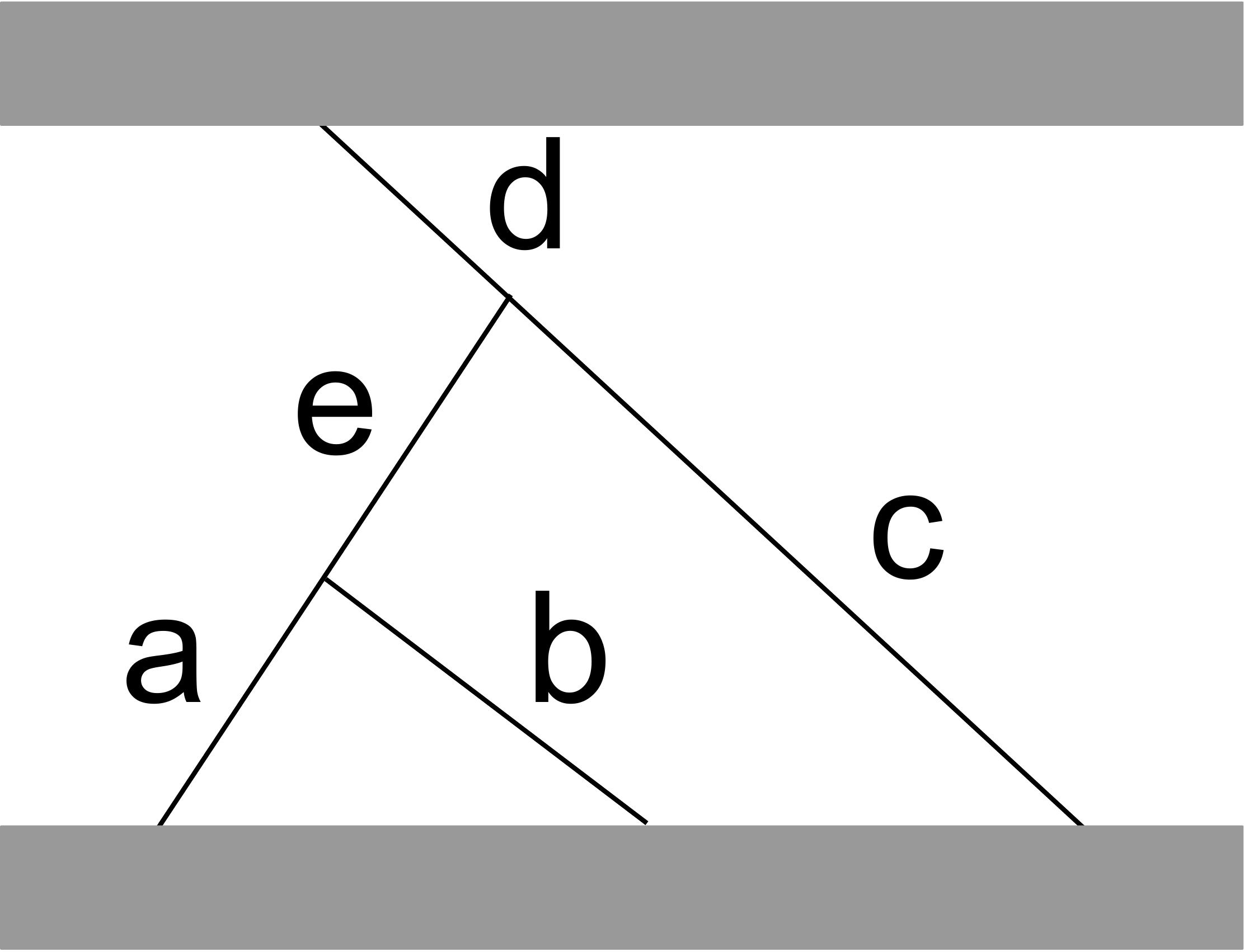}} \right)&=\sum_f (F^{abc}_{d})_{fe}^{-1} \Phi \left(\raisebox{-0.22in}{\includegraphics[height=0.5in]{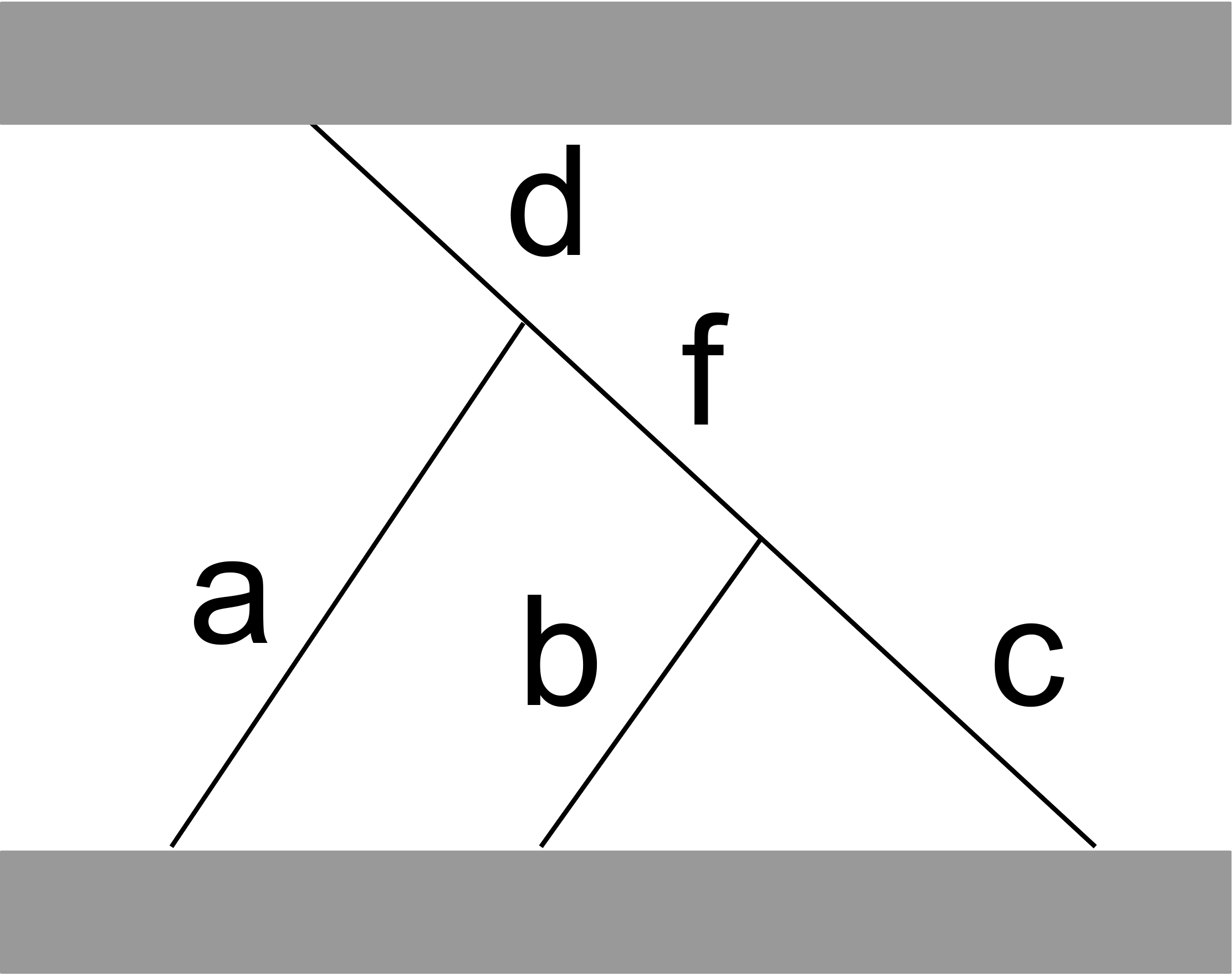}} \right) \label{1d}
\end{align}
\label{localrules}
\end{subequations}
Here $a,b,c,\dots$ are arbitrary string types (including the null string) and the shaded regions represent arbitrary string-net configurations which are not changed from on side of the equation to the other. The symbol $\delta_{c,d}=1$ if $c=d$ and $\delta_{c,d}=0$ otherwise.
 
The relative amplitudes in Eqs. (\ref{localrules}) are unchanged by horizontal bendings of the strings, but are not invariant under vertical bending.  Indeed, we do not allow smooth vertical bends in our string-net graphs at all; only kinks, which are equivalent to vertices $(a, \bar{a}; 0)$ or $(\bar{a},a; 0)$.  These can be added or removed using the appropriate $F$ symbols.

The $F$-symbols and quantum dimensions are not free parameters.  Rather, 
to have a well-defined wave function $\Phi$, the parameters $\{F^{abc}_{cde},\tilde{F}^{abc}_{cde},
d_a\} $ need to satisfy:
\begin{subequations}
	\begin{align}
		F^{fcd}_{egl}F^{abl}_{efk}&=\sum_h F^{abc}_{gfh} F^{ahd}_{egk} F^{bcd}_{khl} \label{3a}\\
		F^{abc}_{def}&=1 \quad \text{if }a\text{ or }b \text{ or }c=0  \label{3e}\\
		d_a&=d_{\bar{a}}=1 \quad \text{ if }a=0 \label{3f} \ \ .
	\end{align}
	\label{consistency}
\end{subequations}
These constraints are in fact quite limiting; solutions are described mathematically by a pivotal fusion category.  

In addition, to ensure that the string-net Hamimltonian is \emph{Hermitian}, we require:
\begin{subequations}
	\begin{align}
		(F^{abc}_{d})^{-1}_{ef} &= (F^{abc}_{dfe})^* \\
		|F^{ab\bar{b}}_{ac0}| &= \sqrt{\frac{d_c}{d_a d_b}} \delta^{ab}_c \\
		d_a & = d_{\bar{a}} 
	\end{align}
		\label{hermicity}
\end{subequations}
where $(F^{abc}_{d})^{-1}$ is the matrix inverse of $(F^{abc}_d)$, whose matrix elements are $(F^{abc}_{d})_{ef} \equiv F^{abc}_{def}$.

The conditions (\ref{hermicity}) also imply that the quantum dimensions obey:
\begin{equation} \label{Eq:Dsum}
d_a d_b = \sum_{c} d_c
\end{equation}
where the sum runs over all values of $c$ that satisfy the branching rules.

Local unitary transformations of the string net wave function 
result in new coefficients $\{\hat{F}, \hat{d}\}$, which are related to the original coefficients $\{F, d\}$ via the gauge transformation: 
\begin{equation}
	\begin{split}
		\hat{F}^{abc}_{def}&=F^{abc}_{def}\cdot \frac{f^{ab}_ef^{ec}_d}{f^{bc}_f f^{af}_d}  \\
		\hat{d}_a&=d_a \ .
	\end{split}
	\label{gaugef}
\end{equation}
Here $f^{ab}_c$ parametrize the local unitary transformation; they are complex functions defined on upward vertices, with the downward vertices transformed by $1/f^{ab}_c$.  To preserve the constraints listed above, we require
\begin{equation}
	\begin{split}
		|f^{ab}_c|&=1,
		f^{ab}_c=1 \text{ if }a \text{ or }b=0.
	\end{split}
	\label{gt}
\end{equation}

It is convenient to note that the local rules (\ref{localrules}) imply the following identities:
\begin{subequations}
	\begin{align}
\Phi\left(\raisebox{-0.22in}{\includegraphics[height=0.5in]{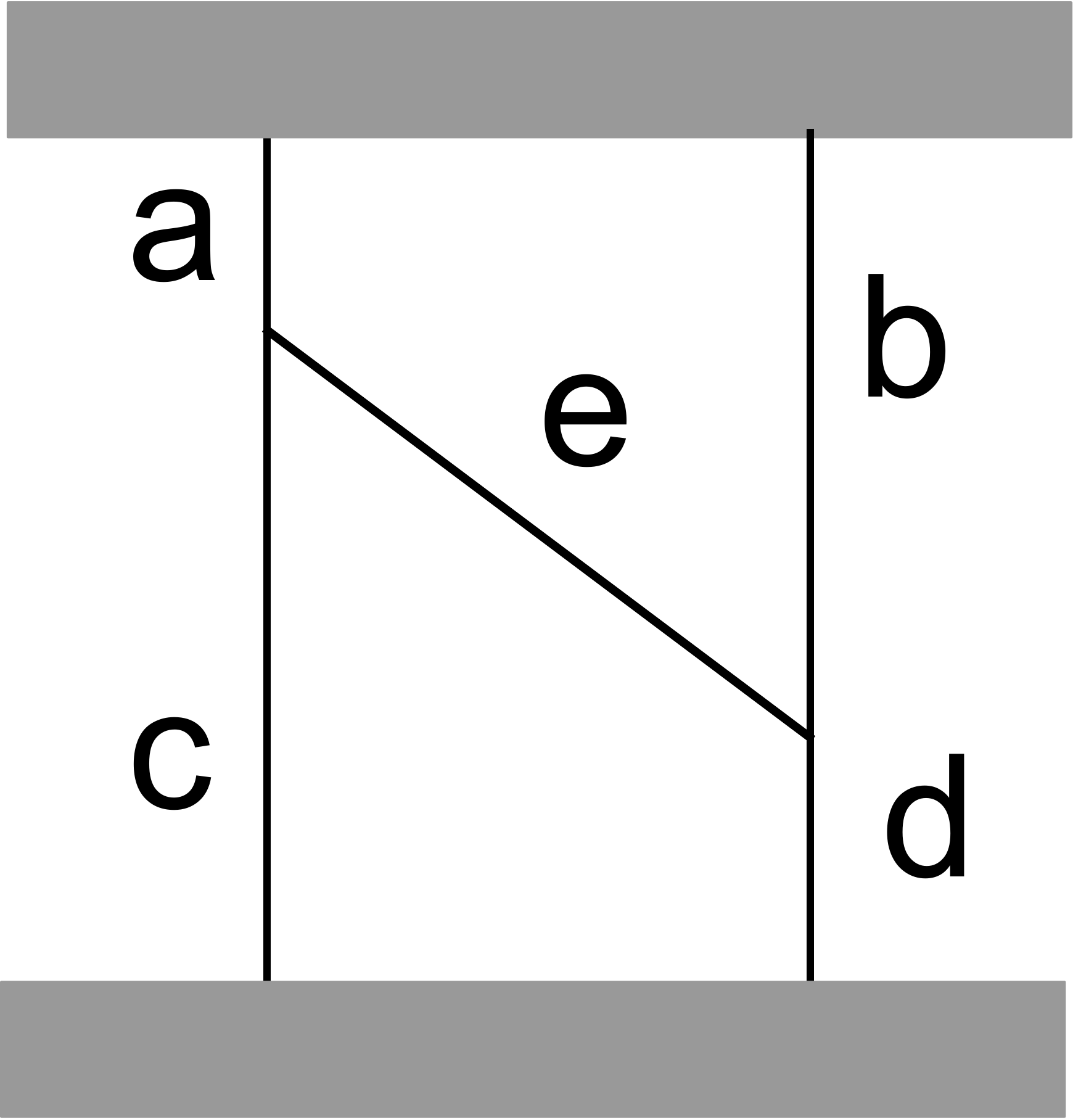}} \right)&=\sum_f [F^{ab}_{cd}]_{ef}\Phi\left(\raisebox{-0.22in}{\includegraphics[height=0.5in]{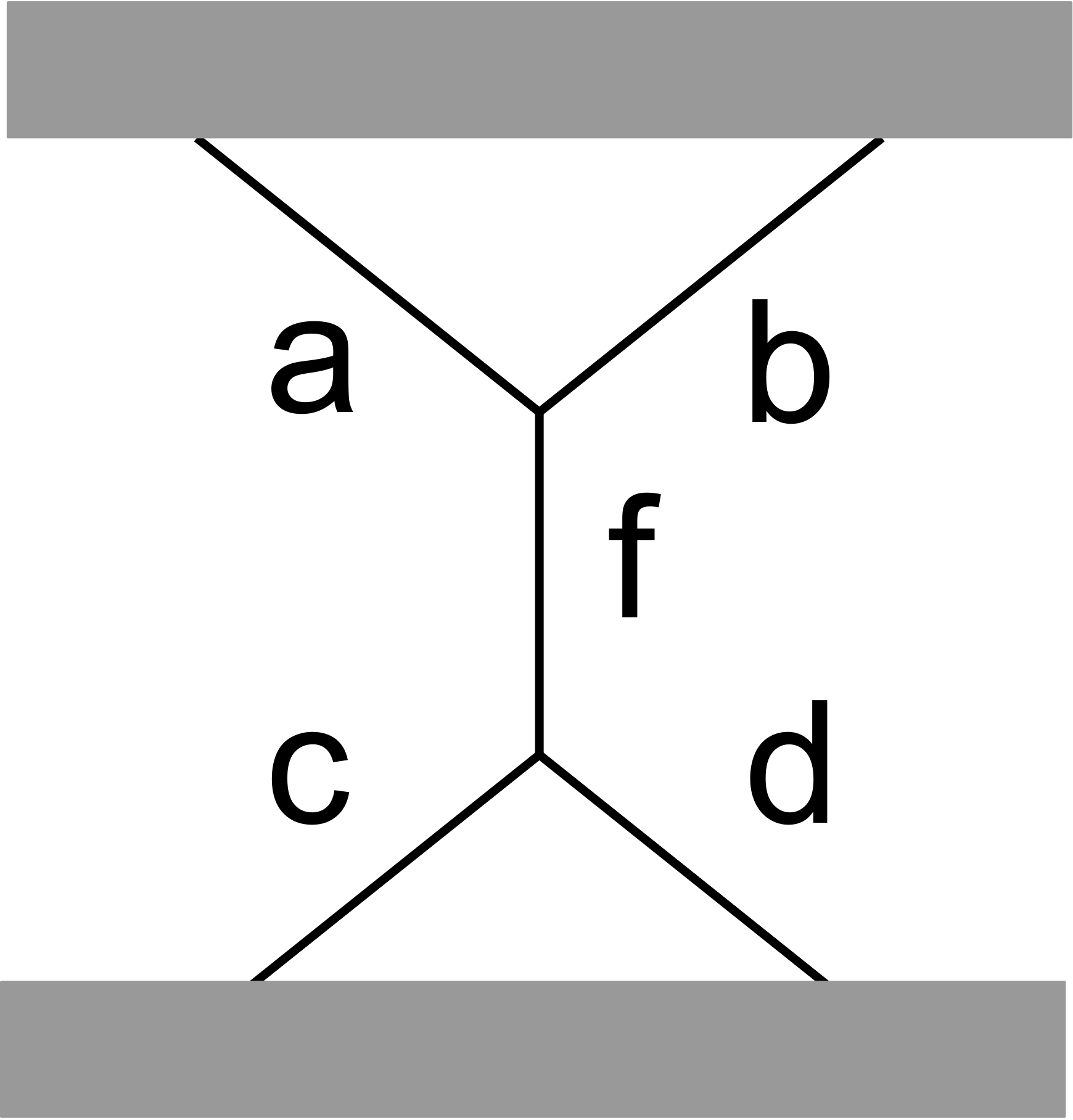}} \right) \label{1e}\\
	\Phi\left(\raisebox{-0.22in}{\includegraphics[height=0.5in]{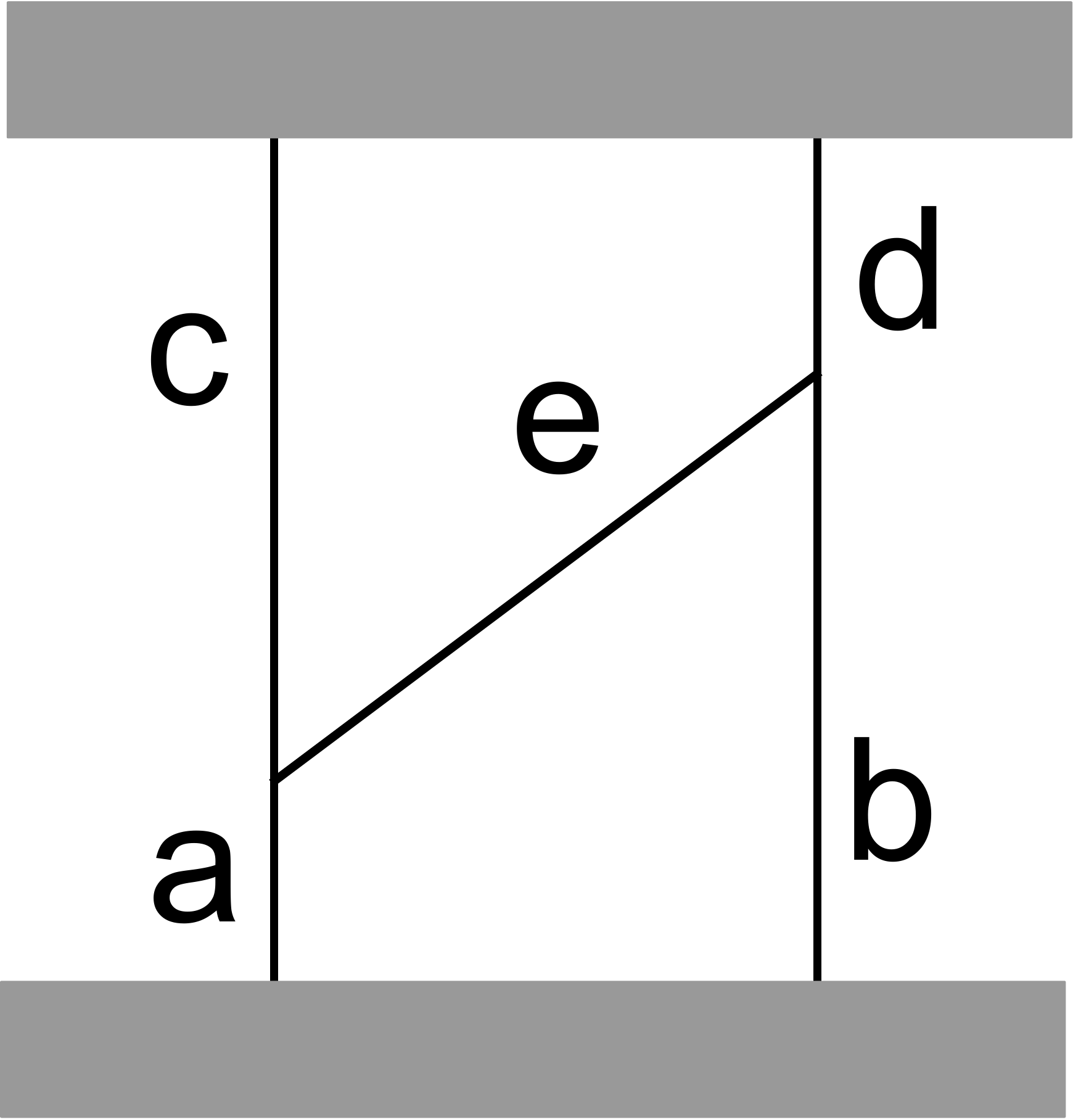}} \right)&=\sum_f [\tilde{F}^{ab}_{cd}]_{ef} \Phi\left(\raisebox{-0.22in}{\includegraphics[height=0.5in]{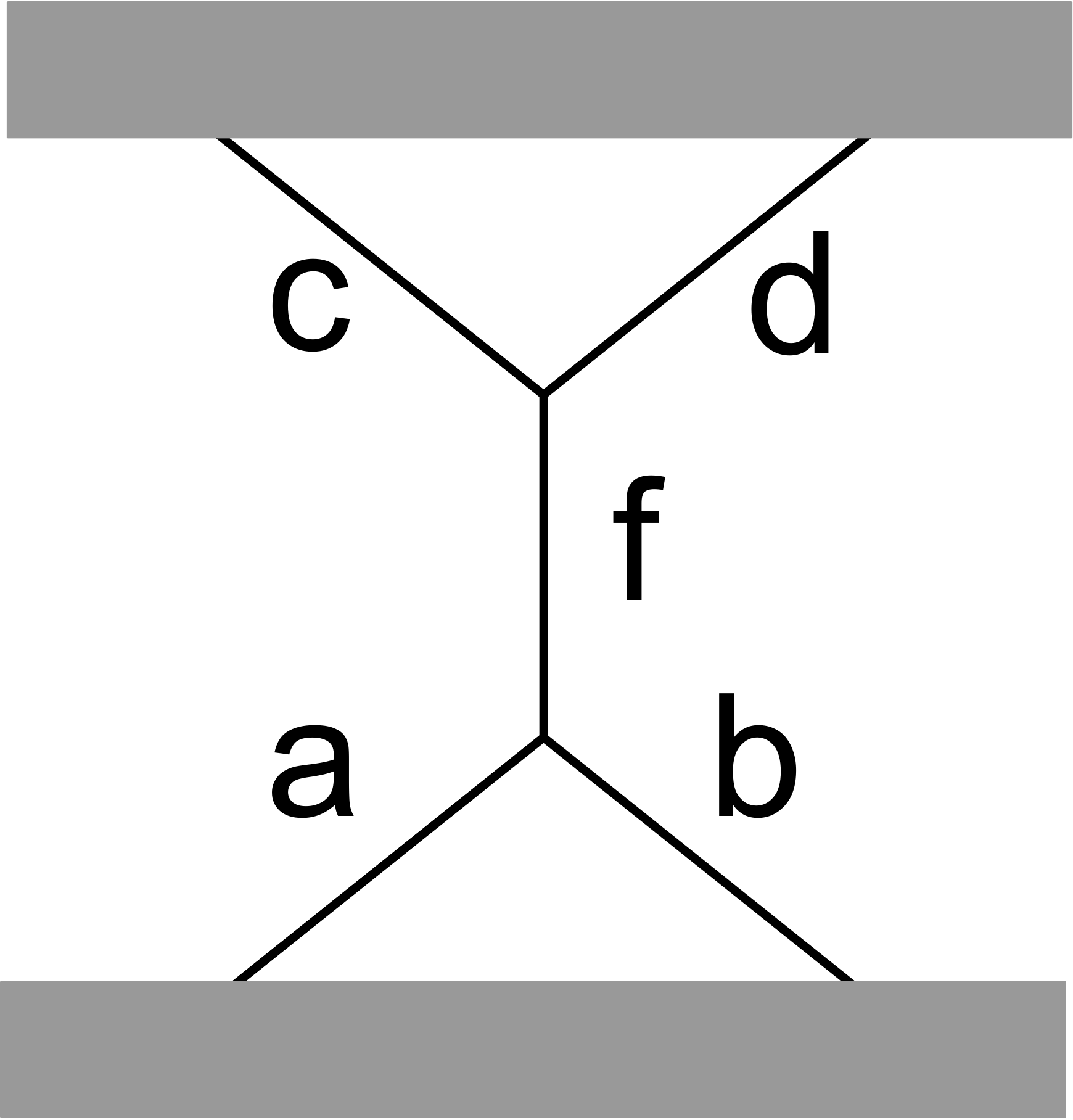}} \right) \label{1f}		
	\end{align}
	\label{localrules1}
\end{subequations}
with 
\begin{subequations}
	\begin{align}
		[F^{ab}_{cd}]_{ef} &
		=(F^{ceb}_{f})_{da}^{-1} \frac{ d_e d_f }{d_d d_a}
		\label{3c}\\
	[\tilde{F}^{ab}_{cd}]_{ef} &= F^{ceb}_{fad}  \frac{d_e  d_f }{d_a  d_d } \label{3d}
	\end{align}
	\label{consistency1}
\end{subequations}

\subsubsection{Abelian string operators} \label{Sec:Strops}

Next, we review the string operators that create point-like anyon excitations when acting on the ground state.  Here we focus on the case where these anyons  are \emph{abelian bosons}, since these are the excitations we wish to condense.  (For a discussion of more general string operators, see Ref. \cite{LinLevinBurnell}.)
Recall that an abelian anyon is defined by the fact that it has a unique fusion product with any other anyon in the theory; it is a boson if it has trivial statistics with itself.

To create a particle-antiparticle pair $(a, \bar{a})$ at two points in our lattice, we act on the string-net ground state with a string operator $W_a(P)$ along an oriented path $P$.  This creates $a$ at the final endpoint of $P$, and its antiparticle $\bar{a}$ at the initial endpoint.  On a given string-net state $\<X|$, we depict this action by drawing an $a$-labeled string along the path $P$ \emph{under} the string-net graph.  The string label $a$ specifies both a choice of one or more string types, and some extra data required to resolve crossings between the path $P$ and 
the string-net graph.  

If $a \equiv \phi$ is an abelian anyon, the label $\phi$ corresponds to a single string type $s$,
meaning that in regions where $P$ does not cross any edges of the string-net, we replace the label $\phi$ with $s$ on upward-oriented segments of $P$, and $\bar{s}$ on  downward-oriented segments of $P$.  Further, $s$ (and $\bar{s}$) must have a unique fusion product with all other string types, meaning that for each $a$, the branching rules contain $(a,s; a')$ (and also $(s,a; a')$) for only one $a'$, which we will sometimes denote as $a' \equiv a \times s$.    It follows that 
\begin{equation}
d_s = d_{\bar{s}} = 1,
\end{equation}
and thus $d_a = d_{a'}$ by Eq. (\ref{Eq:Dsum}).
In this case the coefficients associated with the moves (\ref{1b}) and (\ref{1c}) are unity.  

For abelian anyons, the crossings are resolved using the rules:
\begin{align}
	\begin{split}
	\left<\raisebox{-0.22in}{\includegraphics[height=0.5in]{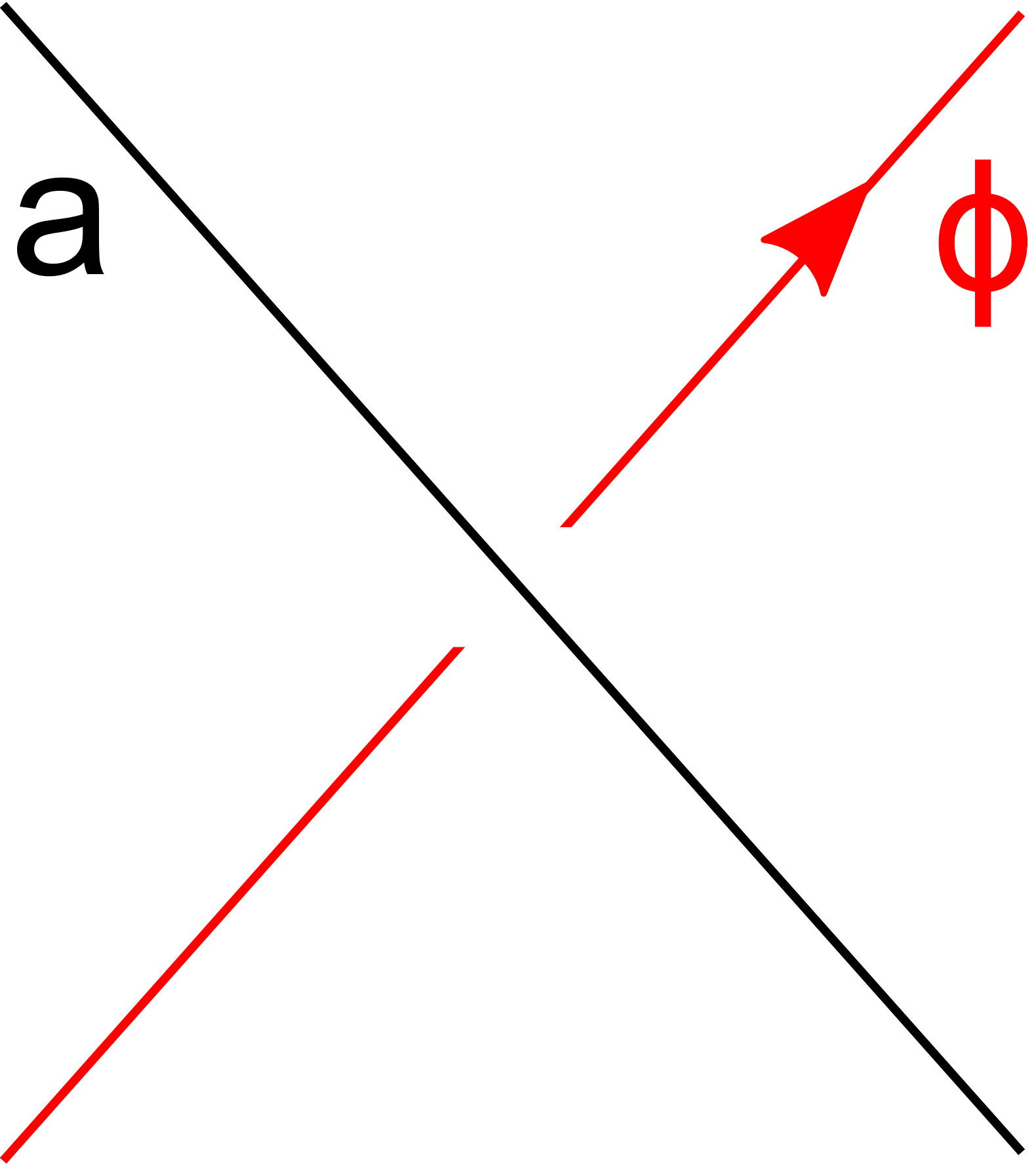}} \right|=w_\phi(a)\left<\raisebox{-0.22in}{\includegraphics[height=0.5in]{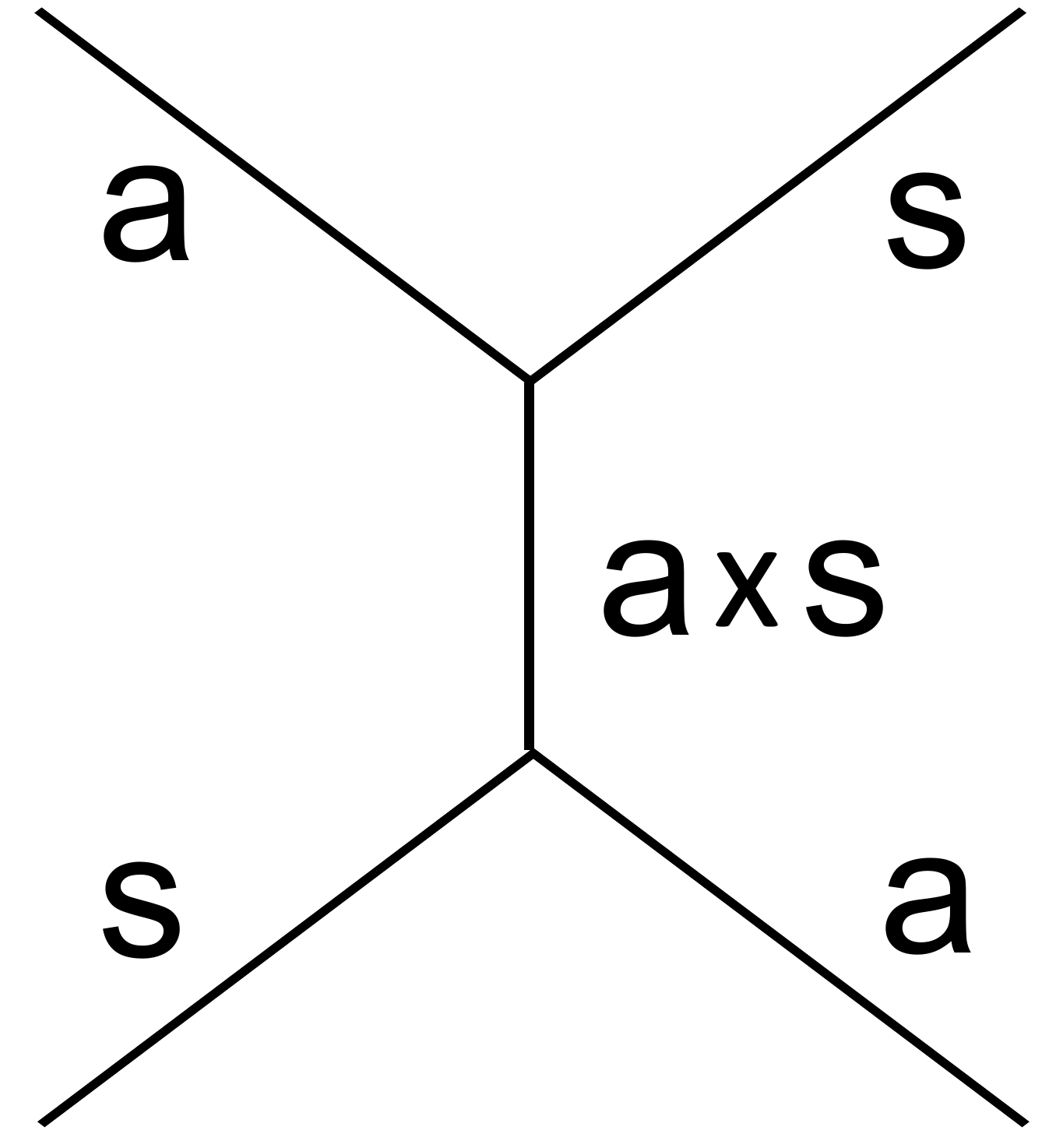}} \right|\\
	\left<\raisebox{-0.22in}{\includegraphics[height=0.5in]{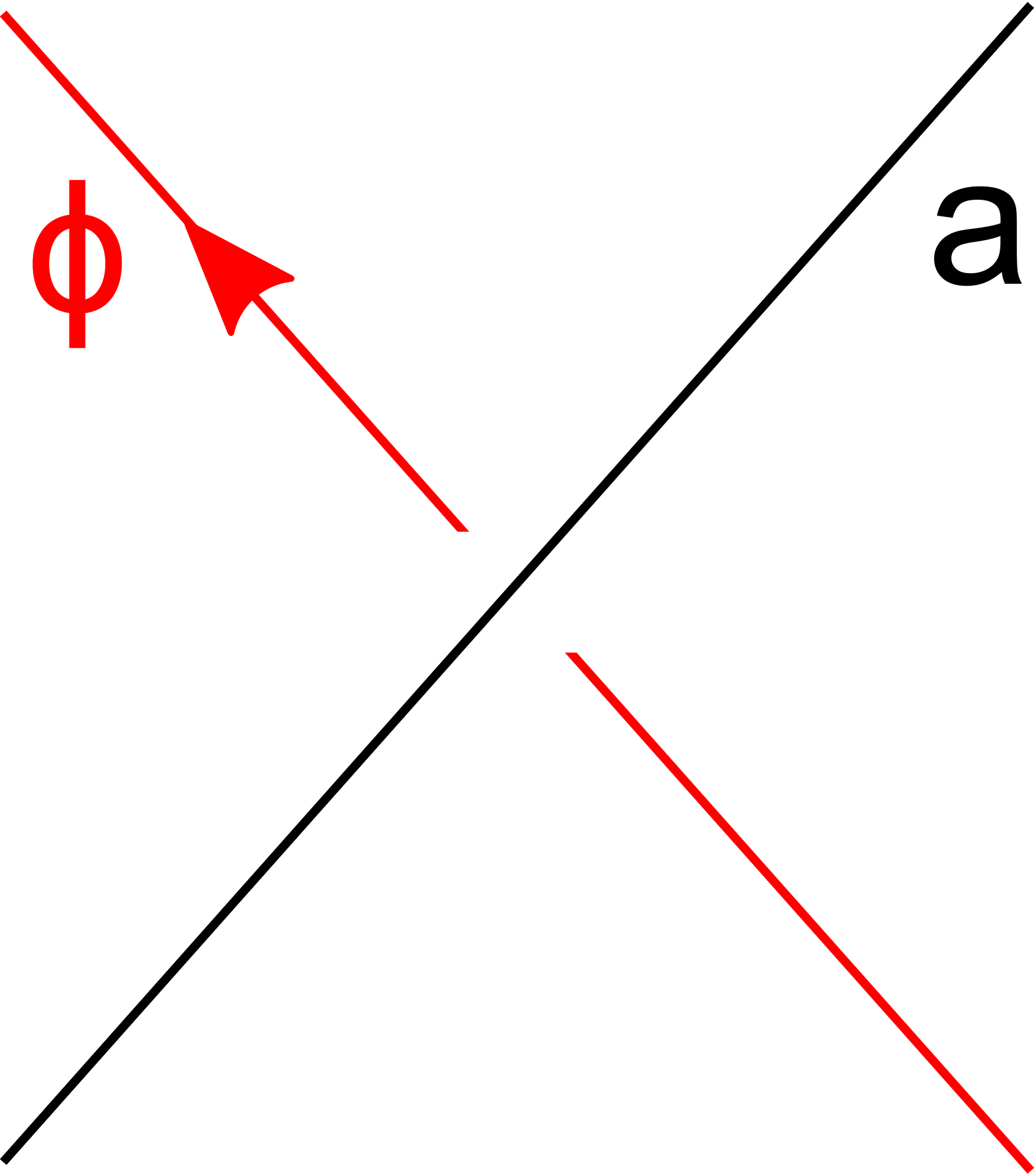}} \right|=\bar{w}_\phi(a)\left<\raisebox{-0.22in}{\includegraphics[height=0.5in]{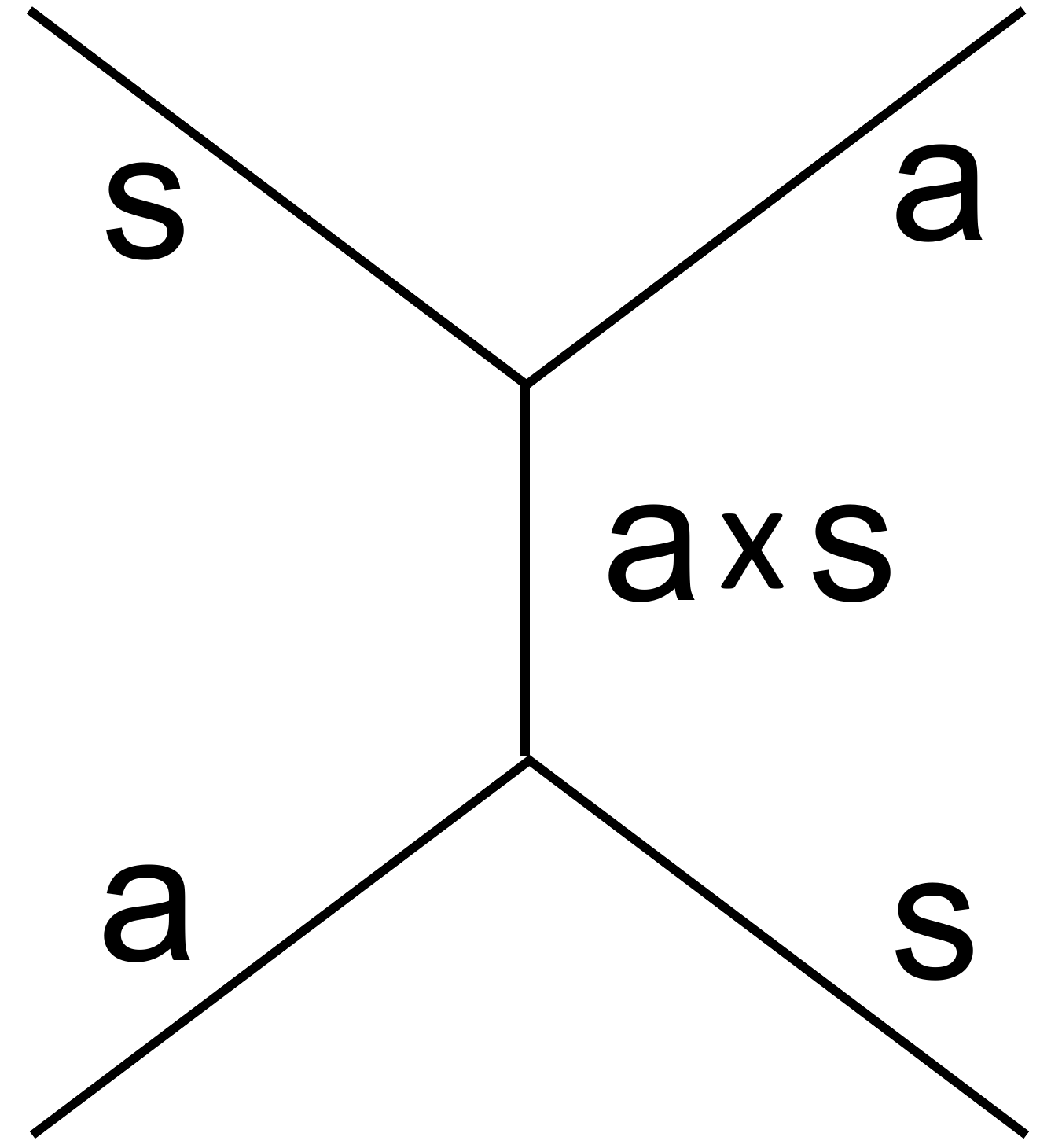}} \right|\\
\end{split}
	\label{stringrules}
\end{align}
Here $w_\phi(a),\bar{w}_\phi(a)$ 
are complex-valued functions of the string type $a$, with $w_\phi(0)=\bar{w}_\phi(0)
=1$.  
These rules, together with the local string-net rules (\ref{localrules}), dictate how to fuse the $\phi$-string into the string-net graph,
giving a new string-net states $\<X'|$, multiplied by a product of $w_\phi,\bar{w}_\phi$. 
This defines the action of $W_\phi(P)$ in terms of the parameters $(s,w,\bar{w})$.  For every abelian anyon $\phi$, there is an inverse anyon $\bar{\phi}$, obeying $\phi \times \bar{\phi} = \mathbf{1}$, where $\mathbf{1}$ denotes the identity anyon.

To ensure that $W_\phi(P)$ creates point-like excitations in the string-net ground state, we choose the parameters $(s,w,\bar{w})$ such that when acting on the string-net ground state $|\Phi \rangle$, the path independence condition:
\begin{equation}
	W_\phi(P)|\Phi\>=W_\phi(P')|\Phi\>
	\label{Eq:PathInd}
\end{equation}
is satisfied for any two upward paths $P,P'$ with the same end points.
Eq. (\ref{Eq:PathInd}) will be satisfied if the corresponding parameters $(s,w,\bar{w})$ obey
\begin{subequations}
\begin{align}
		w_\phi(a)w_\phi(b)&=C_s(a,b,c) w_\phi(c), \label{weq1}\\
		\bar{w}_\phi(a)&=w_\phi(a)^{-1} \label{weq2}\\
		w_\phi(a)w_{\bar{\phi}}(a)&=C_a(s,\bar{s},0)^{-1} \label{weq5}
\end{align}
\label{weqs}
\end{subequations}
with
\begin{equation}
	C_s(a,b,c)=\frac{F^{sab}_{c'a'c}F^{abs}_{c'cb'}}{F^{asb}_{c'a'b'}}
	\label{weqlabel}
\end{equation}
where $x' = x\times s$  for $x=a,b,c$, and $(a,b;c)$ is allowed by the branching rules.  Given a set of $F$-symbols satisfying (\ref{consistency},\ref{hermicity}) and a choice of the string $s$, in general we will find multiple solutions to Eqs. (\ref{weqs}) for $w_\phi$.  We label these by $m$, and the corresponding anyon by $\phi=(s,m)$ where $s$ is the string type created by the corresponding string operator $W_\phi$ and $m$ labels distinct solutions for a given $s$.

For example, the $\mathbb{Z}_N$ string-net model has $N$ string types $a\in\{0,1,\dots,N-1\}$ with $\mathbb{Z}_N$ branching rules $(a,b;c=a+b(\text{ mod }N))$.
There are $N$ distinct solution to (\ref{consistency})\cite{MooreSeiberg,PropitiusThesis}
\begin{equation}
	F(a,b,c)=e^{2\pi i \frac{pa}{N^2}(b+c-[b+c]_N)}.
	\label{Fsabelian}
\end{equation}
labeled by $p=0,\dots,N-1$. The arguments $a,b,c$ take values in $0,\dots,N-1$ and $[b+c]_N$ denotes $b+c$ (mod $N$) with values also taken in $0,\dots,N-1$.  
Each $\mathbb{Z}_N$ string-net model has $N^2$ topologically distinct quasiparticle excitations labeled by $\phi=(s,m)$ where $s,m=0,1,\dots,N-1$. The corresponding string operators $W_\phi$ are defined by the string parameters
\begin{align}
	w_{\phi}(a)=e^{2\pi i(\frac{psa}{N^2}+\frac{ma}{N})}.
	\label{}
\end{align}

The braiding statistics of quasiparticles can be extracted from the commutation algebra of the corresponding string oeprators. (see Ref.  \onlinecite{LinLevinstrnet}, \onlinecite{LevinWenstrnet}, \onlinecite{LinLevinBurnell} for details). Specifically, the exchange statistics of $\phi=(s,m)$ is
\begin{equation}
	e^{i\theta_\phi} = w_{\phi}(s).
	\label{}
\end{equation}
Thus self-bosons satisfy
\begin{equation}
	w_{\phi}(s)=1.
	\label{}
\end{equation}
If $\phi = (s,m)$ and $\chi = (r,n)$ are two abelian bosons that we wish to condense simultaneously, then they must have trivial braiding.  This requires that\cite{LinLevinBurnell}
\begin{equation}
w_\phi(r) w_\chi(s) = 1 .
\end{equation}

\subsection{Extended string-net model}

In the usual string-net construction, if $s \neq 0$, $W_\phi(P)$ creates states outside of the string net Hilbert space,  since  near the endpoints of $P$ there is no way to fuse an $s$-labeled string into the string net graph without creating vertices that violate the branching rules.  
When we are only interested in the topological nature of the excitations, the resulting ambiguity in the action of $W_\phi(P)$ near the endpoints is unimportant, since it affects only the immediate vicinity of the excitation and hence cannot impact its topological properties.  In order to condense $\phi$, however, we require a more careful treatment of these endpoints.  We achieve this by extending the string-net Hilbert space.

\subsubsection{Extended string-net Hilbert space}

The extended string-net Hilbert space, $\mathcal{H}_{\{\phi\}}$, is defined with respect to a set $\{ \phi \}$  of abelian bosons that we wish to condense.   
Since every finite abelian group is isomorphic to a direct product of cyclic groups, we can assume without loss of generality that the group is $G=\mathbb{Z}_{N_1}\times \dots \times \mathbb{Z}_{N_k}$.   To understand how to condense all bosons in $G$, it is therefore sufficient to understand how to condense bosons in a single $\mathbb{Z}_{N_j}$ factor; thus in what follows, for simplicity we will often restrict ourselves to the case that the set of bosons to be condensed comprise a cyclic group.
  
\begin{figure}[ptb]
\begin{center}
\includegraphics[width=0.8\columnwidth]{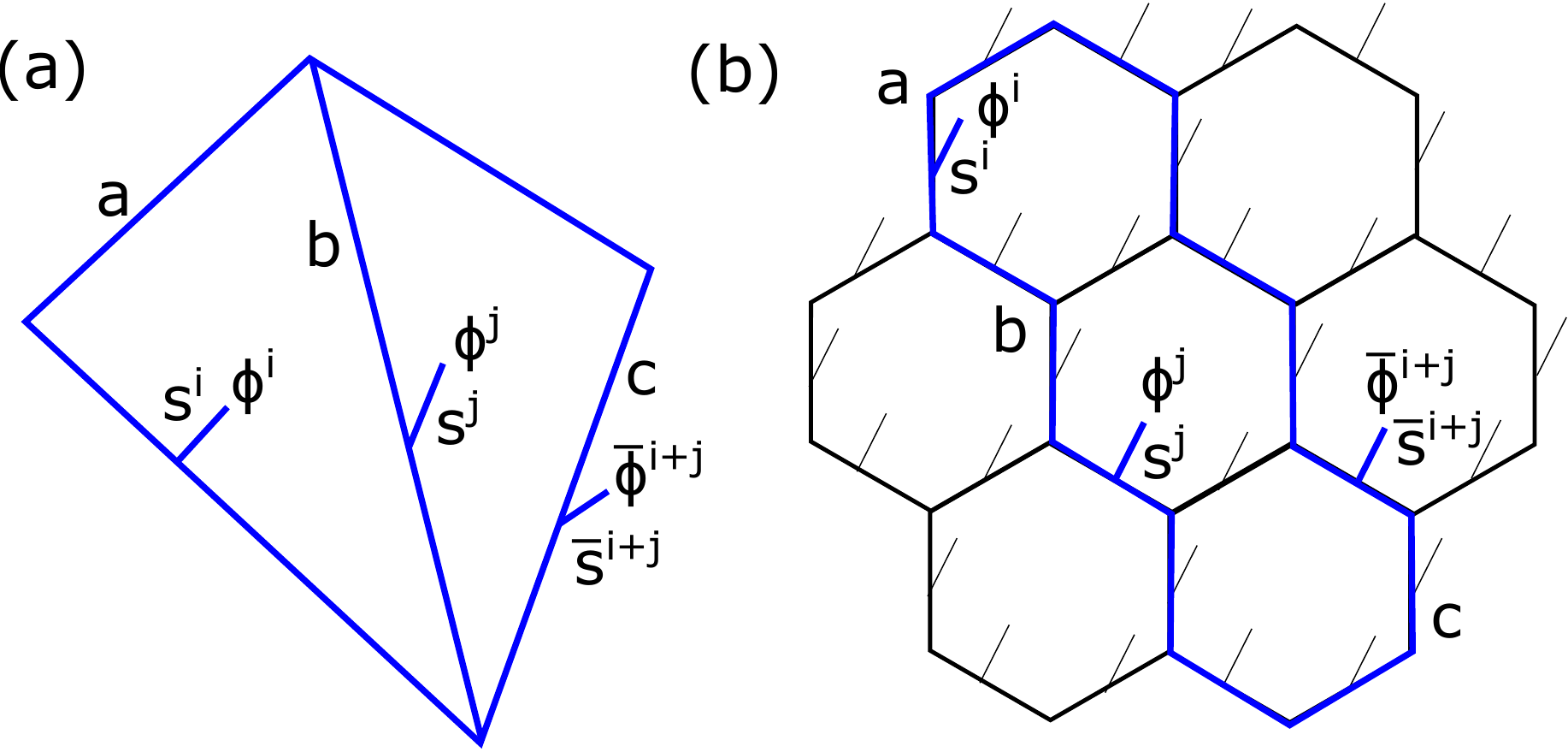}
\end{center}
\caption{A typical string-net configuration in $\mathcal{H}_\phi$ in continnum (a) and on the decorated honeycomb lattice (b).  Regular edges connecting two trivalent vertices can host any string label $a \in \mathcal{C}$.  Edges connecting to only one trivalent vertex, which we call sticks, may {\it only} host edge labels $\{ s\} $ associated with the string operators that generate the set of condensing bosons $\{\phi \}$.  These edge labels necessarily have abelian fusion rules, $a \times s =s \times a =  a'$ for any $a \in \mathcal{C}$.  Sticks also carry a label from the set $\{\phi \}$ at their end-points. 
} 
\label{fig:openstringnet}
\end{figure}

The string-nets in $\mathcal{H}_{\{\phi\}}$ are  oriented trivalent graphs with \emph{two} types of edges, as shown in Fig. \ref{fig:openstringnet}. The first type, which we will simply call edges, are edges connecting two trivalent vertices. Each such edge carries a string label as defined in $\mathcal{H}$.  
The second type of edge, which we will call sticks, has one end-point at a trivalent vertex, and one open endpoint.  A stick carries a $|G|$-spin label, which takes values in the set of abelian bosons $\{\phi\}$.  This spin label $\phi=(s,m)$ also dictates a string label $s$ associated with the stick; we require the labels at each trivalent vertex to satisfy the branching rules, and the total $G$-spin label (i.e. the sum of spin labels of all sticks) must be trivial.  
An orthonormal basis for the extended Hilbert space $\mathcal{H}_\phi$ is thus given by the set of all oriented trivalent graphs with sticks which (1) satisfy the branching rules at each trivalent vertex, and (2) have a net trivial $G$-spin.

To describe these extended string-nets on the lattice, we work on a decorated honeycomb lattice: at the center of each edge of the honeycomb lattice, we add an upward-pointing stick (see Fig. \ref{fig:openstringnet} (b)). We introduce two types of spins on the decorated lattice: link spins, which live on its edges, and end spins, which live at the endpoints of each stick.
The link spins take values in the string types $\{0, a, b, c, ...\}$ of the standard string-net model, and end spins take values in excitation labels $\{\phi\}$.  We require a stick carrying a label $\phi=(s,m)$ to have the string label $s$, and that all trivalent vertices satisfy the branching rules.

\subsubsection{$W_\phi(P)$ in the extended string-net Hilbert space}

In the following, we will use the extended string-net Hilbert space in two ways.  First, we may use it to describe a system whose ground state is the original string-net ground state, but which can also describe certain excited states that are not allowed in the original string-net Hilbert space.  In this case, sticks with non-trivial labels appear only in excited states, and the string-net ground state is exactly as described in Sec. \ref{Sec:StrOld}.  
Second, in order to describe the condensed phase, we can view all sticks as part of the ground-state Hilbert space.  This will allow us to describe a modified set of local rules capturing the condensed phase, as we discuss in Sec. \ref{Sec:newsnmodel}.

Here, we take the first perspective, and describe the action of the string operator $W_\phi(P)$ in the extended Hilbert space.   The action of $W_{\phi}(P)$ on the string net ground state $|\Phi \rangle$ is exactly as specified in Sec. \ref{Sec:Strops} away from the end-points of $P$.  However, we now require $P$ to begin and end on two sticks.  In adddition to its action on the edge labels, $W_\phi(P)$ acts by raising the end spin at the final and initial end-point of the path $P$ by $\phi$ and $\overline{\phi}$, respectively.

\begin{figure}[ptb]
\begin{center}
\includegraphics[width=1\columnwidth]{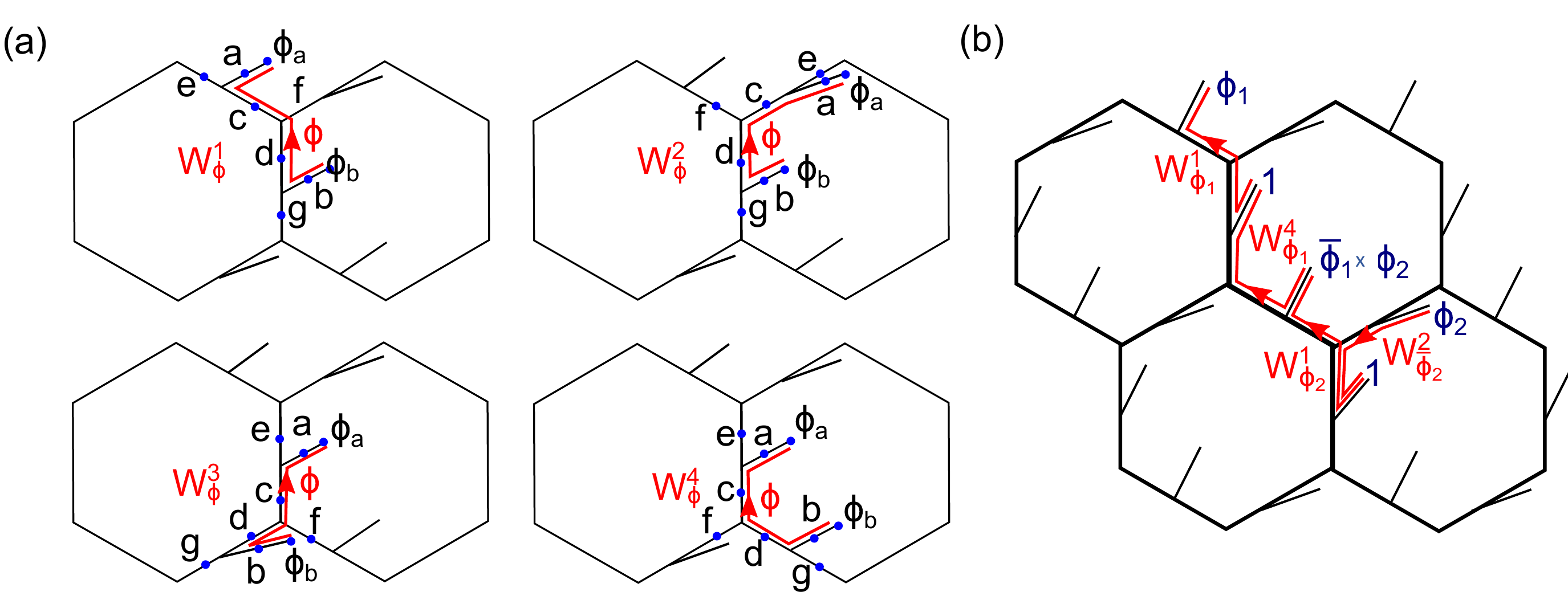}
\end{center}
\caption{(a) Four building blocks of any string operator defined on the decorated honeycomb lattice. $W_{\phi}^1,W_{\phi}^2,W_{\phi}^3,W_{\phi}^4$ act along four different paths (the red line) connecting two nearest neighboring sticks. Here $a,b,c,d$ and $e,f,g$ denote the initial link spin states along the path and on the external legs of the path, respectively and $\phi_a,\phi_b$ are end spin states at two ends of the path. Their matrix elements are given in (\ref{stringmatrix}). (b) A typical open string operator $W(P)$ along the path $P$ can be decomposed into product of basic string blocks acting on each vertex along $P$.
} 
\label{fig:stringoperator}
\end{figure}

When $\{ \phi\}$ is a set of abelian bosons with trivial mutual statistics,  we can describe any string operator $W_\phi(P)$ as a product of ``basic string operators" $W^i_\phi$, each of which connects a pair of sticks on adjacent edges.  
The four basic string operators on the decorated honeycomb lattice act along the four paths $p_1,p_2,p_3,p_4$ shown in Fig. \ref{fig:stringoperator}(a). 
The operators $W_\phi^1,W_\phi^2$ act on paths $p_1$ and $p_2$ centered at upward vertices, while $W_\phi^3,W_\phi^4$ act on paths $p_3$ and $p_4$, centered at the downward vertices. 
Their action is defined as follows.  Let $a,b,c,d$ and $e,f,g$ denote the initial link spin states along $p_i$ and on the external legs of $p_i$ respectively, and let $\phi_a,\phi_b$ be the initial end spin states at stick $a,b$ respectively (see Fig. \ref{fig:stringoperator}). 
The matrix elements of $W^{i}_\phi$ between an initial state $a,b,c,d,e,f,g,\phi_a,\phi_b$ and a final state $a',b',c',d',e,f,g,\phi_{a'},\phi_{b'}$ are then given by
\begin{align}
	\begin{split}
		W_{\phi,a'b'c'd';\phi_{a'}\phi_{b'}}^{1,abcd;\phi_a\phi_b}(efg)&=\bar{w}_{\phi}(f) \delta_{\phi_a\times \phi,\phi_{a'}}\delta_{\phi_b\times\bar{\phi},\phi_{b'}} \times	\\	
		&F^{eas}_{c'ca'}
		F^{cfs}_{d'df'}
		(F^{csf}_{d'c'f'})^*
		F^{s\bar{s}b}_{b 0 b'}
		(F^{d s b'}_{gd'b})^*\\
	W_{\phi,a'b'c'd';\phi_{a'}\phi_{b'}}^{2,abcd;\phi_a\phi_b}(efg)& =
	\delta_{\phi_a\times \phi,\phi_{a'}}\delta_{\phi_b\times \bar{\phi},\phi_{b'}}\times\\
	&F^{eas}_{c'ca'}
	F^{fcs}_{d'dc'}
	F^{s\bar{s}b}_{b 0 b'}
	(F^{d s b'}_{gd'b})^*  \\
	W_{\phi,a'b'c'd';\phi_{a'}\phi_{b'}}^{3,abcd;\phi_a\phi_b}(efg)&={w}_{\phi}(f) \delta_{\phi_a\times \phi,\phi_{a'}}\delta_{\phi_b\times \bar{\phi},\phi_{b'}} \times	\\	
		&F^{eas}_{c'ca'}
		(F^{dfs}_{c'cf'})^*
		F^{dsf}_{c'd'f'}
		F^{s\bar{s}b}_{b 0 b'}
		(F^{d s b'}_{gd'b})^* \\
	W_{\phi,a'b'c'd';\phi_{a'}\phi_{b'}}^{4,abcd;\phi_a\phi_b}(efg)&=
	\delta_{\phi_a\times \phi,\phi_{a'}}\delta_{\phi_b\times \bar{\phi},\phi_{b'}}\times\\
	&F^{eas}_{c'ca'}
	(F^{fds}_{c'cd'})^*
	F^{s\bar{s}b}_{b 0 b'}
	(F^{d s b'}_{gd'b})^*
	\end{split}
	\label{stringmatrix}
\end{align}
Here $x' = x \times s$ (or $x \times \bar{s}$, if $x=b)$, where 
we use multiplicative notation for the abelian group operation on both edge and end spins.

Notice that the matrix elements of open string operators are not invariant under local unitary transformations of the form (\ref{gt}), and thus are gauge dependent.
When $\phi$ are cyclic abelian bosons with trivial mutual statistics, there exists a convenient gauge
\begin{equation}
	F(s^i,s^j,s^k)\equiv F^{s^is^js^k}_{s^{(i+j+k)}s^{(i+j)}s^{(j+k)}}=1
	\label{f1}
\end{equation}
where $s^i,s^j,s^k$ are any string types associated with condensing bosons (see Appendix \ref{app:string}).  We will work in the gauge (\ref{f1}) in the rest of the paper.

In the gauge (\ref{f1}), the basic string operators have the following important properties, which we derive in Appendix \ref{app:string}.
First, all basic string operators commute with each other:
\begin{equation}
	[W_\phi^i,W_{\phi'}^j]=0.
	\label{wpcommute}
\end{equation}
Second, one can show that
\begin{equation}
	\begin{split}
		W^{i\dagger}_\phi =W^{i}_{\bar{\phi}}	
\end{split}
	\label{wdagger}
\end{equation}
for $i=1,2,3,4$.  

\begin{figure}[ptb]
\begin{center}
\includegraphics[width=1\columnwidth]{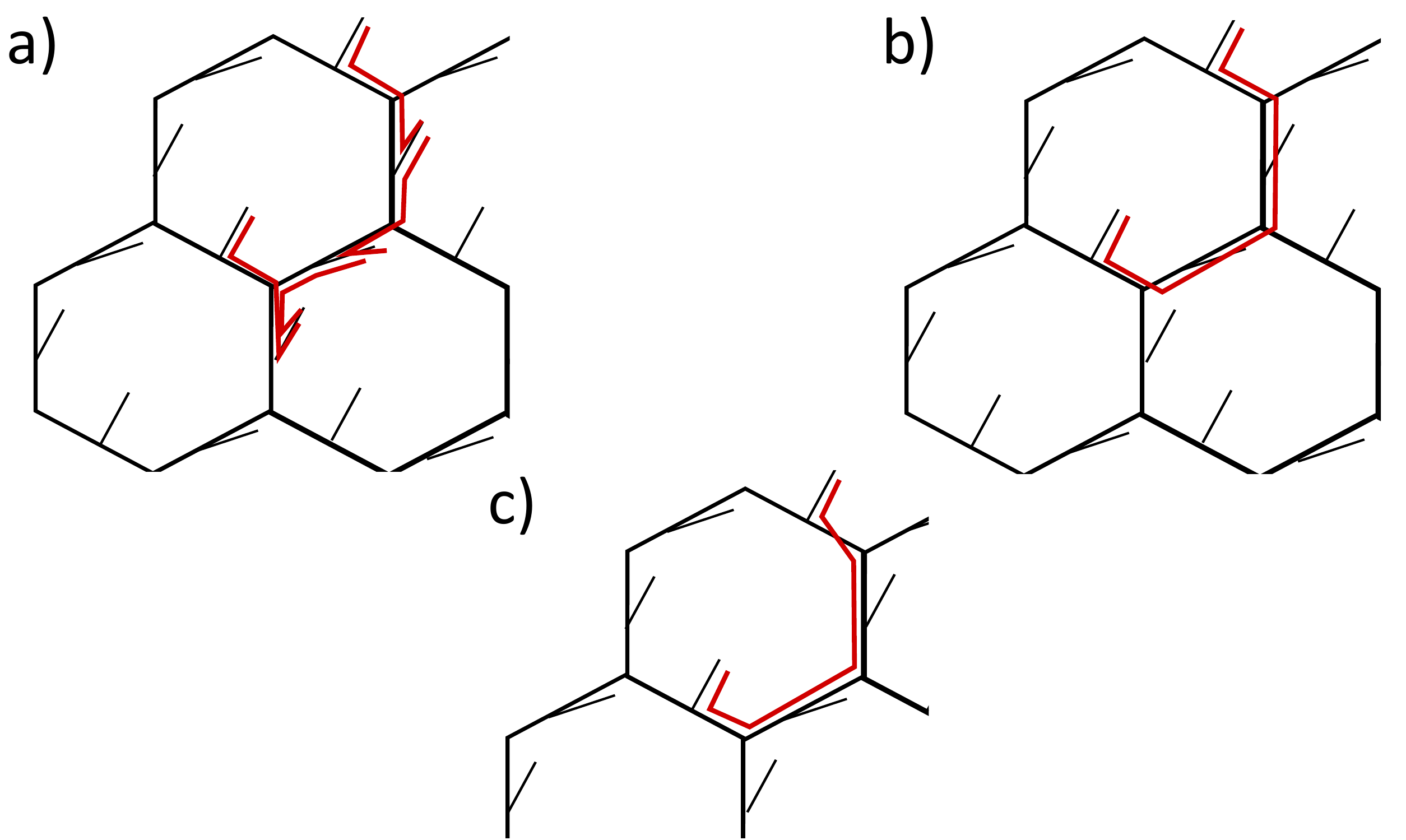}
\end{center}
\caption{(a) An example of a  product of the basic string operators creating a $(\phi, \bar{\phi})$ pair on two neighbouring plaquettes.  (b) The action of these products is equivalent to the action of a string $W_{\phi}(P)$ starting on a stick in one plaquette, and ending on a stick in the other plaquette, which visits no other sticks in between.   (c) By path independence (see Eq. (\ref{Eq:PathInd})), such a string can be deformed to cross only the edge separating the plaquette pair.  
} 
\label{StringCombineFig}
\end{figure}

Finally, a general string operator can be expressed as a product of simple string operators.
First, $W_{\phi_1}^i, W_{\phi_2}^i$ along the same path $p_i$ can be combined as
\begin{equation}
	W_{\phi_1}^i \cdot W_{\phi_2}^i =W_{\phi_1\times \phi_2} ^i
	\label{w1w20}
\end{equation}
where the $\cdot$ operation is defined by
\begin{equation}
	\begin{split}
	&W_{\phi_3=\phi_1\times \phi_2,a_3b_{\bar{3}}c_3d_3;\phi_a\times \phi_3,\phi_b\times \bar{\phi}_3}^{i,abcd;\phi_a\phi_b}(efg) = \\
	&\qquad \qquad W_{\phi_1,a_1b_{\bar{1}}c_1d_1;\phi_{a}\times\phi_1,\phi_{b}\times \bar{\phi}_1}^{i,abcd;\phi_a\phi_b}(efg) \times \\
	&\qquad \qquad W_{\phi_2,a_3b_{\bar{3}}c_3d_3;\phi_a\times\phi_3,\phi_b\times \bar{\phi}_3}^{i,a_1b_{\bar{1}}c_1d_1,\phi_a\times \phi_1,\phi_b\times \bar{\phi}_1}(efg)
	\end{split}
	\label{w1w2}
\end{equation}
for $i=1,\dots,4$.  Thus if the set of condensing bosons is cyclic and generated by $\phi$, we can express all basic string operators as products of the basic string operator $W^i_\phi$.  
Second, 
let $P$ be a path obtained from a union of two basic paths, $ p_i(r_1,r_2)$, which begins on a stick at positions $r_1$, and ends on a stick at position $r_2$, and $ p_j(r_2,r_3)$, which begins on the stick at $r_2$ and ends on a stick at $r_3$ (see Fig. (\ref{StringCombineFig})).   (Note that it is because $w_\phi(s) =1$ that we can combine string end-points into a single string that crosses the sticks).  
Then we have:
\begin{equation}
	W_{\phi}(P) =  W_{\phi}^{i, (12)} W_\phi^{j,(23)}
	\label{}
\end{equation}
and similarly for paths composed of more than two concatenated segments, as shown in the Figure.  Here we define $W_\phi^{i, (12)} = W^i_{\phi}$ if $p_i(r_1,r_2)$ is oriented upwards, and $(W^i_{{\phi}})^\dag$ otherwise.  By joining string operators along multiple basic paths in this way, we can thus express $W_{\phi}(P)$ as a product of basic string operators for any path $P$.  (Note that since the basic string operators commute, the order in which we apply them is unimportant.)
It follows that {\it any} product of $\phi$-string operators for $\phi$ in our chosen set of abelian bosons can be expressed as a product of basic string operators.

\section{Lattice Hamiltonians for condensing abelian bosons} \label{Sec:CondHam}

Next, we identify a lattice Hamiltonian $H(J)$ within the extended string-net Hilbert space that, by tuning a parameter $J$, can bring a system through a transition in which a set of abelian bosons is condensed.  
Our lattice Hamiltonian has the general form
\begin{equation}
	H(J)= H_\mathcal{C} - J H_1 \ .
	\label{hsnc}
\end{equation}
Here $H_\mathcal{C}$ is a Hamiltonian in the extended string-net Hilbert space whose ground state is exactly the original string-net ground state; it can be viewed as a modification of the original string-net Hamiltonian (see Refs. \cite{LevinWenstrnet,LinLevinBurnell}) appropriate to the extended string-net Hilbert space.  $H_1$ is a term which creates particle- antiparticle pairs of anyons in the set $\{ \phi\}$ of condensing bosons.  Here, for simplicity, we take this set to be a cyclic group of order $p$, which we denote $\< \phi \> = \{  \phi^i, i = 1, ... , p\}$. with $\phi^p = 1$.

We will show that $H(J)$ has the following properties.  First,  $H(J=0)$ is identical to the original string-net Hamiltonian when acting on states where all stick labels are trivial, and states with sticks carrying non-trivial labels $\phi^i$ have a finite energy cost.  Thus in this limit
 string-net eigenstates with sticks carrying non-trivial labels $\phi^i$ correspond to gapped excited states, and the ground state is the original string-net ground state $|\Phi \rangle$.
Second,  $H(J = \infty)$ is a commuting projector model with a frustration-free ground state $|\Psi \rangle$, in which excitations in the set $\< \phi \>$ have condensed, in the sense that they are present in arbitrary number in the ground state.  Third, $|\Psi \rangle$ can be obtained by applying a certain projector to the $J=0$ ground-state $| \Phi \rangle$.  Thus we can describe the $J= \infty$ ground state explicitly in terms of the string types and local rules associated with $|\Phi \rangle$, and use this description to investigate the topological data of the condensed phase.

It is worth noting that, as we show below, the ground state of $H(J)$ for any $J$ contains only excitations on the sticks, and no plaquette defects.  Thus for $\phi^p = 1$, the critical point separating condensed and uncondensed phases is always of the Potts or clock variety, depending on the specific choice of $H_1$.  Here we have chosen a Potts- like version, resulting in first order transitions for $p \geq 3$.

\subsection{The Hamiltonian $H_{\mathcal{C}}$ \label{modifiedstring}}

\begin{figure}[ptb]
\begin{center}
\includegraphics[width=0.5\columnwidth]{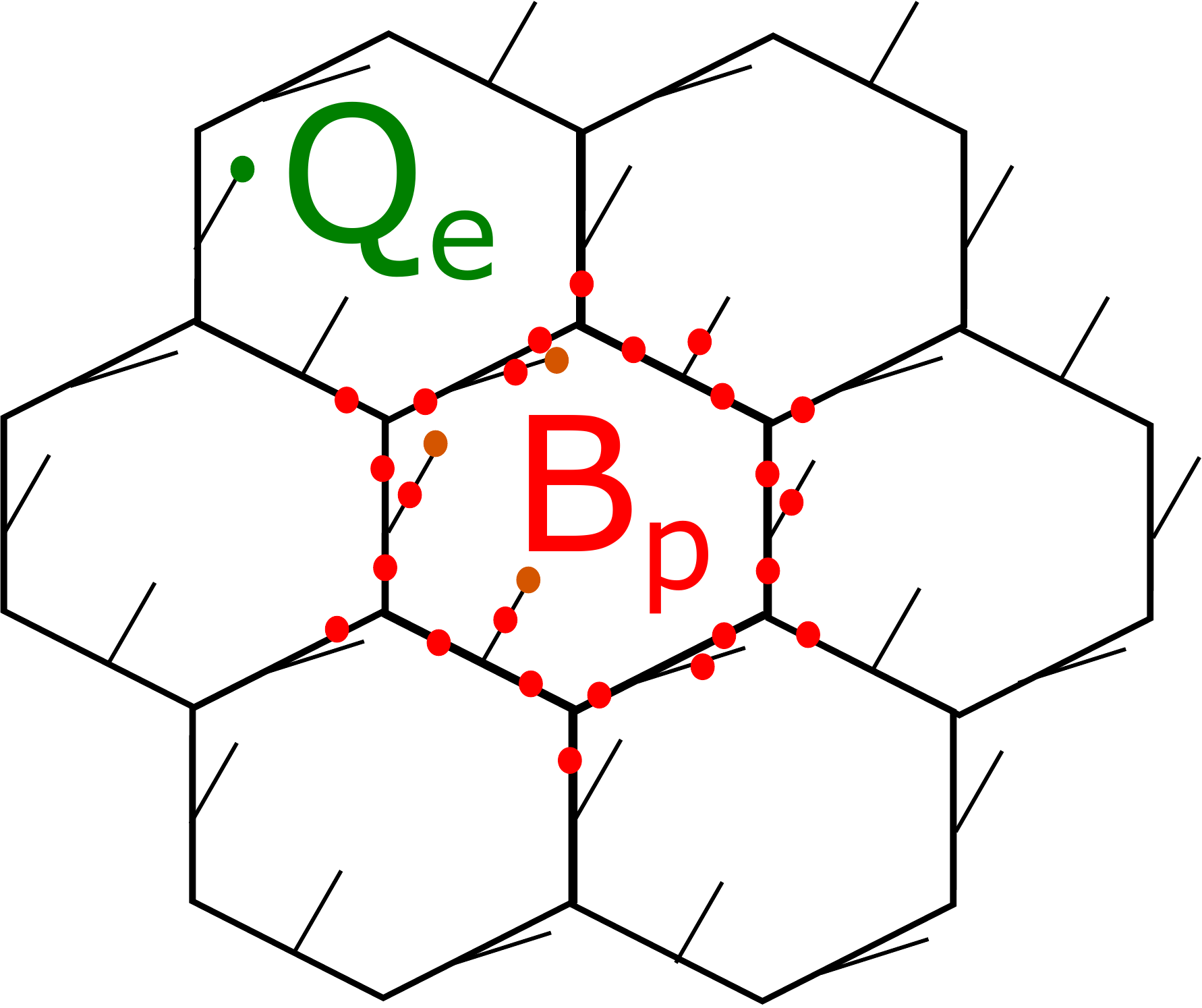}
\end{center}
\caption{Decorated honeycomb lattice with an upward stick on each link of the honeycomb lattice. The $Q_e$ operator acts on the end spin. The $B_p$ operator acts on 27 spins adjacent to the plaquette $p$.
} 
\label{fig:lattice}
\end{figure}

We first define the Hamiltonian $H_{\mathcal{C}}$ in the extended string-net Hilbert space $\mathcal{H}_\phi$ of the honeycomb lattice (see Fig. \ref{fig:openstringnet}).  
$H_{\mathcal{C}}$ is of the form 
\begin{equation}
	H_{\mathcal{C}}=-\sum_e Q_e -\sum_p B_p^\phi 	
	\label{hc1}
\end{equation}
The two sums run over end spins $e$ and plaquettes $p$ of the decorated honeycomb lattice.
The operator $Q_e$ acts on the end spins
\begin{equation}
	Q_e=\delta_{e,\mathbf{1}}
	\label{}
\end{equation}
where $\delta_{e,\mathbf{1}}=1$ if $e=\mathbf{1}$ (no excitation) and $\delta_{e,\mathbf{1}}=0$ otherwise ($\phi$ excitations).
The operator $Q_e$ penalizes the states with $\phi$ excitations at ends of sticks.  Note that unlike in the usual string-net Hamiltonian, we have not included a term imposing the branching rules at each vertex; instead, we will work exclusively in the string-net Hilbert space, where these are necessarily satisfied.

The operator ${B}_p^\phi$ on the decorated honeycomb lattice is more complicated, but the main idea is as follows.  First, $[B_p^\phi, B_{p'}^{\phi}] = [{B}_p^\phi,Q_e] =0$, ensuring that $H_{\mathcal{C}}$ is a sum of commuting projectors.  Second, analogous to the plaquette term in the usual string-net models\cite{LevinWenstrnet,LinLevinBurnell},$B_p^\phi$ maps between different string-net configurations in the extended string-net Hilbert space, ensuring that the ground states (for which all stick labels are trivial) obey the local rules (\ref{localrules}).   Indeed, when acting on states where all stick labels are trivial, our plaquette term is identical to that of the generalized string net models \cite{LinLevinBurnell}. Third, unlike the plaquette term of the usual string-net models, ${B}_p^\phi$ commutes with the string operators $W_{\phi^k}(P)$ even for paths $P$ ending or beginning on the plaquette $p$.   

We note that here we use a prescription that ensures that ${B}_p^\phi$ commutes with $W_{\phi}(P)$ for {\em any} choice $\phi = (s,m)$; this allows us to discuss all abelian anyon condensation transitions on the same footing.  For some classes of models, however (those for which the fusion category describing the string types is {\em braided}), an alternative and potentially computationally simpler formulation of the Hamiltonian resulting in the same condensed phase exists; this is discussed in Appendix \ref{app:bps}.  

We now describe the operator ${B}_p^\phi$ in detail.  ${B}_p^\phi$ has the form:
\begin{equation}
	{B}_p^\phi=\sum_{s=0}^{N-1}\frac{d_s}{D} {B}_p^{\phi,s}
	\label{bp}
\end{equation}
where $D=\sum_{s=0}^{N-1}d_s^2$ and ${B}_p^{\phi,s}$ describes a 27 spin interaction involving the 24 link spins around $p$ and 3 end spins inside $p$ (see Fig. \ref{fig:lattice}). 
Its action can be understood as a sequence of three operations:
\begin{equation}
	{B}_p^{\phi,s}=\sum_{\phi_{10},\phi_{11},\phi_{12}} W^{\dagger}_{\phi_{10},\phi_{11},\phi_{12}} B_p^s W_{\phi_{10},\phi_{11},\phi_{12}}
	\label{}
\end{equation}
where the sums run over the possible spin labels of the three sticks inside $p$, with
\begin{equation}
	W_{\phi_{10},\phi_{11},\phi_{12}}=
	  (W_{\phi_{10}}^1)^\dag \mathcal{P}_{\phi_{10}}  \cdot
	W_{\phi_{11}}^1 \mathcal{P}_{\phi_{11}}  \cdot
	 W_{\phi_{12}}^3 \mathcal{P}_{\phi_{12}}
	\label{}
\end{equation}
Here $\mathcal{P}_{\phi_a}=|\phi_a\>\<\phi_a|$ projects the end spin label of stick $a$ onto $\phi_a$, and  $(W_{\phi_{10}}^1)^\dag,W_{\phi_{11}}^1,W_{\phi_{12}}^3$ are basic string operators (see Eq.  (\ref{stringmatrix})) that lower the spin label on sticks $10, 11$, and $12$ by $\phi_{10}, \phi_{11}$, and $\phi_{12}$ respectively.  
The operator $W_{\phi_{10},\phi_{11},\phi_{12}}$ therefore moves any excitations on the sticks 
inside the plaquette $p$ to sticks outside of $p$:
\begin{equation}
	\left<\raisebox{-0.32in}{\includegraphics[height=0.7in]{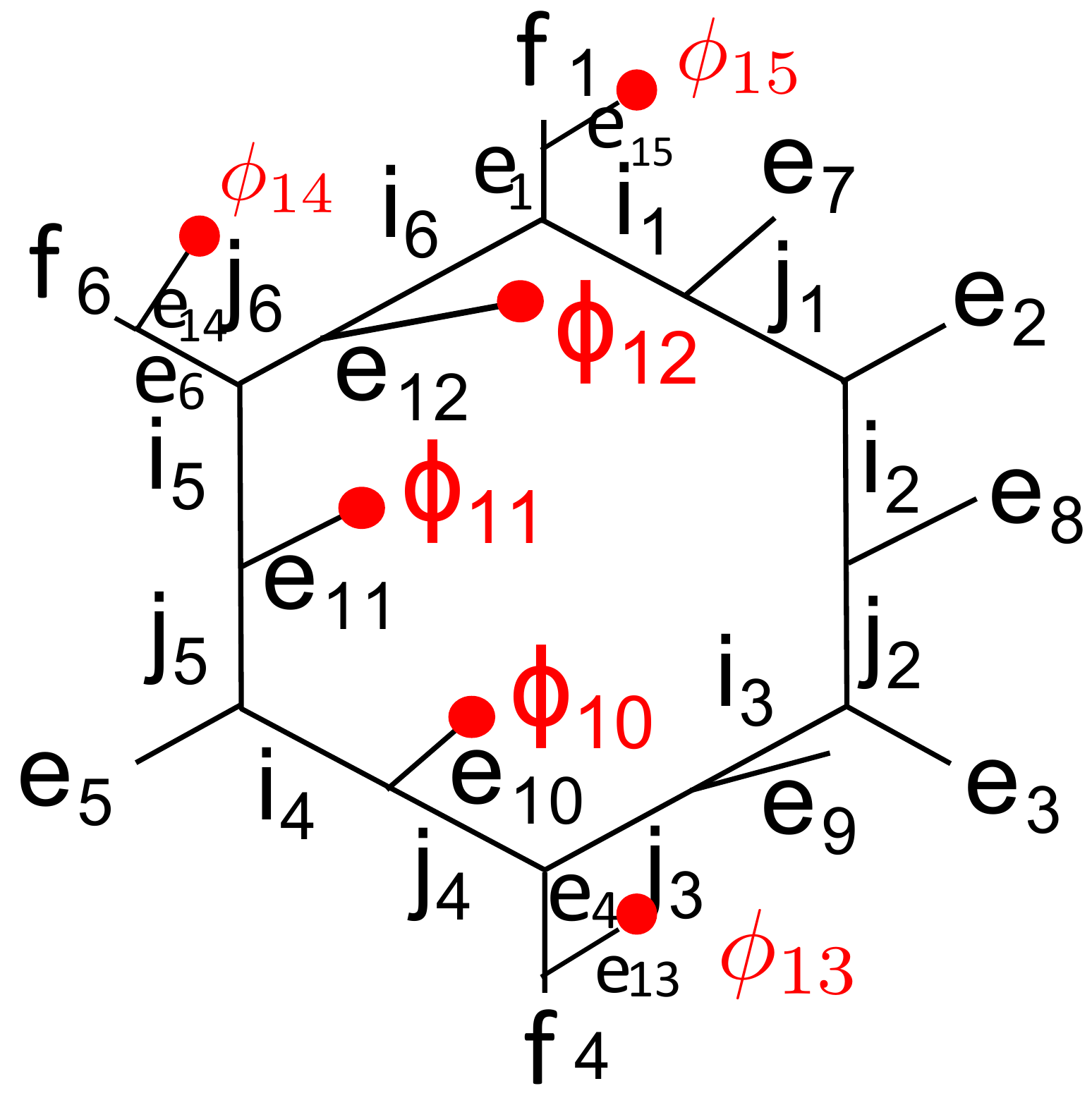}} \right|
	W_{\phi_{10},\phi_{11},\phi_{12}}=C_1 \cdot
	\left<\raisebox{-0.32in}{\includegraphics[height=0.7in]{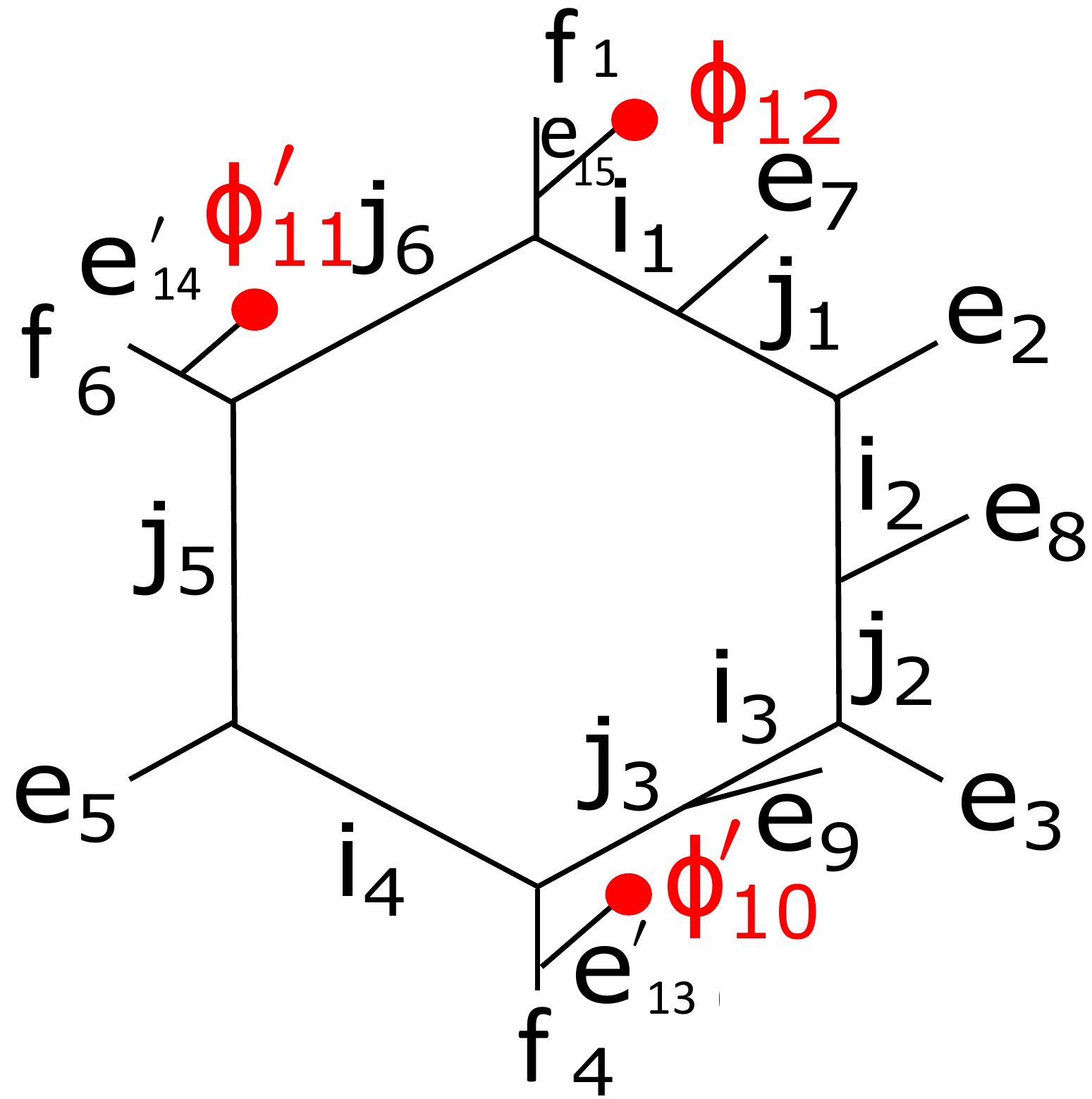}}\right|.
	\label{}
\end{equation} 
This action is nontrivial only if $\{\phi_{10},\phi_{11},\phi_{12}\}$ contains non-trivial end spin labels.  In particular, it is trivial when acting on  ground states of $H_{\mathcal{C}}$.  

The second operation $B_p^s$ is the same as the plaquette operator defined in the Ref. \onlinecite{LevinWenstrnet} which adds a loop of type-$s$ string around the boundary of $p$:
\begin{equation}
	\begin{split}
	\left<\raisebox{-0.32in}{\includegraphics[height=0.7in]{bp2b-eps-converted-to.pdf}} \right
	|B_p^s&=
	\left<\raisebox{-0.32in}{\includegraphics[height=0.7in]{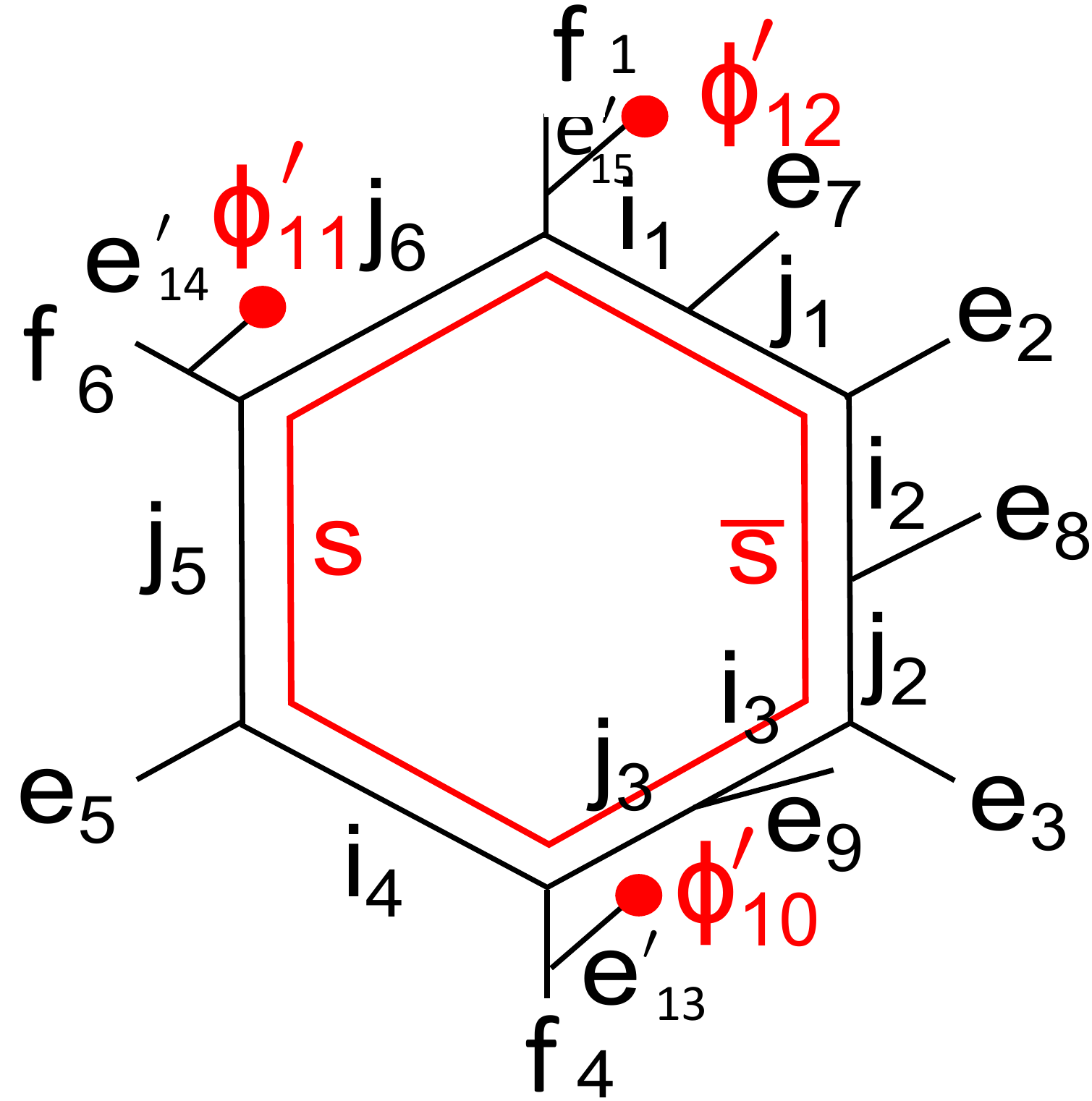}} \right| \\
	&=\sum_{i_1'\dots i_6'} C_2  \cdot \left<\raisebox{-0.32in}{\includegraphics[height=0.7in]{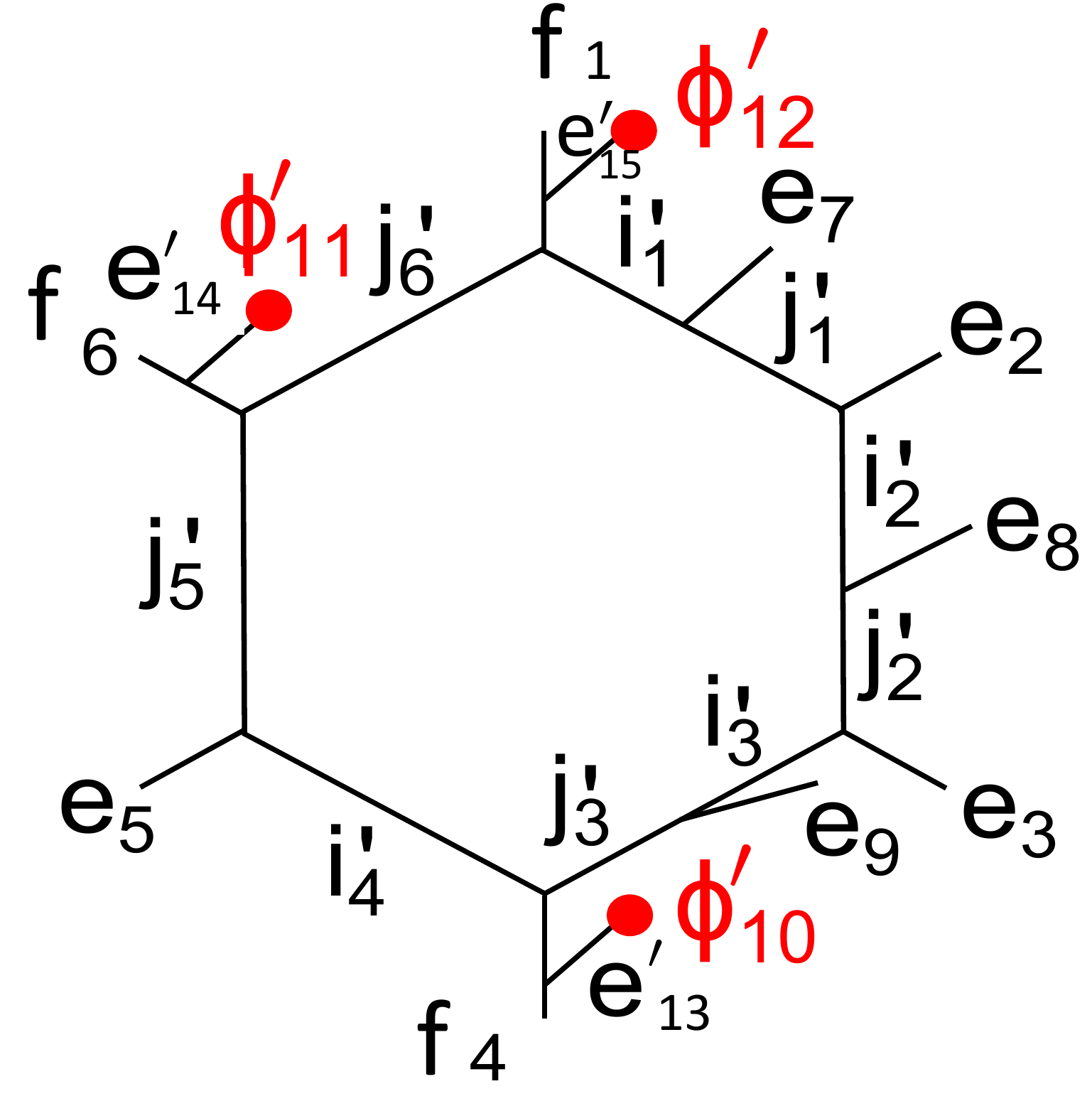}} \right|.
\end{split}
	\label{}
\end{equation}
Finally, the operation $W^{\dagger}_{\phi_{10},\phi_{11},\phi_{12}}$ moves the excitations $\{\phi^{10},\phi^{11},\phi^{12}\}$ back to the appropriate sticks in $p$:
\begin{equation}
	\left<\raisebox{-0.32in}{\includegraphics[height=0.7in]{bp2g-eps-converted-to.pdf}} \right
	|W_{\phi_{10},\phi_{11},\phi_{12}}^{\dagger}=C_3  \cdot
	\left<\raisebox{-0.32in}{\includegraphics[height=0.7in]{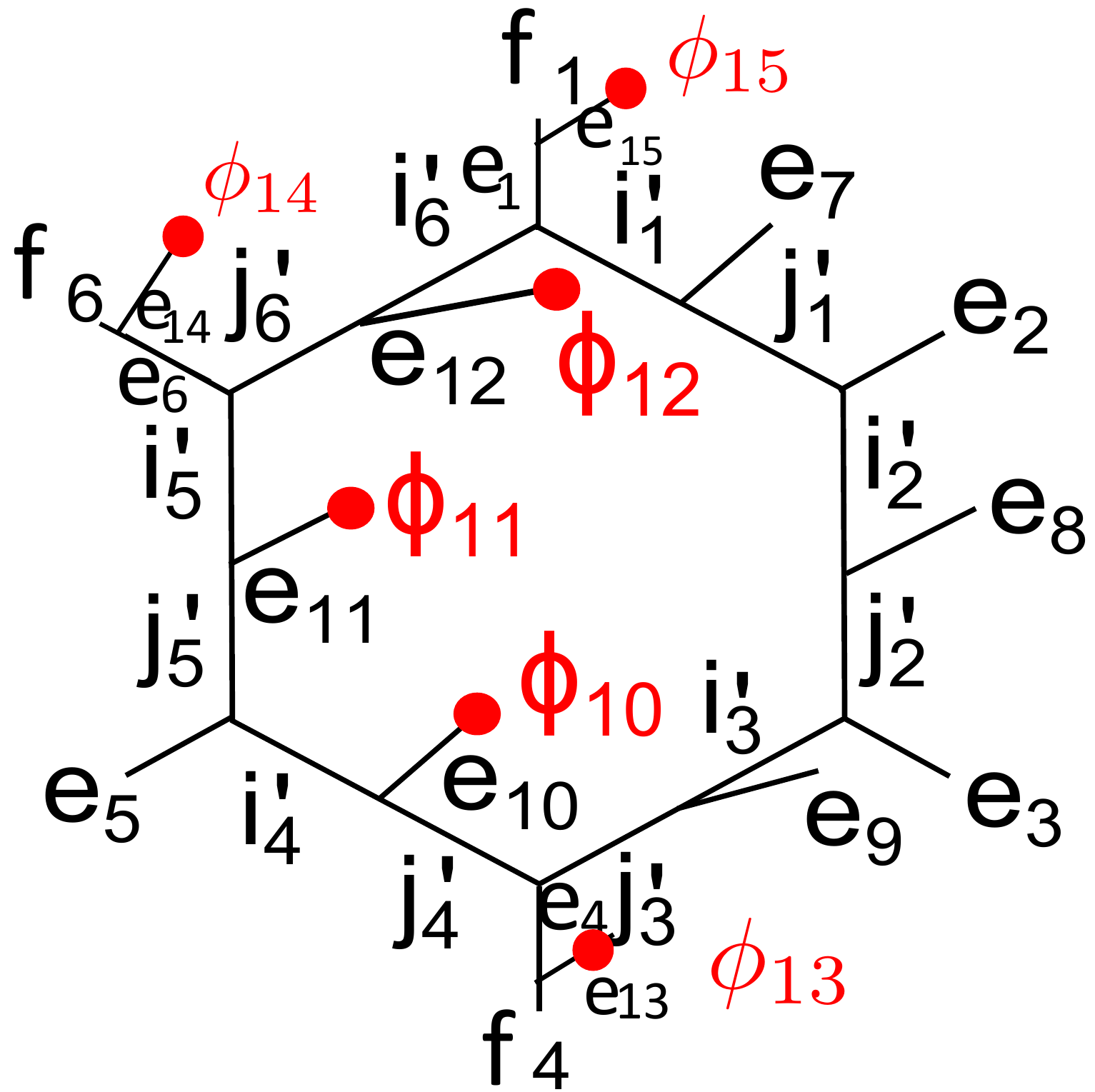}} \right|.
	\label{}
\end{equation}
Here $C_1,C_2,C_3$ are the corresponding matrix elements of the three operations. The product $C_1 \cdot C_2 \cdot C_3$ gives the matrix elements of $B_p^{\phi,s}$.

More precisely, the matrix elements of ${B}_p^{\phi,s}$ are defined by
\begin{equation}
	\begin{split}
	&\left<\raisebox{-0.32in}{\includegraphics[height=0.7in]{bp2a-eps-converted-to.pdf}} \right| {B}_p^{\phi,s}  \\
	&=\sum_{i'_1\dotsi'_{6}}
	\left\<\raisebox{-0.32in}{\includegraphics[height=0.7in]{bp2h-eps-converted-to.pdf}} \right|
	B^{s,i_1\dots i_{6}j_1\dots j_6}_{p,i'_1\dots i'_{6}j'_1\dots j'_6}(e_1\dots e_{12};\phi_{10},\phi_{11},\phi_{12})
\end{split}
	\label{}
\end{equation}
where 
\begin{equation}
	\begin{split}
		&B^{s,i_1\dots i_{6}j_1\dots j_6}_{p,i'_1\dots i'_{6}j'_1\dots j'_6}(e_1\dots e_{12};\phi_{10},\phi_{11},\phi_{12})=\\
		&d_s\sqrt{\frac{d_{i_1}d_{j'_2}d_{i_3}d_{i_4}d_{j'_5}d_{j_6}}{d_{i'_1}d_{j_2}d_{i'_3}d_{i'_4}d_{j_5}d_{j'_6}}}\times \\
		&(F^{\bar{s}i_1 e_7}_{j'_1i'_1j_1})^*
		(F^{\bar{s}j_1e_2}_{i'_2j'_1i_2})^*
		(F^{\bar{s}i_2e_8}_{j'_2i'_2j_2})^*
		F^{\bar{s}i_3e_3}_{j'_2i'_3j_2}
		(F^{\bar{s}i_3e_9}_{j'_3i'_3j_3})^* 
		(F^{e_5i_4s}_{j'_5j_5i'_4})^* \times \\
		&F^{e'_6j_6s}_{j'_5j_5j'_6} 
		F^{j_6s\bar{s}}_{j_6j'_60}
		(F^{i_4s\bar{s}}_{i_4i'_40})^*
		(F^{i'_6\bar{s}i_1}_{e'_1j_6i'_1})^*
		F^{i'_4\bar{s}j_3}_{e'_4i_4j'_3} \times \\
		&W^{1,e_{10}e_{13} j_4e_4;\phi_{10}\phi_{13}}_{\bar{\phi}_{10},0e_{13}' i_4e_4';\mathbf{1}\phi_{13}'}(i_4j_3f_4)
		W_{\phi_{10},e_{10}e_{13} j'_4e_4;\phi_{10}\phi_{13}}^{1 ,0e_{13}'i'_4e_4';\mathbf{1}\phi_{13}'}(i'_4j'_3f_4) 
		 \times\\
		 &W^{1,e_{14} e_{11}e_6i_5;\phi_{14}\phi_{11}}_{\phi_{11},e'_{14}0e_6'j_5;\phi_{14}'\mathbf{1}}(f_6j_6j_5)
		 W_{\bar{\phi}_{11},e_{14} e_{11}e_6i'_5;\phi_{14} \phi_{11}}^{1 ,e_{14}'0e_6'j'_5;\phi_{14}' \mathbf{1}}(f_6j'_6j'_5) \times \\
		 &W^{3,e_{15} e_{12}e_1i_6;\phi_{15}\phi_{12}}_{\phi_{12},e_{15}'0e_1' j_6;\phi_{15}'\mathbf{1}}(f_1i_1j_6)
		 W_{\bar{\phi}_{12},e_{15} e_{12}e_1i'_6;\phi_{15}\phi_{12}}^{3 ,e_{15}'0e_1'j'_6;\phi_{15}'\mathbf{1}}(f_1i'_1j'_6)
	\end{split}
	\label{bp1}
\end{equation}
Here $e_7,e_8\dots e_{12}$ take values in abelian string types and thus $j_p=i_p\times e_{p+6}$ for $p=1\dots6$ while $e_1'=e_1\times e_{12},e_4'=e_4\times \bar{e}_{10}$ and $e'_6=e_6\times e_{11}$.
The matrix elements of $W^1_\phi,W^3_{\phi},W_\phi^{1\dagger},W_\phi ^{3 \dagger}$ are defined in (\ref{stringmatrix}) and (\ref{wdagger}).

From this explicit form, one can check that the plaquette operator $B_p^{\phi,s}$ commutes with any basic string operator $W_{\phi^i}^{j}$ (and hence any string operator $W_{\phi^i}(P)$): 
\begin{equation}
	[B_p^{\phi,s},W_{\phi^i}^j]=0.
	\label{bpw}
\end{equation}
We leave the derivation to Appendix \ref{app:wbcommute}. 

 We can now show that 
$H_{\mathcal{C}}$ (\ref{hc1}) has the following properties.  First, it is a sum of commuting projectors:
clearly $[Q_e, Q_{e'}] = [Q_e, B_p^{\phi}] = 0$, since $B_p^{\phi}$ does not alter the value of the spin-label on any stick, and hence preserves the eigenvalue of $Q_e$.  
Moreover, in Appendix \ref{app:commute} we show that $[B_p^{\phi}, B_{p'}^{\phi} ] =0$.  Essentially, this results from the fact that the two plaquette operators commute in the absence of excitations on the sticks, and also that the string operator used to move a stick excitation off the shared edge between two adjacent plaquettes commutes with $B^{\phi}_p$, where $p$ is the plaquette that the stick points outward from.  

It follows that, like the conventional string-net Hamiltonians, $H_{\mathcal{C}}$ is exactly solvable. 
Second, there exists at least one state that satisfies $Q_e=B_p^\phi=1$ for all $e,p$; this state is therefore a ground state.  Clearly, states with only trivial stick labels satisfy $Q_e = 1$ at every vertex; when restricted to theese states, $H_{\mathcal{C}}$ reduces to the original string-net model, whose ground state $|\Phi\>$ is an eigenstate of the plaquette term of the corresponding Hamiltonian with eigenvalue $1$ on every plaquette\cite{LinLevinBurnell}.   When $Q_e \equiv 1$, $|\Phi \rangle$ is therefore also an eigenstate satisfying $B_p^{\phi}\equiv 1$.  In other words, $H_{\mathcal{C}}$  is exactly solvable, and its ground state(s) are exactly the string-net ground states defined by the local rules (\ref{localrules}).  

Though they have the same ground state(s), the excited states of our extended string-net model differ from those of conventional string-net Hamiltonians.   
In conventional string nets, where sticks are not included, 
excited eigenstates are either string-net states with $B_p^\phi=0$ on some plaquettes, or states that  violate the branching rules and hence are outside of the string-net Hilbert space (for which necessarily we also have $B_p^{\phi}=0$ on some plaquettes). 
In our models, however, there are $\phi^j$-type excitations of  $H_{\mathcal{C}}$ satisfying $B_p^\phi=1$ everywhere, with $Q_e = 0$ on some sticks.\footnote{We note that if we allow states outside of the string net Hilbert space, this leads to a redundancy,  since in our extended Hilbert space $\phi$ can also be realized by an eigenstate with $Q_e \equiv  1$, with either some $B_p^\phi=0$  or a violation of the branching rules.}
  As a consequence, the ground state of $H(J)$ satisfies $B_p^\phi\equiv 1$ {\em for every positive $J$}.

\subsection{The Hamiltonian $H_1$}

To define $H_1$, we begin by defining the projector along the path $p_i,i=1,2,3,4$:
\begin{equation}
	P_{\phi}^i({\bf r}) =\frac{1}{p}\sum_{k=1}^{p}W_{\phi^k}^i ({\bf r})
	\label{pij}
\end{equation}
where the sum runs over basic open string operators $W_{\phi^k}^i$ with $\phi^k \in \<\phi\>$, and ${\bf r}$ indexes a unit cell of the honeycomb lattice.
The set of operators $\{P_{\phi}^i\}$ form commuting projectors.
The operator $H_1$ is defined as a sum of commuting projectors over all neighboring sticks
\begin{equation}
	H_1=\sum_{i, {\bf r} }P_{\phi}^i ({\bf r})
	\label{h11}
\end{equation} 
By (\ref{bpw}), we have $[H_1, B_P^{\phi}] =0$. Thus, $H_1$ creates excitations only of $Q_e$ in $H_{\mathcal{C}}$, while leaving the operator $B_P^\phi$ in its ground state on every plaquette.

\subsection{Condensed phase and the $J \rightarrow \infty$ limit}

For $J$ sufficiently large, the string net describes a new topological phase, in which the anyons $\{ \phi^j, 1 \leq j < q \}$ have condensed.   
That this is so can be most easily understood by considering the limit $J \rightarrow \infty$. 

Since $H_1$ is a sum of commuting projectors, in the $J\rightarrow \infty$ limit, the low-energy Hilbert space $\tilde{\mathcal{H}}$ consists of states in the image of the projector:
\begin{equation}
	P_{\phi}=\prod_{i, {\bf r}}P_{\phi}^i ({\bf r}) \ .
	\label{pphi}
\end{equation}
These states have eigenvalue $1$ under all terms in $H_1$.  

To leading order in $1/J$, the effective Hamiltonian, which acts within the low energy Hilbert space $\tilde{\mathcal{H}}$, is 
\begin{equation}
	H_{\mathcal{\tilde{C}}}=P_\phi H_{\mathcal{C}} P_\phi=-\sum_p P_\phi B_p^{\phi}+ \text{const} .
	\label{effH}
\end{equation}
In the second equality, we use (\ref{bpw}) and the fact that $H_\phi \equiv \sum_{e} P_\phi Q_e P_\phi $ is simply the number of ways to combine operators in $P_\phi $ to obtain a trivial label on every vertex, which is a system size independent constant.
Note that since $P_{\phi}$ and $B_p^{\phi}$ are projectors, and $[P_{\phi}, B_p^{\phi} ]=[B_p^{\phi} , B_{p'}^{\phi} ]=0$, $P_\phi B_p^{\phi}$ are also commuting projectors. 
Moreover, if $|\Phi\>$ is the ground state of $H_{\mathcal{C}}$, we have $P_\phi B_p^{\phi} (P_\phi |\Phi\> ) =P_\phi |\Phi\> $ for every $p$.  Hence
the ground state $|\Psi\>$ of $H_{D}$ is given by\footnote{In fact, projecting any state $|\Phi_{ex}\>$ satisfying $B_p^{\phi}|\Phi_{ex}\> =|\Phi_{ex}\>$ in this way gives the ground state $|\Psi \>$ of $H_{D}$.  This is because such states have the form $|\Phi_{ex}\>=W_\phi(P)|\Phi\>$, and as $P_\phi W_\phi(P)=P_\phi$, we have $P_\phi|\Phi_{ex}\>=P_\phi W_\phi(P)|\Phi\>=P_\phi |\Phi\>$. } 
\begin{equation}
	|\Psi\>=P_\phi |\Phi\>.
	\label{psiD1}
\end{equation}

To show that $|\Psi \>$ is indeed a state in which the bosons $\< \phi \>$ have condensed, we expand the projector 
$P_{\phi}$ according to:
\begin{equation}
	P_\phi=\frac{1}{p^{2 N_V}} \sum_{\{\phi_j, r_{ij} \}}W_{\{\phi_j, r_{ij}\}}
	\label{pphi1}
\end{equation}
Here $N_V$ is the number of vertices on our honeycomb lattice; for each such vertex there are two  simple string operators.  (Note that throughout this paper, we will assume boundary conditions where this is the case).  $W_{\{\phi_j, r_{ij}\}}$ is the composite string operator which creates excitations $\{\phi_j\}$ using string operators along the paths $\{ r_{ij} \}$ on the lattice, and the sum runs over all possible configurations  $\{\phi_j, r_{ij}\}$ on  the lattice, subject to the constraint that $\prod_j \phi_j = 1$. 
We can use (\ref{pphi1}) to expand the new ground state (\ref{psiD1}) as
\begin{equation}
	|\Psi\> =\frac{1}{p^{2 N_V}}  \sum_{\{\phi_j, r_{ij} \}} W_{\{\phi_j, r_{ij}\}} |\Phi  \> \ .
	\label{psiD}
\end{equation}
In other words, the ground state $|\Psi\>$ is a superposition of all possible configurations of $\phi$ excitations -- a $\phi$ condensed state.

We can also make some educated guesses about  the topological order in the condensed phase by studying the effect of $P_{\phi}$ on low-lying excited states of $H_{\mathcal{C}}$.  These are created by generalized versions of the string operators $W_{\phi}(P)$, which we deonte $W_{\alpha}(P)$, where $P$ describes a path on the lattice, and $\alpha$ is the anyon type.  The data associated with $W_{\alpha}(P)$ is essentially the same as that for $W_{\phi}(P)$, except that resolving string crossings requires a matrix $\Omega_\alpha(a)$, rather than a scalar $w_\phi(a)$; a detailed description can be found in Ref. \cite{LinLevinBurnell}.  Unless $\alpha = \phi^j$, here we require that $P$ starts and ends at vertices of the lattice, rather than on sticks.

Consider how the operators $P_{\phi}^i ({\bf r})$ act on the string operators $W_{\alpha}(P)$.   
The latter can suffer one of three possible fates.  First, if 
\begin{equation}
W_{\phi}(P)  W_{\alpha}(P) = W_{\alpha'}(P)
\end{equation}
then the  operators $W_{\alpha}(P)$, $W_{\alpha'}(P)$ have identical actions on states in the image of $P_{\phi}$.  This suggests that in the limit $J \rightarrow \infty$, the two anyons $\alpha$ and $\alpha'$ have been {\it identified}, meaning that they comprise a single anyon type in the condensed phase. For example, all of the condensing bosons $\{ \phi^k \}$ are identified with the vacuum in the condensed phase.
This conclusion agrees with the expectations of other approaches to anyon condensation\cite{SlingerlandBais,Eliens}.  

Note that if
\begin{equation}
W_{\phi^r}(P)  W_{\alpha}(P) = W_{\alpha}(P)
\end{equation}
for $r| p$, then in the condensed phase $W_{\alpha}(P)$ becomes identified with $r-1$ distinct anyon string operators, rather than with $q-1$.  For example, if $r=1$,  $W_{\alpha}(P)$ does not become identified with any other anyon string operators.  Though this statement seems rather innocuous here, in fact in such cases $\alpha$ {\it splits} into multiple anyon types after condensation\cite{SlingerlandBais,Eliens}.  We will not discuss the splitting at the level of anyons in detail here; however it is closely related to the splitting of string net labels which, as we show in Section \ref{Sec:VLC}, arises in the ground states of our condensed string net model.

Second, if $\alpha$ braids non-trivially with one of the condensing bosons $\phi$, then when the path $P_1$ crosses $P_2$, and $\phi$ is an abelian boson,\cite{LinLevinBurnell}
\begin{equation}
W_{\bar{\phi}}(P_1) W_{\alpha}(P_2) W_{\phi}(P_1) = \frac{ S_{\alpha \phi}}{S_{\alpha \mathbf{1}}} W_{\alpha}(P_2)   W_{\phi}(P_1)
\end{equation}
In this case, the string operator $W_{\alpha}(P)$  maps states in the image of $P_{\phi}$ (for which $W_{\phi}(P_1) |\Psi \rangle = |\Psi \rangle$ for every choice of $P_1$) to states outside of this image.   This suggests that $\alpha$ anyons are no longer a point-like excitations in the condensed phase, and become confined, again agreeing with expectations based on other approaches to anyon condensation.

In the following sections, rather than pursue the analysis of anyon string operators, we will instead focus on the fate of the string net ground state in the condensed phase. We will show how to describe the ground state of $H_{\mathcal{C}}(J \rightarrow \infty)$  
as a conventional string net of the type described in Ref. \cite{LinLevinBurnell}.  Such string net ground states can always be associated with a commuting projector string net Hamiltonian\cite{LinLevinBurnell}, whose topological order can be inferred directly from the string net data.  (Specifically, it is the Drinfeld center of the fusion category comprising the string net).  Thus this approach allows us to identify the topological order of the condensed phase without requiring an explicit discussion of anyon string operators in the condensed phase.

\section{The condensed Hilbert space}  \label{Sec:newH}

In order to understand the condensed phase, 
we begin by studying the effective Hilbert space
\begin{equation}
	\tilde{\mathcal{H}} = \text{span}\{\<\tilde{X}|:\<\tilde{X}|P_\phi = \<\tilde{X}|\}.
	\label{hspaceeff}
\end{equation}
that describes states of finite energy in the limit $J \rightarrow \infty$, which we refer to as the {\em condensed Hilbert space}. Our goal is to show that $\tilde{\mathcal{H}}$ can be thought of as a new (non-extended) string-net Hilbert space,  whose basis states are string-net states with new string labels and new branching rules.

To construct a basis state $\<\tilde{X}|$ in the effective Hilbert space defined by Eq. (\ref{hspaceeff}), we begin with a reference state $\<X|$ in the uncondensed string-net Hilbert space $\mathcal{H}$.  The corresponding  basis state in $\tilde{\mathcal{H}}$ is: 
\begin{equation}
	\<\tilde{X}|=\<X|P_\phi
	\label{}
\end{equation}
Since $W^i_{\phi^k} ({\bf r} ) P_{\phi}^i({\bf r} ) =  P_{\phi}^i({\bf r} )$ for every $i$ and ${\bf r}$, we have
\begin{equation} \label{Eq:Pmult}
W_{\{\phi_j, r_{ij}\}}  P_{\phi} = P_{\phi} W_{\{\phi_j, r_{ij}\}}  =P_{\phi}  \ .
\end{equation}
Consequently, if $\< X' | = \< X | W_{\{ \phi_j, r_{ij} \} }$, then $\<X|P_\phi = \<X'|P_\phi$.  Thus, to construct a basis 
 of $\tilde{\mathcal{H}}$, we must find a suitable basis $\{ \<X_0|\}$ of $\mathcal{H}$ such that $\<X^{i}_0| W_{\{ \phi_j, r_{ij} \} } | X_0^{j} \rangle = \delta_{ij}$. 
 Since we are interested in identifying a set of string-net labels appropriate to the condensed phase, we take $\{ \<X_0| \}$ to be states in the string-label basis -- i.e., $\< X_0^j |$ has a fixed string label for every edge.

To find the new string-label basis $\{\<\tilde{X}|\}$, we consider two classes of condensing bosons. The first class is $\phi = (0,m)$ -- i.e. the string operator $W_{ \phi}(P)$ does not change the string labels of edges that it acts on.  This type of condensation, which does not require the extended Hilbert space, has been discussed in detail in Ref. \cite{TSBLong}. The second class, with $\phi = (s,m)$, condenses bosons whose string operators do change the string net labels.  These condensation transitions do require the extended Hilbert space that we introduce here.  

We start with the first case, $\phi=(0,m)$. For any edge in the lattice, $P_\phi$ contains an equal contribution from $\phi^j$-labeled strings that cross that edge, for every $j$ (see Fig. \ref{StringCombineFig}).  Thus,
\begin{equation}
	\left<\raisebox{-0.22in}{\includegraphics[height=0.5in]{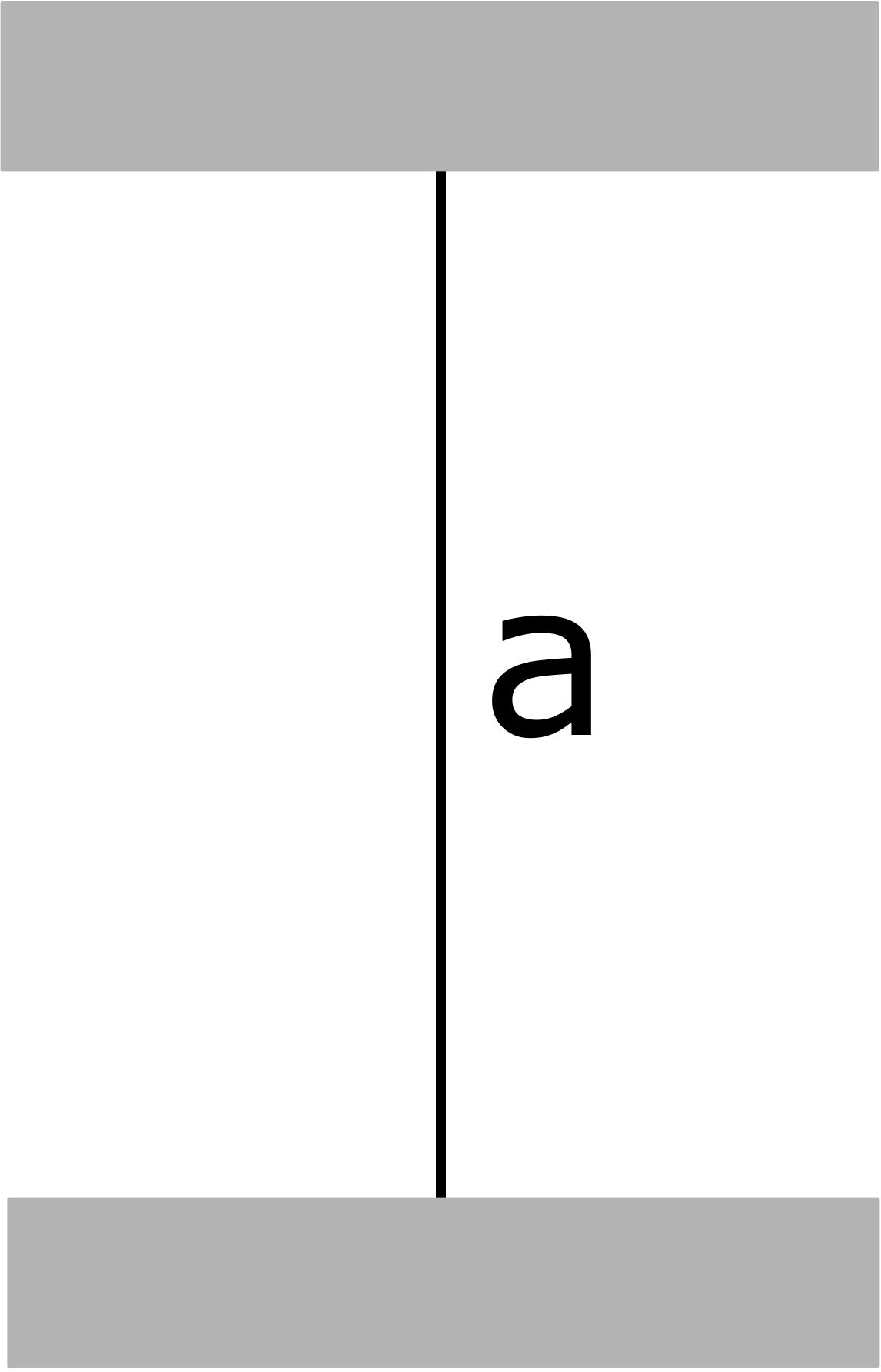}} \right| P_\phi= 
	\left<\raisebox{-0.22in}{\includegraphics[height=0.5in]{line-eps-converted-to.pdf}} \right|  \frac{1}{p}\sum_{i=1}^p w_{\phi^i}(a) =
	\left<\raisebox{-0.22in}{\includegraphics[height=0.5in]{line-eps-converted-to.pdf}} \right|  \delta_{w_\phi(a),1}
	\label{}
\end{equation}
where $p$ is the order of $\<\phi\>$.
Thus, only string types $a$ with 
\begin{equation}
	w_\phi(a)=1
	\label{wphi1}
\end{equation}
remain after condensation of $\phi=(0,m)$ bosons. 
Hence, the new string-label basis $\{\<\tilde{X}|\}$ is simply the subset of states in $\< X_0 |$ containing only string types $a$ satisfying Eq. (\ref{wphi1}).  We say that the remaining string labels, which do not appear in the low-energy Hilbert space after condensation, are {\it confined}.

Now, we consider the second case, $\phi=(s,m)$. If $\<\phi\>$ contains a subgroup $\<\phi_{m^j}\>\subset \<\phi\>$ which is generated by $\phi_{m^j} =(0,m^j)$, then by the same reasoning as above, the string types $a$ that appear in the condensed Hilbert space must satisfy 
\begin{equation}
	w_{\phi_{m^j}}(a)=1,
	\label{wm1}
\end{equation}
with other string types being confined.  
We find it useful to reorganize the deconfined string types into new string labels as follows. 
First, we define the new null string label via:
\begin{equation}
	\tilde{0}= \oplus_{j=0}^{q-1} s^j \ .
	\label{new0}
\end{equation}
where we have assumed the condensing bosons form a cyclic group generated by $\phi = (s,m)$, with $s^q= 1$ and $q|p$.  Here the symbol $\oplus$ means that, in the original string net basis, an edge carrying the label $\tilde{0}$ carries a superposition of labels in the set $\{s^j \}$.  
Similarly, other  condensed string types are given by superpositions of the form:
\begin{equation}
	\tilde{a}=\oplus_{j=0}^{q-1} (a \times s^j ).
	\label{newa}
\end{equation}
It is convenient to pick a particular representative for $\tilde{a}$ in the original label set, which we will denote $a$.  We denote the remaining terms on the right-hand side of Eq. (\ref{newa}) as:
\begin{equation}
a^j  \equiv  a \times s^j  \ , \ \ a^0 \equiv a
\end{equation}
Then all $a^j$ project to the same condensed string type, while if $b \neq a \times s^j$, then $a$ and $b$ project to different string types.  As we discuss in detail below, if one or more of the labels obey $a^r = a$ for some $r |q$, in the condensed phase the single string type $a$ splits into multiple string types $\tilde{a}_1, ... \tilde{a}_{q/r}$.  
Finally, 
the branching rules for the new string labels can be deduced from the branching rules of the original string labels.   In the absence of splitting, given the branching rules $(a,b;c)$, the new branching rules are $(\tilde{a},\tilde{b};\tilde{c})$.  We discuss the new branching rules in the presence of splitting in Sec. \ref{Sec:Splitting}.

The condensed string labels, together with the associated branching rules, define the string-label basis in the condensed Hilbert space.
Specifically, a string-net state $\< \tilde{X} | \in \tilde{\mathcal{H}}$ has edges labeled with the condensed string types $\{ \tilde{0}, \tilde{a}, \tilde{b}, ... \}$, and satisfies the new branching rules $(\tilde{a},\tilde{b};\tilde{c})$ at each vertex.  Note that $\tilde{\mathcal{H}}$ should be viewed as a conventional string-net Hilbert space, since after condensation all sticks are effectively in the trivial vacuum state.

\subsection{Mapping between new and old string net labels}

Since $\< \tilde{X}| = \< X | P_{\phi}$, any state in $\tilde{\mathcal{H}}$ can also be expressed as a superposition of string-net states in our original Hilbert space $ \mathcal{H}_{\phi}$.  This superposition contains states in which each edge label is replaced by an appropriate superposition of original string net labels, with the branching rules obeyed at every trivalent vertex, and arbitrary allowed labels on the sticks.  Notationally, we represent the resulting string-net configurations in the original label set by $X\in\tilde{X}$, where $X$ represents a labeling of edges in the original string-net basis, and $\tilde{X}$ represents the corresponding labels in the condensed basis.  
Explicitly, we may write
\begin{equation}
	\<\tilde{X}|= \frac{1}{p^{2 N_V}} \sum_{X\in\tilde{X}} C(X)\<X|
	\label{basisX}
\end{equation}
where 
$C(X)$ are numerical coefficients. 

The coefficients $C(X)$ are highly constrained.  
For any $X_1,X_2\in\tilde{X}$, $\< X_1|$ is related to $\< X_2|$ by the action of some composite string operator $W(\{ \phi_j, r_{ij} \})$:
\begin{align}
	&\< X_1|W(\{ \phi_j, r_{ij} \})=W(\{ \phi_j, r_{ij} \})_{X_2}^{X_1}\< X_2 | 
	\label{cx2cx2}
\end{align}
where $W(\{ \phi_j, r_{ij} \})_{X_2}^{X_1}$ is the relevant matrix element of $W(\{ \phi_j, r_{ij} \})$, and for a fixed configuration $\phi_j, r_{ij}$ the state $X_2$ is unique as the condensing anyons are abelian.  Since the composite string operator is unitary, we equivalently have 
$ W(\{ \phi_j, r_{ij} \})|X_2 \> = W(\{ \phi_j, r_{ij} \})_{X_2}^{X_1}|X_1 \>$.  
Therefore, the coefficients satisfy:
\begin{align}
C(X_2)&=\< \tilde{X} | X_2 \> = \< \tilde{X} |W(\{ \phi_j, r_{ij} \})| X_2 \>  \nonumber \\
 &= W(\{ \phi_j, r_{ij} \})_{X_2}^{X_1} \< \tilde{X} | X_1 \>  \nonumber \\
& =W(\{ \phi_j, r_{ij} \})_{X_2}^{X_1} C(X_1)  \ . 
	\label{cx2cx1}
\end{align}  
 where in the first line, we have used Eq. (\ref{Eq:Pmult}).  Eq. (\ref{cx2cx1}) allows us to determine the coefficients $C(X)$, up to an  overall coefficient $C(X_0)$ for each distinct reference configurations $\{X_0\}$.

\subsection{Vertex coefficients} \label{Sec:VLC}

Solutions $\{ C(X) \}$ to Eq. (\ref{cx2cx1}) can be expressed $C(X) = \prod_v C_v(X)$, where the product runs over all trivalent vertices in the extended string-net cofiguration $X$, and $C_v(X)$ is a coefficient that depends only on the three string labels surrounding the vertex $v$.  
This is because the action of any string operator can be broken up into a product of actions of simple string operators, with  each simple string operator acting at a single honeycomb vertex and the vertices associated with nearby sticks.  
Thus for each simple string operator, Eq. (\ref{cx2cx1}) can be reduced to a set of equations relating products of at most three of the vertex coefficients $C_v(X)$ to at most three of the vertex coefficients $C_v(X')$.   We will show that the resulting equations are self-consistent, and sufficient to fully determine the coefficients of any configuration $X \in \tilde{X}$ from that of a reference configuration $X_0$.  

To parametrize the vertex coefficients $C_v(X)$, we  define a set of \emph{root vertices}, which contain one representative vertex $(a,b; c)$ in the original string label set for each condensed vertex $(\tilde{a}, \tilde{b}; \tilde{c})$.  
Then any condensed string net state $\< \tilde{X} |$ can be obtained by projecting a reference string-net state $\< X_0|$ for which all vertices are root vertices.   Conversely, any two states that differ by at least one root vertex project to two distinct states in $\tilde{\mathcal{H}}$. 
We then define two types of vertex coefficients: $\{A^{ab}_c \}$  associated with the root vertices $(a,b; c)$, and $\{B^{a^jb^k}_{c^{j+k}} \}$, associated with the remaining vertices $(a^j, b^k; c^{j+k})$, where $a^j \equiv a \times s^j$.  The coefficients $\{ B^{a^jb^k}_{c^{j+k}} \}$ can be expressed in terms of $\{A^{ab}_c \}$ using Eq. (\ref{cx2cx1}).     On the other hand,  $\{A^{ab}_c \}$, which are associated with the vertex coefficients of our reference configuration, are not fully determined, and in some cases admit multiple, physically distinct solutions.   

We begin by defining the root vertices. Again, we have two cases to consider.
The first case is the $\phi=(0,m)$ condensed phases. In this case, we define the root vertices by
\begin{equation}
	\raisebox{-0.22in}{\includegraphics[height=0.5in]{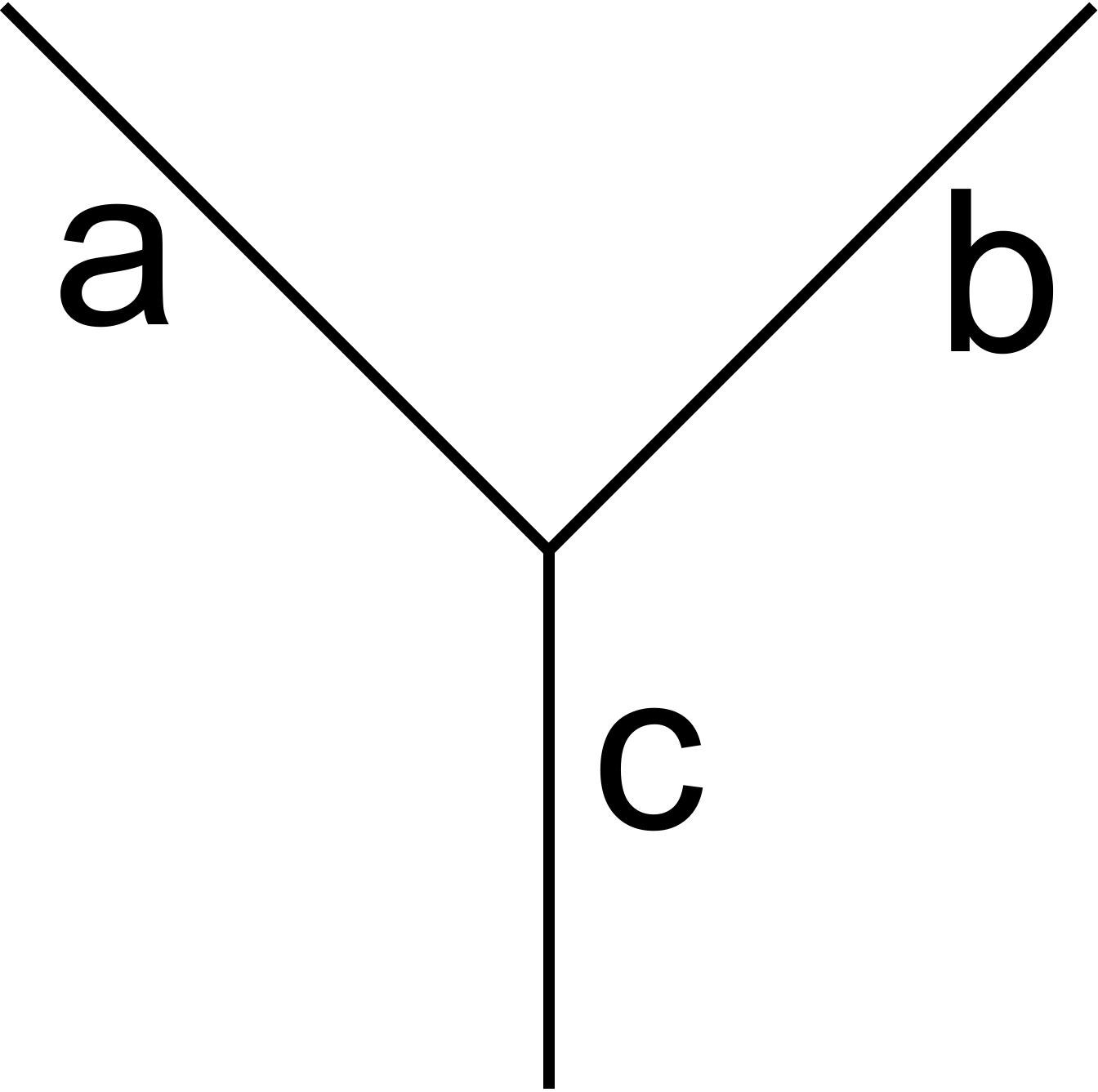}},\quad 
	\raisebox{-0.22in}{\includegraphics[height=0.5in]{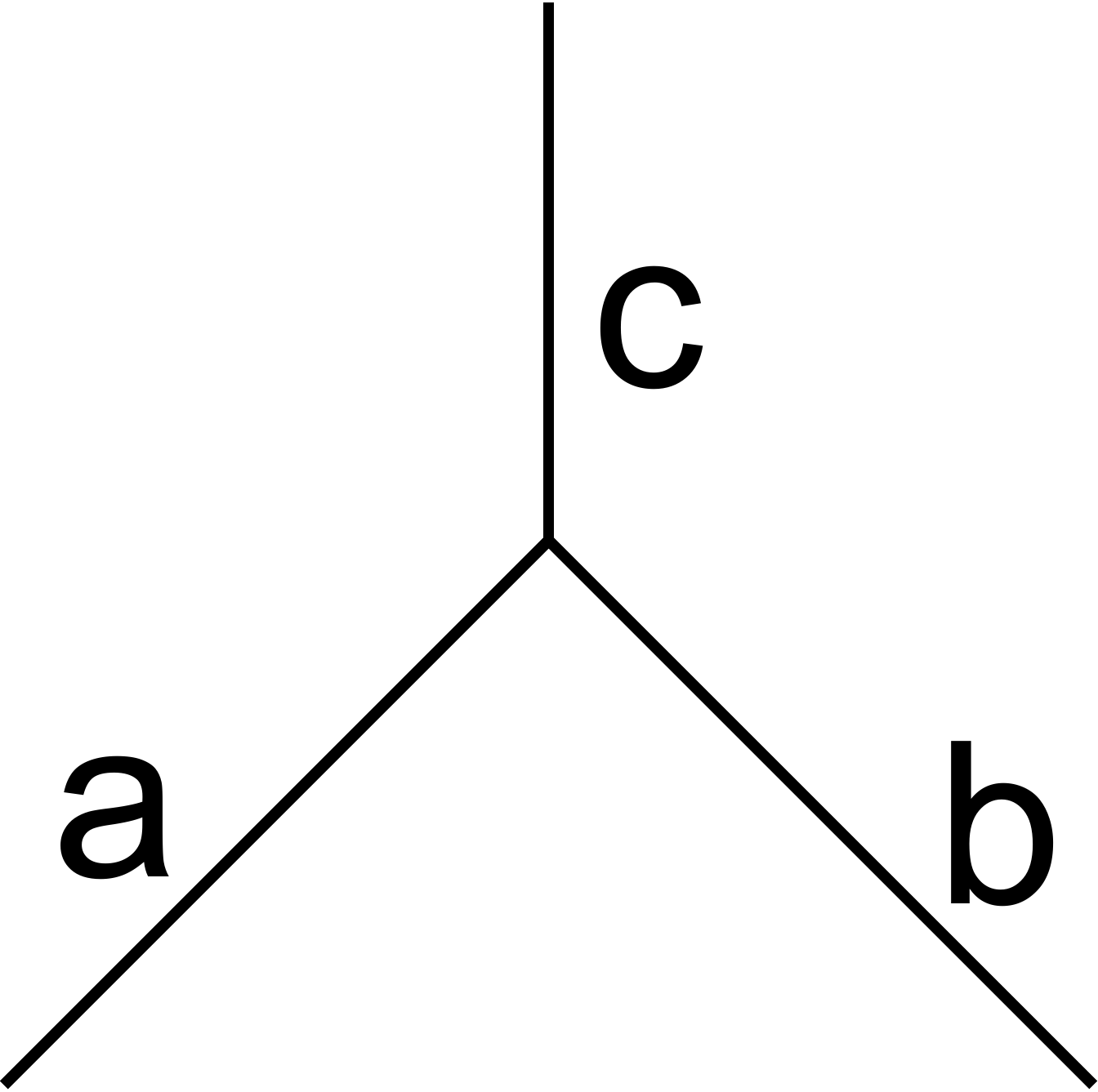}}
	\label{rv2a}
\end{equation}
where $a,b,c$ are the deconfined string types which satisfy (\ref{wphi1}), and $(a,b;c)$ satisfies the branching rules.  In this case all vertices are root vertices.

The second case is the $\phi=(s,m)$ condensed phase. In this case, for each new string label $\tilde{a}$, we choose a representative label $a \in \tilde{a}$.  We define two classes of root vertices.  First, we have the root vertices:
\begin{equation}
	\raisebox{-0.22in}{\includegraphics[height=0.5in]{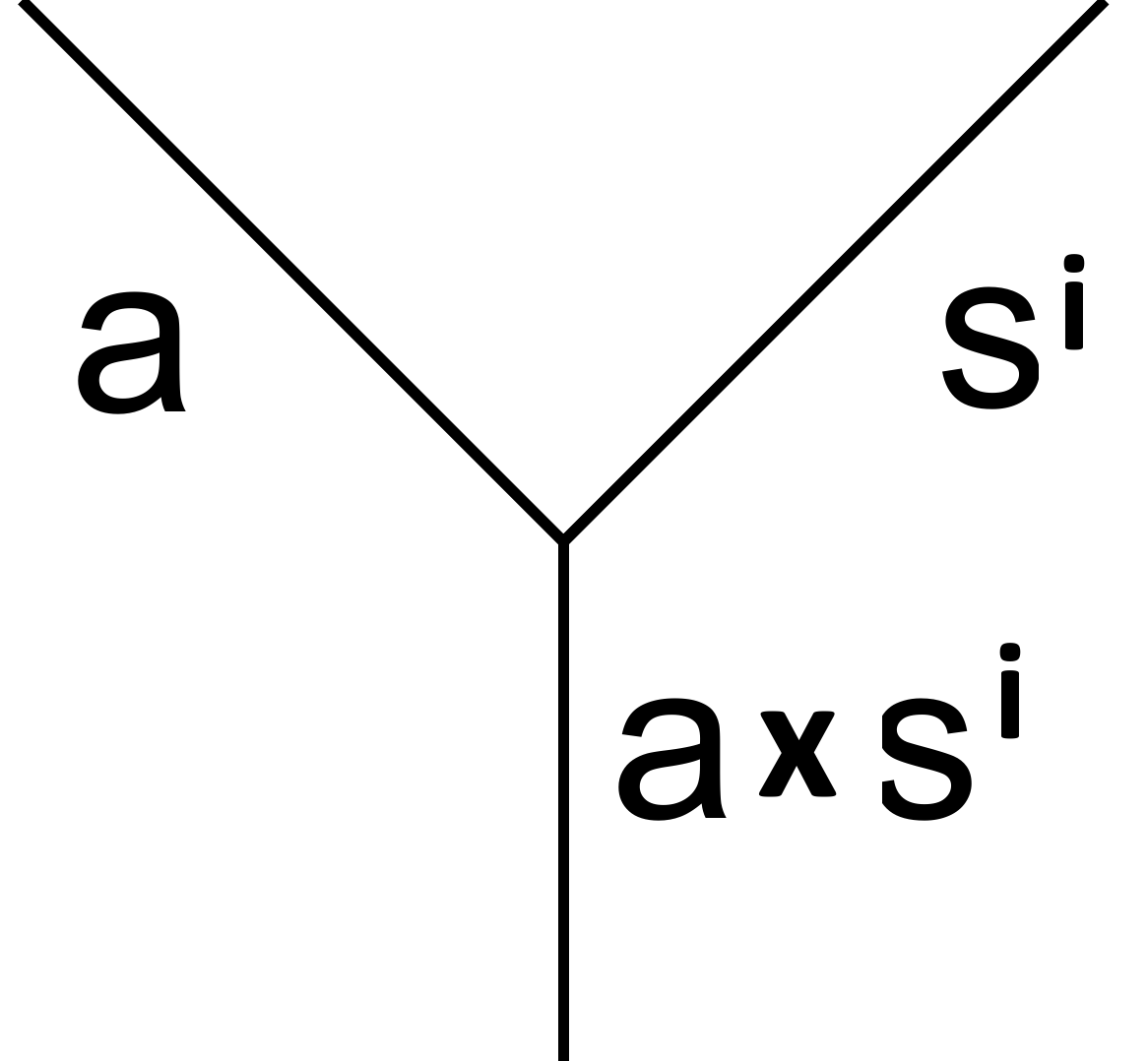}},\quad 
	\raisebox{-0.22in}{\includegraphics[height=0.5in]{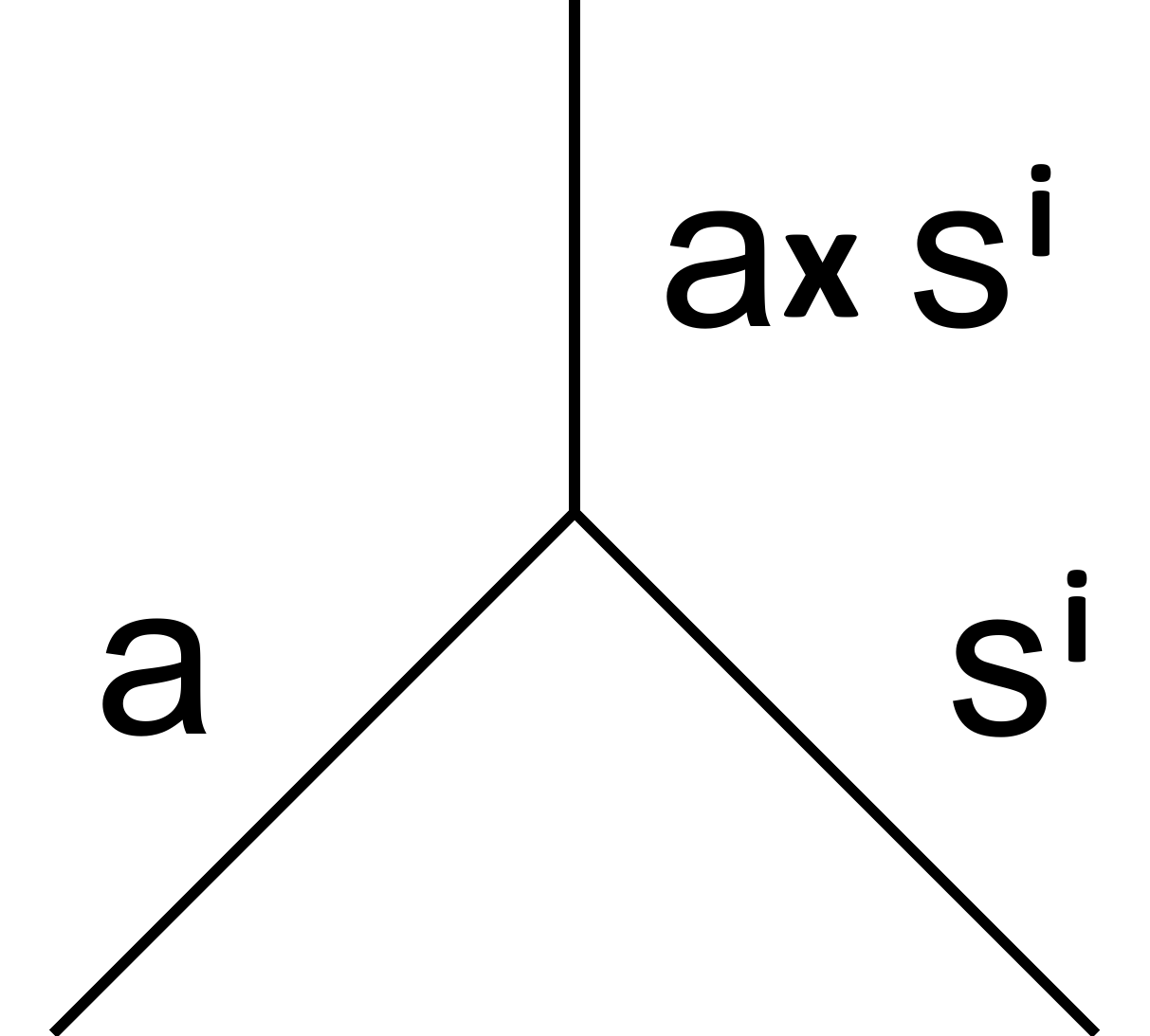}},\quad
	\label{rv2b0}
\end{equation}
where $s^i$ denotes the string type associated with the boson $\phi^{j q + i}$, and $0 \leq i < q$.  (Here, as above, we take $\phi = (s,m)$, with $\phi^p = 1$ and $s^q = 0$, where $q|p$.)  In this case, for reasons that will become apparent shortly, it is convenient to consider all different powers $i$ to be root vertices, in spite of the fact that all of these correspond to the same projected label $\tilde{0}$.  
Second, we have root vertices with two non-null string types in the two upper (lower) legs of upward (downward) vertices:
\begin{equation}
	\begin{split}
	\raisebox{-0.22in}{\includegraphics[height=0.5in]{abc-eps-converted-to.pdf}},\quad 
	\raisebox{-0.22in}{\includegraphics[height=0.5in]{abcin-eps-converted-to.pdf}} \text{ with }a,b \not\in \tilde{0} \ ,
	\end{split}
	\label{rv2b}
\end{equation}
where $(a,b;c)$ satisfies the original branching rules.

The vertex coefficients associated with root vertices for general $\phi=(s,m)$ are defined by:
\begin{equation}
	\begin{split}
	C\left(\raisebox{-0.22in}{\includegraphics[height=0.5in]{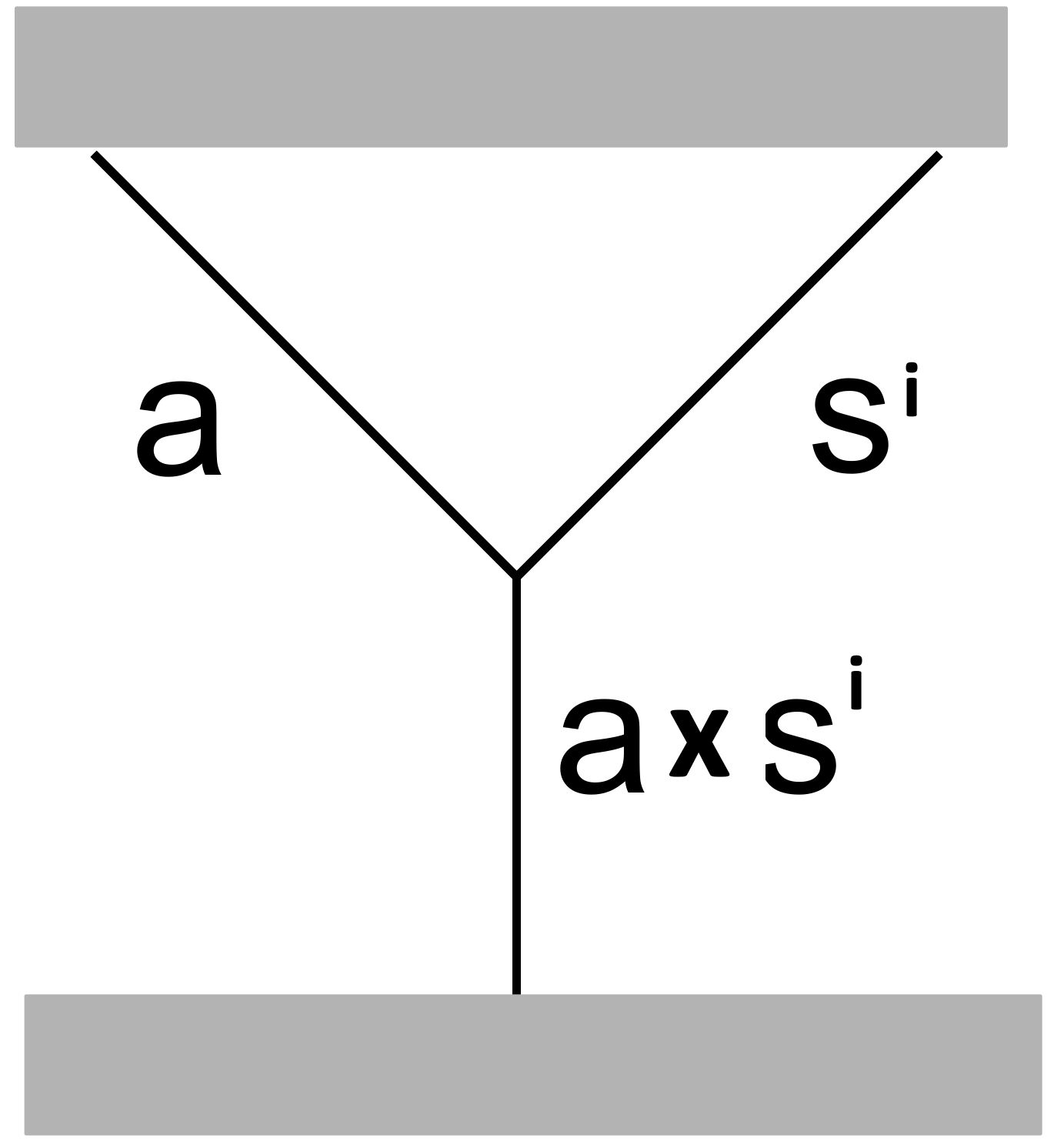}}\right)\sim A^{as^i}, 
	\quad 	
	C\left(\raisebox{-0.22in}{\includegraphics[height=0.5in]{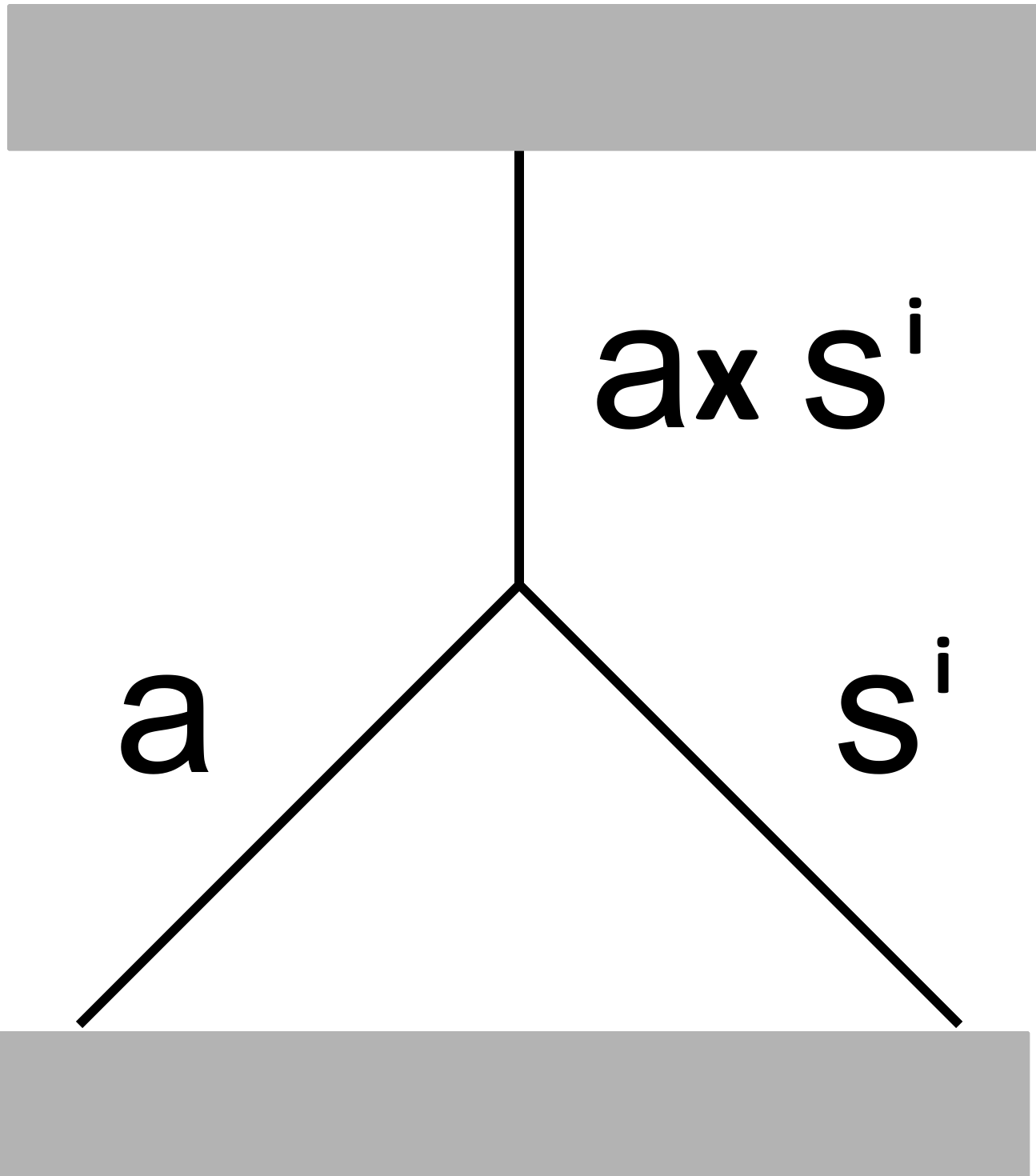}}\right)\sim \frac{1}{A^{as^i}}
	\end{split}
	\label{basic1}
\end{equation}
and 
\begin{equation}
	C\left(\raisebox{-0.22in}{\includegraphics[height=0.5in]{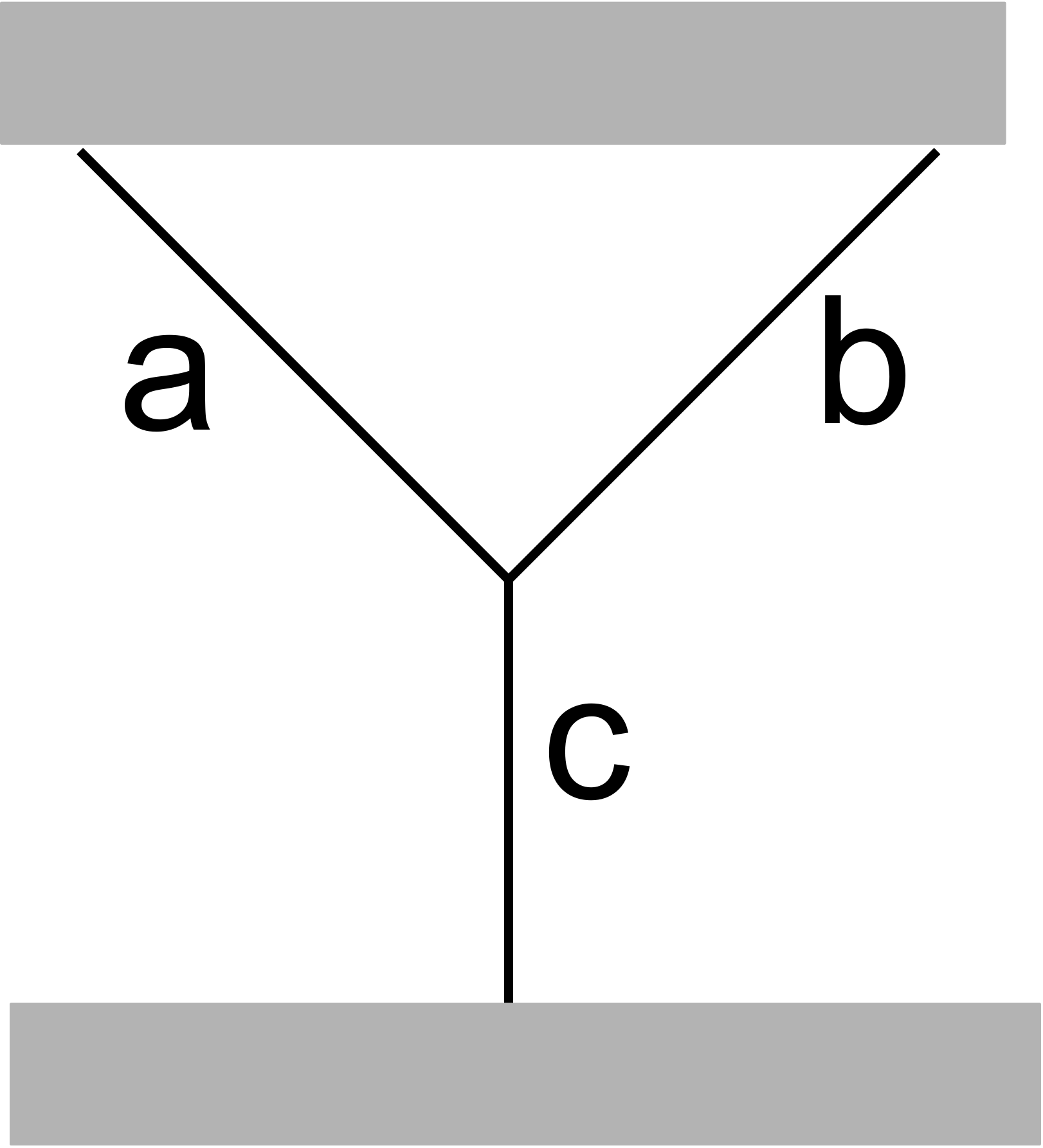}}\right)\sim A^{ab}_c, 
	C\left(\raisebox{-0.22in}{\includegraphics[height=0.5in]{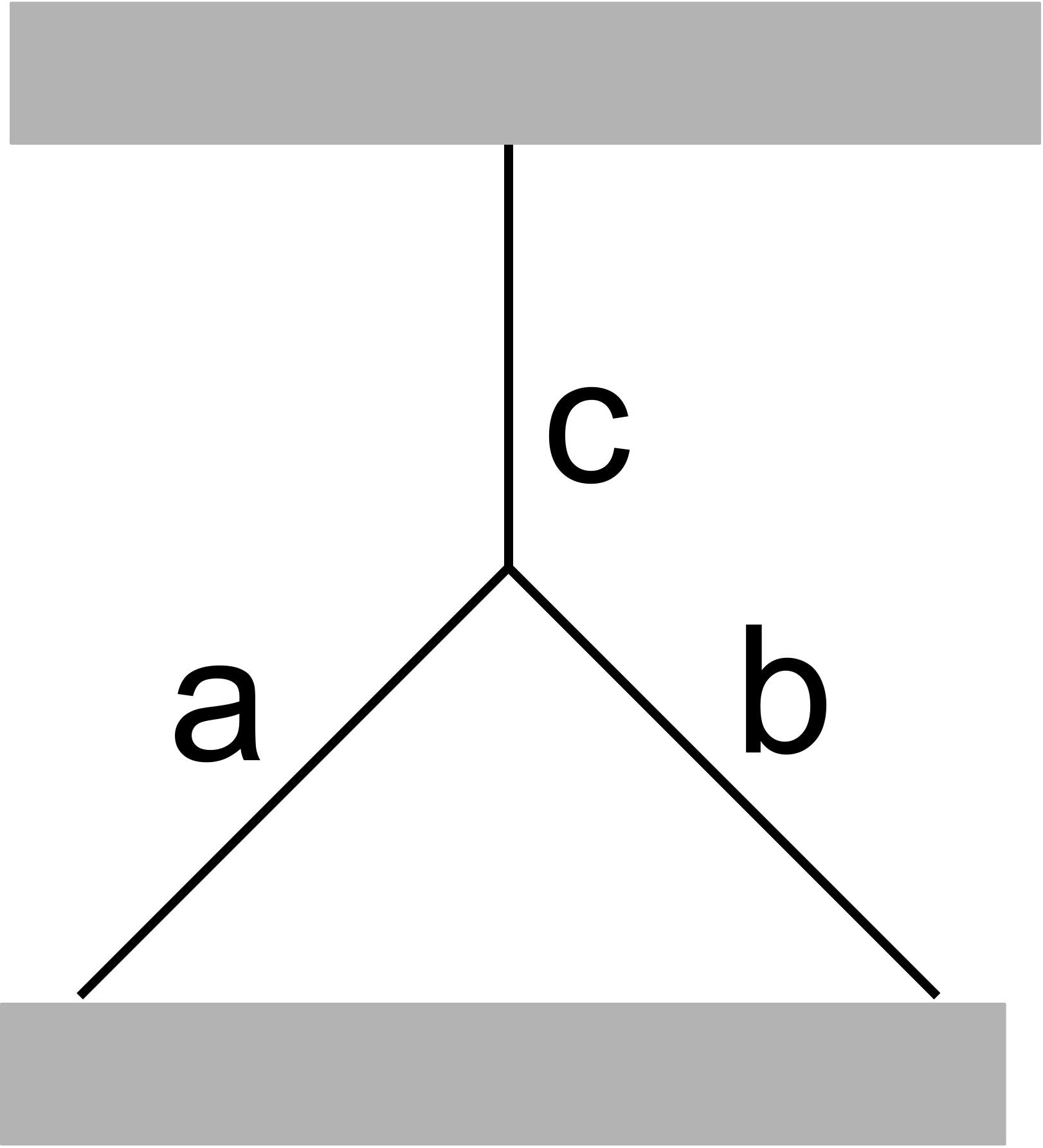}}\right)\sim \frac{1}{A^{ab}_c}
	\text{ with }a,b\not \in \tilde{0}.
	\label{basic2}
\end{equation}
for root vertices (\ref{rv2b0}) and (\ref{rv2b}) respectively.
Here $\sim$ means equality up to factors associated with other vertices in the grey area. The vertex coefficients $A^{ab}_c,A^{as^i}$ are complex numbers which satisfy that
\begin{equation}
	A^{ab}_c=A^{a s^i}=1 \text{ if }a \text{ or }b=0.
	\label{Eq:Achoice}
\end{equation}
In the absence of splitting, these are not constrained, and can be any complex number of unit modulus; in particular, we may choose them all to be $1$.

The $\phi=(0,m)$ condensed phases can be thought of special cases of the $\phi=(s,m)$ condensed phases where $s=0$ and thus $\tilde{0}=\{0\}$. In this case, the vertex coefficients (\ref{basic2}) associated with root vertices (\ref{rv2a}) correspond to a gauge choice for our string net model\cite{LinLevinBurnell}. 

When $s \neq 0$, to find the coefficients $C(X)$ in Eq. (\ref{basisX}), we define a set of vertex coefficients associated with non-root vertices via:
\begin{equation}
	\begin{split}
	C\left(\raisebox{-0.22in}{\includegraphics[height=0.5in]{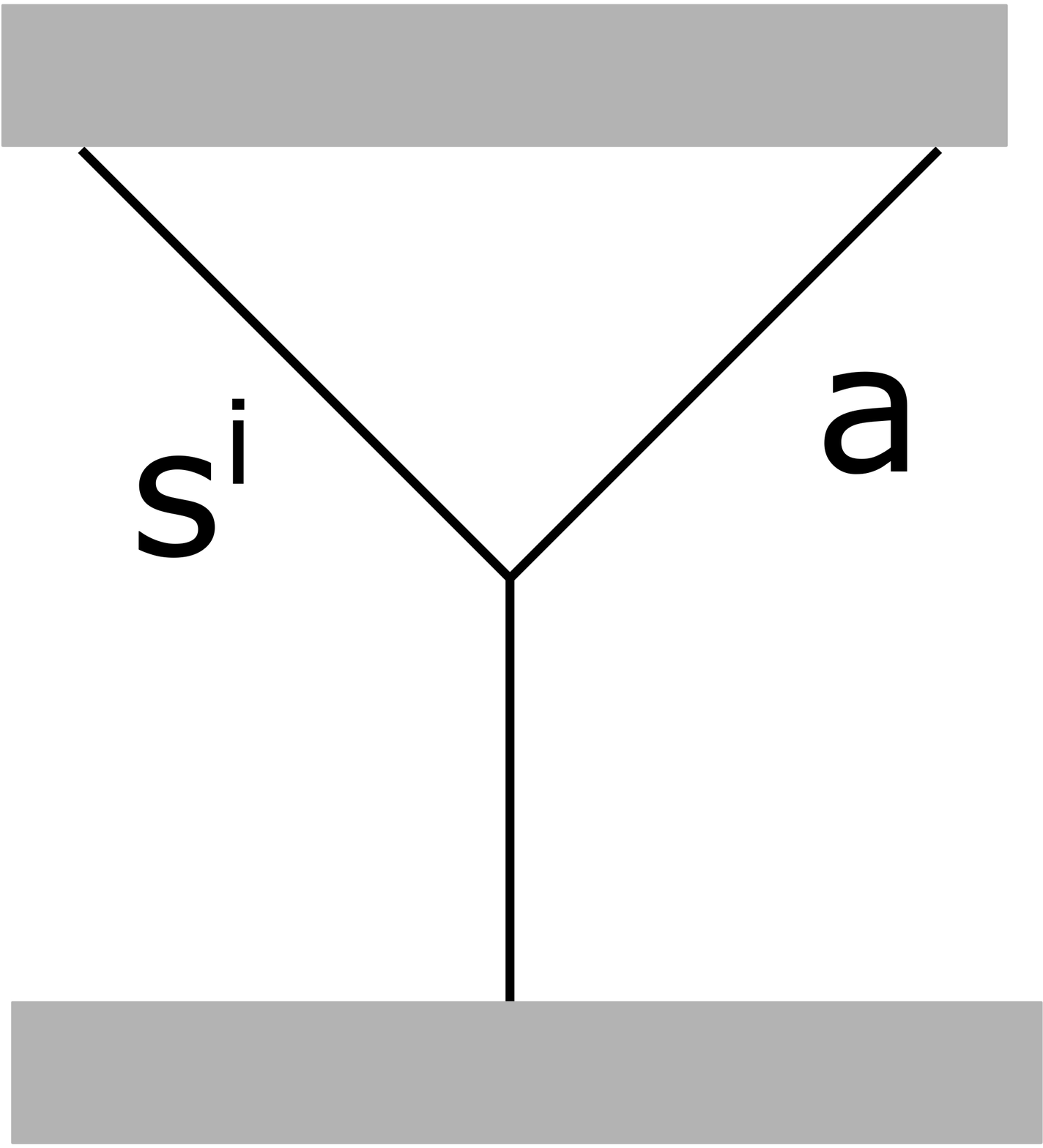}}\right)\sim B^{s^ia}, 
	\quad 
	C\left(\raisebox{-0.22in}{\includegraphics[height=0.5in]{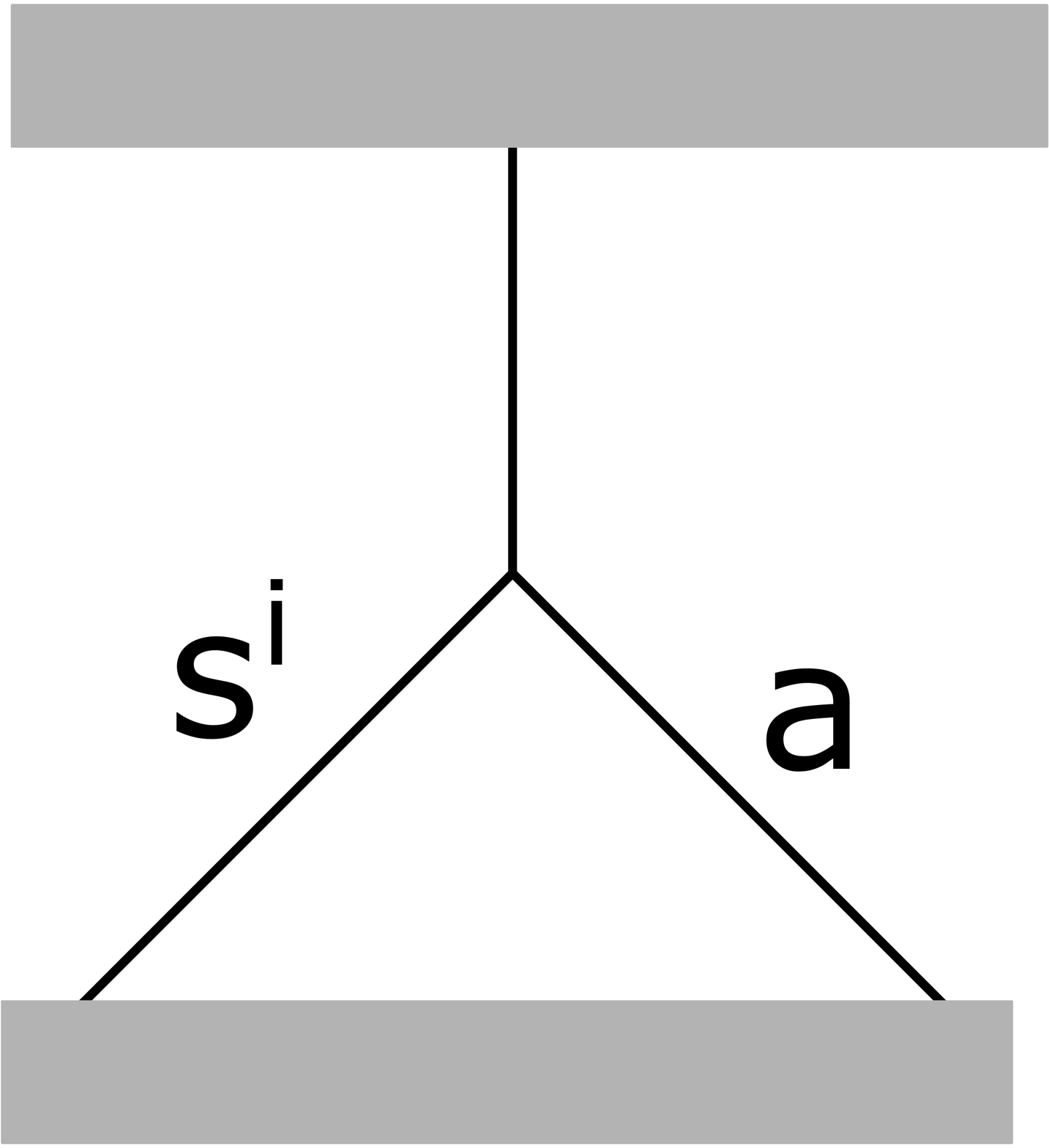}}\right)\sim \frac{1}{B^{s^ia}}%, 
		\end{split}
	\label{basic3}
\end{equation}
and 
\begin{equation}
	\begin{split}
	C\left(\raisebox{-0.22in}{\includegraphics[height=0.5in]{abc1-eps-converted-to.pdf}}\right)\sim B^{a^jb^k}_{c^{i+j}}, \quad
	C\left(\raisebox{-0.22in}{\includegraphics[height=0.5in]{abc1in-eps-converted-to.pdf}}\right)\sim \frac{1}{B^{a^jb^k}_{c^{i+j}}}.	
	\end{split}	
	\label{Bab}
\end{equation}
where $B^{a^jb^k}_{c^{i+j}}$ are complex numbers, and at least one of ($k,j)$ are non-zero, such that at least one of $a^j, b^k$ are not in our chosen set of reference labels.

The division into $A$-type and $B$-type vertex coefficients is useful since the latter are fully determined by the root vertex coefficients $A^{ab}_c,A^{a s^i}$ using Eq. (\ref{cx2cx1}).
The coefficients $A^{ab}_c$ and $A^{a s^i}$, on the other hand, are not fixed by Eq. (\ref{cx2cx1}) provided that $a \times s^r \neq a$ for any $r< q$.  
In this case, coefficients $C(X)$ in (\ref{basisX}) can be parametrized as 
\begin{equation}
	C(X) =C(X_0) \prod_{v\in X} B_v
	\label{}
\end{equation}
where the product runs over all vertices $v$ in $X$ and $B_v$ is the corresponding vertex coefficient.  

The coefficient $C(X_0)$ associated with the given reference configuration $X_0$ is determined by the root vertex coefficients via:
\begin{equation}
	C(X_0) = \prod_{v\in X_0} A_v
	\label{}
\end{equation}
where $v$ runs over all root vertices in $X_0$ and $A_v$ is the corresponding root vertex coefficient.   When $a \times s^r \neq a$ for any $r< q$ we will find that all choices of root vertex coefficients are equivalent, and the freedom to choose $C(X_0)$ amounts to a gauge choice.

If $a^r = a$ for some $r | q$, the parametrization of $C(X)$ is similar.  However, in this case Eq. (\ref{cx2cx1}) imposes additional constraints on the root vertex coefficients $A^{a s^k}$.  In this case we find that only $A^{a s^k}$ for $k \leq r$ are free parameters, and that there are $q/r$ distinct solutions for each of these coefficients.  These distinct solutions correspond to the fact\cite{SlingerlandBais} that after condensation the label $a$ {\it splits} into $q/r$ distinct labels; correspondingly we also obtain multiple vertex coefficients $A^{a b}_{c}$.
We now discuss each of these cases in turn.

\subsubsection{Case 1: $a^k \neq a$}

First, let us verify that a solution to Eq. (\ref{cx2cx1}) can be expressed as a product of vertex coefficients -- i.e. that any mapping between two configurations with the same sets of initial and final vertices has the same numerical coefficient.  If $a^k \neq a$ for any $a$ or $k$, then it suffices to consider sequences of simple string operators connecting the same inital and final vertex configurations.  The properties of basic string operators outlined in Eqs. (\ref{wpcommute}) - (\ref{w1w20}), as well as the consistency conditions (\ref{consistency}) and (\ref{weqs}), ensure that all combinations of basic string operators relating a given initial and final set of vertices will have the same numerical coefficient.  

Second, we use Eq. (\ref{cx2cx1}) to solve for the $B$-type vertex coefficients in terms of $\{ A^{a s^i}, A^{ab}_c \}$.  
First, consider vertices where both $a$ and $b$ legs are powers of the condensing label $s$.  In this case, in the gauge  (\ref{f1}), and using $w_{\phi}(s^j) = 1$, all non-vanishing string operator matrix elements are simply $+1$, and we have
\begin{equation}
\begin{split}
&A^{0 s^j} A^{0 s^{-j} }B^{s^{j}s^{-j}}= A^{00} = 1 \\
 &B^{s^i s^j} A^{0 s^i} A^{0 s^j} B^{s^{i+j} s^{-i-j} } = A^{00} = 1  .
 \end{split}
	\label{}
\end{equation}
It follows that, given the condition (\ref{Eq:Achoice}), 
\begin{equation} \label{Bss1}
B^{s^i s^j } = 1
\end{equation}
for any $i,j$.  

Next suppose $a \notin \tilde{0}$, with $b=s^i$, and $a^j \neq a$.   In this case, we have
\begin{equation}
\label{Aeq}
A^{a s^k} B^{a^k, s^{l-k} } B^{s^l,s^{-k}} = A^{a s^l}  (F^{as^k s^{l-k}}_{a^{l} a^k s^{l }} )^*
\end{equation}
where the coefficient is obtained by acting on the vertex $(a, s^l; a^l)$ with the product $W^1_{\phi^k} W^2_{\phi^{-k}} $.  
Given Eq. (\ref{Bss1}), this fixes $B^{a^k, s^{l-k}}$ in terms of $A^{a s^l}$ and $A^{a s^k}$.

If $a,b \neq s^k$, we have 
\begin{equation}
\label{Eq:VEq0}
A^{a s^j} A^{b s^k} B^{a^j, b^k}_{c^{j+k}} B^{c^{j+k} \bar{s}^{j+k} }= W_{jk} (abc) A^{ab}_c \ ,
\end{equation}
 The matrix element is given by acting with the product $W^2_{\phi^k} W^1_{\phi^j}$ on the vertex $(a,b; c)$ (see Fig. \ref{fig:stringoperator}):
\begin{align}
\label{Eq:Wabc}
W_{jk} (abc) =\bar{w}_{\phi^j}(b)  F^{a b s^j}_{c^j c b^j} (F^{a s^j b}_{c^j a^j b^j})^* F^{a^j b s^k}_{c^{j+k} c^j b^k} (F^{c s^j \bar{s}^j}_{c c^j 0} F^{c^j s^k \bar{s}^{j+k}}_{c c^{j+k} \bar{s}^j})^*
\end{align}
Given Eq. (\ref{Aeq}), this fixes $B^{a^j, b^k}_{c^{j+k}}$ in terms of $A^{ab}_c$ and $\{ A^{a s^j} \}$.

Finally, using string paths of the form
\begin{equation}
	\raisebox{-0.22in}{\includegraphics[height=1.in]{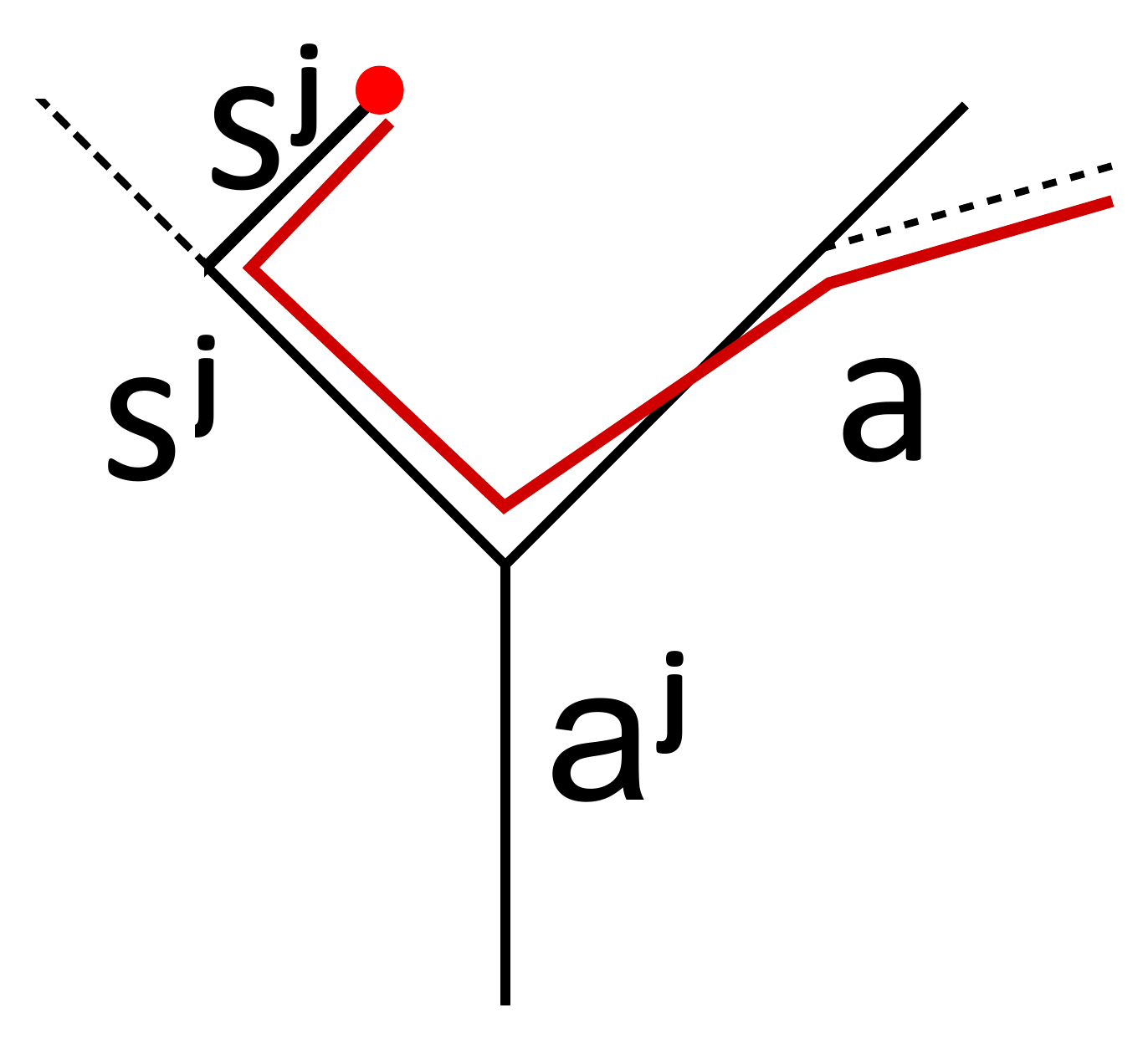}}
\end{equation}
we have
 \begin{subequations} \label{Eq:Cross}
	\begin{align}
	B^{s^j a^i} W_{\bar{j} j} (s^j, a^i, a^{i+j})&= B^{a^j s^i} 	\label{ex0a}\\
	B^{s^j a}  W_{\bar{j} j} (s^j, a, a^j) 
	&= A^{a s^i} 	\label{ex0b} ,
	\end{align}
\end{subequations}
which shows that both $B^{s^i a}$ and $B^{s^k a^j}$ can be expressed in terms of $A^{a s^j}$-type vertex coefficients.  
These relations have a particularly simple form: From Eq. (\ref{Eq:Wabc}), we can show that
\begin{equation}
	\begin{split}
		W_{\bar{j} j} (s^j, a, a^j)
		=& w_{\phi^j}(a)       
	\end{split}
	\label{Eq:CrossCoeff}
\end{equation}
where we have used  Eqs. (\ref{weq2},\ref{weq5}), as well as the identity
\begin{equation}
F^{a s^j \bar{s}^j}_{a a^j 0} F^{s^j \bar{s}^j s^j}_{s^j 0 0} ( F^{ a^j, \bar{s}^j, s^j}_{a^j a 0})^* = 1
\end{equation}
which follows from Eq. (\ref{consistency}).  

This leaves us with the vertex coefficients $A^{ab}_c$ and $A^{a s^j}$ (where $a, b \neq s^k$).  The former are clearly free parameters, since by definition there is no $\phi$ string operator that takes such a root vertex to another root vertex.  Indeed, they represent a choice of gauge for the $F$ matrices describing the condensed phase (see Eq. (\ref{gt})), and can be set to $1$.  
To see that $A^{a s^j}$ are also free parameters, we note that there is a residual gauge freedom when solving (\ref{cx2cx1}).   
Specifically, given a set of vertex coefficients that satisfy Eq. (\ref{cx2cx1}),  
we can construct an infinite class of other solutions $\tilde{A}^{a b}_c, \tilde{B}^{a^j b^k}_{c^{j+k}}$ via:
\begin{equation}
	\tilde{A}^{a b}_c=A^{a b}_c \cdot \frac{g(a)g(b)}{g(c)} \ , \ \ \tilde{B}^{a^j b^k}_{c^{j+k}}=B^{a^j b^k}_{c^{j+k}} \cdot \frac{g(a^j)g(b^k)}{g(c^{j+k})} \ .
	\label{ggauge}
\end{equation}
Here $a,b$ and $c$ are any string labels (including $s^j$), and
 $g(a)$ is any function with
\begin{equation*}
	g(s^i)g(s^j)=g(s^{i+j}),\quad g(0)=1.
	\label{}
\end{equation*}
It is straightforward to verify that the transformations (\ref{ggauge}) do not alter the equalities dictated by the action of any of the basic string operators at a vertex.

When $a^i\neq a^j$ for $i\neq j$ (mod $q$), the gauge transformation (\ref{ggauge}) fully fixes the coefficients $A^{a s^i}$: we can always choose the ratio $g(a)/g(a^j)$ to set
\begin{equation}
	A^{a s^j} = 1 \ .
	\label{asa1}
\end{equation}
Accordingly, we have
\begin{equation}
	B^{s^j a} = w_{\phi^j}(a) \ , \ \ \ B^{a^{j}  s^{i-j}}=  (F^{as^js^{i-j}}_{a^{i}a^j s^{i}})^*
	\label{}
\end{equation}
by Eqs. (\ref{Aeq})  and  (\ref{ex0b}).

\subsubsection{Case 2: $a^r = a$} \label{Sec:Splitting}

If $a^r=a$ for some $r|q$, 
there are additional constraints relating the coefficients  $A^{a, s^{nr}}$, and we cannot set these to $1$ using transformations of the form (\ref{ggauge}).

We begin by considering configurations involving only vertices of the form $(a, s^{jr}; a)$, $j = 1,... q/r$, and their cyclic permutations.  
We first show that for such configurations, there exists a solution to Eq. (\ref{cx2cx1}) that can be expressed as a product of vertex coefficients.  To show this, we must establish that the coefficients relating configurations with the same sets of initial and final vertices do not depend on the relative positions of these vertices -- an issue that did not crop up in case $1$.  For example, using a $W^1$-type simple string acting on a vertex $(a, s^j, a)$ with sticks on the two $a$ legs carrying labels $s^i$ and $s^k$, we can derive:
\begin{align} \label{Eq:Afind}
A^{a s^{(i+l) r}} A^{a s^{(k-l) r}}  =&A^{a s^{i r}} A^{a s^{k r}}  F^{ a s^{j r} s^{ l r}}_{a a s^{(j+l)r}} (F^{ a s^{ l r} s^{j r} }_{a a s^{(j+l)r}} )^* \nonumber \\ 
  & \times 
    F^{a s^{i r} s^{l r}}_{a a s^{(i+l) r}} F^{ s^{lr} \bar{s}^{lr} s^{k r}}_{s^{kr} 0 s^{(k-l)r}} ( F^{ a s^{lr} s^{(k-l)r }}_{a a s^{kr} } )^*
\end{align}
where we have removed a common factor of $A^{a s^{j r}} $ (which is non-zero) from both sides, and used $w_{\phi^j}(s^k) =1$ for all $j,k$.     
This can be true only if the coefficient does not depend on $j$.
Similar consistency requirements arise from acting with $W^2_{\phi^{lr}}$ on a vertex $(s^{jr}, a; a)$ and with $W^2_{\bar{\phi}^{lr}}W^1_{\phi^{lr}}$ on a vertex $(a, \bar{a}; s^{jr})$.  
 In Appendix \ref{ConsistencyApp}, using the conditions (\ref{3a}, \ref{weq1}), we show that in all three cases, in the gauge (\ref{f1}), the coefficients are indeed independent of $j$.   (A similar result holds for vertices of the form $(a, s^{i + j r}; a^i)$ with $i<r$, for which the coefficient is also indepedent of $j$).  
Thus we see that, for configurations with only vertices involving $a^j$, $\bar{a}^k$ ($j,k < r)$, and powers of $s^{ i}$, the simple string operators at each vertex yield a consistent set of equations for the $A^{a s^j}$.

Having established that a consistent solution exists, let us solve for the coefficients $A^{a s^{nr}}$.  (As above, the coefficients $A^{a s^i}$ for $i < r$ can be consistently set to $1$ by a gauge transformation).   Taking $k=l=1$ in Eq. (\ref{Eq:Afind}),
 we find:
\begin{equation} \label{Eq:Afind3}
 A^{a s^{(i+1) r}}  = A^{a s^{i  r}} A^{a s^{ r}}  F^{a s^{i r} s^{ r}}_{a a s^{(i+1) r}}   F^{ s^{r} \bar{s}^{r} s^{ r}}_{s^{kr} 00} F^{a s^{j r} s^{r}}_{ a a s^{(j+1)r} }  (F^{ a s^{r} s^{jr} }_{a a s^{(j+1)r}} )^* 
\end{equation}
As shown in Eq. (\ref{Eq:EFprod}), in fact 
\begin{equation} \label{Eq:Afind2}
F^{a s^{j r} s^{lr}}_{ a a s^{(j+l)r} } = F^{ a s^{lr} s^{jr} }_{a a s^{(j+l)r}} 
\end{equation} 
It follows that in our gauge of choice,
\begin{equation} \label{Eq:Acond1s}
A^{a s^{n  r}}   = (A^{a s^{ r}} )^n \prod_{k=1}^{n-1} ( F^{a s^{k r} s^{ r}}_{a a s^{(k+1) r}} )
\end{equation}
Thus of the $q/r$ vertex coefficients $A^{a s^{n r}}$, we can freely choose only one, which we take to be $A^{a, s^{r}}$.  

Moreover, the coefficient $A^{a s^{ r}} $ is not unconstrained: taking $n = q/r$ in Eq. (\ref{Eq:Acond1s}), and noting that $s^{(q/r) r} = 0$, we see that 
\begin{equation} \label{Eq:Acond2s}
(A^{a s^{ r}} )^{q/r}  = \prod_{k=1}^{q/r-1} ( F^{a s^{k r} s^{ r}}_{a a s^{(k+1) r}} )^*
\end{equation}
Thus, we see that $A^{a s^{ r}}$ must be a $q/r^{th}$ root of the product on the right-hand side, and we have exactly $q/r$ possible choices for this coefficient, which we label $(A^{a s^{ r}})_i$, $i = 1, ... q/r$.  
We note that the product on the right-hand side (and hence also $(A^{a s^{ r}})_i$) has modulus $1$, since by unitarity $( F^{a s^{k r} s^{ r}}_{a a s^{(k+1) r}} )^* F^{a s^{k r} s^{ r}}_{a a s^{(k+1) r}} =1$.  
.

Physically, the fact that we obtain multiple, physically inequivalent choices of $A^{a s^{n r}}$ implies that in the condensed phase, the string label $a$ 
``splits" into $q/r$ distinct label types, which we denote
\begin{equation}
	\tilde{a}_\mu =(a, \mu)  \ , \ \ \mu=1,\dots,q/r
	\label{newa1}
\end{equation}
 Here $(a, \mu) $ indicates that any vertex associated with the label $\tilde{a}_\mu$ is assigned a vertex coefficient consistent with the choice $(A^{a s^r} )_\mu= [ \prod_{k=1}^{q/r-1} ( F^{a s^{k r} s^{ r}}_{a a s^{(k+1) r}} )^*]^{ \mu r /q}$.

Armed with this knowledge of splitting, we may return to scrutinize other types of vertices.   Vertex coefficients for vertices $(a,b;c)$ where none of the three string labels split can be solved for as above; this includes all vertices of the form $(b, s^{j}; b^{j})$ for $0 \leq j < q$ where $b \times s^j \neq b$ for any $j < q$.   Thus consider a vertex of the form $(a,b; c)$ (or one of its cyclic permutations), where $a^r = a$, but $b$ and $c$ do not split.  Attaching a stick carrying the label $s^{rl}$ to the $a$ edge, the analog of Eq. (\ref{Eq:VEq0}) is:
\begin{align}
\label{Eq:ABSplit1}
& (A^{a s^{(j+l) r}})_\mu A^{b s^k} B^{a, b^k}_{c^{j r+k}} B^{c^{jr+k} \bar{s}^{jr+k} }= \nonumber \\
 &(A^{a s^{l r}})_\mu F^{a s^{l r} s^{jr}}_{a a s^{(l+j)r}} W_{jk} (abc) A^{ab}_c  \ ,
\end{align}
 This set of equations allows for a consistent definition of $B^{a, b^k}_{c^{jr +k}}$ in terms of $A^{ab}_c$ only if the $l$-dependence of the two sides cancels.  Indeed, (see Eq. \ref{E21}) 
\begin{equation} \label{Afracs}
\frac{(A^{a s^{(j+l) r}})_\mu }{(A^{a s^{l r}})_\mu} (F^{a s^{l r} s^{jr}}_{a a s^{(l+j)r}})^* = (A^{a s^{jr}})_\mu
\end{equation}  
so the $l$ dependence is indeed trivial, and a consistent definition is possible.  Note that the coefficient $B^{a, b^k}_{c^{jr +k}}$ on the left-hand side of Eq. (\ref{Eq:ABSplit1})
will depend on the choice of $\mu$; correspondingly, we define the $q/r$ vertex coefficients $(B^{a b^k}_{c^{jr+k}})_\mu$.  
 Similar considerations apply for the cyclic permutations $(\bar{c}, a; \bar{b})$, $(b, \bar{c}; \bar{a})$.

Finally, consider a vertex of the form $(a,b; c)$ where at least two of the labels split.   For example, suppose that $a^r = a, c^r = c$, and consider applying a $W^1_{\phi^{lr}}$ string operator to the vertex $(a,b; c)$, with sticks on the $a$ and $c$ edges initially labeled by $s^{ir}$ and $s^{kr}$ respectively.  This gives the relation: 
\begin{align} \label{Eq:Ac1}
&A^{a s^{(i+l) r}} A^{c s^{(k-l) r}} A^{a b}_c  \nonumber \\ 
&=A^{a b}_c A^{a s^{i r}} A^{c s^{k r}}   \bar{w}_{\phi^{lr}}(b) F^{ a b s^{ l r}}_{c c b^{lr}} (F^{ a s^{ l r} b}_{c a b^{lr}} )^* \nonumber \\ 
  & \times F^{a s^{i r}  s^{l r}}_{a a s^{(i+l) r}} ( F^{ c s^{lr} s^{(k-l)r} }_{c c s^{k r }} )^* F^{s^{lr} \bar{s}^{lr} s^{ k r}}_{ s^{ k r} 0  s^{ (k-l) r}}
\end{align}
Note that in this case, $A^{ab}_c$ appears on both sides of the equation.  If the coefficient on the right-hand side does not depend on $b$, then we can simply cancel these factors and we recover an equation that is satisfied by solutions to (\ref{Eq:Acond1s}) (see Eq. (\ref{Afracs})).  However, in general the coefficient is invariant only under replacing $b \rightarrow b^{jr}$ for some integer $j$, and may be different for distinct choices of $b$ (see Appendix (\ref{ConsistencyApp})).  The resolution to this is that we must replace Eq (\ref{Eq:Ac1}) with the equation: 
\begin{align} \label{Eq:Ac2}
\frac{(A^{a s^{l r}})_\mu }{(A^{c s^{l r}})_\nu} (A^{a b}_c)^\mu_\nu  =&(A^{a b}_c)^\mu_\nu  M_l(a,b,c) 
\end{align}
where we have used Eqs. (\ref{Eq:Afind2}) and (\ref{Afracs}) to simplify the factors associated with vertices $A^{a s^{ir}}, A^{c s^{jr}}$, and 
\begin{equation}
M_l(a,b,c)  =  \bar{w}_{\phi^{lr}}(b) F^{ a b s^{ l r}}_{c c b^{lr}} ( F^{ a s^{ l r} b}_{c a b^{lr}} )^*
\end{equation}
 encodes the dependence on the label $b$.  
Thus either $(A^{a b}_c)^\mu_\nu = 0$, or 
\begin{equation} \label{Eq:munuconsist}
 (A^{a s^{ l r}})_\nu = (A^{c s^{lr}})_\mu M_l(a,b,c) . 
\end{equation}
 In general, this gives us  a condition that fixes the values of $(\mu,\nu)$ for which $(A^{a b}_c)^\mu_\nu \neq 0$, and hence specifies the fusion rules of the new, split anyon labels.  
 Note that the conditions for $(A^{a b}_c)^\mu_\nu $, $(A^{\bar{c} a}_{\bar{b}})^\mu_\nu $, and $(A^{b \overline{c} }_{\overline{a}})^\mu_\nu $ to be non-vanishing involve different coefficients in general.     

In Appendix \ref{ConsistencyApp}, we show that when $(a,b;c)$ is allowed by the branching rules, there is necessarily at least one choice of $(\mu,\nu)$ such that $(A^{a b}_c)^\mu_\nu  \neq 0$, and hence at least one choice of $(\mu,\nu)$ for which $(\tilde{a}_\mu, \tilde{b}; \tilde{c}_\nu)$ is allowed by the new branching rules.   (We also show that the same is true for the cyclic permutations $(A^{\bar{c} a}_{\bar{b}})^\mu_\nu $, and $(A^{b \overline{c} }_{\overline{a}})^\mu_\nu $  of this vertex).  Indeed, provided that $\tilde{b}$ does not split, generically there are $q/r$ such solutions.  
This allows us to partially characterize the fusion rules of the new theory.  For example, suppose that only vertices of the form $(\bar{ c}, a; \bar{b})$ are allowed by the branching rules, where $b \times s^v \neq b$ for any $0 < v < q$.  
In the condensed Hilbert space, we have
\begin{equation}
\left (\sum_{\nu =1}^{q/r} \tilde{\bar{c}}_\nu  \right )  \times \left (\sum_{\mu =1}^{q/r} \tilde{a}_\mu \right ) =  \frac{q}{r} \sum_{ \tilde{\bar{b}}}  \tilde{\bar{b}} 
\end{equation} 
where the sum on the right-hand side runs over all distinct choices of $\tilde{\bar{b}}$ that are compatible with the original branching rules.  
Since $a \times s^r = a, c \times s^r = c$, we have $s^r \times( \bar{ c}\times a) =  \bar{ c}\times a$; hence if  $(\bar{ c}, a; \bar{b})$ is allowed by the fusion rules, then so is  $(\bar{ c}, a; \bar{b}^{j r})$ for any $j$.   Consequently, provided that $(\bar{ c}, a; \bar{b}^{k})$ is not allowed by the branching rules of the original theory for any $0<k < r$, in the condensed phase there are $q/r$ copies of $\tilde{\bar{b}}$ in the fusion product $ \tilde{\bar{c}} \times \tilde{a} $. 
This corresponds exactly to the number of distinct choices of $(\mu, \nu)$ that solve Eq. (\ref{Eq:munuconsist}) -- i.e. the number of choices of $(\mu, \nu)$ for which $(\tilde{\bar{c}} , \tilde{a}_\mu; \tilde{\bar{b}})$ is allowed by the new branching rules.  
In this case, each copy of $\tilde{\bar{b}}$ can be associated with a distinct solution, such that typically the Hilbert space at the vertex $(\tilde{\bar{c}} , \tilde{a}_\mu; \tilde{\bar{b}})$  is one dimensional (i.e. the new theory does not have fusion multiplicity).   In particular, there is no fusion multiplicity associated with the vacuum $\tilde{0}$, since the cyclic property of the fusion rules ensures that only vertices of the form $(a, \overline{a}; s^{jr})$ are allowed.

The possible fusion rules for other types of vertices, such as vertices $(\tilde{a}_\mu,  \tilde{b}_\lambda; \tilde{c}_\nu )$ where all three labels split, are discussed in Appendix \ref{ConsistencyApp}.

In summary, the new Hilbert space $\tilde{\mathcal{H}}$  consists of string-net states with  both unsplit string types of the form (\ref{new0}), (\ref{newa}), whose branching rules are fixed by those of the original labels, and \emph{split} string types given by (\ref{newa1}), whose branching rules are fixed by a combination of those of the original theory, and the solutions to Eq. (\ref{Eq:munuconsist}).

\section{String net model of the condensed phase} \label{Sec:newsnmodel}

We now show that the ground state $|\Psi \rangle$ of our extended string-net model as $J \rightarrow \infty$ can be expressed as an ordinary string net ground state using the new label set $\{ \tilde{a}_i, \tilde{b}, \tilde{c}, ... \}$.  
In particular, we show how to use the vertex coefficients $\{ A^{ab}_c , B^{a^j, b^k}_{c^{j+k}} \}$ 
described in the previous section to obtain the fusion data describing the string net in the condensed phase.  If there is no splitting, we find that the vertex coefficients, together with the fusion data of the original category, fully fix the fusion data for the condensed string net.  With splitting, these do not fully fix the fusion data; the remaining freedom can be eliminated by imposing the consistency conditions (\ref{consistency}).   

\subsection{The topological data for the condensed phases}

Deep in the condensed phase, the basis states in $\tilde{\mathcal{H}}$ allow us to express the condensed ground state $|\Psi\> = P_{\phi} |\Phi \>$ as a new string-net condensed state with amplitudes
\begin{align}
\Psi \left (\tilde{X} \right )&\equiv  \<\tilde{X}|\Psi\> 
	\nonumber \\
	&= \frac{1 }{p^{2 N_V}}   \sum_{X\in\tilde{X}} C_{\tilde{X}} (X)  \< X  |\Psi\> \nonumber \\
	&\equiv\frac{1 }{p^{2 N_V}}  \sum_{X\in\tilde{X}}   C_{\tilde{X}}(X) \Psi(X).
	\label{psiamp}
\end{align}
Here 
the sum over $X\in\tilde{X}$ is over all configurations of uncondensed string labels compatible with the configuration $\tilde{X}$.  Note that when one or more labels split there are multiple distinct solutions for the vertex coefficients, associated with the multiple distinct split string labels; in this case the coefficients $C(X)$ depend not only on $X$ but on the choice of which label in each set $\{ \tilde{a}_{\mu} \}$ is in the configuration $\tilde{X}$; to indicate this dependence, we have added a subscript, denoting the coefficient $C(X)$  as $C_{\tilde{X}}(X)$. 

\subsubsection{Topological data in theories without splitting}

We first describe how to use Eq. (\ref{psiamp}) to obtain the topological data associated with the condensed string net in theories where none of the original labels split.  
We begin by simplifying Eq. (\ref{psiamp}), using the relation:
\begin{equation}
	C_{\tilde{X}}(X_1)\Psi(X_1)= C_{\tilde{X}}(X_2)\Psi(X_2)	\text{ for any }X_1,X_1\in \tilde{X}.
	\label{lemma}
\end{equation}
To see that these are equal, observe that on the one hand, we have
\begin{equation}
	C_{\tilde{X}}(X_2)=C_{\tilde{X}}(X_1)W(P)^{X_1}_{X_2}
	\label{eq1}
\end{equation}
for any $X_1,X_2 \in \tilde{X}$ by (\ref{cx2cx2},\ref{cx2cx1}).
On the other hand, the new ground state $|\Psi\>$ satisfies
\begin{equation}
	\Psi(X_2) = \<X_2|\Psi\>=\frac{\<X_1|W(P)|\Psi\>}{W(P)^{X_1}_{X_2}}=\frac{\Psi(X_1)}{W(P)^{X_1}_{X_2}}.
	\label{eq2}
\end{equation}
Here we use (\ref{cx2cx2}) in the second equality and (\ref{psiD1}) in the third equality.
Putting (\ref{eq1},\ref{eq2}) together, we establish (\ref{lemma}).

By using (\ref{lemma}), we can rewrite (\ref{psiamp}) as
\begin{equation}
	\Psi\left (\tilde{X} \right )=  C(X_0)\Psi(X_0)
	\label{psiamp1}
\end{equation}
were $X_0$ denotes a reference configuration of our choice from the set $X \in \tilde{X}$; in the following it will be convenient to choose $X_0$ to have only trivial labels on all sticks. 
Observe that in the absence of splitting,  each vertex in $\tilde{X}$ corresponds to $q^2$ configurations $X$, obtained by acting with $W^1_{\phi_j} W^2_{\phi^k}$ for $0 \leq j,k < q$ at each vertex.   Each such configuration appears $p/q$ times when we act with $P_\phi$ on $X_0$. 
Thus summing over $X \in \tilde{X}$ and expressing all terms in terms of $\Psi(X_0)$ gives a factor of $p^2$ for each vertex, which exactly cancels the normalization pre-factor.

We can use the amplitudes of this new ground state to define the new F-symbols $F^{\tilde{a}\tilde{b}\tilde{c}}_{\tilde{d}\tilde{e}\tilde{f}}$ and quantum dimensions $d_{\tilde{a}}$ by
\begin{subequations}
	\begin{gather}
	\Psi \left( \raisebox{-0.22in}{\includegraphics[height=0.5in]{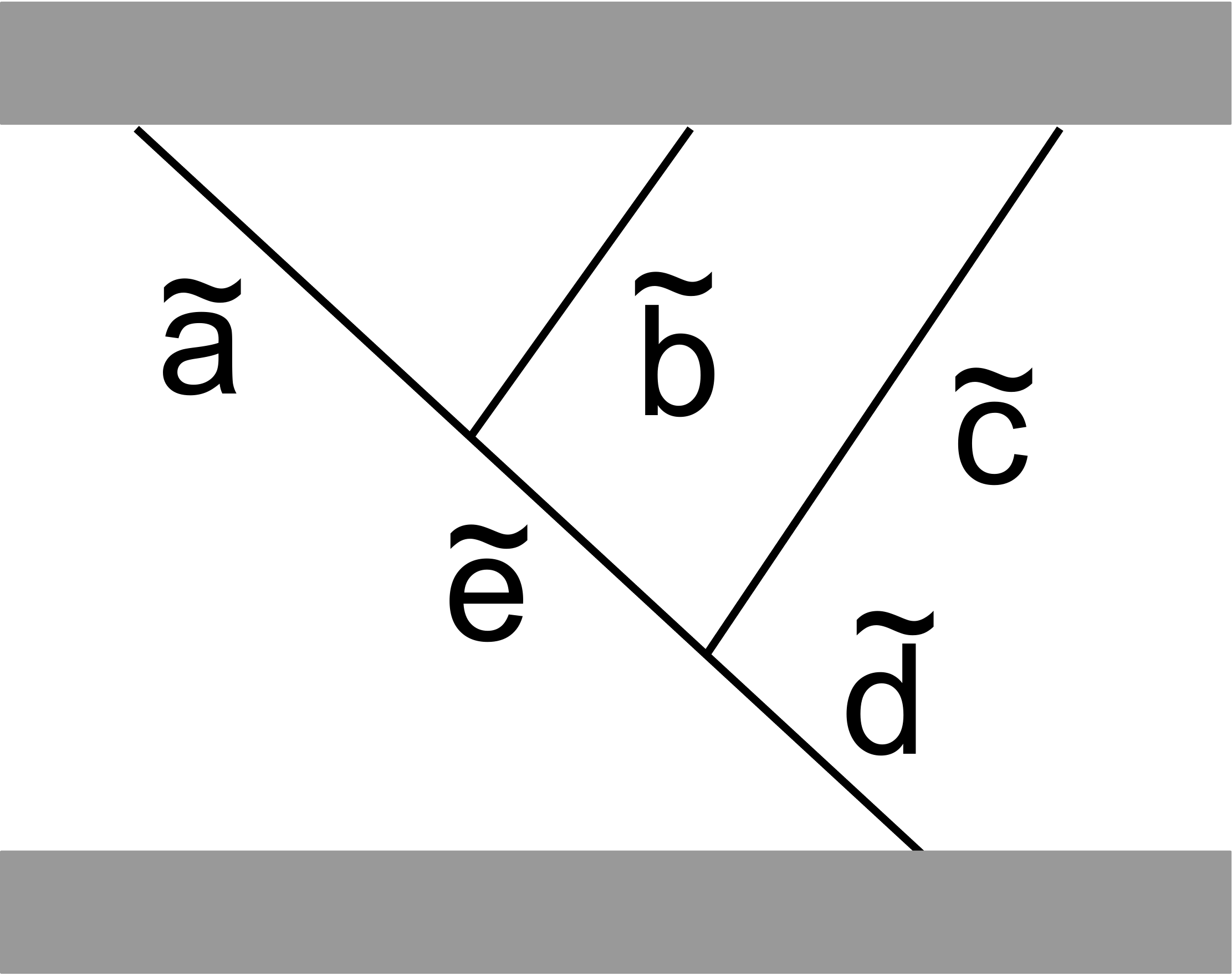}}
	\right) =\sum_{\tilde{f}}F^{\tilde{a}\tilde{b}\tilde{c}}_{\tilde{d}\tilde{e}\tilde{f}} \Psi \left(\raisebox{-0.22in}{\includegraphics[height=0.5in]{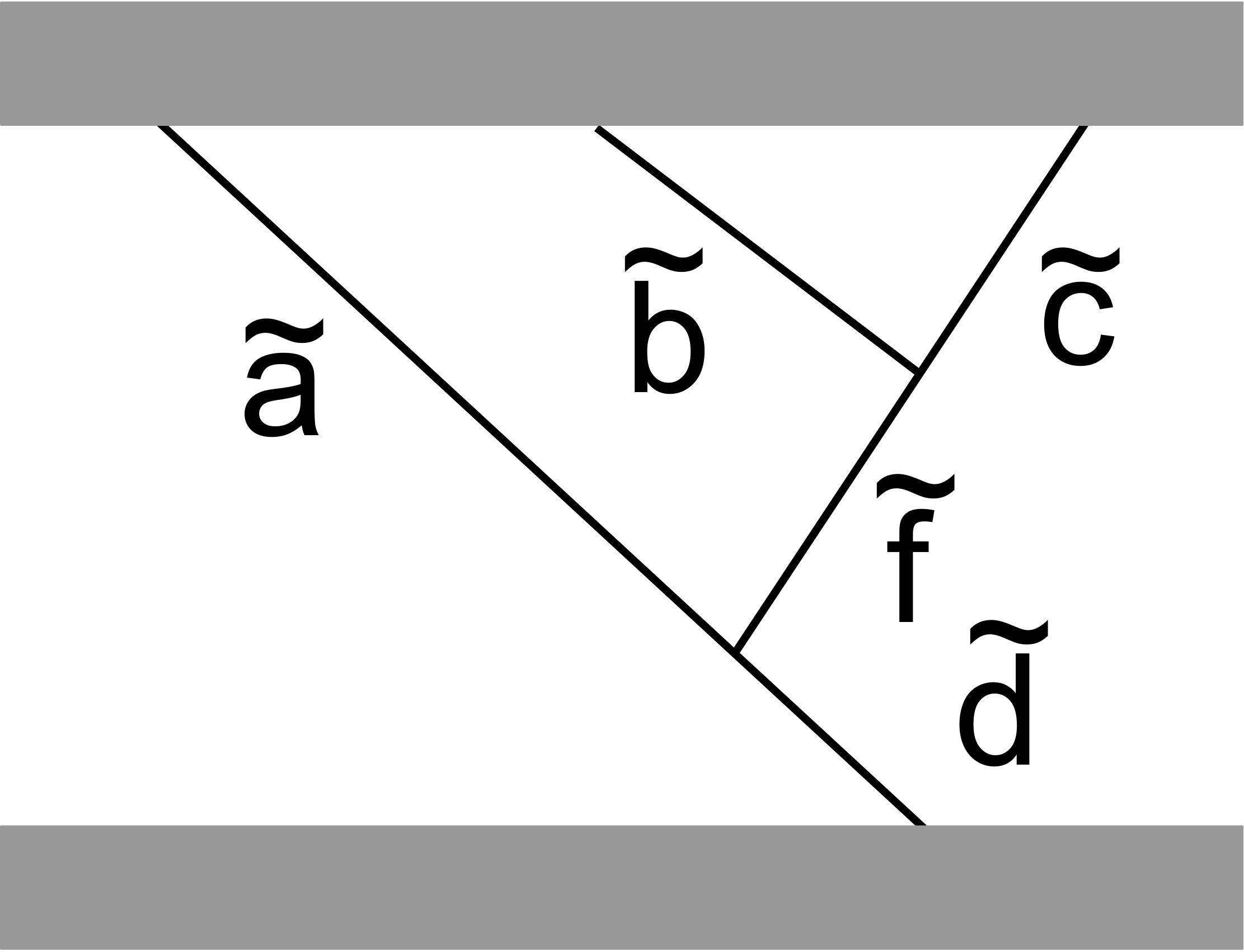}} \right)
\label{newf} \\
	\Psi \left( \raisebox{-0.22in}{\includegraphics[height=0.5in]{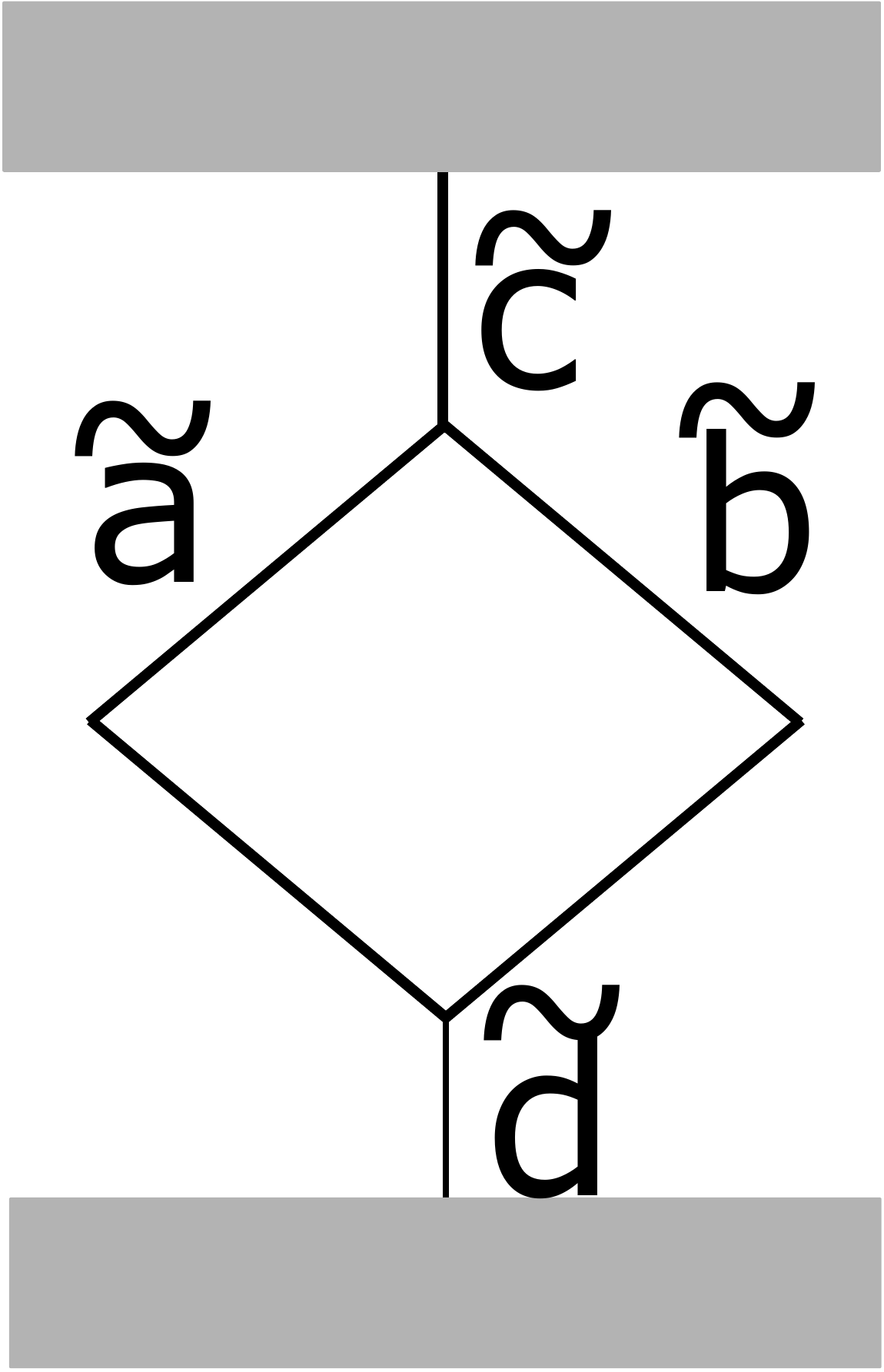}}
	\right) =  \delta_{\tilde{c},\tilde{d}} \sqrt{\frac{d_{\tilde{a}} d_{\tilde{b}}}{d_{\tilde{c}}}} \Psi \left(\raisebox{-0.22in}{\includegraphics[height=0.5in]{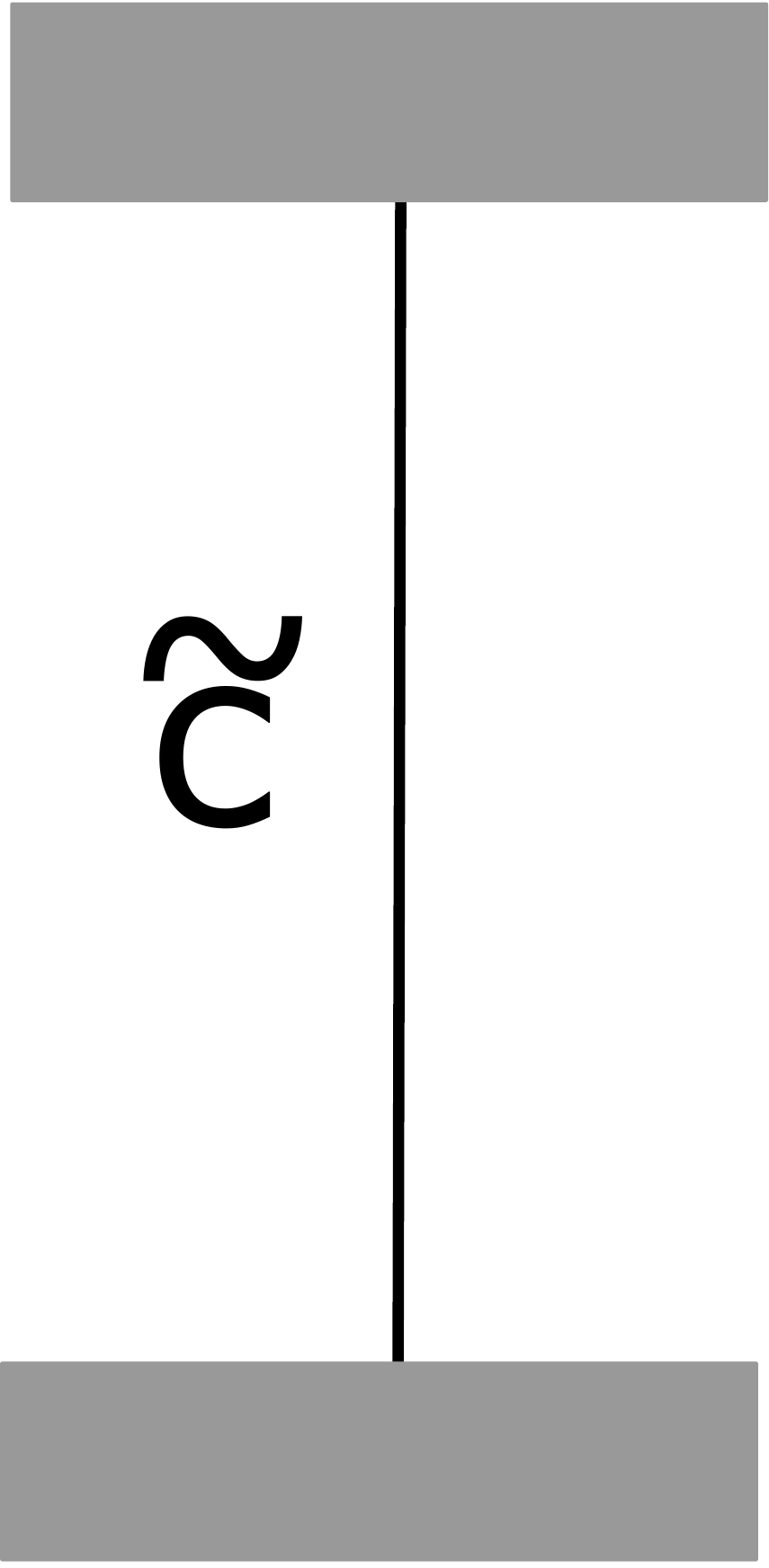}} \right). \label{newd}
	\end{gather}
	\label{newrules}
\end{subequations}%
Here the grey regions denote the part of the configuration that is identical on both sides of the equation.

To relate the new coefficients to the old ones,
consider a pair of reference configurations $X_0, X_0'$ related by one of the local moves in Eq. (\ref{localrules}). For convenience, we choose all reference configurations to be closed configurations in which all sticks carry the trivial label.  
These closed configurations are generated by those terms in $P_{\phi}$ containing only closed loops of simple string operators, all of which act as the identity on the ground state $|\Phi\rangle$ of the original string net.  (Recall that $|\Psi \> = P_{\phi} |\Phi \>$). 
Thus when $X_0, X_0'$ are closed configurations, $\Psi(X_0) \propto \Phi(X_0)$, and similarly for $X_0'$.  Note that the constant of proportionality here depends only on the number of closed loop string operators in $P_{\phi}$, and is the same for all reference configurations. 

Using the fact that $\Phi(X_0), \Phi(X_0')$ are related by the original local rules, and applying (\ref{psiamp1}) to both sides of (\ref{newrules}),  
we conclude that when none of the labels $\tilde{a}, ... \tilde{f}$ split, the old data and the new data are related by
\begin{subequations}
	\begin{gather}
		\frac{ B^{ab}_e B^{ec}_d  }{B^{bc}_f B^{af}_d } F^{abc}_{def}=F^{\tilde{a}\tilde{b}\tilde{c}}_{\tilde{d}\tilde{e}\tilde{f}} \label{main1}\\
		 d_a = d_{\tilde{a}} \label{main2}
\end{gather}
	\label{main}
\end{subequations}
(The local moves (\ref{1b}) and (\ref{1c}) lead to the same definition (\ref{main2}) of $d_{\tilde{a}}$).
Here $B^{ab}_c$ (which can also be root vertex coefficients $A^{ab}_c$) are the vertex coefficients defined in (\ref{basic1}) -(\ref{Bab}) and $\{ F^{abc}_{def},d_a \}$ are the original $F$-symbols and quantum dimensions for the ground state $\Phi$.
The labels $a,b,c,d$ in $F^{abc}_{def}$ are chosen such that they are compatible with the branching rules of the old theory, and such that $a\in \tilde{a},b \in \tilde{b}, c\in \tilde{c}, d\in \tilde{d}$.    
This expression thus fully fixes the new $F$-symbols in terms of the old $F$-symbols and the vertex coefficients.  Further, comparing this to the expression (\ref{gaugef}) for gauge transformations of the $F$-symbols, we can see that the root vertex coefficients $A^{ab}_c$ with $a,b \neq s^i$ are simply gauge transformations of the new $F$'s.  (The remaining vertex coefficients, which are fully fixed by the choice of $A^{ab}_c$, ensure that the left-hand side of the equation is independent of the specific choice $a,b,c,d,e,f$ used in the calculation.)

\subsubsection{Topological data in theories with splitting}

In theories with splitting, instead of expressing amplitudes in terms of a single reference configuration, we replace Eq. (\ref{psiamp1}) with: 
\begin{equation}
	\Psi\left (\tilde{X} \right )= \frac{1}{N(X_0)}\sum_{X_0 \in \tilde{X}} C_{\tilde{X}}(X_0)\Psi(X_0)
	\label{psiamp1s}
\end{equation}
were $X_0$ denotes the set of reference configurations that are compatible with the choice of condensed labels $\tilde{X}$, {\it and} contain only trivial labels on the sticks.  The reason for this replacement is that in theories with splitting, a single reference configuration $X_0$ may not be sufficient to uniquely fix $\tilde{X}$ via the choice of vertex coefficients entering $C_{\tilde{X}}(X_0)$.  We therefore instead keep the minimum number of configurations in our sum necessary to ensure that the right-hand side describes coefficients associated with a specific condensed label set, which is a sum over all configurations compatible with $\tilde{X}$ for which all sticks carry the trivial label.  

Unlike in the unsplit case, the number of configurations associated with each reference configuration $X_0$ in the sum does not, in general, fully cancel the pre-factor of $1/p^{2 N_V}$ in Eq. (\ref{psiamp}).  
Here $N(X_0)$ counts the number of distinct products of simple string operators that leave $X_0$ unchanged -- meaning that they change neither the labels on the sticks, nor any of the edge labels.  This number depends on the number of closed loops in $X_0$ along which all labels split.  Relative to the unsplit case, the number of distinct configurations $X$ in the sum (\ref{psiamp}) is reduced by $N(X_0)$.

To find $F^{\tilde{a}\tilde{b}\tilde{c}}_{\tilde{d}\tilde{e}\tilde{f}}$,
we note that:
\begin{eqnarray}
	\Psi \left( \raisebox{-0.22in}{\includegraphics[height=0.5in]{newfa-eps-converted-to.pdf}}
	\right) 
	&=& \sum_{e\in\tilde{e}}(B^{ab}_e)^{\mu \nu}_{\lambda}  (B^{ec}_d)^{\lambda \rho}_{\sigma}  \Phi \left(\raisebox{-0.22in}{\includegraphics[height=0.5in]{rule1a-eps-converted-to.pdf}}\right)  \nonumber \\
	& = & \sum_{e\in\tilde{e}}(B^{ab}_e)^{\mu \nu}_{\lambda}  (B^{ec}_d)^{\lambda \rho}_{\sigma} \nonumber \\
	&  & \times \sum_f  F^{abc}_{def} \Phi \left(\raisebox{-0.22in}{\includegraphics[height=0.5in]{rule1b-eps-converted-to.pdf}}\right)  
\label{newfsplit}
\end{eqnarray}
In the last equality, the labels $\mu, \nu, \rho, \sigma$ identify the external split legs as $\tilde{a}_{\mu}, \tilde{b}_{\nu}, \tilde{c}_{\rho},$ and $\tilde{d}_{\sigma}$ respectively.  However, it can happen that there is more than one solution $\tilde{f}_{\kappa}$ compatible with both the old label $f$, and the new fusion rules.  In other words, there may be more than one choice of $\kappa$ for which $(B^{bc}_f)^{\nu \rho}_{\kappa}  (B^{af}_d)^{\mu \kappa}_{\sigma} \neq 0$.  
We conclude that the old data and the new data are related by
\begin{equation}
		\sum_{e:e\in\tilde{e}}\frac{ (B^{ab}_e)^{\mu \nu}_{\lambda}  (B^{ec}_d)^{\lambda \rho}_{\sigma}   }{ (B^{bc}_f)^{\nu \rho}_{\kappa}  (B^{af}_d)^{\mu \kappa}_{\sigma} \ } F^{abc}_{def}=\sum_{\tilde{f}: f\in \tilde{f}} F^{\tilde{a}_\mu \tilde{b}_\nu \tilde{c}_\rho}_{\tilde{d_\sigma}\tilde{e_\lambda }\tilde{f_\kappa}} \label{main1split} \ .
\end{equation}

To find $d_{\tilde{a}}$, we observe that we also have:
\begin{equation}
\Psi \left (\sum_{\tilde{X} \in X_0} \tilde{X} \right )
=\frac{1 }{N(X_0) }\sum_{\tilde{X} \in X_0}    C_{\tilde{X}} (X_0) \Psi(X_0).
\end{equation} 
where in this case, we can choose a single reference configuration.  Letting $\tilde{X}$ be a configuration with single closed loop carrying the label $\tilde{a}$, and $X_0$ to have a single loop carrying the label $a$, we find that the number of terms in the sum is precisely $N(X_0)$, and that all terms in the sum contribute equally.  From this, we conclude that 
\begin{equation}
 d_a = \sum_{\tilde{a}: a\in \tilde{a}}d_{\tilde{a}} \label{main2split} \ .
 \end{equation}

Thus, we see that in theories with splitting, the original fusion data and vertex coefficients do not fully fix all  of the new $F's$.  In this case, the remaining freedom must be used to ensure that the new $F$'s satisfy the consistency conditions (\ref{consistency}), as well as the unitarity conditions (\ref{hermicity}).

\subsection{Consistency conditions for $F^{\tilde{a}\tilde{b}\tilde{c}}_{\tilde{d}\tilde{e}\tilde{f}}$ \label{app:pi}}

We now show that the new data $\{ F^{\tilde{a}\tilde{b}\tilde{c}}_{\tilde{d}\tilde{e}\tilde{f}} \}$ satisfy the consistency conditions (\ref{consistency}).   
We begin with the condition (\ref{3e}), which requires $F^{\tilde{a}\tilde{b}\tilde{c}}_{\tilde{d}\tilde{e}\tilde{f}}=1$ if $\tilde{a}$ or $\tilde{b}$ or $\tilde{c}=\tilde{0}$.  
We wish to show that the right-hand side is equal to $1$ if $a$, $b$, or $c$ are powers of $s$. 
Indeed,  using Eq. (\ref{cx2cx1}), one can show:
\begin{equation} \label{e98}
		\frac{B^{a b^i}_{c^i} B^{b s^i} }{B^{c s^i} B^{a b}_c } F^{a b s^i}_{c^i c b^i } = 
		\frac{B^{a s^i} B^{a^i b}_{c^i} }{B^{a b^i}_{c^i}  B^{s^i b}} F^{a s^i b}_{c^i a^i b^i } = 
		\frac{B^{a^i b}_{c^i} B^{ s^i a} }{B^{a b}_{c}  B^{s^i c}} F^{s^i a  b}_{c^i a^i c } = 1 
\end{equation}
If $a$, $b$, or $c$ are powers of $s$, then there are no sums in Eq. (\ref{main1split}); thus Eq. (\ref{e98}) ensures that the new $F$'s satisfy Eq. (\ref{3e}).  

To see that the first term in Eq. (\ref{e98}) is equal to unity, apply a $W^2_{\phi^i}$-type string to the vertex $(a,b;c)$, with the stick on the $c$ edge carrying the label $s^i$.  From Eq. (\ref{stringmatrix}), in the gauge (\ref{f1}), the matrix element associated with this string operator is $ F^{a b s^i}_{c^i c b^i }$, and Eq. (\ref{cx2cx1}) implies that $B^{c s^i} B^{a b}_c = B^{a b^i}_{c^i} B^{b s^i}F^{a b s^i}_{c^i c b^i} $.  
The second equality is obtained by applying a product of the form $W^2_{\phi^i}W^1_{\phi^{-i}}$ to a configuration with the two vertices $(a, s^i; a^i)$ and $(a^i, b; c^i)$, and using Eqs. (\ref{consistency}), (\ref{weqs}), and  (\ref{Eq:Cross}). 
The third equality can be obtained by acting with a $W^1_{\phi^i}$ string on the vertex $(a,b; c)$, with an $s^i$ labeled stick on the $c$ edge.  In our gauge of choice, the corresponding matrix element is $\overline{w}_{\phi^i}(b) F^{a b s^i}_{c^i c b^i} (F^{a s^i b}_{c^i a^- b^i})^*$, which by Eq. (\ref{weqs}) is equal to $w_{\phi^i}(a) \bar{w}_\phi^i(c) (F^{s^i a b}_{c^i a^i c})^*$.  
This gives 
\begin{equation}
A^{ab}_c A^{c s^i} \bar{\omega}_{\phi^i}(c) = F^{s^i a b}_{c^i a^i c} \bar{\omega}_{\phi^i}(a)A^{a s^i} B^{a^i b}_c
\end{equation}
Using Eq. (\ref{Eq:Cross}), we obtain the stated result.

Next, we turn to the pentagon identity (\ref{3a}).
Multiplying both sides of (\ref{3a}) by $B^{fc}_g B^{gd}_e B^{ab}_f$, and summing over $f\in\tilde{f}$ and $g\in\tilde{g}$, gives
\begin{equation}
	\begin{split}
	\sum_{f\in \tilde{f},g\in\tilde{g}}F^{fcd}_{egl}F^{abl}_{efk} & B^{fc}_g B^{gd}_e B^{ab}_f \\
	&=\sum_{f\in \tilde{f},g\in\tilde{g},h}F^{abc}_{gfh}F^{ahd}_{egk}F^{bcd}_{khl} 
	B^{fc}_g B^{gd}_e B^{ab}_f
	\end{split}
	\label{3aa}
\end{equation}
We fist consider the right hand side of (\ref{3aa}).  Using Eq. (\ref{main}), we have:
\begin{equation}
	\begin{split}
	&\sum_{h,g\in\tilde{g}}(\sum_{f\in\tilde{f}} F^{abc}_{gfh}B^{ab}_fB^{fc}_g)F^{ahd}_{egk}F^{bcd}_{khl}B^{gd}_e \\
	&=\sum_{h,\tilde{h}\ni h}F^{\tilde{a}\tilde{b}\tilde{c}}_{\tilde{g}\tilde{f}\tilde{h}}(\sum_{g\in\tilde{g}}F^{ahd}_{egk}B^{ah}_gB^{gd}_e)F^{bcd}_{khl}B^{bc}_h \\
	&=\sum_{\tilde{h},\tilde{k}\ni k}F^{\tilde{a}\tilde{b}\tilde{c}}_{\tilde{g}\tilde{f}\tilde{h}} F^{\tilde{a}\tilde{h}\tilde{d}}_{\tilde{e}\tilde{g}\tilde{k}}(\sum_{h\in\tilde{h}}F^{bcd}_{khl}B^{hd}_kB^{bc}_h)B^{ak}_e\\
	&=\sum_{\tilde{h},\tilde{k}\ni k,\tilde{l}\ni l}F^{\tilde{a}\tilde{b}\tilde{c}}_{\tilde{g}\tilde{f}\tilde{h}} F^{\tilde{a}\tilde{h}\tilde{d}}_{\tilde{e}\tilde{f}\tilde{h}} F^{\tilde{b}\tilde{c}\tilde{d}}_{\tilde{k}\tilde{h}\tilde{l}} B^{cd}_lB^{bl}_kB^{ak}_e
\end{split}
	\label{3aar}
\end{equation}
(In the third line, we exploit the fact that $\sum_{h,\tilde{h}\ni h}=\sum_{\tilde{h},h\in \tilde{h}}$).  A similar treatment of the right hand side of (\ref{3aa})
gives $\sum_{\tilde{l}\ni l,\tilde{k}\ni k}F^{\tilde{f}\tilde{c}\tilde{d}}_{\tilde{e}\tilde{g}\tilde{l}}F^{\tilde{a}\tilde{b}\tilde{l}}_{\tilde{e}\tilde{f}\tilde{k}}B^{bl}_kB^{ak}_eB^{cd}_l$.  
Thus the new $F$-symbols satisfy:
\begin{equation}
	\sum_{\tilde{l}\ni l,\tilde{k}\ni k} F^{\tilde{f}\tilde{c}\tilde{d}}_{\tilde{e}\tilde{g}\tilde{l}}F^{\tilde{a}\tilde{b}\tilde{l}}_{\tilde{e}\tilde{f}\tilde{k}}=\sum_{\tilde{l}\ni l,\tilde{k}\ni k} \sum_{\tilde{h}}F^{\tilde{a}\tilde{b}\tilde{c}}_{\tilde{g}\tilde{f}\tilde{h}}F^{\tilde{a}\tilde{h}\tilde{d}}_{\tilde{e}\tilde{g}\tilde{k}}F^{\tilde{b}\tilde{c}\tilde{d}}_{\tilde{k}\tilde{h}\tilde{l}}.
	\label{e102}
\end{equation}
If $a^r \neq a$ for any $r<q$, then the sums over $\tilde{l}$ and $\tilde{k}$ can be dropped, and the new $F$-symbols automatically satisfy the pentagon identity (\ref{3a}).  
Otherwise, Eq. (\ref{main}) only constrains certain sums of the new $F$'s, and we must use the remaining freedom to choose the new $F$'s to satisfy Eq. (\ref{3a}).

\subsection{Effective Hamiltonian \label{app:effH}}

We have seen that the ground state $|\Psi\rangle$ of the condensed phase is a string net state, described in terms of the new labels $\{\tilde{a} \}$, with new $F$ symbols and quantum dimensions given by Eqs. (\ref{main1split}), (\ref{main2split}), together with the consistency conditions (\ref{consistency}).  Since $|\Psi\rangle$ is also a ground state of the effective Hamiltonian in the condensed phase, this suggests that our effective Hamiltonian acts on the labels in the new basis as a (conventional) string net Hamiltonian.  

In the absence of splitting, it is relatively straightforward to see that this is indeed the case.  Consider the action of $P_{\phi} B^\phi_p$ on a state $\langle \tilde{X} |$ in our new string-net basis.  As above, we can use the fact that $\langle \tilde{X} | P_{\phi} B^\phi_p = \langle \tilde{X} | B^\phi_p =  C(X_0)  \langle X_0| B^\phi_p $, where $X_0$ is a configuration compatible with $\tilde{X}$, and for which all stick labels are trivial.   
We thus have 
\begin{equation}
\begin{split}
\langle \tilde{X} | B^{\phi,t}_p& 
 = \sum_{X_0'} C(X_0)  \langle X'_0| B^{t, i_1... i_6j_1...j_6}_{p, i_1'... i_6'j_1'...j_6'}(e_1,... e_{12};\mathbb{I},\mathbb{I},\mathbb{I})
\end{split}
\end{equation}
where $j_k= i_k, j_k'= i_k'$ for $k=4,5,6$, and $e_{10}= e_{11} = e_{12} = 0$.  Here $\langle X'_0|$ is identical to $\langle X_0|$ except on the boundary of the plaquette $p$, where edges labeled $i_1 ... i_6, j_1... j_6$ in $\langle X_0|$ now carry labels $i'_1 ... i'_6, j'_1... j'_6$.  
Applying Eq. (\ref{main1}) repeatedly, we find that the matrix element can be expressed as a product of new $F$ symbols:  
 \begin{equation}
\begin{split} 
 \sum_{X'}C(X_0)  \langle X'| B^{t, i_1... i_6j_1...j_6}_{p, i_1'... i_6'j_1'...j_6'}(e_1,... e_{12};\mathbb{I},\mathbb{I},\mathbb{I}) =\\
 \sum_{X' } C(X')  \langle X'|   B^{\tilde{t},\tilde{i}'_1\dots \tilde{i}'_6 \tilde{i}'_1\dots \tilde{i}'_6}_{p,\tilde{i}'_1\dots \tilde{i}'_{6} \tilde{i}'_1\dots \tilde{i}'_{6}}(\tilde{e}_1\dots \tilde{e}_6 \tilde{0} \dots \tilde{0};\mathbb{I},\mathbb{I},\mathbb{I})  \\
 = \langle \tilde{X}'|  \tilde{B}_{p,\tilde{i}'_1\dots \tilde{i}'_6}^{\tilde{t},\tilde{i}_1\dots\tilde{i}_6}(\tilde{e}_1\dots \tilde{e}_6) 
	\end{split}
\end{equation}
where the matrix elements of $B_p^t$ are defined in (\ref{bp1}), with the old $F$-symbols in the first line replaced by the new $F$-symbols in the second and third lines.  Matrix elements of the plaquette string operator $B^{\tilde{t}}_p$ acting on states with all sticks carrying the trivial label are exactly the matrix elements of the conventional string-net plaquette operator (see Ref. \cite{LinLevinBurnell}) which we denote $\tilde{B}_{p,\tilde{i}'_1\dots \tilde{i}'_6}^{\tilde{t},\tilde{i}_1\dots\tilde{i}_6}(\tilde{e}_1\dots \tilde{e}_6)$ in the third line.
Thus our effective Hamiltonian in the string-net phase is exactly the new string net Hamiltonian.  

The situation in theories with splitting is similar, though more subtle due to the fact that a single label $t$ in the original theory can represent multiple labels $\tilde{t}_\mu$ in the new theory.

\section{Examples} \label{section:exp}
In this section, we work out some illustrative examples.
We begin by considering condensation  in the abelian
$\mathbb{Z}_2, \mathbb{Z}_4, \mathbb{Z}_6$, $\mathbb{Z}_4\times \mathbb{Z}_4$ string-net models.  
In abelian theories there is never any splitting, and the new fusion data follows directly from the coefficients $B^{a_i, b^j}_{c^{k+k}}$.  
We also describe condensation of abelian bosons in two non-abelian string-net models, based on the fusion categories $\text{Rep}(S_3)$ and $SU(2)_4$.
In these models, we also fully construct the new Hilbert space $\tilde{\mathcal{H}}$ after condensation and compute new $F$-symbols and quantum dimensions for the condensed phases.

Throughout our discussion of abelian string-net models, we will use 
\begin{equation}
	F(a,b,c)=F^{abc}_{def}
	\label{}
\end{equation}
for brevity, since other indices can be deduced from the abelian branching rules.  
Moreover, for string nets based on the group $G=\mathbb{Z}_N$, there are $N$ distinct solutions to (\ref{localrules}), with the explicit form\cite{MooreSeiberg,PropitiusThesis}
\begin{equation}
	F(a,b,c)=e^{2\pi i \frac{pa}{N^2}(b+c-[b+c])}.
	\label{Fsol}
\end{equation}
The integer parameter $p=0,\dots,N-1$ labels the $N$ distinct solutions. The arguments $a,b,c$ take values in $0,\dots,N-1$ and $[b+c]$ denotes $b+c$ (mod $N$) with values also taken in $0,\dots,N-1$.  
For each of the $N$ distinct solutions, we can construct a corresponding string-net model. 
Each such string net model has $N^2$ topologically distinct quasiparticle excitations labeled by $\phi=(s,m)$ where $s,m=0,1,\dots,N-1$. The string operator $W_\phi(P)$ which creates $\phi=(s,m)$ is defined by (\ref{stringrules}) with the string parameters 
\begin{align}
	w_{\phi}(a)=e^{2\pi i(\frac{psa}{N^2}+\frac{ma}{N})}
	\label{}
\end{align}

\subsection{$\mathbb{Z}_2$ string-net model}

To set the stage, we begin with the $\mathbb{Z}_2$ string-net model, whose condensation transitions and phase diagram have been studied extensively in the literature \cite{Wegner,FradkinShenker,VidalToric,VidalToric2,KitaevPhase,TrebstTC,TSBLong,Schuler23}.
Here, we briefly review how our construction replicates these results.

The $\mathbb{Z}_2$ string-net model has two types of strings $\{0,1\}$ with dual strings $\bar{0}=0,\bar{1}=1$. The branching rules are $\{(a,b;c)\text{ with }a+b=c\text{ mod }2 \}$.  
There are two distinct solutions $F(1,1,1)=\pm1$ to (\ref{consistency}). The corresponding models are the Toric code\cite{KitaevToric} and the double semion model respectively. 
The Toric code has two $\mathbb{Z}_2$ bosons $\phi=(1,0)$ and $\phi=(0,1)$, while the double semion model has one $\mathbb{Z}_2$ boson $\phi=(0,1)$.

We first consider the condensation of $\phi=(0,1)$ in the two models, as the two condensed phases are identical. After condensation, only string type $a$ which satisfies $w_\phi(a)=(-1)^a=1$ remains, namely the remaining string type is $\tilde{0}=\{0\}$ and thus the Hilbert space $\tilde{\mathcal{H}}$ is the vacuum state which is the same as the vacuum state in $\mathcal{H_\phi}$. Hence, there is no string-net topological order after $\phi$ condensation.

Next, we consider the $\phi=(1,0)$ condensation in the Toric code. After condensation, the new string type is $\tilde{0}=\{0,1\}$ and thus $\tilde{\mathcal{H}}$ is the vacuum state which is the equal superposition of all states in $\mathcal{H}_\phi$. Thus there is no string-net topological order after $\phi$ condensation.

With the Hamiltonian described here, all of these phase transitions are in the $2+1$D Ising universality class.

\subsection{$\mathbb{Z}_4$ string-net model}

We next show how our construction allows us to construct certain condensed phases of the $\mathbb{Z}_4$ string net model.  The full phase diagram of this model was studied in detail in Ref. \cite{IqbalZ4}.

The $\mathbb{Z}_4$ string-net model has four types of strings $\{0,1,2,3\}$ with dual strings $\bar{0}=0,\bar{1}=3,\bar{2}=2,\bar{3}=1$. The branching rules are $\{(a,b;c) \text{ with }a+b=c \text{ mod }4\}$. The Hilbert space consists of all possible string-nets with the above string types and branching rules.

There are four distinct solutions to the self-consistency conditions (\ref{consistency})
\begin{equation}
	F(a,b,c)=e^{i \frac{2\pi p a(b+c-[b+c]_4)}{4^2}}
	\label{}
\end{equation}
labeled by $p=0,1,2,3$. Here $[x]_n=x$ mod $n$.
The corresponding $\mathbb{Z}_4$ string-net models realize the topological order described by the Chern-Simons theory with the $K$-matrix
\[K= \left( \begin{array}{cc}
0 & 4\\
4 & -2p
 \end{array} \right).\]

 All four models have a $\mathbb{Z}_4$ boson $\phi=(0,1)$ and a  $\mathbb{Z}_2$ boson $\phi=(0,2)$. In addition, the $p=0$ model has other two $\mathbb{Z}_2$ bosons $\phi=(2,0)$ and $\phi=(2,2)$.
We consider the topological order after condensation of each of these bosons.

As in the $\mathbb{Z}_2$ case, condensing $\phi=(0,1)$ leads to a trivial order.  Thus, 
we begin with the condensation of $\phi=(0,2)$.  In the condensed phase, the remaining string types $a$ are those which satisfy $w_{\phi}(a)=e^{i 2\pi \frac{2a}{4}}=1$, namely, the remaining string types are $a\in\{0,2\}$. 
Thus the Hilbert space $\tilde{\mathcal{H}}$ after $\phi$-condensation contains string-nets with the new string labels $\{ \tilde{0}=\{0\},\tilde{1}=\{2\} \}$ and the $\mathbb{Z}_2$ branching rules.  
As discussed in Eq. (\ref{Eq:Achoice}), in this case all non-vanishing vertex coefficients can be set to $1$.  Solving Eq. (\ref{main}), 
we then find that the $F$ symbols of the condensed phase are simply a subset of those of the uncondensed phase.  Specifically:
\begin{equation}
	\text{all }A=1\quad ,F(\tilde{1},\tilde{1},\tilde{1})=1,\quad  d_{\tilde{1}}=1
	\label{}
\end{equation} for the $p=0,2$ models, and
\begin{equation}
	\text{all }A=1,\quad F(\tilde{1},\tilde{1},\tilde{1})=-1,\quad d_{\tilde{1}}=1
	\label{}
\end{equation} for the $p=1,3$ models.
Thus, the $\phi=(0,2)$ condensed phase in the $p=0,2$ models is described by the Toric code while the $\phi=(0,2)$ condensed phase in the $p=1,3$ models is described by the double semion model.

Condensing $\phi=(1,0)$ also leads to a trivial topological phase; thus we
next turn to the condensation of the $\phi=(2,0)$ and $\phi=(2,2)$ bosons in $p=0$ model.
The Hilbert space $\tilde{\mathcal{H}}$ for both cases contains string-nets with the new string types $\{\tilde{0}=\{0,2\},\tilde{1}=\{1,3\}\}$ and the $\mathbb{Z}_2$ branching rules.
To find the topological order for after condensation, we solve for the vertex coefficients, and use Eq. (\ref{main}) to deduce the topological data of the condensed phase.
When the condensing boson is $\phi=(2,0)$, we find that
\begin{equation}
	\text{all }A=1,\quad F(\tilde{1},\tilde{1},\tilde{1})=1,\quad d_{\tilde{1}}=1.
	\label{}
\end{equation}
In this case, the condensed $F$-symbols are simply a subset of the uncondensed ones; this is always the case when condensing $(q,0)$-type bosons in untwisted abelian lattice gauge theories.  Thus, the $\phi=(2,0)$ condensed phase is described by the Toric code.  

When condensing boson is $\phi=(2,2)$, in contrast, not all vertex coefficients can be chosen to be unity.  In this case, we can choose:
\begin{equation}
	A^{t2}=(-1)^t,\text{ other }A=1, \quad F(\tilde{1},\tilde{1},\tilde{1})=-1,\quad d_{\tilde{1}}=1.
	\label{}
\end{equation}
with $t={0,1,2,3}$. Thus, the $\phi=(2,2)$ condensed phase is described by the double semion model.

\subsection{$\mathbb{Z}_6$ string-net models}
The $\mathbb{Z}_6$ string-net model has six types of strings $\{0,1,2,\dots,5\}$. The dual string type is defined by $\bar{a}=6-a \text{ mod }6$ while the branching rules are the triplets $(a,b;c)$ that satisfy $a+b=c$ (mod $6$). By using the general solution
\begin{equation}
		F(a,b,c)=e^{i \frac{2\pi p a(b+c-[b+c]_6)}{6^2}},
	\label{}
\end{equation}
we can construct six distinct string-net models labeled by $p=0,1,\dots,5$. The corresponding topological order can be described by the Chern-Simons theory with the K-matrix
\[K= \left( \begin{array}{cc}
0 & 6\\
6 & -2p
 \end{array} \right).\]

Analogously to the previous examples, condensing a $\mathbb{Z}_6$ boson
results in a trivial topological phase.  Thus we  focus on condensing 
the $\mathbb{Z}_2$ and $\mathbb{Z}_3$ abelian bosons, which are summarized for the 6 distinct $\mathbb{Z}_6$ string-net models in Table \ref{table1}.
  
We  first consider condensing $\mathbb{Z}_3$ bosons.
Condensing $\phi=(0,2)$, which is a boson for any $p$, leaves the string types  $\{\tilde{0}=\{0\},\tilde{1}=\{3\}\}$ with $\mathbb{Z}_2$ branching rules.
In this case, the condensed phase is described by:
\begin{equation}
	\text{all }A=1,\quad F(\tilde{1},\tilde{1},\tilde{1})=1,\quad d_{\tilde{1}}=1
	\label{}
\end{equation} for the $p=0,2,4$ models
and 
\begin{equation}
	\text{all }A=1,\quad F(\tilde{1},\tilde{1},\tilde{1})=-1,\quad d_{\tilde{1}}=1
	\label{}
\end{equation} 
for the $p=1,3,5$ models.
Thus, the $\phi=(0,2)$ condensed phase in the $p=0,2,4$ models is described by the Toric code while the $\phi=(0,2)$ condensed phase in the $p=1,3,5$ models is described by the doubled semion model.

Second, we condense the $\mathbb{Z}_3$ boson $\phi=(2,0)$ in the $p=0$ model
and the $\mathbb{Z}_3$ boson $\phi=(2,2)$ in the $p=3$ model.
The new string labels are $\{\tilde{0}=\{0,2,4\},\tilde{1}=\{1,3,5\}\}$ with $\mathbb{Z}_2$ branching rules after condensation. 
Analogous to the $\mathbb{Z}_4$ case, we find
\begin{equation}
	\text{all }A=1,\quad F(\tilde{1},\tilde{1},\tilde{1})=d_{\tilde{1}}=1 
	\label{}
\end{equation}
for condensation of $\phi=(2,0)$ in the $p=0$ model and
\begin{equation}
	\begin{split}
	A^{12}=A^{32}=A^{52}=-1,\text{other } A=1,\\
	F(\tilde{1},\tilde{1},\tilde{1})=-1, d_{\tilde{1}}=1.
	\end{split}
	\label{}
\end{equation}
for condensation of $\phi=(2,2)$ in the $p=3$ model.
Thus the two condensed phases are described by the Toric code and the doubled semion model respectively.

Next, we consider condensing $\mathbb{Z}_2$ bosons.
When the condensing boson is $\phi=(0,3)$, the remaining string types are $\{\tilde{0}=\{0\},\tilde{1}=\{2\},\tilde{2}=\{4\}\}$ with $\mathbb{Z}_3$ branching rules.
After solving (\ref{main}), we find
\begin{equation}
	\text{all }A=1,\quad F(\tilde{a},\tilde{b},\tilde{c})=e^{i2\pi \frac{q\tilde{a}(\tilde{b}+\tilde{c}-[\tilde{b}+\tilde{c}]_{\tilde{3}})}{9}},\quad d_{\tilde{a}}=1
	\label{}
\end{equation}
with $q=0$ for the $p=0,3$ models and $q=1$ for the $p=1,4$ models and $q=2$ for the $p=2,5$ models. 
Thus, the $\phi=(0,3)$ condensed phase in the $p=0,3$ models is described by the $\mathbb{Z}_3$ string-net model with $q=0$ while
the $\phi=(0,3)$ condensed phase in the $p=1,4$ and $p=2,5$ models is described by the $\mathbb{Z}_3$ string-net model with $q=1$ and $q=2$ respectively.

Finally, we condense the $\phi=(3,0),(3,1),(3,2)$ bosons in the $\mathbb{Z}_6$ string-net models with $p=0,2,4$ respectively.
The new string types after condensation are $\{\tilde{0}=\{0,3\},\tilde{1}=\{1,4\},\tilde{2}=\{2,5\}\}$ with $\mathbb{Z}_3$ branching rules.
Condensing $\phi=(3,0)$ in the $p=0$ model, all vertex coefficients can be taken to be $1$, and we obtain:
\begin{equation}
	\text{all }A=1,\quad F(\tilde{a},\tilde{b},\tilde{c})=1,\quad d_{\tilde{a}}=1.
	\label{}
\end{equation}
To condense $\phi=(3,1)$ in the $p=2$ model, we may take:
\begin{equation}
	\begin{split}
		&A^{13}=A^{43}=e^{i\frac{2\pi}{3}},A^{23}=A^{53}=e^{-i\frac{2\pi}{3}},\quad \text{other }A=1\\
	&F(\tilde{1},\tilde{1},\tilde{2})=F(\tilde{1},\tilde{2},\tilde{1})=F(\tilde{1},\tilde{2},\tilde{2})=e^{-i\frac{2\pi}{3}}, \\
	&F(\tilde{2},\tilde{2},\tilde{1})=F(\tilde{2},\tilde{1},\tilde{2})=F(\tilde{2},\tilde{2},\tilde{2})=e^{i\frac{2\pi}{3}},\quad
	\text{other }F =1\\
	&d_{\tilde{a}}=1.
	\end{split}
	\label{}
\end{equation}
Finally, to condense $\phi=(3,2)$ in the $p=4$ model, we obtain:
\begin{equation}
	\begin{split}
		&A^{13}=A^{43}=e^{-i\frac{2\pi}{3}},A^{23}=A^{53}=e^{i\frac{2\pi}{3}},\quad\text {other }A=1\\
	&F(\tilde{1},\tilde{1},\tilde{2})=F(\tilde{1},\tilde{2},\tilde{1})=F(\tilde{1},\tilde{2},\tilde{2})=e^{i\frac{2\pi}{3}}, \\
	&F(\tilde{2},\tilde{2},\tilde{1})=F(\tilde{2},\tilde{1},\tilde{2})=F(\tilde{2},\tilde{2},\tilde{2})=e^{-i\frac{2\pi}{3}},\quad
	\text{other }F=1 \\
	&d_{\tilde{a}}=1.
	\end{split}
	\label{}
\end{equation}
Thus the three condensed phases are described by the $\mathbb{Z}_3$ string-net models labeled by $p=0,2,1$ respectively.

 We summarize the condensed phases after condensing abelian bosons in the $\mathbb{Z}_6$ models in Table \ref{table1}.
\begin{table}
\begin{tabular}{|p{1.6cm}|p{0.8cm} | p{2.05cm} || p{0.8cm} | p{2.3cm}|}
	 \hline
	 $\mathbb{Z}_6$ Models  $p$ & 
	 $\mathbb{Z}_2$ $\phi$ & 
	 condensed phase & 
	 $\mathbb{Z}_3$ $\phi$ & 
	 condensed phase \\
	  \hline
	 $0,\dots,5$ & $(0,3)$ & $K_{3,0}$, $p=0,3$ 
	 $K_{3,1}$, $p=1,4$
	 $K_{3,2}$, $p=2,5$ & $(0,2)$ & $K_{2,0}$, $p=0,2,4$
	 $K_{2,1}$, $p=1,3,5$ \\
	 \hline
	 $0$ & $(3,0)$ & $K_{3,0}$ & $(2,0)$ & $K_{2,0}$ \\
	 \hline
	 $2$ & $(3,1)$ & $K_{3,2}$ & & \\
	 \hline
	 $3$ & & & $(2,2)$  & $K_{2,1}$ \\
	 \hline
	 $4$ & $(3,2)$ & $K_{3,1}$ & &\\
	 \hline
\end{tabular}
\caption{Six $\mathbb{Z}_6$ string-net models labeled by $p=0,1,\dots,5$. Column 2, 4 show the $\mathbb{Z}_2$ and $\mathbb{Z}_3$ bosons labeled by $(s,m)$ in the models. Column 3, 5 show the $K$-matrix for the condensed phases after condensing $\phi$ bosons in Column 2, 4 respectively. Here $K_{a,b}= \left( \begin{array}{cc}
0 & a\\
a & -2b
 \end{array} \right).$ }
\label{table1}
\end{table}

\subsection{$\mathbb{Z}_4\times \mathbb{Z}_4$ string-net model}
The $\mathbb{Z}_4\times \mathbb{Z}_4$ string-net model has 16 types of strings labeled by $a\in\{(a_1,a_2),a_1,a_2\in\{0,1,2,3\}\}$ with dual strings $\bar{a}=(\bar{a}_1,\bar{a}_2)=(4-a_1,4-a_2).$ The branching rules are $\{(a_i,b_i;c_i) \text{ with } a_i+b_i=c_i \text{ mod }4\}$ with $i=1,2$. The Hilbert space consists of all possible string-nets with the above string types and branching rules.

The general form of solutions to the self-consistency conditions (\ref{consistency}) for $\mathbb{Z}_N \times \mathbb{Z}_N$ string net models is known\cite{MooreSeiberg,PropitiusThesis}.
Here, we consider one such solution, for which: 
\begin{equation}
	F(a,b,c)=e^{i 2\pi a^T \mathbf{N}^{-1} \mathbf{P} \mathbf{N}^{-1} (b+c-[b+c]) }
	\label{fz4z4}
\end{equation}
with 
\begin{equation}
	\mathbf{N}= \left( \begin{array}{cc}
4 & 0\\
0 & 4
 \end{array} \right),
 \mathbf{P}= \left( \begin{array}{cc}
0 & 2\\
0 & 0
 \end{array} \right).
	\label{}
\end{equation}
Here the square bracket $[b+c]$ denotes a 2-component vector whose $i$-th component is $b_i+c_i$ (mod $4$).
From the solution (\ref{fz4z4}), we can construct the $\mathbb{Z}_4\times \mathbb{Z}_4$ string-net Hamiltonian.

We focus on the four $\mathbb{Z}_2$ bosons in the model and we denote them by
\begin{equation*}
	\begin{split}
	\phi_1=(2,0,0,3),\quad \phi_2=(2,0,0,1),\\
	\phi_3=(2,0,2,1),\quad \phi_4=(2,0,2,3).
	\end{split}
	\label{}
\end{equation*}
Here the bosons are labeled by $(s_1,s_2,m_1,m_2)$ with $s_1,s_2$ being the flux and $m_1,m_2$ being the charge carried by the particle.
Now, we consider the condensation of the four bosons $\phi_i$ in the $\mathbb{Z}_4\times \mathbb{Z}_4$ model in order.
In the $\phi_i$ condensed phase, we define the 2-component new string labels by
\begin{equation*}
	\begin{split}
		\tilde{a}=(\tilde{a}_1,\tilde{a}_2)=\{(a_1,a_2),(2+a_1,a_2)\} \\
		\text{ with }a_1\in\{0,1\},a_2\in\{0,1,2,3\}
	\end{split}
	\label{}
\end{equation*}
To find the topological order for the $\phi_i$ condensed phase, we have to solve for the vertex coefficients.
First, we find that
\begin{equation}
	\begin{split}
	\text{all }A&=1 \text{ for $\phi_1$ condensed phase} \\
	A^{a,(2,0)}&=(-1)^{a_2}, \text{other $A=1$ for $\phi_2$ condensed phase}.
	\end{split}
	\label{}
\end{equation}
For the $\phi_1$ condense phase, we then solve Eq.  (\ref{main}) to find:
 \begin{equation}
	 F(\tilde{a},\tilde{b},\tilde{c})=e^{i 2\pi \tilde{a}^T \tilde{\mathbf{N}}^{-1} \tilde{\mathbf{P}} \tilde{\mathbf{N}}^{-1} (\tilde{b}+\tilde{c}-[\tilde{b}+\tilde{c}]) } 
	 \label{newfz2z4}
 \end{equation}
with 
\begin{equation}
	\tilde{\mathbf{N}}= \left( \begin{array}{cc}
2 & 0\\
0 & 4
 \end{array} \right),
 \tilde{\mathbf{P}}= \left( \begin{array}{cc}
0 & 1\\
0 & 0
 \end{array} \right).
	\label{newfz2z41}
\end{equation}
For the $\phi_2$ condensed phase, the new $F$-symbol is gauge equivalent to (\ref{newfz2z41}). 
Thus the topological order in $\phi_1$ or $\phi_2$ condense phase is described by the Chern-Simons theory with $K$-matrix\cite{LinLevinstrnet}
\begin{equation}
	K= \left( \begin{array}{cccc}
0 & 0 & 2 & 0\\
0 & 0 & 0 & 4\\
2 & 0 & 0 & -1\\
0 & 4 & -1 & 0
 \end{array} \right).
	\label{}
\end{equation}

Second, we find that
\begin{equation}
	\begin{split}
	A^{a,(2,0)}&=(-1)^{a_1+a_2}, \text{other $A=1$ for $\phi_3$ condensed phase} \\
	A^{a,(2,0)}&=(-1)^{a_1}, \text{other $A=1$ for $\phi_4$ condensed phase}.
	\end{split}
	\label{}
\end{equation}
For the $\phi_4$ condensed phase, we find that $F(\tilde{a},\tilde{b},\tilde{c})$ is given by (\ref{newfz2z4})
with 
\begin{equation}
	\tilde{\mathbf{N}}= \left( \begin{array}{cc}
2 & 0\\
0 & 4
 \end{array} \right),
 \tilde{\mathbf{P}}= \left( \begin{array}{cc}
1 & 1\\
0 & 0
 \end{array} \right).
	\label{newfz2z42}
\end{equation}
The new $F$-symbol for the $\phi_3$ condensed phase is gauge equivalent to (\ref{newfz2z42}). 
Thus the topological order in $\phi_3$ or $\phi_4$ condense phase is described by the $K$-matrix
\begin{equation}
	K= \left( \begin{array}{cccc}
0 & 0 & 2 & 0\\
0 & 0 & 0 & 4\\
2 & 0 & -2 & -1\\
0 & 4 & -1 & 0
 \end{array} \right).
	\label{}
\end{equation}

\subsection{$S_3$ string-net model \label{s3exp}}

Our last two examples concern condensation transitions involving splitting.  
We begin with the $S_3$ string-net model (constructed from the fusion category Rep$(S_3)$, which has three types of strings $\{0,1,2\}$ with dual strings $\bar{0}=0,\bar{1}=1,\bar{2}=2$. The branching rules are 
\begin{equation}
	\begin{split}
	&\{(0,0;0),(1,0;1),(2,0;2), \\
	&(1,1;0)(1,1;1),(1,1;2),(1,2;1),(2,2;0)\}.
	\end{split}
	\label{s3rules}
\end{equation} 
Here the triplets $(a,b;c)$ are understood as the fusion $a\times b=c$ and are symmetric in the first two labels $a,b$. 
The nontrivial F-symbols and $d$ to self-consistency conditions (\ref{consistency}) are
\begin{equation}
	\begin{split}
	F^{111}_{1ef}= 
	\left( \begin{array}{ccc}
		\frac{1}{2} & -\frac{1}{\sqrt{2}} & \frac{1}{2} \\
		-\frac{1}{\sqrt{2}} & 0 & \frac{1}{\sqrt{2}} \\
		\frac{1}{2} & \frac{1}{\sqrt{2}} & \frac{1}{2}
	\end{array} \right) \\
	F^{111}_{211}=F^{112}_{111}=F^{121}_{111}=F^{211}_{111}=-1 \\
	d_0=d_2=1, d_1=2
	\end{split}
	\label{}
\end{equation}
where the matrix indices $e,f$ can be $0,1,2$. 

The model has 8 quasiparticles. Among them, there is a $\mathbb{Z}_2$ abelian boson, which we denote $\phi=(2,0)$. The corresponding string operator is defined by the string parameter
\begin{equation}
	w_\phi(a)=(-1)^a.
	\label{}
\end{equation}
Since $2 \times 1 = 1\times 2 = 1$, condensing $\phi$ will cause the original string label $1$ to split into two distinct labels, which we denote $\tilde{1}_1, \tilde{1}_2$.  

To describe the Hilbert space $\tilde{\mathcal{H}}$ after condensation, we first
solve (\ref{Eq:Acond2s}) for $A^{s^ia}$.  The 
two distinct solutions are: 
\begin{equation}
	(A^{21})_1= (A^{12})_2 =  1,\quad (A^{12})_1 =(A^{21})_2 =- 1.
	\label{asols3}
\end{equation}
Thus, the new string labels for $\tilde{\mathcal{H}}$ are 
\begin{equation}
	\tilde{0}=\{0,2\},\quad \tilde{1}_1=\{1\},\quad \tilde{1}_2=\{1\}.
	\label{labels3}
\end{equation}
The branching rules can be deduced from the branching rules for the old string labels and are given by 
\footnote{Before condensation, we have $1\times1=0+1+2$. After condensation, the fusion becomes $(\tilde{1}_1+\tilde{1}_2)\times(\tilde{1}_1+\tilde{1}_2)=\tilde{0}+\tilde{1}_1+\tilde{1}_2+\tilde{0}$.	Thus, $\tilde{1}_1,\tilde{1}_2$ can be either self-dual or not self-dual. However, from the associativity of the fusion $\tilde{1}_1\times (\tilde{1}_2\times \tilde{1}_1)=(\tilde{1}_1\times \tilde{1}_2)\times \tilde{1}_1$, we conclude $\tilde{1}_1\times \tilde{1}_2=\tilde{0},\tilde{1}_1\times \tilde{1}_1=\tilde{1}_2,\tilde{1}_2\times \tilde{1}_2=\tilde{1}_1$.  This can also be deduced directly from Eq. (\ref{Eq:Ac2}).}
\begin{equation}
	\{(\tilde{0},\tilde{1}_1;\tilde{1}_1),
	(\tilde{0},\tilde{1}_2;\tilde{1}_2),
	(\tilde{1}_1,\tilde{1}_2;\tilde{0}),
	(\tilde{1}_1,\tilde{1}_1;\tilde{1}_2),
	(\tilde{1}_2,\tilde{1}_2;\tilde{1}_1)\}
	\label{branchs3}
\end{equation}
Thus the condensed phase has $\mathbb{Z}_3$ (abelian) branching rules. 

Next, we want to find the topological order in the $\phi$ condensed phase.
A solution for the full vertex coefficients is given by Eq. (\ref{asols3}), together with:
\begin{equation}
\begin{split}
	&(A^{1 1}_{0})^{1,2} =(A^{1 1}_{0})^{2,1}=(A^{11}_2)^{1,2}=-(A^{11}_2)^{2,1}=1 \nonumber \\ 
	& (A^{11}_{1})^{1,1}_2 (A^{11}_{1})^{2,2}_1=\sqrt{2},
	\label{}
	\end{split}
\end{equation} 
where $(A^{1 1}_{0})^{1,2} $ is the coefficient that is relevant to the $(\tilde{1}_1, \tilde{1}_2; 0)$ vertex in the condensed phase, $ (A^{11}_{1})^{1,1}_2$ is relevant to the $(\tilde{1}_1, \tilde{1}_`; \tilde{1}_2)$ vertex, and so on.  
Using this data, it is possible to solve (\ref{main}) for the fusion data in the condensed phase:
\begin{equation}
	\begin{split}
		F^{\tilde{1}_a \tilde{1}_a \tilde{1}_a}_{0 \tilde{1}_b \tilde{1}_b}
		=F^{\tilde{1}_a \tilde{1}_a \tilde{1}_b}_{\tilde{1}_a\tilde{1}_b 0}
		=F^{\tilde{1}_a \tilde{1}_b \tilde{1}_a}_{\tilde{1}_a 0 0}
		=F^{\tilde{1}_b \tilde{1}_a \tilde{1}_a}_{\tilde{1}_b 0 \tilde{1}_b}=1
	\end{split}
	\label{}
\end{equation}
for $a,b\in\{1,2\}$.
With the data, we can construct ground states and lattice Hamiltonian for the condensed phase. It turns out the topological order in the condensed phase is described by the $\mathbb{Z}_3$ string-net model charactered by the $K$-matrix
\[K= \left( \begin{array}{cc}
0 & 3\\
3 & 0
 \end{array} \right).\]
In other words, this is an untwisted ($p=0$) $\mathbb{Z}_3$ string net model.  

\subsection{$SU(2)_4$ string-net model \label{section:su2_4}}
The $SU(2)_4$ string-net model has five string types $\{0,1,2,3,4\}$ with all strings being self dual. The branching rules are
\begin{equation}
	\begin{split}
	&\{(a,0;a) \text{ for }a=0,1,2,3,4,\\
	&(a,a;0),(a,a;2),(a,2;1),(a,2;3) \text{ for }a=1,3,\\
	&(1,3;2),(1,3;4),(1,4;3),(3,4;1) \\
	&(2,2;0),(2,2;2),(2,2;4),(2,4;2),(4,4;0) \}
	\end{split}
	\label{su2rules}
\end{equation}
If we only keep the even labels, the above branching rules (\ref{su2rules}) are the same as the branch rules for the $S_3$ model (\ref{s3rules}). The data $\{F,d\}$ satisfying (\ref{consistency}) are known and we refer the readers to Ref. \onlinecite{BondersonThesis} for details.

The model has 25 particles. Among these, there is one $\mathbb{Z}_2$ abelian boson, which we denote $\phi=(4,0)$. The corresponding string operator is defined by the string parameter
\begin{equation}
	w_\phi(a)=(-i)^a.
	\label{}
\end{equation}
We consider the string net obtained by condensing $\phi=(4,0)$.

Since the string labels obey $2 \times 4 = 4 \times 2 = 2$, the label $2$ will split into two distinct string types after condensation, which we denote $\tilde{2}_1$ and $\tilde{2}_2$.  
 We first define $\tilde{\mathcal{H}}$ after condensation. Specifically, solving for the vertex coefficients $A^{4,a}$, we obtain:
\begin{equation}
	\begin{split}
	A^{41}=1,\quad A^{43}=-1,\quad A^{14}=A^{34}=-i,\\
	(A^{42})_1 = (A^{24})_2 = 1,\quad (A^{24})_1 =(A^{42})_2 =- 1.
	\end{split}	
	\label{asolsu2}
\end{equation}
Thus, the new string labels for $\tilde{\mathcal{H}}$ are
\begin{equation}
	\tilde{0}=\{0,4\},\quad
	\tilde{1}=\{1,3\},\quad
	\tilde{2}_1=\{2\},\quad
	\tilde{2}_2=\{2\}.
	\label{labelsu4}
\end{equation}
The new branching rules can be deduced from the old branching rules, together with Eq. (\ref{Eq:Ac2}), and are given by
\begin{equation}
	\begin{split}
		\{(\tilde{1},\tilde{1};\tilde{0}),(\tilde{1},\tilde{1};\tilde{2}_1),(\tilde{1},\tilde{1};\tilde{2}_2),\\
	(\tilde{1},\tilde{2}_1;\tilde{1}),(\tilde{1},\tilde{2}_2;\tilde{1}),\\
	(\tilde{2}_1,\tilde{2}_1;\tilde{2}_2),(\tilde{2}_1,\tilde{2}_2;\tilde{0}),(\tilde{2}_2,\tilde{2}_2;\tilde{2}_1).
	\}
	\end{split}
	\label{branchsu4}
\end{equation}
Thus, $\tilde{\mathcal{H}}$ is the string-net Hilbert space with new string labels (\ref{labelsu4}) and branching rules (\ref{branchsu4}).

Next, we want to find $\{F,d\}$ in the $\phi$ condensed phase.
A valid choice of the full vertex coefficients is given by (\ref{asolsu2}), together with: 
\begin{equation}
	\begin{split}
	(A^{11}_{2})_1&=-\sqrt{2}, \quad (A^{12}_1)^{1}=-\frac{1}{\sqrt{2}}, \quad (A^{22}_{2})^{1 1}_2=\frac{1}{\sqrt{2}},\\
	(A^{11}_{2})_2&=1, \qquad (A^{12}_1)^2=1, \qquad (A^{22}_{2})^{2 2}_1=-2,
	\end{split}
	\label{}
\end{equation}
where $ (A^{12}_1)^2=1$ pertains to the vertex $(\tilde{1}, \tilde{2}_2; \tilde{1})$ and so on.  
Using these, and Eq. (\ref{main1split}), 
we find the nontrivial new $F$-symbols are 
\begin{equation}
	\begin{split}
		&F^{\tilde{1}\tilde{1}\tilde{1}}_{\tilde{1}00}=
		F^{\tilde{1}\tilde{1}\tilde{1}}_{\tilde{1}0\tilde{2}_a}=
		F^{\tilde{1}\tilde{1}\tilde{1}}_{\tilde{1}\tilde{2}_a0}=-\frac{1}{\sqrt{3}},    
		\quad F^{\tilde{1}\tilde{1}\tilde{1}}_{\tilde{1}\tilde{2}_a\tilde{2}_b}=-\frac{1}{\sqrt{3}}e^{-i \frac{2\pi ab}{3}},\\
		&F^{\tilde{1}\tilde{2}_a \tilde{1}}_{\tilde{2}_b\tilde{1}\tilde{1}}=F^{\tilde{2}_a\tilde{1}\tilde{2}_b}_{\tilde{1}\tilde{1}\tilde{1}}=e^{-i \frac{2\pi ab}{3}},
		\qquad F^{\tilde{1}\tilde{2}_a\tilde{2}_b}_{\tilde{1}\tilde{1}0}=F^{\tilde{2}\tilde{2}_b\tilde{1}}_{\tilde{1}0\tilde{1}}=-1.
	\end{split}
	\label{fsu2}
\end{equation}
Here $a\neq b=1,2$. Interestingly, the data (\ref{fsu2}) are exactly the $\mathbb{Z}_3$ Tambara-Yamagami category\cite{TYcat} ($TY_3$). The $TY_3$ category has 4 labels $[0]=\tilde{0},[1]=\tilde{2}_1,[2]=\tilde{2}_2,\sigma=\tilde{1}$. The first 3 labels have $\mathbb{Z}_3$ fusion rules. The last label $\sigma$ represents the symmetry defect:
\begin{equation}
	\begin{split}
		[a]\times \sigma &=\sigma \\
		\sigma \times \sigma &=[0]+[1]+[2].
	\end{split}
	\label{}
\end{equation}
With the data (\ref{fsu2}), we can construct the ground state and effective string-net model for the $\phi$ condensed phase. The braiding data of excitations, the S, T matrices, are known in Ref. \onlinecite{TYcatST}.

Thus, condensing the $\mathbb{Z}_2$ boson in the SU(2)$_3$ string net, we obtain the $TY_3$ string net.  In this case, because the input fusion category is modular, this transition is analogous to condensing the $\mathbb{Z}_2$ boson in the top layer of an SU(2)$_4 \times \overline{\text{SU(2)}}_4$ bilayer system.  The resulting topological order is SU(3)$_1 \times \overline{\text{SU(2)}}_4$, which is exactly that of the $TY_3$ string net.

\section{Discussion}

In this paper, we have systematically studied condensation of arbitrary abelian bosons in string-net models.  We have introduced a Hamiltonian that tunes the system through a condensation transition, and given a detailed description of the string net in the condensed phase.   
We have shown how, in the low-energy Hilbert space of the condensed phase, the input fusion category $\mathcal{C}$ of the uncondensed string net becomes a new fusion category $\mathcal{\tilde{C}}$, with both the effective Hamiltonian and the  ground state in the condensed phase being $\mathcal{\tilde{C}}$ string nets. Finally, we have shown how both the labels and the fusion data for $\tilde{\mathcal{C}}$ can be calculated directly from the data of the string operators of the condensing bosons, together with the fusion data of $\mathcal{C}$.  

Because the transitions discussed here involve condensation of abelian bosons, the degrees of freedom that become gapless at the critical point can all be mapped onto variables in a Potts model, using a method similar to that described in Ref. \cite{TSBLong}.  By modifying $H_1$, one could also achieve phase transitions in the clock universality class.  

One useful result of our construction is the possibility of systematically extracting not only the label set, but also the fusion data of $\tilde{\mathcal{C}}$, by solving for the vertex coefficients implied by the string operators $W^a_{\phi^j}$.  
We note that Ref. \onlinecite{Eliens} similarly introduced vertex coefficients when studying the effect of anyon condensation on the fusion and braiding data of the UMTC describing the topological order, and used these to determine the $F$ and $R$ symbols for the condensed theory. 
The vertex coefficients that we introduce here can be viewed as analogs of Ref. \onlinecite{Eliens}'s vertex coefficients, albeit for the fusion category underpinning the string net, rather than for the UMTC associated with the anyon model itself.  

One potential application of our construction, illustrated in the last example, is to obtain the fusion data for string nets of lower symmetry by condensing anyons in string nets with higher symmetry.  For example, we can begin with a string net that has explicit time-reversal symmetry, such as SU(2)$_4 \times \overline{\text{SU(2)}}_4$, and condense a chiral abelian boson in one of the two copies, to obtain a string net that does not have time reversal symmetry.  This is useful because the data for many high-symmetry string nets, such as those constructed from rational conformal field theories, is known.  

A second potential application is to string nets realizing symmetry enriched topological phases, where the enriching symmetry is abelian.  Specifically, condensing $\mathbb{Z}_p$ abelian anyons can be viewed as ``un-gauging" a $\mathbb{Z}_p$ symmetry, and a modification of the construction here can lead to condensed phases in which the models exhibit a global $\mathbb{Z}_p$ symmetry, similar to the constructions of Refs. \onlinecite{Heinrich16,ChengAPS}.  Such a construction may enable a simpler string-net realization of many of these symmetries than in the existing literature.  It also gives a framework that could be used to construct similar models with anyon-permuting symmetry at the boundary of a three-dimensional Walker-Wang string net.

{\bf Note added} Shortly before completing this work, we became aware of Ref. \cite{HustonCond}, which also discusses anyon condensation in string net models, including some non-abelian examples.    

\section*{acknowledgments}
C.-H. Lin thanks Lan Tian, Chenjie Wang, and Yidun Wan for helpful discussions.  FJB is grateful for the support of NSF DMR 1928166.

\appendix

\section{Properties of Abelian string operators \label{app:string}}
In this section, we prove the basic properties of our abelian string operators that we use in the main text.

\subsection{Finding a gauge where $F$-symbols are trivial}

Many of the properties of abelian string operators that we will use are valid only in the gauge where $F (s^i, s^j, s^k) =1$, where we use the notation $F(a,b,c) = F^{a b c}_{abc, ab, bc}$ appropriate to $F$-symbols involving only abelian string labels.  To see that such a gauge exists, we use the fact that if $\phi = (s,m)$ to be a boson, then  $w_\phi(s) =1$.  (We note that while $w_\phi(a)$ is not gauge invariant for general $a$, $w_\phi(s)$, which represents the self-twist of the particle, is a gauge-invariant quantity). 
From Eq. (\ref{weqs}), we see that
\begin{equation}
w_\phi(s) w_\phi(s^j) = F(s, s^j,s) w_\phi (s^{j+1})
\end{equation}
If $s^2 = 0$ and $j=1$, $F(s, s^j,s)$ is gauge invariant, and this tells us that only if $F(s, s^j,s) =1$ can $(s,m)$ be a boson.  Otherwise, under gauge transformations, we have
\begin{equation}
\hat{F}(s, s^j,s) = F(s, s^j,s)  \frac{ f^{s s^j}_{s^{j+1}} }{f^{s^j s}_{s^{j+1}}} \frac{f^{s^{j+1} s}_{s^{j+2}}}{f^{s s^{j+1}}_{s^{j+2}}}
\end{equation}
where our string net construction requires $f^{s 0}_s = f^{0 s}_s = 1$.  
For $1 \leq j < p-1$, we can use the ratio  $ f^{s^j+1s }_{s^{j+2}} / (f^{ss^{j+1} }_{s^{j+2}})$ to fix $F(s, s^j, s) = 1$, where $s^p = 0$.  
 Further, we have $w_\phi(s) w_\phi(s^{p-1}) = F(s, s^{p-1},s) w_\phi (0)$, and hence also $F(s, \bar{s}, s) = 1$, where $\bar{s} = s^{p-1}$.  
It follows that in this gauge, for all $i,j$, we have 
\begin{equation}
F(s, s^i, s) = 1 \ , \ \ \ F(s,s^i, s^j ) F(s^i, s^j, s) = F(s^i, s, s^j) \ .
\end{equation}
 In this gauge, we see that $w_\phi(s^j) = 1$ for all $j$. 
 
Next, consider $F(s, s^j, s^k)$ with $k>1$.  
Under gauge transformations, we have:
\begin{equation}
\hat{F}(s, s^j,s^k) = F(s, s^j,s^k)  \frac{ f^{s s^j}_{s^{j+1}} }{f^{s s^{j+k}}_{s^{j+k+1}}} \frac{f^{s^{j+1} s^k}_{s^{j+k+1}}}{f^{s^j s^k}_{s^{j+k}}}
\end{equation}
For $k>1$, and a fixed choice of $f^{s s^i}_{s^{i+1}}$ for each $i$, we can set all of these to $1$ by fixing the ratio $\frac{f^{s^{j+1} s^k}_{s^{j+k+1}}}{f^{s^j s^k}_{s^{j+k}}}$.  (In this case, this also works for $j= p-1$).  

Thus, if $w_\phi(s) = 1$, we have enough gauge freedom to simultaneously set $F(s, s^j, s^k) = 1$ for all $j,k$.  Using the pentagon relation, we also have
\begin{align}
F(s,s^i, s^j) F(s,s^{i+j}, s^k) F(s^i, s^j, s^k)  \nonumber \\
= F(s^{i+1}, s^j, s^k) F(s, s^i, s^{j+k})
\end{align}
In the gauge where $F(s, s^j, s^k) = 1$ for all $j,k$, we find that 
\begin{equation}
F(s^i, s^j, s^k) = F(s^{i+1}, s^j, s^k) 
\end{equation}
from which it follows that $F(s^i, s^j, s^k) = 1$ for all $i,j,k$.

\subsection{Basic string operators in a general gauge}

The gauge choice $F(s^i, s^j, s^k) = 1$ is convenient, because the action of the operator $W_\phi(P)$ is identical to acting with a string operator with a fixed end-point that is located away from the stick, and fusing it into the lattice appropriately.  With a different gauge choice, the difference between the operator $W_\phi(P)$ and such open string operators can be described by a gluing operator $O_l$,  
whose action is defined by
\begin{equation}
	\begin{split}
	\left<\raisebox{-0.22in}{\includegraphics[height=0.5in]{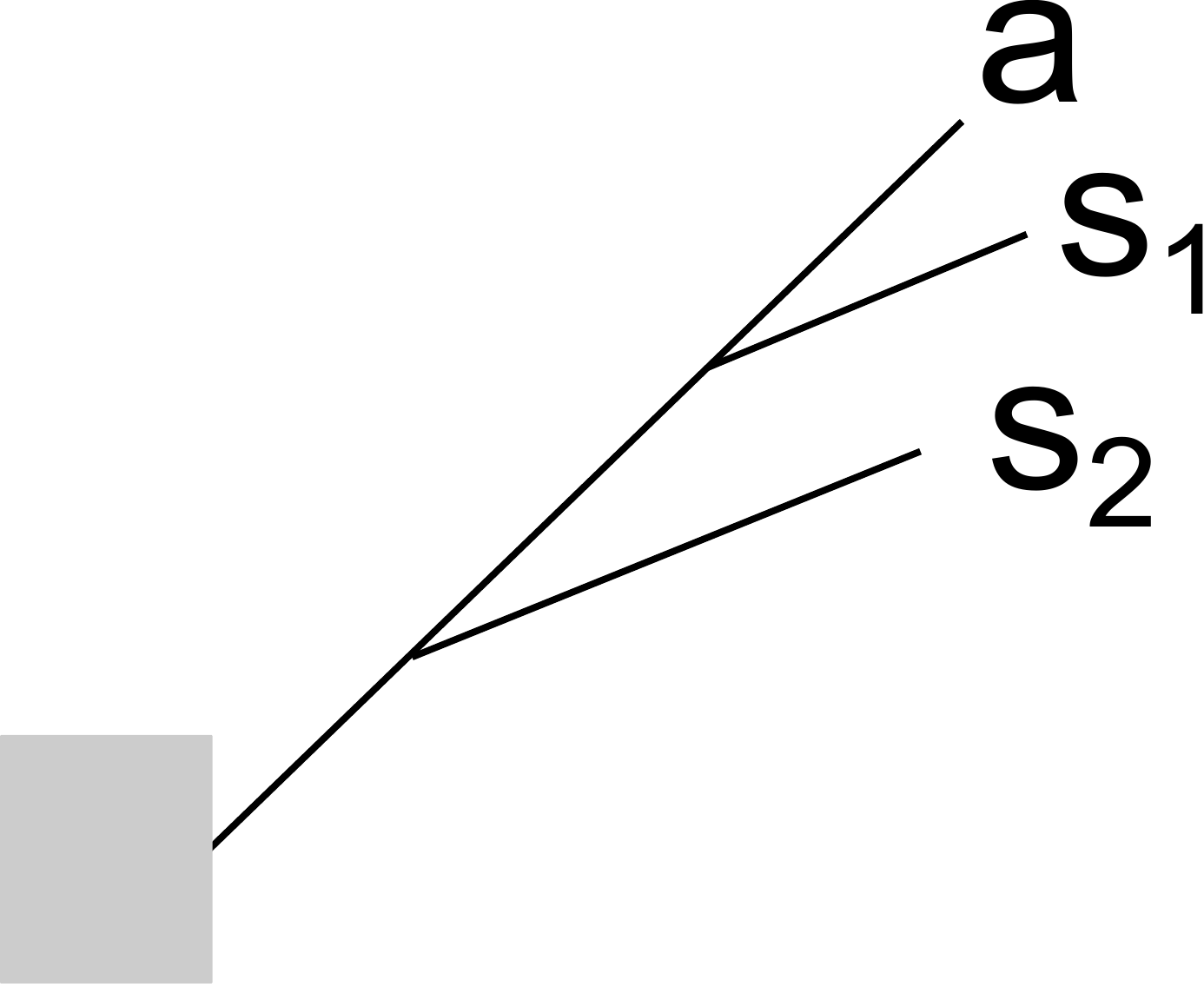}} \right|O_l
	&=F(a,s_1,s_2)\left<\raisebox{-0.22in}{\includegraphics[height=0.5in]{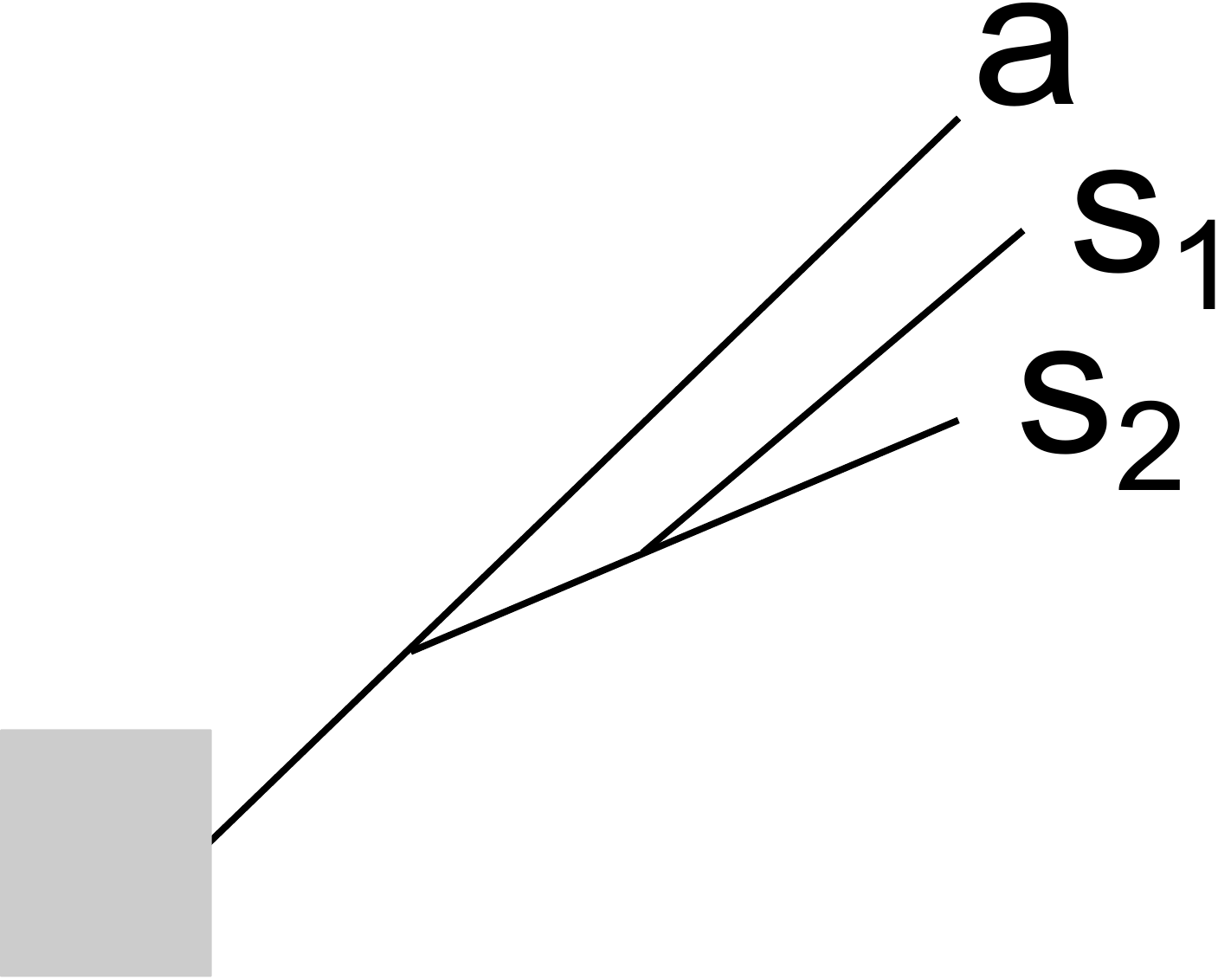}} \right| \\
	\left<\raisebox{-0.22in}{\includegraphics[height=0.5in]{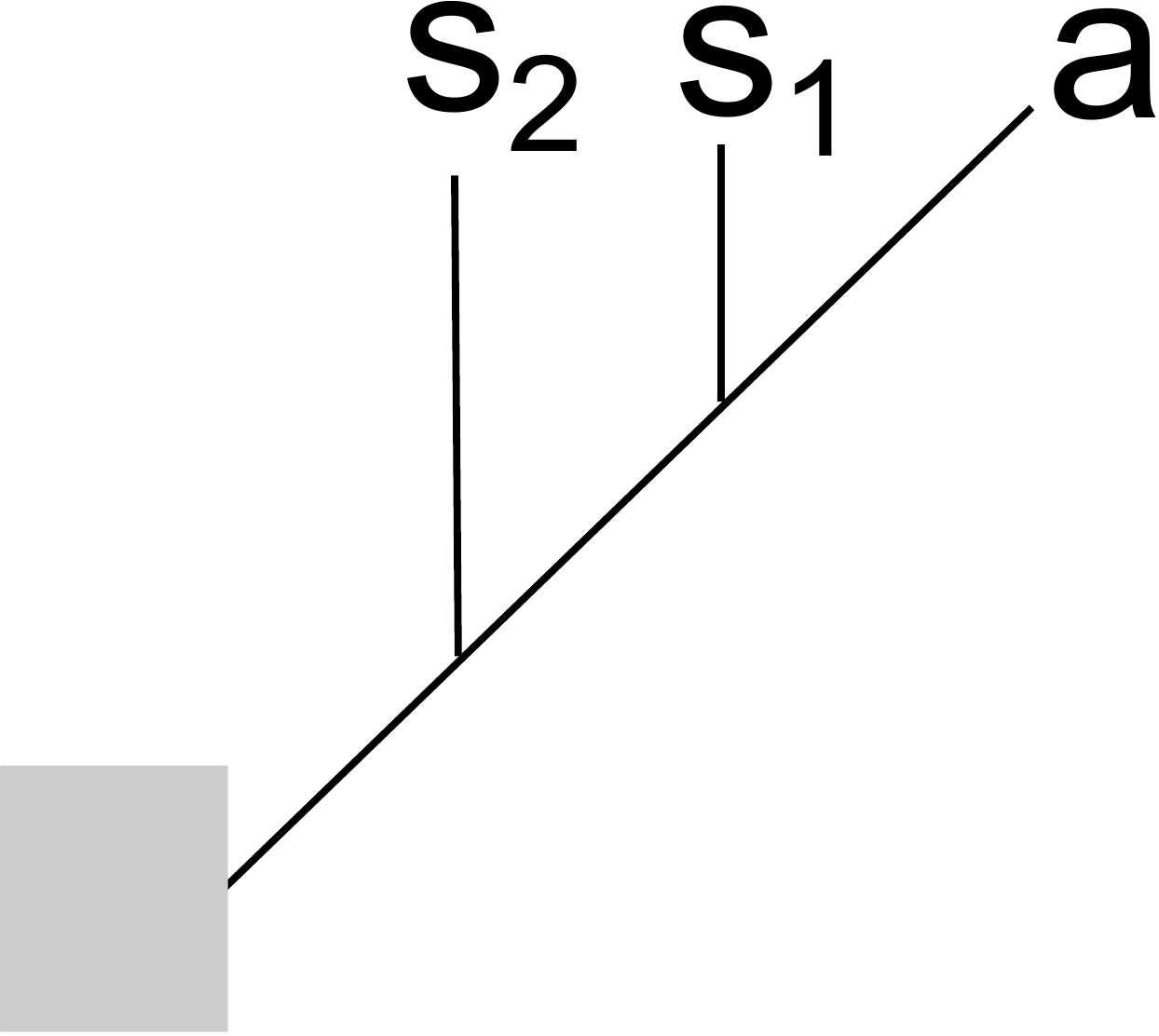}} \right|O_l
	&=F(s_2,s_1,a)^{-1}\left<\raisebox{-0.22in}{\includegraphics[height=0.5in]{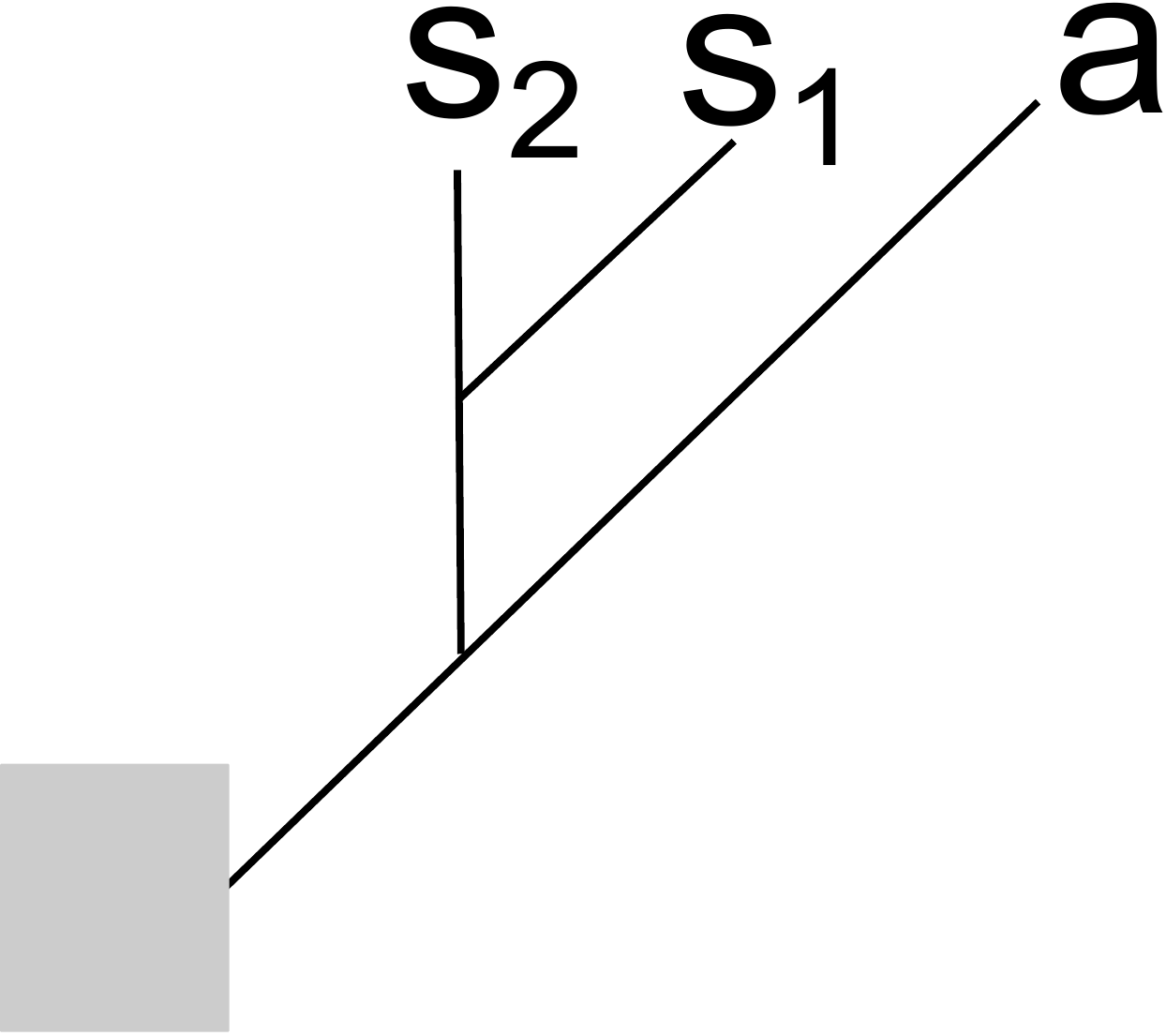}} \right| \\
	\left<\raisebox{-0.22in}{\includegraphics[height=0.5in]{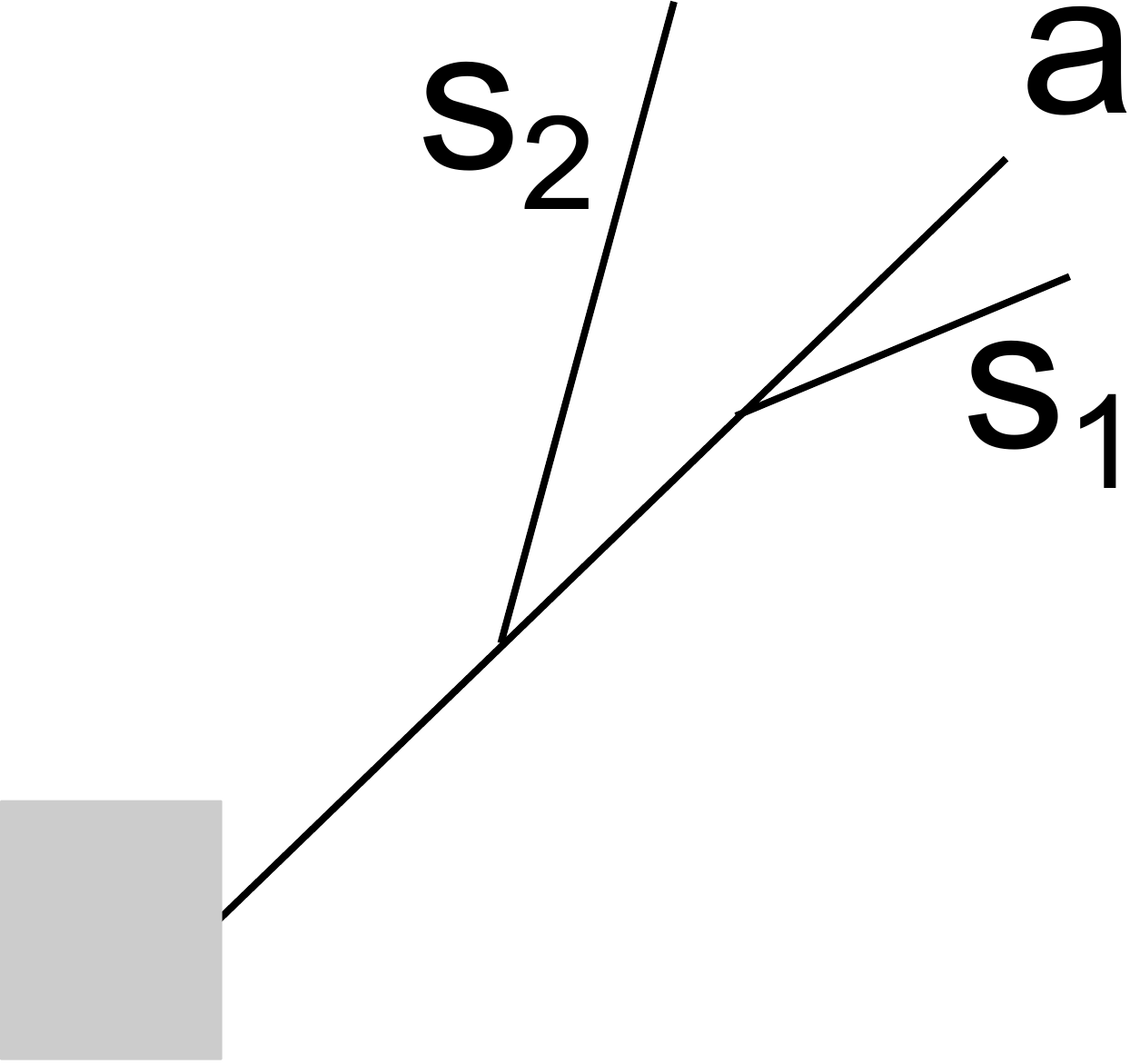}} \right|O_l
	&=\left<\raisebox{-0.22in}{\includegraphics[height=0.5in]{stick3a-eps-converted-to.pdf}} \right| \\
	\left<\raisebox{-0.22in}{\includegraphics[height=0.5in]{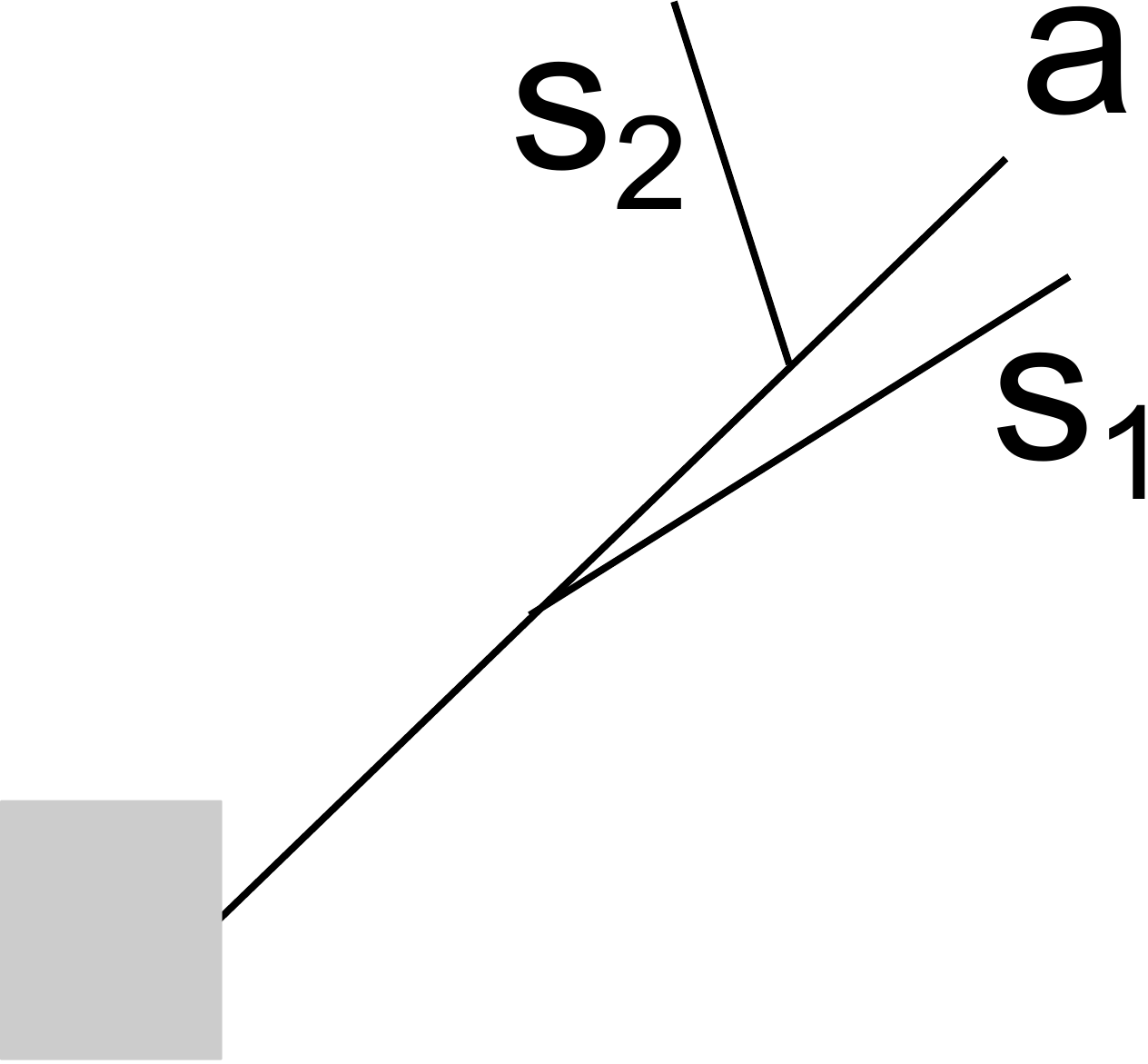}} \right|O_l
	&=F(s_2,a,s_1)\left<\raisebox{-0.22in}{\includegraphics[height=0.5in]{stick3a-eps-converted-to.pdf}} \right| 
	\end{split}
	\label{glue}
\end{equation}
Here $a$ denotes the string label of the stick,  and 
the grey region denotes the configuration which does not change.

In addition, in this gauge, when taking the product $W_{\phi^1}^i \cdot W_{\phi^2}^i$, we may ignore the vertical bendings of the jointed path $p^i\cup p^j$. For example, consider two basic string operators $W_{\phi^1}^i,W_{\phi^2}^i$ along the same path $p_i$. When acting the composite operator $W_{\phi^1}^i \cdot W_{\phi^2}^i$ on the vacuum state, we have
\begin{equation}
	\begin{split}
	\<vac|W_{\phi^1}^i\cdot W_{\phi^2}^i&=
	\left<\raisebox{-0.16in}{\includegraphics[height=0.4in]{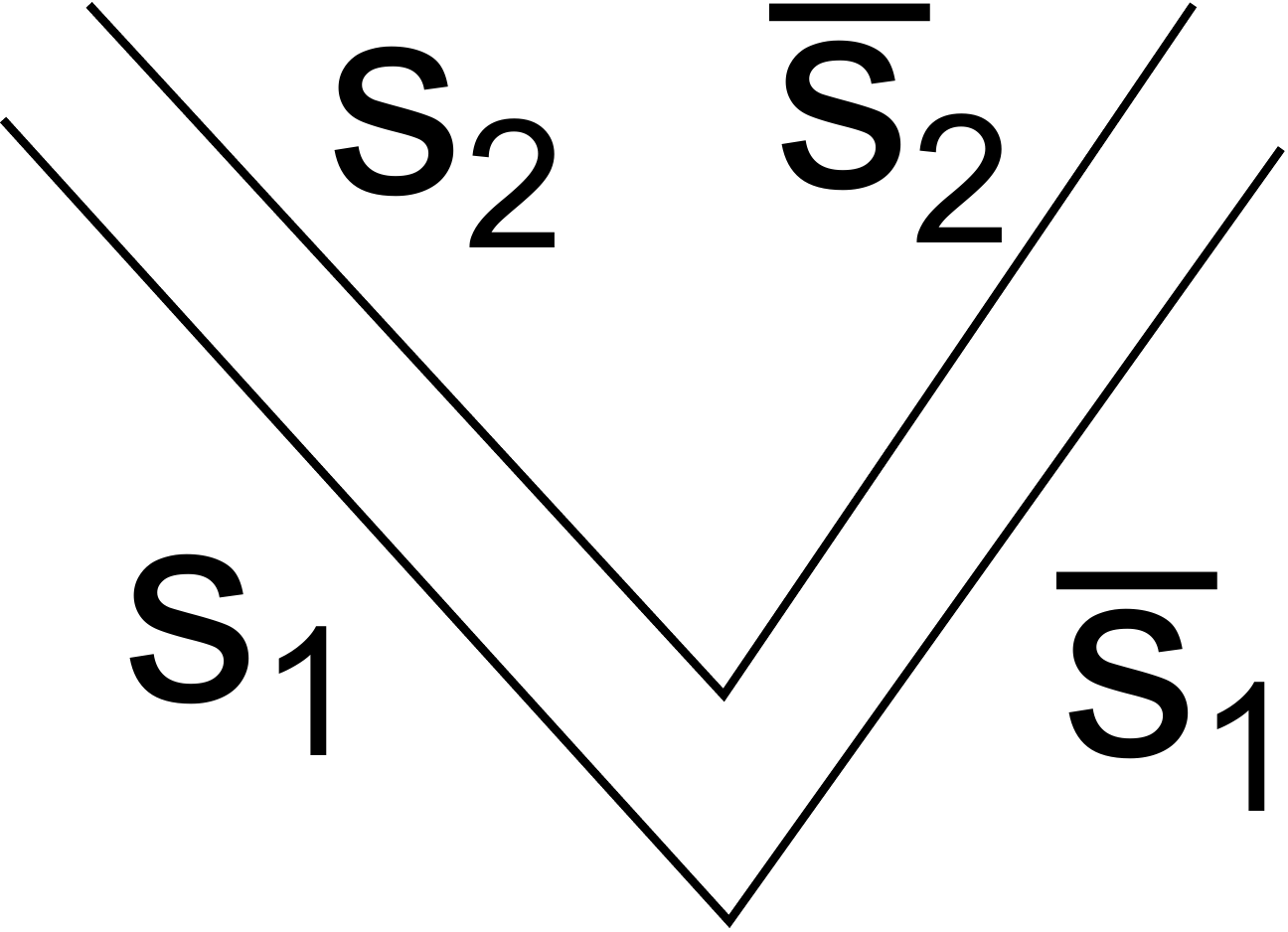}} \right|\\
	&=\left<\raisebox{-0.16in}{\includegraphics[height=0.4in]{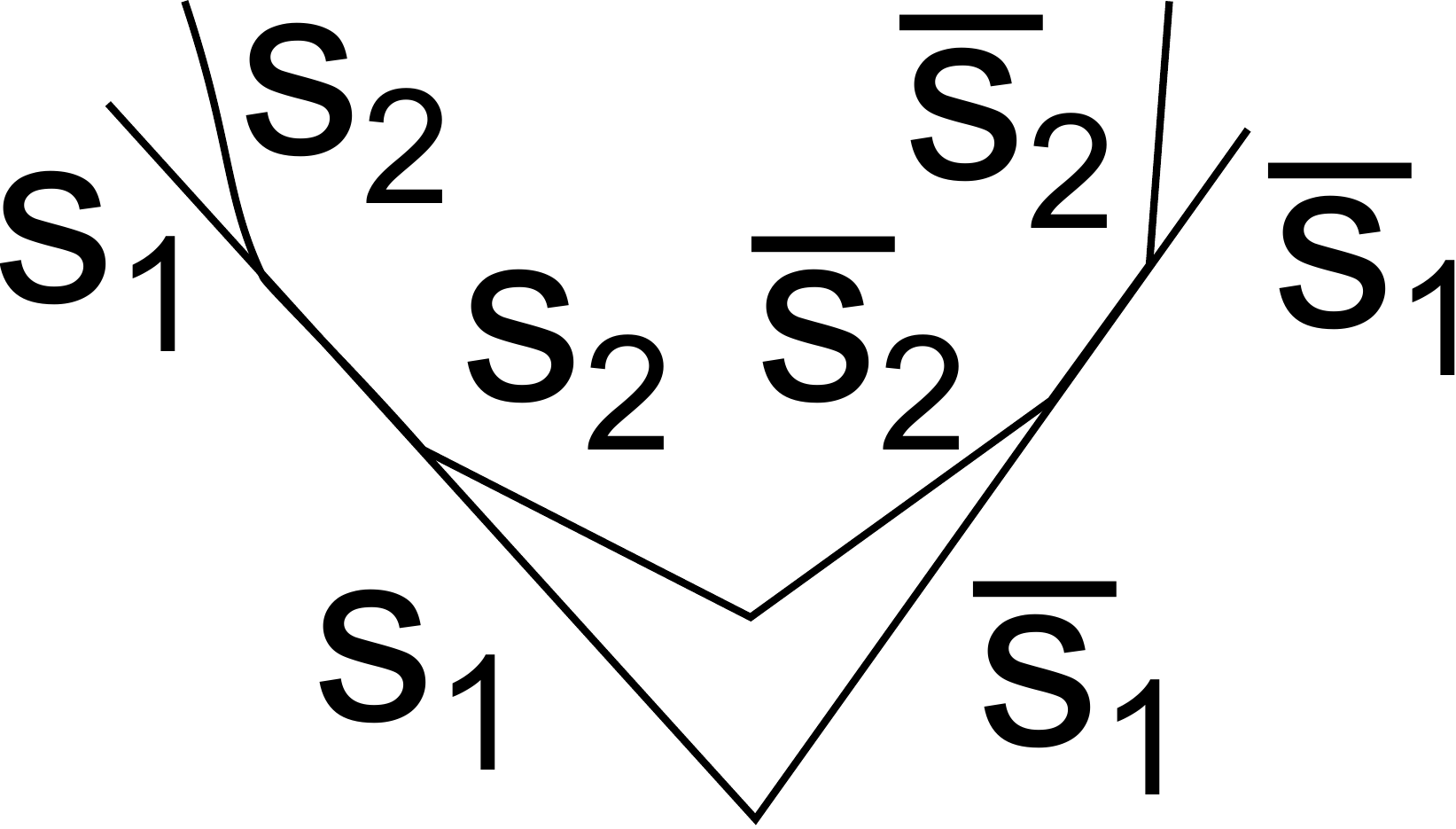}} \right|\\
	&=\left<\raisebox{-0.16in}{\includegraphics[height=0.4in]{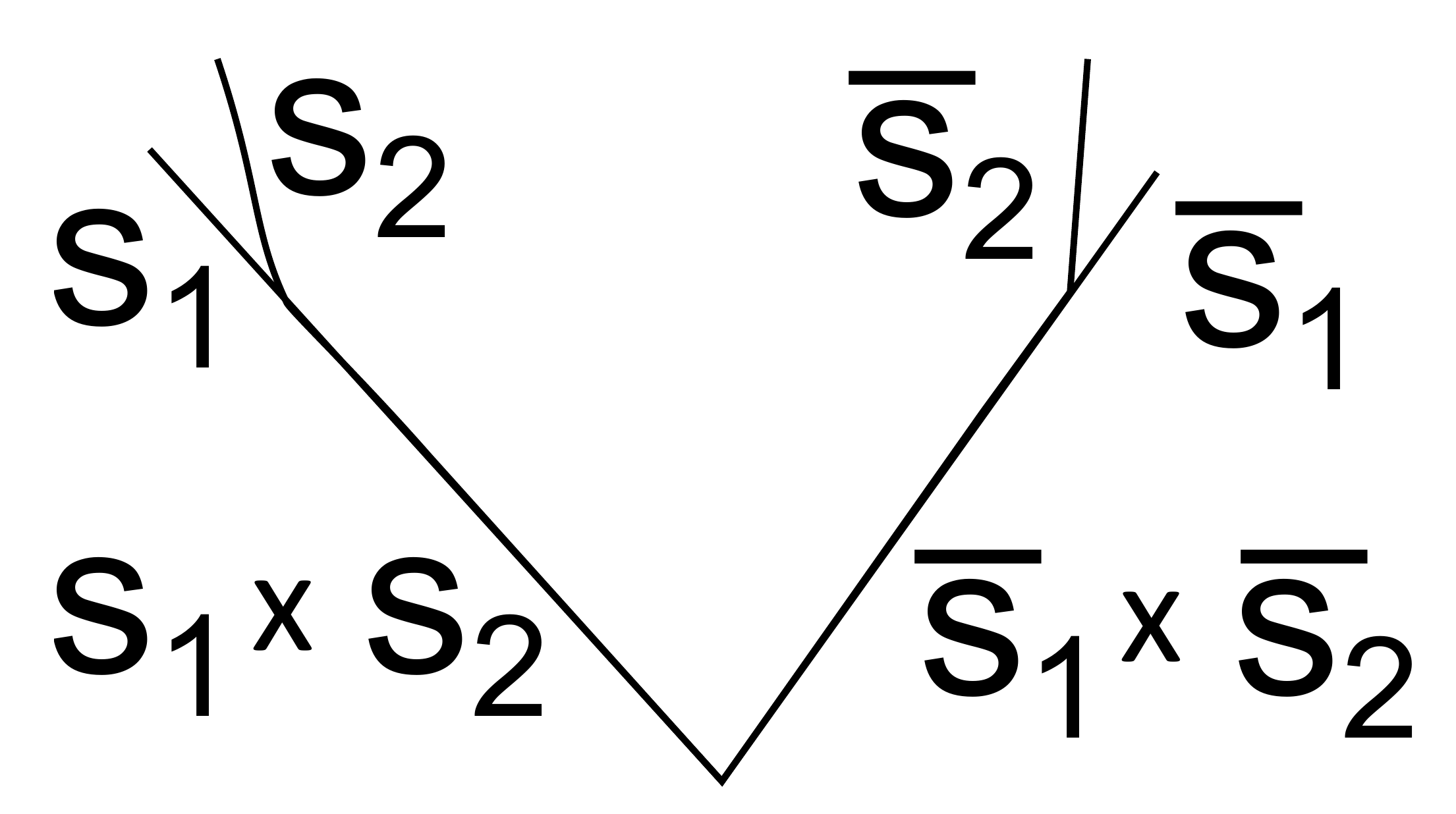}} \right| \theta_{s_1,s_2}\\
	&=\<vac|W_{\phi^1\times \phi^2}^i \cdot \theta_{s_1,s_2}
	\end{split}
	\label{}
\end{equation}
with
\begin{equation}
	\theta_{s_1,s_2}=\frac{F(s_2,\bar{s}_2,\bar{s}_1)}{F(s_1,s_2,\bar{s}_1\times \bar{s}_2)}.
	\label{}
\end{equation}
Here we use (\ref{localrules}) to remove the loop in the second line at the expense of the factor $\theta_{s_1,s_2}$.

Thus, in a general gauge, we do not have $W_{\phi^1}^i\cdot W_{\phi^2}^i=W_{\phi^1\times \phi^2}^i$.
Instead, we find:
\begin{equation}
	\begin{split}
	&W_{\phi^3=\phi^1\times \phi^2,a_3b_{\bar{3}}c_3d_3;\phi^a\times \phi^3,\phi^b\times \overline{\phi}^3}^{i,abcd;\phi^a\phi^b}(efg)= \\
	&\qquad \qquad W_{\phi^1,a_1b_{\bar{1}}c_1d_1;\phi^{a}\times \phi^1,\phi^{b}\times \overline{\phi}^1}^{i,abcd;\phi^a\phi^b}(efg) \times \\
	&\qquad \qquad W_{\phi^2,a_3b_{\bar{3}}c_3d_3;\phi^a\times \phi^3,\phi^b\times \overline{\phi}^3}^{i,a_1b_{\bar{1}}c_1d_1,\phi^a\times \phi^1,\phi^b\times \overline{\phi}^1}(efg) \times \theta_{s_1,s_2,a,b} ^{-1}
	\end{split}
	\label{w3w1w2}
\end{equation}
with 
\begin{equation}
	\theta_{s_1,s_2,a,b}=\frac{F(s_2,\bar{s}_2,\bar{s}_1)F(\bar{s}_2,\bar{s}_1,b)}
	{F(s_1,s_2,\bar{s}_1\times \bar{s}_2)F(a,s_1,s_2)}
	\label{theta0}
\end{equation}
for $i=1,\dots,4$.

In addition, one can show that
\begin{equation}
	\begin{split}
	&W^{i \dagger, a'b'c'd';\phi^{a'}\phi^{b'}}_{\phi,abcd;\phi^a\phi^b}  (efg)= \\
	&\qquad \qquad W^{i,a'b'c'd';\phi^{a'}\phi^{b'}}_{\bar{\phi},abcd;\phi^a\phi^b}(efg) \cdot 
	\theta_{s,\bar{s},a,b}^{-1}
\end{split}
	\label{wdagger1}
\end{equation}
for $i=1,2,3,4$.

\subsection{Properties of basic string operators in the gauge $F(s^i, s^j, s^k ) = 1$}

Next, we establish the properties of basic string operators in the gauge 
\begin{equation}
	F(s^i,s^j,s^k)=1  \ .
	\label{f1a}
\end{equation}
First, from equations (\ref{w3w1w2},\ref{wdagger1}), we see that in this gauge,
\begin{equation}
	W_{\phi^1}^i\cdot W_{\phi^2}^i =W_{\phi^1+\phi^2}^i
	\label{eq:Wprods}
\end{equation}
and
\begin{equation}
	W_{\phi}^{i\dagger}=W_{\bar{\phi}}^i.
	\label{}
\end{equation}

Second, all basic string operators commute:
\begin{equation}
	[W_{\phi}^i,W_{\phi'}^j]=0.
	\label{}
\end{equation}
This follows from Eq. (\ref{w3w1w2}) if $i=j$ (i.e. if the two paths are the same).  If the two paths intersect only on one stick (for example, $i=1, j=3$), this follows from the fact that using Eq. (\ref{consistency}), one can show that the two operator products differ by a factor of $F(s^i, b\times s, s^j)$, which is unity if $b$ labels a stick.  If $i=1, j=2$, we can use the identity
\begin{equation}
w_\phi(f) w_\phi(s) = F^{s f s}_{f'', f', f'} w_\phi(f \times s)
\end{equation}
to show that $\bar{w}_\phi(f) F^{s f s}_{f'', f', f'} = \bar{w}_\phi(f \times s)$, which shows that $[W_\phi^1, W_\phi^2] = 0$.  We can use this, together with Eq. (\ref{eq:Wprods}),  to show that $[W_{\phi^i}^1, W_{\phi^j}^2] = 0$.  A similar argument shows $[W_{\phi^i}^3, W_{\phi^j}^4] = 0$.

Moreover, in this gauge we have $F^{s \bar{s} b}_{b' 0 b'} = w_{\phi}(b) = 1$ when $b$ is a string label associated with the condensing boson.  It follows that $W_{\phi}^1 W_{\phi}^3 = W_{\phi} (p_1 \cup p_3)$, where the path $p_1 \cup p_3$ crosses straight under the $b$-labeled stick.  Similar results hold for other products of simple string operators with paths that overlap only on a single stick.  Using the identity (derived from Eq. (\ref{consistency})) $F^{f \bar{s} s}_{f, f\times \bar{s},0}  F(\bar{s} s \bar{s}) = F^{f \times \bar{s}, s, \bar{s}}_{f \times \bar{s}, f, 0}$, the product $W_\phi^1 W_{\bar{\phi}}^2$  can similarly be shown to be equal to an operator running along the path $(p_1 \cup p_2)$, which directly connects the two sticks.  (The consistency relations ensure that any deformation of this path which does not change the end-points yields the same operator).   

Thus, in this gauge, we may express a general string operator by concatenating string operators along a series of adjacent basic paths.

\section{Diagrammatical representation of the $B_p^{\phi,s}$ operator \label{app:bps}}

In this section, we present the graphical representation of $B_p^{\phi,s}$ in $H_{\mathcal{C}}$ (\ref{hc1}) which lead to the matrix elements in Eq. (\ref{bp1}), as well as an alternative (simpler) formulation

The action of ${B}_p^{\phi,s}$ in $H_{\mathcal{C}}$ is defined by
\begin{equation}
	{B}_p^{\phi,s}=\sum_{\phi_{10},\phi_{11},\phi_{12}} {W}_{\phi_{10},\phi_{11},\phi_{12}} B_p^s {W}^{\dagger}_{\phi_{10},\phi_{11},\phi_{12}}.
	\label{}
\end{equation}
with
\begin{equation}
	{W}_{\phi_{10},\phi_{11},\phi_{12}}=\mathcal{P}_{\phi_{10}} W^{1}_{\phi_{10}} \cdot \mathcal{P}_{\phi_{11}} W^{1}_{\phi_{11}} \cdot \mathcal{P}_{\phi_{12}} W^{3}_{\phi_{12}}.
	\label{}
\end{equation}
Here the sums run over three end spins states $\phi_{10},\phi_{11},\phi_{12}$ in $p$. The $\mathcal{P}_{\phi_{i}}=|\phi_{i}\>\<\phi_{i}|$ is the projector to the end spin state $|\phi_{i}\>$ and $W^{1}_{\phi_{10}},W^{1}_{\phi_{11}},W^{3}_{\phi_{12}}$ are three basic string operators defined in (\ref{stringmatrix}). The $B_p^s$ is defined to add a loop-$s$ to the boundary of $p$ after $W_{\phi_{10},\phi_{11},\phi_{12}}$ moves the excitations $\{\phi_{10},\phi_{11},\phi_{12}\}$ to the exterior of $p$. Finally, after fusing the loop-$s$ to the boundary of $p$, ${W}^{\dagger}_{\phi_{10},\phi_{11},\phi_{12}}$ moves back the excitations to $p$.

Diagrammatically, the matrix elements of $B_p^{\phi,s}$ can be obtained by
\begin{widetext}
\begin{equation}
	\begin{split}
	&\left<\raisebox{-0.32in}{\includegraphics[height=0.7in]{bp2a-eps-converted-to.pdf}} \right|B_p^{\phi,s}
	=\left<\raisebox{-0.32in}{\includegraphics[height=0.7in]{bp2a-eps-converted-to.pdf}} \right|
	{W}_{\phi_{10},\phi_{11},\phi_{12}} B_p^s {W}^{\dagger}_{\phi_{10},\phi_{11},\phi_{12}}
	=C_1 \left<\raisebox{-0.32in}{\includegraphics[height=0.7in]{bp2b-eps-converted-to.pdf}} \right| B_p^s {W}^{\dagger}_{\phi_{10},\phi_{11},\phi_{12}}\\
	&=C_1 \left<\raisebox{-0.32in}{\includegraphics[height=0.7in]{bp2d-eps-converted-to.pdf}} \right| {W}^{\dagger}_{\phi_{10},\phi_{11},\phi_{12}}
	=C_1 \sum_{i_1'i_2'\dotsi_6'} C_{2_a} \left<\raisebox{-0.32in}{\includegraphics[height=0.7in]{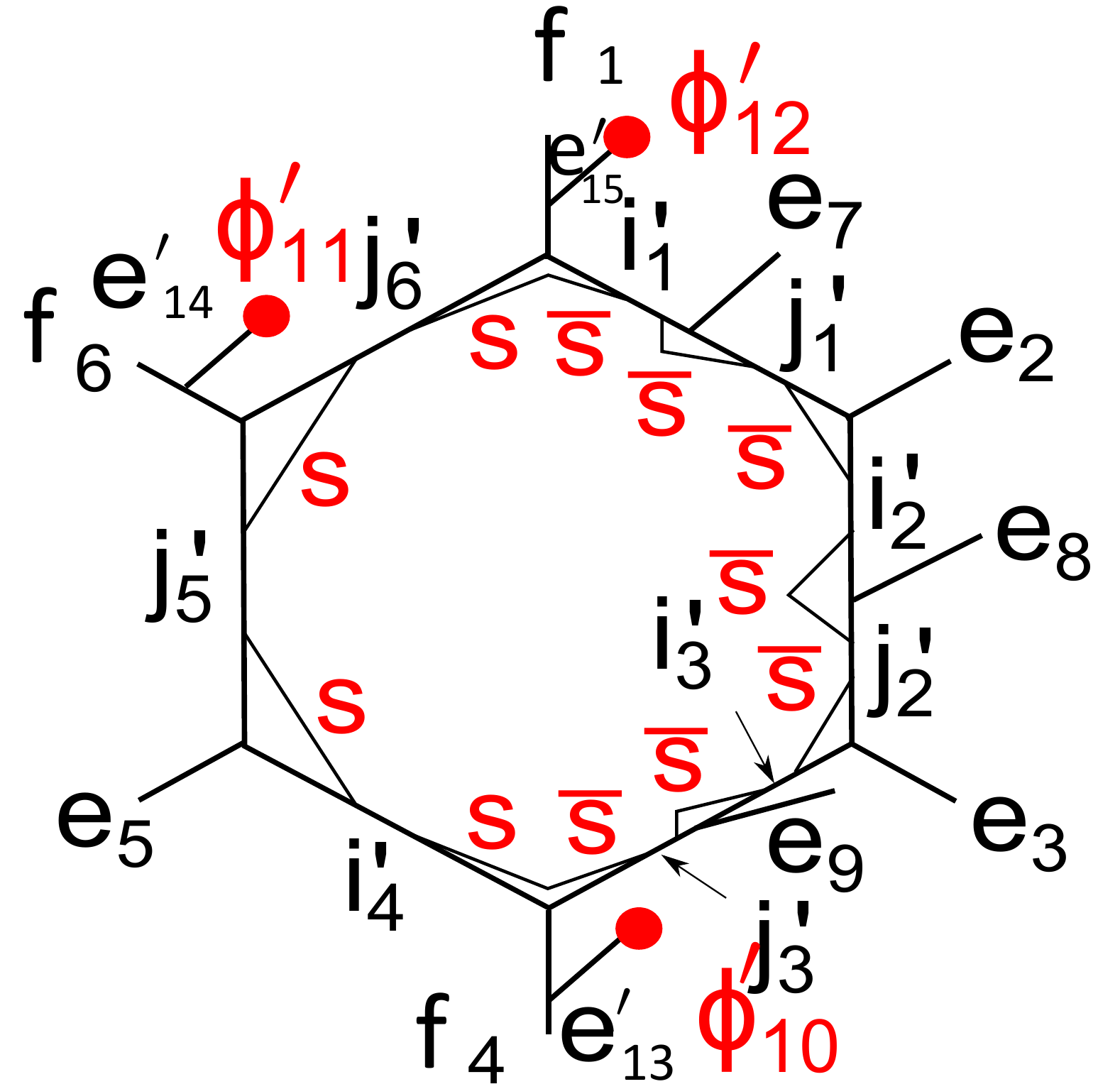}} \right| {W}^{\dagger}_{\phi_{10},\phi_{11},\phi_{12}}\\
	&=C_1 \sum_{i_1'i_2'\dotsi_6'} C_{2_a} C_{2_b}\left<\raisebox{-0.32in}{\includegraphics[height=0.7in]{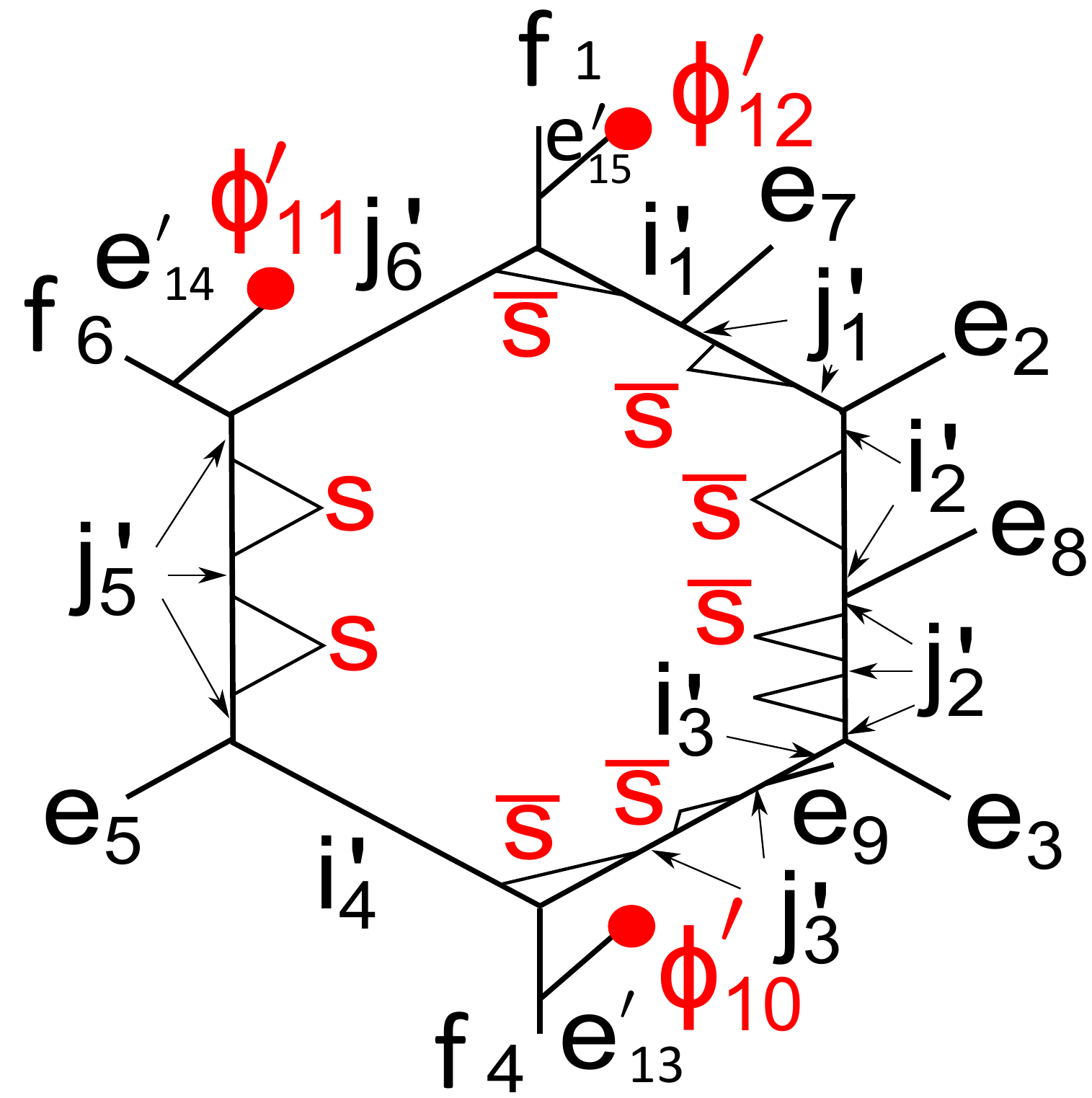}} \right|  {W}^{\dagger}_{\phi_{10},\phi_{11},\phi_{12}}
	=C_1 \sum_{i_1'i_2'\dotsi_6'} C_{2_a} C_{2_b} C_{2_c} \left<\raisebox{-0.32in}{\includegraphics[height=0.7in]{bp2g-eps-converted-to.pdf}} \right| {W}^{\dagger}_{\phi_{10},\phi_{11},\phi_{12}} \\
	&=C_1 \sum_{i_1'i_2'\dotsi_6'} C_{2_a} C_{2_b} C_{2_c} C_3 \left<\raisebox{-0.32in}{\includegraphics[height=0.7in]{bp2h-eps-converted-to.pdf}} \right| 
	\equiv\sum_{i_1'i_2'\dotsi_6'}\left<\raisebox{-0.32in}{\includegraphics[height=0.7in]{bp2h-eps-converted-to.pdf}} \right| B^{s,i_1\dots i_{6}j_1\dots j_6}_{p,i'_1\dots i'_{6}j'_1\dots j'_6}(e_1\dots e_{12};\phi_{10},\phi_{11},\phi_{12})
	\end{split}
	\label{bps}
\end{equation}
\end{widetext}where
\begin{equation}
	B^{s,i_1\dots i_{6}j_1\dots j_6}_{p,i'_1\dots i'_{6}j'_1\dots j'_6}(e_1\dots e_{12};\phi_{10},\phi_{11},\phi_{12})=C_1C_{2_a} C_{2_b} C_{2_c} C_3
	\label{bps1}
\end{equation}
with
\begin{equation}
	\begin{split}
		&C_1=W^{1,e_{10}e_{13} j_4e_4;\phi_{10}\phi_{13}}_{\bar{\phi}_{10},0e_{13}' i_4e_4';\mathbf{1}\phi_{13}'}(i_4j_3f_4)
		W^{1,e_{14} e_{11}e_6i_5;\phi_{14}\phi_{11}}_{\phi_{11},e'_{14}0e_6'j_5;\phi_{14}'\mathbf{1}}(f_6j_6j_5)\\
		&\qquad \quad   \times  W^{3,e_{15} e_{12}e_1i_6;\phi_{15}\phi_{12}}_{\phi_{12},e_{15}'0e_1' j_6;\phi_{15}'\mathbf{1}}(f_1i_1j_6)\\		
		&C_{2_a} C_{2_b} C_{2_c}=
		\frac{1}{d_s}\sqrt{\frac{d_{i'_1}d_{j_2}d_{i'_3}d_{j_3}d_{i'_4}d_{j_5}d_{j'_6}}{d_{e'_1}d_{j'_2}d_{i_3}d_{e'_4}d_{j'_5}}} \times \\
		&\qquad [F^{i'_1e_7}_{\bar{s}j_1}]_{i_1j'_1}[F^{j'_1e_2}_{\bar{s}i_2}]_{j_1i'_2}[F^{i'_2e_8}_{\bar{s}j_2}]_{i_{2}j'_2}
		[\tilde{F}_{\bar{s}j_2}^{i'_3e_3}]_{i_3j'_2}
		[F^{i'_3e_9}_{\bar{s}j_3}]_{i_3j'_3} \times\\
		&\qquad [F^{j_5s}_{e_5i'_4}]_{i_4j'_5}
		[\tilde{F}_{e_6'j'_6}^{j_5s}]_{j_6j'_5}
		[\tilde{F}_{j_60}^{j'_6\bar{s}}]_{sj_6}
		[F^{i'_4\bar{s}}_{i_40}]_{si_4}
		[F^{j_6i_1}_{j'_6i'_1}]_{\bar{s}e_1'}
		[\tilde{F}_{i'_4j'_3}^{i_4j_3}]_{\bar{s}e_4'}\\
		&C_3=
		W_{\phi_{10},e_{10}e_{13} j'_4e_4;\phi_{10}\phi_{13}}^{1 ,0e_{13}'i'_4e_4';\mathbf{1}\phi_{13}'}(i'_4j'_3f_4) \\
		&\qquad \quad \times 
		 W_{\bar{\phi}_{11},e_{14} e_{11}e_6i'_5;\phi_{14} \phi_{11}}^{1 ,e_{14}'0e_6'j'_5;\phi_{14}' \mathbf{1}}(f_6j'_6j'_5) \\
		&\qquad \quad \times   W_{\bar{\phi}_{12},e_{15} e_{12}e_1i'_6;\phi_{15}\phi_{12}}^{3 ,e_{15}'0e_1'j'_6;\phi_{15}'\mathbf{1}}(f_1i'_1j'_6)
	\end{split}
	\label{bps2}
\end{equation}
where $\phi'_{13} = \phi_{13} \times \phi_{10}$, and similarly for $\phi'_{14}, \phi'_{15}$.  Each product is unique because all stick labels have abelian fusion rules.  
Here $e_7,e_8\dots e_{12}$ take values in abelian string types and thus $j_p=i_p\times e_{p+6}$ for $p=1\dots6$ while $e_1'=e_1\times e_{12},f_1' = f_1 \times e_{12}, e_4'=e_4\times \bar{e}_{10}, f_4'=f_4\times \bar{e}_{10}, e'_6=e_6\times e_{11}$ and $f'_6=f_6\times e_{11}$. 
The functions $W^{1,abcd;\phi^a\phi^b}_{\phi,a'b'c'd';\phi^{a'}\phi^{b'}}(efg),W^{3,abcd;\phi^a\phi^b}_{\phi,a'b'c'd';\phi^{a'}\phi^{b'}}(efg)$ are defined in (\ref{stringmatrix}), and we have used the fact that $(W^{j}_{\phi})^\dag = W^{j}_{\bar{\phi}}$. 
By using (\ref{consistency}) to simplify (\ref{bps1}), we obtain $B^{s,i_1\dots i_{6}j_1\dots j_6}_{p,i'_1\dots i'_{6}j'_1\dots j'_6}(e_1\dots e_{12};\phi_{10},\phi_{11},\phi_{12})$ in Eq. (\ref{bp1}).

\subsection{Simplified condensation Hamiltonian for string nets constructed from braided fusion categories}

When the fusion category used to construct the string net is itself an anyon model, a somewhat simpler formulation can be used to describe certain condensation transitions.  We include it here as it may be of interest for, e.g., numerical studies \cite{VidalToric,VidalToric2,SchulzSU2}.   
The existence of a braiding means that in addition to the rules (\ref{consistency}), the string labels also obey rules to determine what happens when an $a$-labeled string crosses over or under a $b$-labeled one:
\begin{equation} \label{BraidRule}
\includegraphics[width=0.32\textwidth]{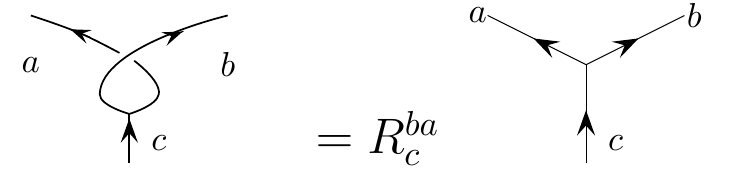}
\end{equation}
In these models, if the string net labels correspond to anyons described by a unitary modular tensor category $\mathcal{C}$, then the particle-like excitations of the string net are anyons in the category $\mathcal{C}\times \overline{\mathcal{C}}$ -- i.e. the string net realizes two copies of the anyon model, with opposite chiralities.  The bosons $a \times \overline{a}$ are plaquette defects\cite{ChainMail}, and can be condensed as described in Ref. \cite{TSBLong}.  Using the original string-net construction, however, a boson $a$ ($\overline{a})$, corresponding to an anyon in the category $\mathcal{C}$ ($\mathcal{\overline{C}})$, violates both vertex and plaquette terms.  However, using a modification of the 
 Walker-Wang construction of 3D string nets\cite{WalkerWang}, it is rellatively straightforward to construct a modified plaquette term for which open string operators creating these anyons commute with all plaquette terms.  
 
 When the condensing anyons are all from $\mathcal{C}$, we can do this in the generalized string-net Hilbert space depicted in Fig. \ref{fig:openstringnet}, and impose the constraint that at each trivalent vertex in the new lattice, the combination of edge labels is allowed by fusion.   As in the construction outlined in the main text, we energetically penalize any sticks carrying a label other than the identity.  
  Finally, we modify the plaquette operator's action on configurations where the sticks carry non-trivial labels, to ensure that $B_p$ commutes with open string operators ending on the sticks.    This can be done by threading the loop carrying the plaquette label under  all sticks, and using the fusion and braiding rules (\ref{localrules},\ref{BraidRule}) to resolve the diagram and obtain the matrix elements, using exactly the same procedure as in the 3D Walker-Wang string net models (see \cite{WalkerWang, VonKey10}).  Intuitively, this construction can be viewed as starting from a single layer of the Walker-Wang Hamiltonian, with a smooth lower boundary (see \cite{VonKey10}),  and open vertical edges extending upwards out of the plane.  In the full 3D construction, our sticks thus correspond to edges of vertical plaquettes, and anyon condensation is achieved by adding ``half-plaquette" operators along these vertical plaquettes.  Commutativity of adjacent plaquette operators, as well as of plaquette operators with the anyon string operators corresponding to adding such vertical plaquettes, follows from commutativity of the full Walker-Wang Hamiltonian.

Similarly, to condense a set of bosons that are all from $\mathcal{\overline{C}}$, we reverse the procedure above, drawing a Walker-Wang model with a smooth upper boundary, and keeping half-plaquettes extending downwards from this plane.  In this case, plaquette operator matrix elements are obtained by drawing the plaquette loop over the sticks, and then using appropriate fusion and braiding rules.  Condensing some anyons in $\mathcal{C}$, and some in $\mathcal{\overline{C}}$, can similarly be achieved by adding two sticks on each edge, one extending above the plane, and one below it.  

The procedure for identifying string net data in the condensed phase using this construction is exactly analogous to that of the more general construction outlined in the main text.  

\section{Showing that $B_{p_1}^{\phi,t_1},B_{p_2}^{\phi,t_2}$ commute \label{app:commute}}
In this section, we will show that the operators $B^{\phi,t_1}_{p_1}$ and $B^{\phi,t_2}_{p_2}$ commute with one another. We only need to consider two cases. One case is when two plaquettes are the same $p_1=p_2$. The other case is when $p_1$ and $p_2$ are adjacent since two operators will commute if $p_1$ and $p_2$ are further apart.

The first case is when two $B_p^{\phi,t}$ operators act on the same plaquette $p_1=p_2=p$. We will show $B_p^{\phi,t_1}$ and $B_p^{\phi,t_2}$ commute if the branching rules $\delta^{t_1,t_2}_u$ are symmetric in $t_1,t_2$.
We note that $B_p^{\phi,t_1}B_p^{\phi,t_2}=\sum_{\phi_{10},\phi_{11},\phi_{12}} W_{\phi_{10},\phi_{11},\phi_{12}} B_p^{t_1}B_p^{t_2} W^\dagger_{\phi_{10},\phi_{11},\phi_{12}}$. Thus, to show that $B^{\phi,t_1}_p,B^{\phi,t_2}_p$ commute, it is sufficient to show that ${B}^{t_1}_p, {B}^{t_2}_p$ commute.

To this end, we compute
\begin{equation}
	\begin{split}
		\left\<\raisebox{-0.22in}{\includegraphics[height=0.5in]{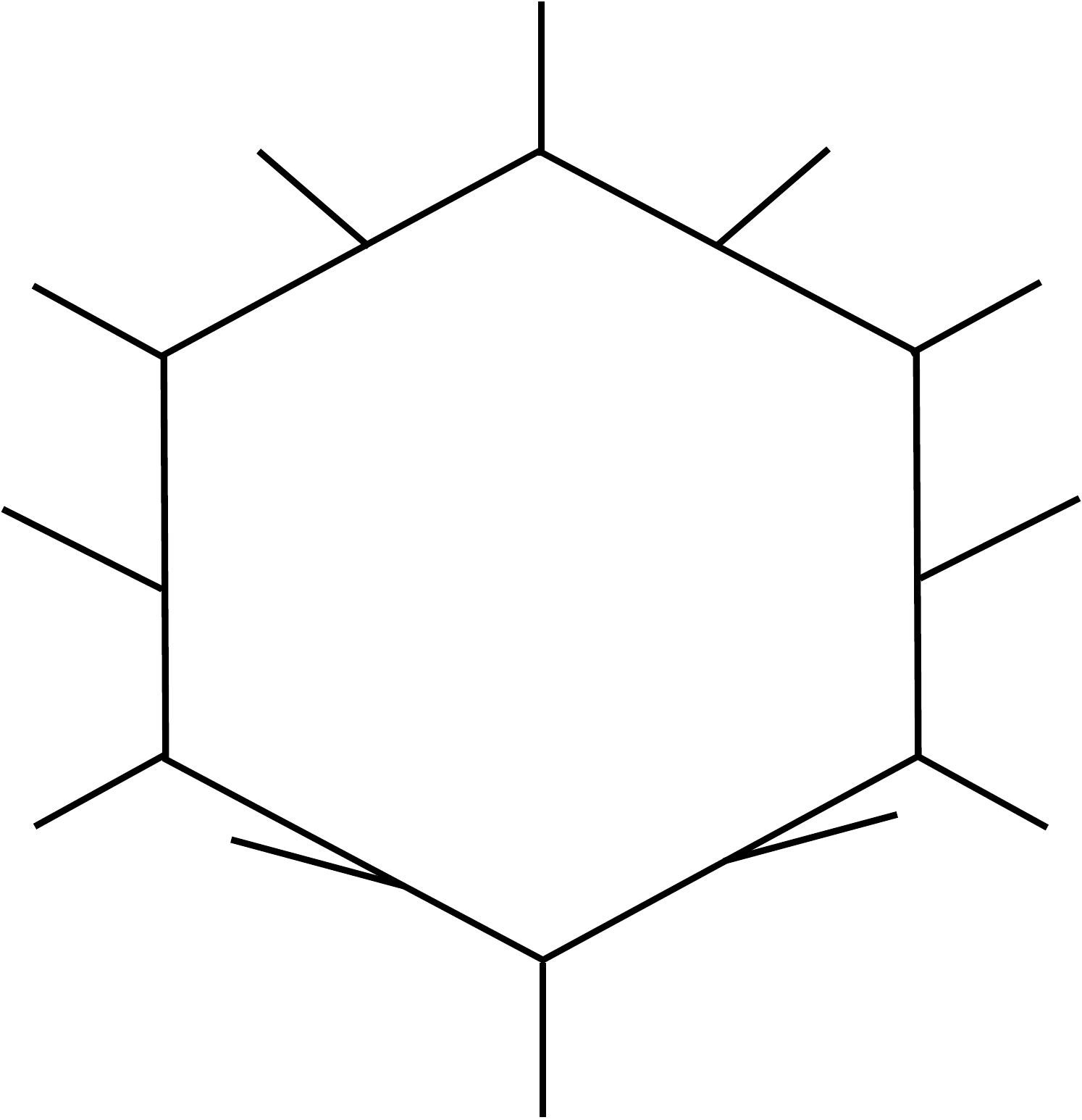}}\right| & {B}_p^{t_1} {B}_p^{t_2}
		=\left\<\raisebox{-0.22in}{\includegraphics[height=0.5in]{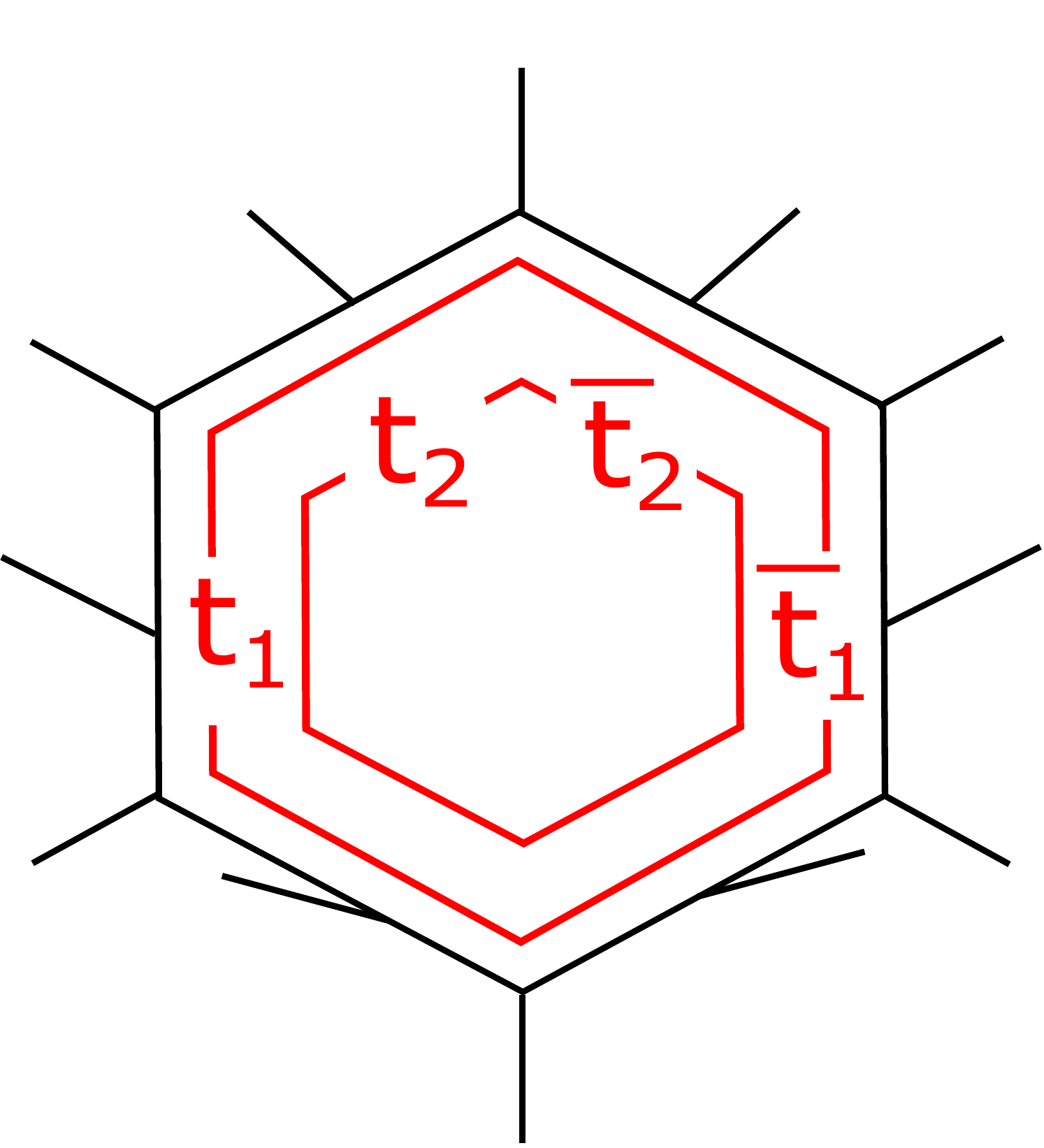}}\right| \\
		&=\sum_{u,u'}\left\<\raisebox{-0.22in}{\includegraphics[height=0.5in]{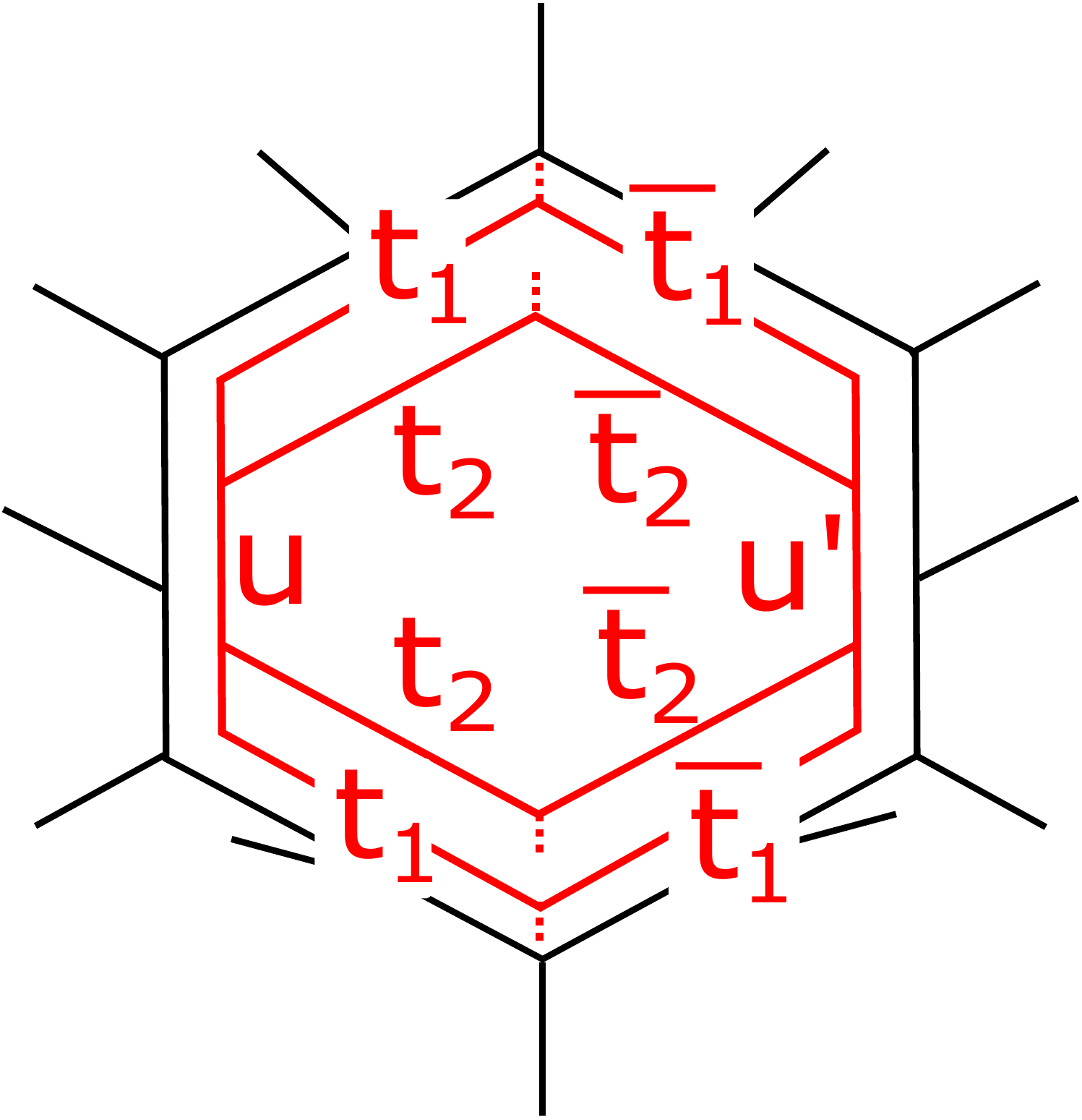}}\right| \frac{\sqrt{d_u d_{u'}}}{d_{t_1}d_{t_2}} \\
		&=\sum_{u} \left\<\raisebox{-0.22in}{\includegraphics[height=0.5in]{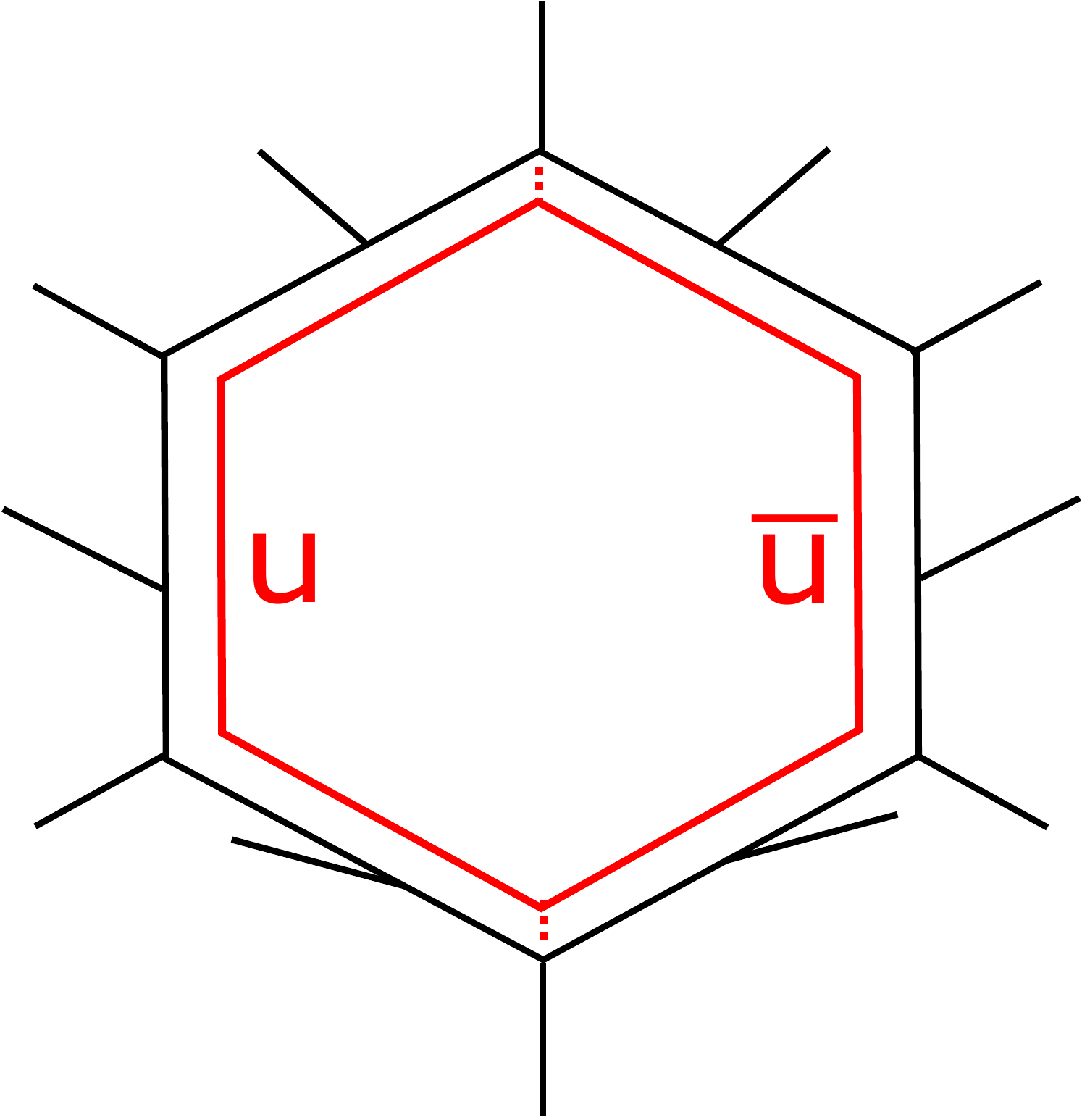}}\right|
		|[F^{t_1\bar{t}_1}_{u\bar{u}}]_{\bar{t}_20}|^2 
		|[F^{u\bar{t}_2}_{t_10}]_{t_2t_1}|^2 
		\frac{d_{t_1} d_u}{d_{t_2}}\\
		&=\sum_{u} \delta^{t_1 t_2}_{u}
		\left\<\raisebox{-0.22in}{\includegraphics[height=0.5in]{bp3d-eps-converted-to.pdf}}\right|
	\end{split}
	\label{}
\end{equation}
Here we use $|[F^{t_1\bar{t}_1}_{u\bar{u}}]_{\bar{t}_20}|^2=\frac{d_{t_2}}{d_{t_1}d_u}$ and $|[F^{u\bar{t}_2}_{t_10}]_{t_2t_1}|^2=|F^{t_1t_2\bar{t}_2}_{t_1u0}|^2\frac{d_{t_1}d_{t_2}}{d_u}=\delta^{t_1t_2}_u $ from (\ref{consistency}). Thus, we have
\begin{equation}
	{B}_p^{t_1} {B}_p^{t_2}=\sum_{u} \delta^{t_1 t_2}_{u} {B}^u_p.
	\label{bp1bp2}
\end{equation}
If $\delta^{t_1 t_2}_u$ is symmetric in $t_1,t_2$, then $B_p^{t_1}B_p^{t_2}=B_p^{t_2}B_p^{t_1}$ and thus ${B}^{\phi,t_1}_p,{B}^{\phi,t_2}_p$ commute.
In general, $\delta^{t_1 t_2}_u$ is \emph{not} symmetric in $t_1,t_2$ and thus ${B}^{\phi,t_1}_p,{B}^{\phi,t_2}_p$ do not commute.

\begin{figure}[ptb]
\begin{center}
\includegraphics[width=0.4\columnwidth]{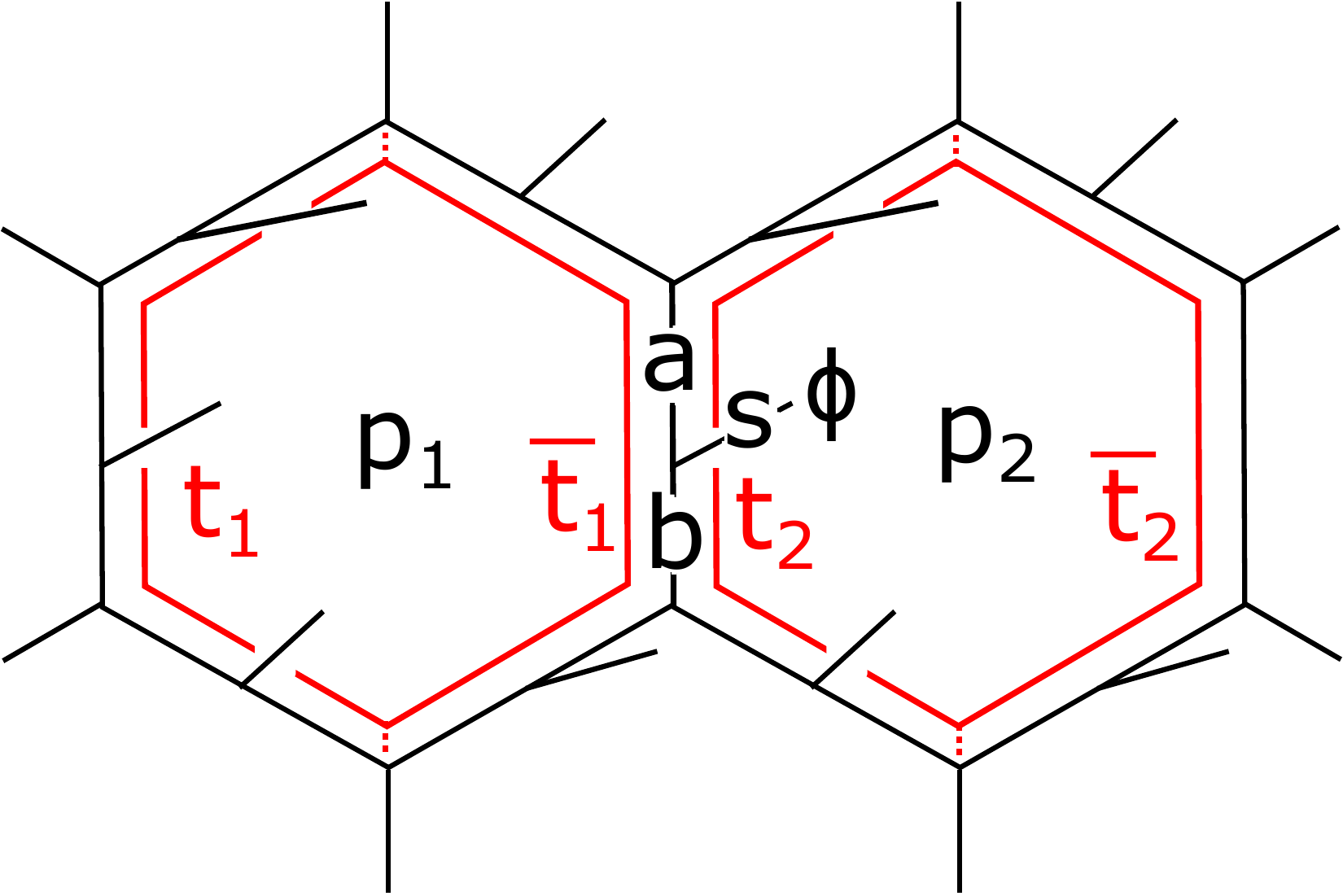}
\end{center}
\caption{Two plaquette operators $B_{p_1}^{\phi,t_1}$ and $B_{p_2}^{\phi,t_2}$ act on two adjacent plaquettes and add two loops. 
} 
\label{fig:2bp}
\end{figure}

The second case is when the two $B_p^{\phi,t}$ operators act on two adjacent plaquettes $p_1,p_2$. We want to show that $B_{p_1}^{\phi,t_1}B_{p_2}^{\phi,t_2}=B_{p_2}^{\phi,t_2}B_{p_1}^{\phi,t_1}$. 
To show this, we write down the matrix elements of the operators on each side by (\ref{bp1}) and then show they are equal. 
In fact, it is sufficient to compare the factors for the two operations which are different. These factors depend on the spin states on the shared boundary between $p_1,p_2$ (see Fig. \ref{fig:2bp}). 

Specifically, we write down the factors which are different. First, the action of $B_{p_1}^{\phi,t_1}B_{p_2}^{\phi,t_2}$ on the shared boundary contributes the factors
\begin{equation}
	\begin{split}
		\sum_{a''b''a'''b'''} \frac{w_\phi(a'')}{w_\phi(a)}[\tilde{F}^{sa''}_{bt_2}]_{ab''}[F^{a'''s}_{t_1b''}]_{a''b''}
\end{split}
	\label{bb1}
\end{equation}
Second, the action of $B_{p_2}^{\phi,t_2}B_{p_1}^{\phi,t_1}$ on the shared boundary contributes the factors
\begin{equation}
	\begin{split}
		\sum_{a'b'a''b''a'''b'''} \frac{w_\phi(a''')}{w_\phi(a')}[\tilde{F}^{sa'''}_{b't_2}]_{a'b'''}[F^{a's}_{t_1b}]_{ab'}
	F^{t_1at_2}_{a'''a'a''}\tilde{F}^{t_1bt_2}_{b'''b'b''}.
\end{split}
	\label{bb2}
\end{equation}
Here $x'=x \times t_1,x''=x \times t_2,x'''=x \times t_1 \times t_2$ for $x=a,b$ and $a+s=b$.
All we need is to show that (\ref{bb1}) = (\ref{bb2}).

To this end, we use (\ref{weq1}) and (\ref{consistency}) to simplify (\ref{bb1}) and (\ref{bb2}) as
\begin{equation}
	\sum_{a''b''a'''b'''} w_\phi(t_2)\frac{F^{ast_2}_{b''bt'_2}(F^{t_1a''s}_{b'''a'''b''})^*}{F^{at_2s}_{b''a''t'_2}}.
	\label{bb1a}
\end{equation}
and
\begin{equation}
	\sum_{a'b'a''b''a'''b'''}w_\phi(t_2) \frac{F^{a'st_2}_{b'''b't'_2} F^{t_1at_2}_{a'''a'a''} (F^{t_1as}_{b'a'b} F^{t_1bt_2}_{b'''b'b''})^*}{F^{a't_2s}_{b'''a'''t'_2}}
	\label{bb2a}
\end{equation}
respectively. Here $t_2'=t_2+ s$.

The next step is to simplify (\ref{bb2a}) further. By using the two pentagon identities
\begin{equation}
	\begin{split}
	F^{a'st_2}_{b'''b't'_2}F^{t_1at'_2}_{b'''a'b''}=F^{t_1as}_{b'a'b}F^{t_1bt_2}_{b'''b'b''}F^{ast_2}_{b''b't'_2} \\
	F^{t_1at'_2}_{b'''a'b''}F^{a't_2s}_{b'''a'''t'_2}=\sum_{h}F^{t_1at_2}_{a'''a'h}F^{t_1 h s}_{b'''a'''b''}F^{at_2s}_{b''ht'_2}
	\end{split}
	\label{}
\end{equation}
and the unitary conditions
\begin{equation}
	\begin{split}
	\sum_{b'}F^{t_1bt_2}_{b'''b'b''}(F^{t_1bt_2}_{b'''b'b''})^*=1 \\
	\sum_{a'}F^{t_1at_2}_{a'''a'a''}(F^{t_1at_2}_{a'''a'h})^*=\delta_{a''h},
	\end{split}	
	\label{}
\end{equation}
we can show (\ref{bb2a})=(\ref{bb1a}). This completes the proof that the $B_{p_1}^{\phi,t_1},B_{p_2}^{\phi,t_2}$ terms commute with one another.

\section{Showing that $[B_p^{\phi,s},W_{\phi^i}]=0$ \label{app:wbcommute}}
In this section, we show that $B_p^{\phi,s}$ and $W_{\phi^i}$ commute with one another. For our purpose, it is sufficient to show they commute in the gauge (\ref{f1}). In fact, we check that $B_p^{\phi,s}$ and $W_{\phi^i}$ commute in \emph{any} gauge.

\begin{figure}[ptb]
\begin{center}
\includegraphics[width=0.5\columnwidth]{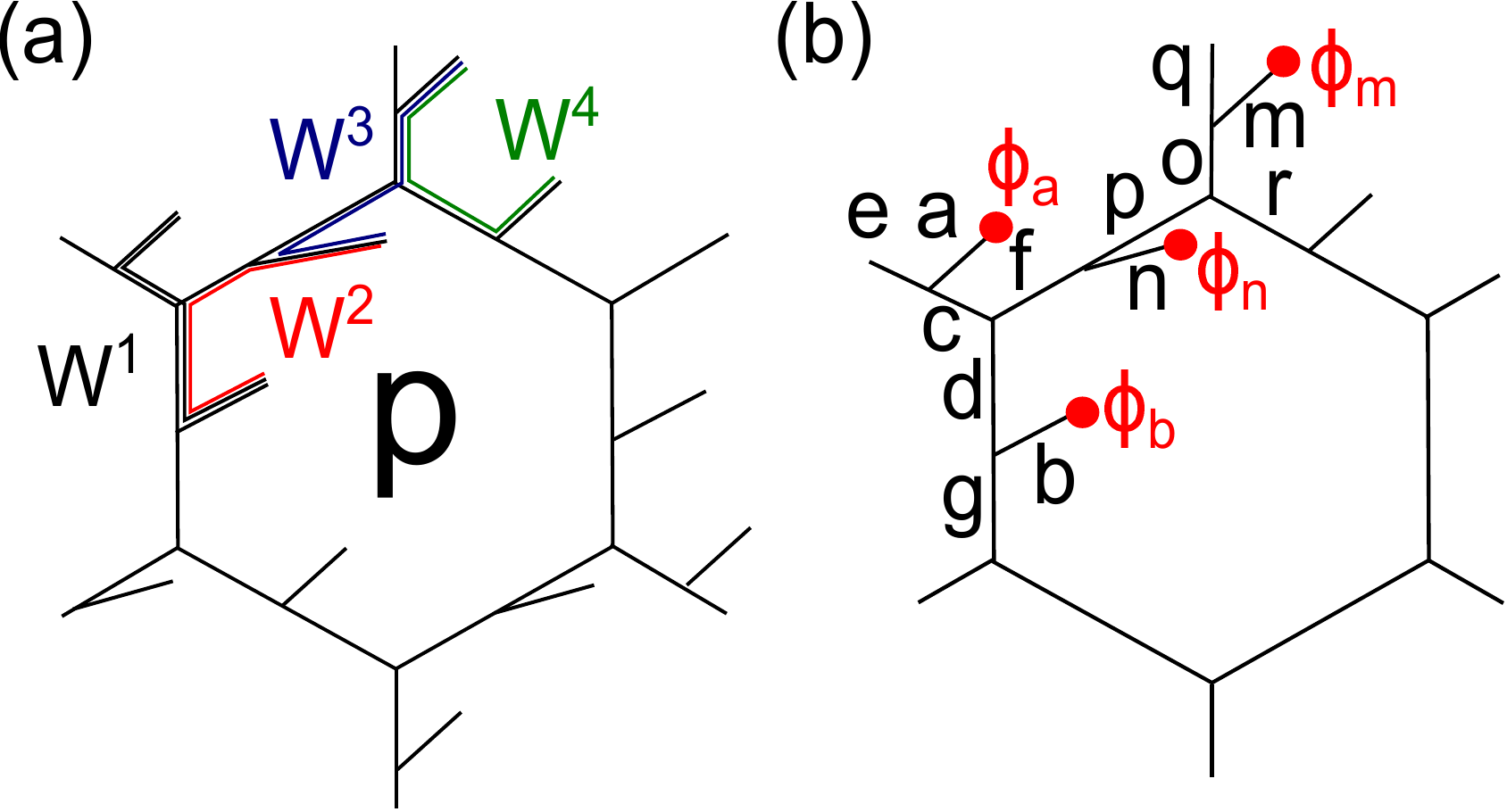}
\end{center}
\caption{(a) Four basic string operators which act around the two upper left vertices of the plaquette $p$. (b) 
} 
\label{fig:wbcommute}
\end{figure}

It suffices to show that the basic string operators $W_{\phi^i}^k$ commute with $B_p^{\phi,s}$ since any $W_{\phi^i}$ can be constructed by gluing the basic string operators along the path.
Thus, we only have to consider the case when the basic string operators $W_{\phi^i}^k$ are around the vertices surrounding the plaquette $p$ since it is clear that two operators commute if they are further apart.

There are two independent basic string operators which act around each vertex surrounding the plaquette $p$ (see Fig. \ref{fig:wbcommute}(a)). We need to show all 12 basic string operators commute with $B_p^{\phi,s}$. Among 12 string operators, there are 6 string operators like $W_{\phi^i}^4$, whose ends lie outside $p$, 4 string operators like $W_{\phi^i}^1$ which intersect $p$ and 2 string operators like $W_{\phi^i}^2$ whose ends lie inside $p$. We will show that $W_{\phi^i}^4,W_{\phi^i}^1, W_{\phi^i}^2$ commute with $B_p^{\phi,s}$. In a similar way, one can show other basic string operators also commute with $B_p^{\phi,s}$.

First, we want to show that $W^4_{\phi^i} B_p^{\phi,s} =  B_p^{\phi,s} W^4_{\phi^i} $ (see Fig. \ref{fig:wbcommute}(b)). To show this, we write out their matrix elements and compare the factors which are different. Specifically, we need to show that the product from the left of the equation
\begin{equation}
	\begin{split}
	&W_{\phi^i,k_ir_{\bar{i}}m_iq_i}^{4,krmq}(pnt) W_{\phi^l,k_{i+l}0m_{i+l}n_l}^{3,k_ilm_in}(pq_if)\times \\
	&W_{\phi^l,k_{i}lm_{i}n_s}^{3\dagger,k_{i+l}0m_{i+l}f_s}(pq_{i+\bar{s}}f_s) (F^{\bar{s}q_ir_{\bar{i}}}_{t_sq_{i+\bar{s}}t}F^{f_s\bar{s}q_i}_{m_{i+l+s}fq_{i+\bar{s}}})^*
	\end{split}
	\label{}
\end{equation}
and the product from the right
\begin{equation}
	\begin{split}
		&W_{\phi^i,k_ir_{\bar{i}}m_iq_{i+\bar{s}}}^{4,krmq_{\bar{s}}}(pn_st_{\bar{s}}) W_{\phi^l,k_{l}0m_{l}n_l}^{3,klmn}(pqf)\times \\
	&W_{\phi^l,klmn_s}^{3\dagger,k_{l}0m_{l}f_s}(pq_{\bar{s}}f_s) (F^{\bar{s}qr}_{t_sq_{\bar{s}}t}F^{f_s\bar{s}q}_{m_{l+s}fq_{\bar{s}}})^*
	\end{split}
	\label{}
\end{equation}
are equal.
Here $x_y=x+y$ and the matrix elements of $W_\phi^k$ are defined in (\ref{stringmatrix}).
By using (\ref{wdagger}) to write $W_{\phi}^{k\dagger}$ in terms of $W_{\bar{\phi}}^k$ and (\ref{weqs},\ref{weqlabel}) to write $w$ in terms of $F$ symbols, we can then show they are equal by (\ref{localrules}).

Second, we want to show that $W^1_{\phi^i} B_p^{\phi,s} =  B_p^{\phi,s} W^1_{\phi^i} $ (see Fig. \ref{fig:wbcommute}(b)). Again, we write down the matrix elements of both sides of the equation and compare the difference between the two. Specifically, from the left is the product
\begin{equation}
	\begin{split}
		W^{1,abcd}_{\phi^i,a_ib_{\bar{i}}c_id_i}(efg)
		W^{1,a_ib_{\bar{i}}c_id_i}_{\phi^b+\bar{\phi}^i,a_b0c_bd_b}(efg)W^{1\dagger,a_b0c_bd_{b+s}}_{\phi^b+\bar{\phi}^i,a_ib_{\bar{i}}c_id_{i+s}}(ef_sg_s)
	\end{split}
	\label{}
\end{equation}
while from the right is the product
\begin{equation}
	\begin{split}
		W^{1,abcd}_{\phi_b,a_b0c_bd_b}(efg)W^{1\dagger,a_b0c_bd_{b+s}}_{\phi_b,abcd_s}(ef_sg_s)
		W^{1,abcd_s}_{\phi^i,a_ib_{\bar{i}}c_id_{i+s}}(ef_sg_s)
	\end{split}
	\label{}
\end{equation}
Writing everything in terms of F symbols by (\ref{wdagger}, \ref{weqlabel}, \ref{weqs}), one can show they are equal by (\ref{localrules}).

Third, we want to show that $W^2_{\phi^i} B_p^{\phi,s} =  B_p^{\phi,s} W^2_{\phi^i} $ (see Fig. \ref{fig:wbcommute}(b)). We write down the their matrix elements and compare the difference. Specifically, from the left we have
\begin{equation}
	\begin{split}
	&W^{2,nbfd}_{\phi^i,n_ib_{\bar{i}}f_id_i}(pcg)
	W^{1,ab_{\bar{i}}cd_i}_{\phi^b+\bar{\phi}^i,a_{b+\bar{i}}0c_{b+\bar{i}}d_b}(ef_ig) \times\\
	&W^{3,mn_i0p}_{\phi^n+\phi^i,m_{n+i}0o_{n+i}p_{n+i}}(qrf_i) 
	W^{1\dagger,a_{b+\bar{i}}0c_{b+\bar{i}}g_s}_{\phi^b+\phi^i,ab_{\bar{i}}cd_{i+s}}(ef_{i+s}g_{s}) \times \\
	&W^{3\dagger,m_{n+i}0o_{n+i}f_{i+s}}_{\phi^n+\phi^i,mn_i0p_s}(qr_{\bar{s}}f_{i+s})
	F^{c_{b+\bar{i}}f_is}_{g_sgf_{i+s}}
	F^{f_is\bar{s}}_{f_if_{i+s}0}
	(F^{f_{i+s}\bar{s}r}_{o_{n+i}f_i r_{\bar{s}}})^*
	\end{split}
	\label{}
\end{equation}
and from the right we have
\begin{equation}
	\begin{split}
		&W^{1,abcd}_{\phi^b,a_b0c_bd_b}(efg)
		W^{3,mnop}_{\phi^n,m_n0o_np_n}(qrf)\\
		&W^{1\dagger,a_b0c_bd_{b+s}}_{\phi^b,abcd_s}(ef_sg_s)
		W^{3\dagger,m_n0o_np_{n+s}}_{\phi^n,mnop_s}(qr_{\bar{s}}f_s)\\
		&W^{2,nbf_sd_s}_{\phi^i,n_ib_{\bar{i}}f_{i+s}d_{i+s}}(p_scg_s)
		F^{c_{b}fs}_{g_sgf_{s}}
		F^{fs\bar{s}}_{ff_{s}0}
		(F^{f_{s}\bar{s}r}_{o_{n}f r_{\bar{s}}})^*.
	\end{split}
	\label{}
\end{equation}
Similarly, by a straightforward but tedious computation, one can show they are equal by (\ref{localrules}).

\section{Consistency of solutions for vertex coefficients in the presence of splitting} \label{ConsistencyApp}

If $a^r = a$ and $c^r = c$, then a $W^1_{\phi^{jr}}$ operator at a vertex $(a, b; c)$ does not change the labels about this vertex; its only effect is to change the labels of the two adjacent sticks.  Thus, we obtain:
\begin{align} \label{Eq:Acon1}
A^{a s^{(i+j) r}} A^{c s^{(k-j) r}} A^{a b}_c  =&A^{a s^{i r}} A^{c s^{k r}} A^{a b}_c  \bar{w}_{\phi^{jr}}(b) F^{ a b s^{ j r}}_{c c b^{jr}} (F^{ a s^{ j r} b}_{c a b^{jr}} )^* \nonumber \\ 
  & \times F^{a s^{i r}  s^{j r}}_{a a s^{(i+j) r}} ( F^{ c s^{jr} s^{(k-j)r} }_{c c s^{k r }} )^*
\end{align}
where we have used the gauge (\ref{f1}).  If $A^{ab}_c$ is non-zero, this can be true only if the coefficient does not depend on $b$.

Similarly, applying a $W^2_{\phi^{jr}}$ operator at a vertex $(b, a; c)$, we obtain:
\begin{align} \label{Eq:Acon2}
A^{a s^{(i+j) r}} A^{c s^{(k-j) r}} A^{ ba}_c  =&A^{a s^{i r}} A^{c s^{k r}} A^{ b a }_c  F^{ b a s^{ j r}}_{c c b^{jr}}  \nonumber \\ 
  & \times F^{a s^{i r}  s^{j r}}_{a a s^{(i+j) r}} ( F^{ c s^{jr} s^{(k-j)r} }_{c c s^{k r }} )^*
\end{align}

Finally, applying $W^2_{\phi^{-jr}} W^1_{\phi^{jr}}$ to a vertex $(a,c; b)$ with $a s^r= a$ and $c s^r = c$ gives:
\begin{align} \label{Eq:Acon3}
A^{a s^{(i+j) r}} A^{c s^{(k-j) r}} A^{ ac}_b  =&A^{a s^{i r}} A^{c s^{k r}} A^{ ac }_b  \bar{w}_{\phi^{jr}}(c) (F^{ c s^{jr} s^{- j r}}_{c c 0})^*  \nonumber \\ 
  & \times (F^{a s^{jr} c}_{b a c})^* F^{a s^{i r}  s^{j r}}_{a a s^{(i+j) r}} F^{ c s^{kr} s^{-jr} }_{c c s^{(k-j) r }} 
\end{align}

Similar equations appear about downward-oriented vertices, involving the string operators $W^{3}_{\phi^{lr}}, W^4_{\phi^{lr}}$; however these do not impose any new conditions required for consistency.  

We now show that the coefficients identified above are independent of $b$ {\it for different choices of $b$ that are related by fusion with $s^{rj}$}.  Iterating this, we see that the coefficients are the same for any $b$ in the fusion orbit of $s^{jr}$.  In particular, for $j=1$ we see that the coefficient is the same for any $b$ in the fusion orbit of $s^r$.  Moreover, the string operator $W^i_{\phi^{jr}} =( W^i_{\phi^r})^j $; hence we conclude that, or any $j$, the coefficients in Eqs. (\ref{Eq:Acon1}- \ref{Eq:Acon3}) are the same for  all $b$ in the fusion orbit of $s^r$, and for any $j$.

 We begin with Eq. (\ref{Eq:Acon1}).  Using Eq. (\ref{consistency} a), we find
\begin{equation} \label{Eq:Afindap2}
F^{ a s^{j r} b}_{c a b^{jr}} F^{a b^{jr} s^{jr}}_{c c b^{2jr}} F^{s^{jr} b s^{jr}}_{b^{2jr} b^{jr} b^{jr}} 
=F^{ab s^{jr} }_{c c b^{jr} } F^{ a s^{jr} b^{jr}}_{c a b^{2jr}}  .
\end{equation}
Next, we multiply  both sides of the equation by $w_{\phi^{jr}}(b^{(jr)})$, and use Eq. (\ref{weqs}) to see that
\begin{equation}
w_{\phi^{jr}}(b) w_{\phi^{jr}}(s^{jr}) = w_{\phi^{jr}}(b^{(jr)}) F^{s^{jr} b s^{jr}}_{b^{2jr} b^{jr} b^{jr}} \ .
\end{equation}
Since $w_{\phi^{jr}}(s^{jr})  = 1$, we thus find: 
\begin{equation} \label{Eq:Afindap3}
w_{\phi^{jr}}(b)F^{ a s^{j r} b}_{c a b^{jr}} (F^{ab s^{jr} }_{c c b^{jr} } )^*
= w_{\phi^{jr}}(b^{(jr)})F^{ a s^{jr} b^{jr}}_{c a b^{2jr}}( F^{a b^{jr} s^{jr}}_{c c b^{2jr}})^*   .
\end{equation}
Iterating this result, we see that the coefficient is the same for all choices of $b$ in the same fusion orbit of $s^{r}$.  

 Next, consider Eq.  (\ref{Eq:Acon2}).  Eq. (\ref{consistency} a) stipulates:
\begin{equation}
F^{s^{jr} b a }_{c b^{jr} c} F^{s^{jr} c s^{kr}}_{c c c} F^{b a s^{kr}}_{cca} = F^{b^{jr} a s^{kr}}_{c c a} F^{s^{jr} b a}_{c b^{jr} c}
\end{equation}
Further, since $s^{jr} c = c$, we have
\begin{equation} \label{Eq:E8}
w_{\phi^{jr}}(c) w_{\phi^{jr}}(s^{kr}) =   \frac{F^{s^{jr} c s^{kr}}_{c c c}  F^{ c s^{kr} s^{jr} }_{c c s^{(k+j)r}}}{F^{c s^{jr} s^{kr}}_{c c s^{(k+j)r}} }w_{\phi^{jr}}(c) 
\end{equation} 
and hence $F^{s^{jr} c s^{jr}}_{c c c} = 1$.  
It follows that $F^{b a s^{jr}}_{cca} = F^{b^{jr} a s^{jr}}_{c c a}$, and the coefficient is the same for any $b$ in the fusion orbit of $s^{jr}$.  

Finally, consider Eq. (\ref{Eq:Acon3}).  We have simplified the coefficient on the right-hand side as follows.  Similar to Eq. (\ref{Eq:Wabc}), the product of $W^2_{\phi^{-jr}} W^1_{\phi^{jr}}$ on the vertex $(a,c; b)$ can be expressed 
\begin{align} \label{W1W2prod}
&  \bar{w}_{\phi^{jr}}(c)  F^{ a c s^{jr}}_{b^{jr} b c} ( F^{a s^{jr} c}_{b^{jr} a c })^* (F^{b s^{jr} s^{t-jr}}_{b^{t} b^{jr} s^{t}})^* (F^{b^{jr} s^{-jr} s^t}_{b^{t} b s^{t - jr}})^* F^{a c s^{-jr}}_{b b^{jr} c} \nonumber \\
  & \times F^{a s^{i r}  s^{j r}}_{a a s^{(i+j) r}} F^{ c s^{kr} s^{-jr} }_{c c s^{(k-j) r }} 
\end{align}
 where the stick on the $a$ edge initially carries the label $s^{ir}$, that on the $c$ edge carries a label $s^{kr}$, and that on the $b$ edge carries $s^t$.  
 
However, we can use Eq. (\ref{consistency}) to show that when $a s^r = a$ and $c s^r = c$, 
\begin{align}
F^{b^{jr} s^{-jr} s^t}_{b^{-t} b s^{t - jr}} F^{b s^{jr} s^{t-jr}}_{b^{-t} b^{jr} s^{t}} = F^{b s^{jr} s^{-jr}}_{b b^{jr} 0} F^{ s^{jr} s^{-jr} s^t}_{s^t 0 s^{t-jr}}
\end{align} 
and
\begin{align}
F^{ a c s^{jr}}_{b^{jr} b c}  F^{a c s^{-jr}}_{b b^{jr} c} F^{c s^{jr} s^{-jr}}_{c c 0} =  F^{b s^{jr} s^{-jr}}_{b b^{jr} 0}
\end{align} 
Thus, in the gauge (\ref{f1}), Eq. (\ref{W1W2prod}) can be simplified to give:
\begin{equation} \label{Eq:SimpleW1W2prod}
  \bar{w}_{\phi^{jr}}(c)   ( F^{a s^{jr} c}_{b^{jr} a c })^*  (F^{c s^{jr} s^{-jr}}_{c c 0})^*
\end{equation}
We can show that the coefficient is the same for any $b$ in the fusion orbit of $s^{jr}$:
\begin{align}
F^{a s^{jr} c}_{b a c} F^{a c s^{jr} }_{ b^{jr} b c} F^{s^{jr} c s^{jr}}_{c c c} = F^{a c  s^{jr} }_{b^{jr}  bc } F^{a s^{jr} c}_{b^{jr} a c}
\end{align} 
Cancelling the redundant factors on both sides, and noting that  by Eq. (\ref{Eq:E8}), $F^{s^{jr} c s^{jr}}_{c c c} =1$, we see that $F^{a s^{jr} c}_{b a c} =  F^{a s^{jr} c}_{b^{jr} a c}$, and again by iterating we find our result.

Now, if $b$ is an abelian particle, and $a \times s^r = a, c \times s^r = c$, then $a \times c = \sum_j b^{jr} + ...$, where $...$ cannot contain $b^l$ for $l \neq jr$.  This follows from the cyclic property of the branching rules.  Suppose $(a,c; b)$ and $(a,c; b^j)$ are allowed by the branching rules.  Then so are $(c, \bar{b}^{-j}; a)$ and $(\bar{b}^{-j}, a; \bar{c})$.  However, if $b^j = b \times s^j = s^j \times b$, then we must also have $(\bar{b}, a; \bar{c}^j)$.  If $j = l r$ then $\bar{c}^j = \bar{c}$, and the outcome of fusing $\bar{b}$ with $a$ is unique.  Otherwise, however, we see that fusing $\bar{b}$ with $a$ can have at least two different outcomes; hence $b$ is not an abelian string label.  In particular, if $b = s^j$, it follows that all equations associated with acting with string operators $s^{jr}$ on the vertices $(a,s^j; a)$, $(s^j, a; a)$, and $(a, \bar{a}; s^j)$ are consistent.  This allows us to solve for the coefficients $A^{a s^{j}}$.

In general, however, vertices of the form (for example) $(b, a;c)$ and $( g,a; c)$, where $g \neq b s^j$, can lead to multiple equations of the form (\ref{Eq:Acon2}), which relate the same pairs of coefficients $A^{a s^{(i+l) r}},A^{c s^{(k-l) r}}$ to  $A^{a s^{i r}} A^{c s^{k r}} $.  If $F^{ b a s^{ l r}}_{c c b^{lr}}  \neq F^{ g a s^{ l r}}_{c c g^{lr}} $, these equations are mutually inconsistent unless we choose either $A^{ba}_c = 0$ or $A^{ga}_c = 0$.  The resolution to this is to recognize that since both $a$ and $c$ split, there are multiple non-zero choices for the coefficient: $A^{ba}_c =(A^{ba}_c)_{ij} $, denoting a choice of $\tilde{a}_i$ and $\tilde{c}_j$.   For a given $b$, we find that the coefficient $F^{ b a s^{ l r}}_{c c b^{lr}}$ takes on at most $p/r$ distinct values (and similarly for other vertices); hence we expect to find at least one non-zero $(A^{ba}_c)_{ij} $ for each $i$.

\subsection{Relations between coefficients $A^{a s^{jr}}$ when $a^{r} = a$} \label{}

Suppose $a^r = a$.  Eq. (\ref{consistency}a) gives:
\begin{align}
F^{a s^{lr} s^{jr}}_{a a s^{(l+j)r} } F^{a s^{(l+j)r} s^{k r}}_{a a s^{(l+j+k)r} } F^{s^{lr} s^{jr} s^{k r}}_{s^{(l+j+k)r}  s^{(l+j)r}  s^{(j+k)r} }  \nonumber \\ = F^{a s^{jr} s^{kr}}_{a a s^{(j+k)r}} F^{a s^{lr} s^{(j+k)r}}_{a a s^{(l+j+k)r} } 
\end{align}
In our gauge of choice, taking $j \rightarrow j-1$ and $k=1$ in the above expression gives
\begin{equation}
 F^{a s^{lr} s^{jr}}_{a a s^{(l+j)r} } 
= \frac{ F^{a s^{lr} s^{(j-1)r}}_{a a s^{(l+j-1)r} } F^{a s^{(l+j-1)r} s^{ r}}_{a a s^{(l+j)r} }  }{F^{a s^{(j-1) r} s^{r}}_{a a s^{jr}} }
\end{equation}

Using this expression repeatedly, we can show that
\begin{align}
F^{a s^{lr} s^{j r}}_{a a s^{(l+j)r} }  &=\left (  \frac{\prod_{k = 0}^{j-1} F^{a s^{(l+k)r} s^{r}}_{a a s^{(l+k+1)r} } }{ \prod_{k = 1}^{j-1} F^{a s^{kr} s^{r}}_{a a s^{(k+1)r} }}\right )  \nonumber \\
&=\left (  \frac{\prod_{k = l}^{l+j-1 } F^{a s^{k r} s^{r}}_{a a s^{(k+1)r} } }{ \prod_{k = 1}^{j-1} F^{a s^{kr} s^{r}}_{a a s^{(k+1)r} }}\right ) 
\end{align}
From this, we see that 
\begin{align}
\frac{ F^{a s^{lr} s^{j r}}_{a a s^{(l+j)r} } } {  F^{a s^{jr} s^{l r}}_{a a s^{(l+j)r} }} 
&=\left (  \frac{\prod_{k = l}^{l+j-1 } F^{a s^{k r} s^{r}}_{a a s^{(k+1)r} } } {\prod_{k = j}^{l+j-1 } F^{a s^{k r} s^{r}}_{a a s^{(k+1)r} } } \right )  \left (  \frac{ \prod_{k = 1}^{l-1} F^{a s^{kr} s^{r}}_{a a s^{(k+1)r} }}{ \prod_{k = 1}^{j-1} F^{a s^{kr} s^{r}}_{a a s^{(k+1)r} }} \right ) 
\end{align}
Without loss of generality, assume that $l>j$.  Then: 
\begin{align} 
\left (  \frac{\prod_{k = l}^{l+j-1 } F^{a s^{k r} s^{r}}_{a a s^{(k+1)r} } } {\prod_{k = j}^{l+j-1 } F^{a s^{k r} s^{r}}_{a a s^{(k+1)r} } } \right ) &=\frac{1}{ \prod_{k = j}^{l-1 } F^{a s^{k r} s^{r}}_{a a s^{(k+1)r} } } \nonumber \\
 \left (  \frac{ \prod_{k = 1}^{l-1} F^{a s^{kr} s^{r}}_{a a s^{(k+1)r} }}{ \prod_{k = 1}^{j-1} F^{a s^{kr} s^{r}}_{a a s^{(k+1)r} }} \right ) & = \prod_{j}^{l-1} F^{a s^{k r} s^{r}}_{a a s^{(k+1)r} } 
\end{align}
so that 
\begin{equation}  \label{Eq:EFprod}
F^{a s^{lr} s^{j r}}_{a a s^{(l+j)r} } =   F^{a s^{jr} s^{l r}}_{a a s^{(l+j)r} }
\end{equation}

Further, from the expression (\ref{Eq:Acond1s}),
\begin{equation} \label{}
(A^{a s^{n  r}})_\mu    = (A^{a s^{ r}} )_\mu^n \prod_{k=1}^{n-1} ( F^{a s^{k r} s^{ r}}_{a a s^{(k+1) r}} )  
\end{equation}
we see that 
\begin{align} 
\frac{ (A^{a s^{(j+l)  r}})_\mu }{(A^{a s^{l  r}})_\nu } & = \frac{ (A^{a s^{ r}} )_\mu^{j+l}}{(A^{a s^{ r}} )_\nu^l } \frac{ \prod_{k=1}^{j+l-1} ( F^{a s^{k r} s^{ r}}_{a a s^{(k+1) r}} )  }{\prod_{k=1}^{l-1} ( F^{a s^{k r} s^{ r}}_{a a s^{(k+1) r}} )  } \\
& = \frac{ (A^{a s^{ r}} )_\mu^{j+l}}{(A^{a s^{ r}} )_\nu^l }  \prod_{k=l}^{j+l-1} ( F^{a s^{k r} s^{ r}}_{a a s^{(k+1) r}} )  
\end{align} 
Hence: 
\begin{align}  \label{E21}
\frac{ (A^{a s^{(j+l)  r}})_\mu }{(A^{a s^{l  r}})_\nu (A^{a s^{j  r}})_\rho  } & = \frac{ (A^{a s^{ r}} )_\mu^{j+l}}{(A^{a s^{ r}} )_\nu^l (A^{a s^{ r}} )_\rho^j} \frac{ \prod_{k=l}^{j+l-1} ( F^{a s^{k r} s^{ r}}_{a a s^{(k+1) r}} )  }{  \prod_{k=1}^{j-1} ( F^{a s^{k r} s^{ r}}_{a a s^{(k+1) r}} )  } \nonumber \\
& = \frac{ (A^{a s^{ r}} )_\mu^{j+l}}{(A^{a s^{ r}} )_\nu^l (A^{a s^{ r}} )_\rho^j} F^{a s^{lr} s^{j r}}_{a a s^{(l+j)r} } 
\end{align} 
where in the last line, we have used Eq. (\ref{Eq:EFprod}).  Evidently, if all factors come from the same solution (i.e. $\mu = \nu = \rho$), then the right-hand side is simply $F^{a s^{lr} s^{j r}}_{a a s^{(l+j)r} } $.  

\subsection{Vertices where multiple labels split}

For vertices where multiple labels split (and none of the labels are a power of $s$), we must address two questions.  First, is it the case that for every choice of $b$, there exists at least one pair $(\mu, \nu)$ for which Eq. (\ref{Eq:munuconsist}) can be satisfied?  If not, we must conclude that at least one of the particles $a,b$, or $c$ must be confined.  

Recall that if $s^q=1$, then if $ w_{\phi^{q}}(b) \neq 1$, n $b$ does not correspond to any label in our effective Hilbert space; hence in this case we must set $(A^{ab}_c)^{\mu}_{\nu} = 0$ for all $\mu,\nu$.   
When $ w_{\phi^{q}}(b) = 1$,  $W^1_{\phi^q}$ acts as the identity operator at the vertex $(a,b;c)$, since it is an excitation of the form $(0,m)$ which does not involve any fusion.  
Applying the operator $W^1_{\phi^{r}}$ $q/r$ times, in the gauge (\ref{f1}) we obtain the matrix element
\begin{equation}
 M_1(a,b,c))^{q/r}  \left( \prod_{k=1}^{q/r-1} F^{a s^{kr} s}_{a a s^{(k+1)r}}) \right) \left( \prod_{k=1}^{q/r-1} (F^{c s^r \bar{s}^{kr}}_{c c \bar{s}^{(k-1)r}})^* \right)
\end{equation}
Since $W^1_{\phi^q} = ( W^1_{\phi^{r}})^{q/r}$, it follows that: 
\begin{equation}
( M_1(a,b,c))^{q/r} =  \left( \prod_{k=1}^{q/r-1} (F^{a s^{kr} s}_{a a s^{(k+1)r}})^*  \right) \left( \prod_{k=1}^{q/r-1} (F^{c s^r \bar{s}^{kr}}_{c c \bar{s}^{(k-1)r}}) \right)
\end{equation}
 hence $M_1(a,b,c)$ is a $q/r^{th}$ root of the product 
on the right-hand side.  Now, $A^{a s^r}$ is a $q/r^{th}$ root of $ \prod_{k=1}^{q/r-1}( F^{a s^{kr} s^r}_{a a s^{(k+1)r}})^*$, while $A^{c s^r}$ is a $q/r^{th}$ root of $\prod_{k=1}^{q/r-1} (F^{c  s^{kr}s^r}_{c c \bar{s}^{(k+1)r}})^* =  \left(  \prod_{k=1}^{q/r-1} F^{c s^r  \bar{s}^{k r} }_{c c s^{(k+1)r}} \right)^{-1}$, where we have used Eq. (\ref{Eq:EFprod}). Thus $A^{a s^r}/A^{c s^r}$ is also a $q/r^{th}$ root of the product on the right-hand side.  It follows that for every $b$, there exists at least one choice of $\mu, \nu$ for which Eq. (\ref{Eq:Ac2}) is satisfied -- in which case it is also satisfied for $b^{kr}$, $0 \leq k < q/r$.  
This suggests that for a fixed $M_1(a,b,c)$ of modulus $1$, we expect $q/r$ distinct solutions $A_\mu^{a s^r}/A_\nu^{c s^r} =M_1(a,b,c)e^{ 2 \pi i n r/q} $, $0 \leq n < q/r$.

At this point, it is worth commenting on the fusion rules of the new theory.  In the most general case, we have
\begin{equation}
(\sum_{\mu =1}^{q/r} \tilde{a}_\mu ) \times (\sum_{\nu =1}^{q/r} \tilde{c}_\nu ) = \frac{q}{r} \sum_{\tilde{b} | b \times s^r \neq b}  c_b \tilde{b}  + \sum_{\tilde{d}| d \times s^v = d} \sum_{\lambda} c_{d,\lambda}   \tilde{d}_\lambda \ .
\end{equation} 
The first sum contains any terms that do not split, and the second contains terms that do).  
As discussed in the main text, if $(a, c; b^{k})$ is not an allowed vertex for any $0 < k < r$, then $c_b = 1$; otherwise, $c_b$ counts the number of distinct $k$, $0 \leq k < r$, for which $(a, c; b^{k})$ is allowed.   The second sum runs over labels $d$ for which $d \times s^v = d$, with $0 < v < q$.  In this case, in the condensed theory a fusion channel $\tilde{d}_\lambda$  appears with a coefficient $c_{d, \lambda}$, whose value depends on $v, r$, and the number of values of $\lambda$ for which the new fusion rules admit solutions.  

As discussed in the main text, when $c_b=1$, the number of distinct values of $(\mu, \nu)$ for which $(\tilde{a}_\mu, \tilde{c}_\nu; \tilde{b})$  is allowed by the branching rules is equal to the number of copies of $\tilde{b}$ on the right, and the new theory need not have fusion multiplicity.  
If $c_b >1$, 
 the label $\tilde{b} = \sum_{j=1}^{q-1} b^j$ at the vertex $(\tilde{a}_\mu, \tilde{c}_\nu; \tilde{b})$ may be associated with multiple different values of the coefficient $ M_l(a,b^j,c)$.  Each distinct coefficient then corresponds to a distinct set of solutions $(\mu,\nu)$ to Eq. (\ref{Eq:Ac2}).  Thus, if $ \{ M_l(a,b^j,c), 0 \leq j < r \}$ are all distinct, then we can find up to $c_b q/r$ distinct solutions to Eq. (\ref{Eq:Ac2}), and again the new theory need not have fusion multiplicities.   On the other hand, if these coefficients are not all distinct, then in general fusion multiplicities are expected, meaning that the coefficients $(A^{ab}_c)^\mu_\nu$ are matrices.
 
The situation for vertices $(\tilde{a}_\mu, \tilde{c}_\nu; \tilde{d}_\lambda)$ is similar, except that in this case if $v$ and $r$ are not mutually prime, additional constraints are imposed which fix which coefficients $(A^{ab}_c)^{\mu,  \lambda}_\nu$ are non-zero.   Again, we cannot rule out the possibility of fusion multiplicities and the need to make these coefficients matrices.

\bibliography{stringnet2}
\end{document}